\documentclass[12pt]{report}
\usepackage{amssymb}
\usepackage{amsmath}

\usepackage{graphicx}
\begin{document}
\voffset = -2.9cm
\hoffset = -1.3cm
\def\itm{\newline \makebox[8mm]{}}
\def\ls{\makebox[8mm]{}}
\def\fra#1#2{\frac{#1}{#2}}
\def\fr#1#2{#1/#2}
\def\frr#1#2{#1 \mbox{\Large $/$} #2}
\def\frl#1#2{\mbox{\large $\frac{#1}{#2}$}}
\def\frh#1#2{\mbox{\Large $\frac{#1}{#2}$}}
\def\frn#1#2{\mbox{\normalsize $\frac{#1}{\rule[-0mm]{0mm}{3.15mm} #2}$}}
\def\frm#1#2{\mbox{\normalsize $\frac{#1}{#2}$}}
\def\leftp{\mbox{$\left( \rule[-1.5mm]{0mm}{5.7mm} \right.$} \hs{-0.2mm}}
\def\leftn{\mbox{$\left( \rule[-1.5mm]{0mm}{4.7mm} \right.$} \hs{-0.2mm}}
\def\rightp{\hs{-0.2mm} \mbox{$\left. \rule[-1.5mm]{0mm}{5.7mm} \right)$}}
\def\rightn{\hs{-0.2mm} \mbox{$\left. \rule[-1.5mm]{0mm}{4.7mm} \right)$}}
\def\hs#1{\mbox{\hspace{#1}}}
\def\p{\hs{0.5mm} + \hs{0.5mm}}
\def\m{\hs{0.5mm} - \hs{0.5mm}}
\def\b{\begin{equation}}
\def\e{\end{equation}}
\def\arccot{\mbox{arccot}}
\def\hn{\hspace{0.7mm}}
\vspace*{6mm}
\makebox[\textwidth][c]
{\large \bf{}}
\vspace{-4mm} \newline
\makebox[\textwidth][c]{\normalsize \bf{Different representations of the
Levi-Civita Bertotti Robinson solution}}
\vspace{4mm} \newline
\makebox[\textwidth][c]
{\normalsize \O yvind Gr\o n$^{*}$ and Steinar Johannesen$^*$}
\vspace{1mm} \newline
\hspace*{12mm} {\scriptsize $*$} \parbox[t]{150mm}
{\scriptsize Oslo and Akershus University College of Applied Sciences,
Faculty of Technology, Art and Design,
\newline P.O.Box 4 St.Olavs Plass, N-0130 Oslo, Norway}
%
\vspace{6mm} \newline
{\bf \small Abstract} {\small The Levi-Civita Bertotti Robinson (LBR)
spacetime is investigated in various coordinate systems. By means of
a general formalism for constructing coordinates in conformally flat
spacetimes, coordinate transformations between the different coordinate
systems are deduced. We discuss the motion of the reference frames in
which the different coordinate systems are comoving. Furthermore we
characterize the motion of the different reference frames by their
normalized timelike Killing vector fields, i.e. by the four velocity
fields of the reference particles. We also deduce the formulae in the
different coordinate systems for the embedding of the LBR spacetime
in a flat 6-dimensional manifold. In particular we discuss a scenario
with a spherical domain wall having LBR spacetime outside the wall and
flat spacetime inside. We also discuss the internal flat spacetime
using the same coordinate systems as in the external LBR spacetime
with continuous metric at the wall. Among the different cases one
represents a Milne-LBR universe model with a part of the Milne universe
inside the wall and an infinitely extended LBR universe outside it. In
an appendix we define combinations of trigonometric and hyperbolic
functions that we call k-functions and present a new k-function calculus.}
%
%
\vspace{10mm} \newline
{\bf 1. Introduction}
\vspace{3mm} \newline
Conformally flat spacetimes have vanishing Weyl tensor. The line
element of such spacetimes can in general be given the form of a
conformal factor times the Minkowski line element. The coordinates
in which the line element takes this form are called conformally flat
spacetime (CFS) coordinates.
\itm The FRW universe models are conformally flat. We have recently given
a systematic description of these universe models in CFS coordinates
[\ref{r_1}-\ref{r_3}].
\itm In the present article we shall give a similar treatment of the LBR
spacetime which was found by T.\hn Levi-Civita [\ref{r_4},\ref{r_5}]
already in 1917, and was rediscovered by B.\hn Bertotti [\ref{r_6}] and
E.\hn Robinson [\ref{r_7}] in 1959. It was proved by N.\hn Tariq and
B.\hn O.\hn J.\hn Tupper [\ref{r_20}] and by N.\hn Tariq and
R.\hn G.\hn McLenaghan [\ref{r_21}], and later emphasized by H.\hn Stephani
et al. [\ref{r_8}] that the LBR spacetime is the only conformally flat
solution of the Einstein-Maxwell equations which is homogeneous and has
a non-null Maxwell field. The physical interpretation of the solution has
been discussed by D.\hn Lovelock [\ref{r_9},\ref{r_10}], P.\hn Doland
[\ref{r_11}] and the present authors [\ref{r_12}].
\itm Our article is organized as follows. In section 2 we present a new
method for finding different coordinates of the LBR spacetime. We give a
general formalism in section 3 for finding coordinate transformations
between the canonical CFS system and an arbitrary coordinate system.
Section 4 is the main part of the article. Here we find the different
coordinate systems and give a thorough discussion of their properties
and of the reference frames in which they are comoving. In section 5 we
discuss a particularly interesting example, a Milne-LBR universe model
where there is LBR spacetime outside a charged domain wall with a radius
equal to the distance corresponding to its charge, and there is a part of
the Milne universe inside the domain wall. The motions of the reference
frames are further characterized in section 6, where we calculate the
four-acceleration of the reference particles from the Killing vectors.
In section 7 we present embedding parametrizations for the different
coordinate representations of the LBR spacetime in a 6-dimensional,
flat spacetime. Our results are summarized in section 8. We define
k-functions, which are combinations of trigonometric and hyperbolic
functions, in an appendix where we also present the k-calculus of
these functions.
%
%
%
\vspace{6mm} \newline
{\bf 2. A new method for finding different representations of the LBR
spacetime}
\vspace{2mm} \newline
By the {\it Levi-Civita Bertotti Robinson} (LBR) {\it spacetime} we
shall mean a conformally flat and static spacetime which is a solution
of the Einstein-Maxwell equations with an electromagnetic field having
a constant energy momentum tensor. This solution has usually been called
the Bertotti Robinson solution, but it was actually discovered by
T.\hn Levi-Civita already in 1917 [\ref{r_4},\ref{r_5}]. Hence we shall
call it the Levi-Civita Bertotti Robinson solution.
\itm In a previous paper [\ref{r_12}] we have given a new interpretation
of the LBR solution. According to our interpretation this solution
describes a static, spherically symmetric and conformally flat spacetime
with a radial electrical field outside a charged domain wall. There is
Minkowski spacetime inside the wall.
\itm It is well known that the LBR solution can be represented by a
spacetime which is the product of a 2-dimensional anti de Sitter
space and a spherical surface [\ref{r_11}]. Hence the line element may
be written in a spherically symmetric form with an angular part which
is $K^2 d \Omega^2$, where $K$ is a constant. According to our
interpretation [\ref{r_12}] the constant $K$ is equal to the radius
$R_{\mbox{\tiny $Q$}}$ of the domain wall. Also the radius of the domain
wall is determined by its charge $Q$ so that $R_{\mbox{\tiny $Q$}} =
[G/(4 \pi {\epsilon}_0 c^4)]^{1/2} Q$, i.e. $R_{\mbox{\tiny $Q$}}$ is
the length corresponding to the charge $Q$. The line element may
then be given the form
\vspace{-1mm} \newline
\begin{equation} \label{e_211}
ds^2 = - e^{2 \alpha(\check{t},\check{r})} d\check{t}^2
+ e^{2 \beta(\check{t},\check{r})} d\check{r}^2
+ R_{\mbox{\tiny $Q$}}^2 d\Omega^2
\mbox{ .}
\end{equation}
%
%
This line element is rather general, and only in the case where
the Weyl tensor vanishes does it describe the LBR spacetime.
M.\hn G\"{u}rzes and \"{O}.\hn Sario\u{g}lu [\ref{r_23}] have shown that
a $D$-dimensional conformally flat LBR spacetime, which is a product of
a 2-dimensional anti de Sitter spacetime and a $(D-2)$-dimensional
spherical surface, permits a cosmological constant proportional to
$1 - (D - 3)^2$. Hence in the 4-dimensional LBR spacetime the
cosmological constant vanishes, which has earlier been noted by
V.\hn I.\hn Khlebnikov and \'{E}.\hn Shelkovenko [\ref{r_31}] and by
J.\hn Podolsk\'{y} and M.\hn Ortaggio [\ref{r_32}].
\itm Using the radial coordinate $\check{r}$ in the line element invites
the interpretation of the spacetime as a spherically symmetric space
in the spacetime ${\bf R}^4$. An alternative interpretation is
also possible. Neglecting the time dimension in the 2-dimensional
anti de Sitter space and a spatial dimension in the spherical surface,
replacing it by a circle, the spacetime can be interpreted as a cylinder.
Then the electrical field is directed along the axis of the cylinder.
We here want to consider both physical interpretations. The spacetime
with a domain wall will be called the WLBR spacetime, and the spacetime
with a product of a 2-dimensional anti de Sitter space and a spherical
surface will be called the PLBR spacetime. We use LBR in statements
concerning both WPBL and PLBR. Note that in the PLBR interpretation
the coordinate $\check{r}$ shall not be interpreted as a radial
coordinate.
\itm In the present case it follows from the geodesic equation that a
free particle instantaneously at rest has an acceleration
\begin{equation} \label{e_341}
\ddot{\check{r}} = - {\Gamma}^{\hs{0.3mm} \check{r}}_{\hs{1.5mm}
\check{t} \check{t}} \hs{0.8mm} \dot{\check{t}}^{\hs{0.5mm} 2}
= - e^{- 2 \beta} \alpha_{,\check{r}}
\mbox{ ,}
\end{equation}
where the dot denotes differentiation with respect to the proper time of
the particle.
Hence there is attractive gravity, i.e. the acceleration of gravity
points in the negative ${\bf e}_{\check{r}}$-direction, if $\alpha$ is
an increasing function of $\check{r}$ and repulsive gravitation if
$\alpha$ is a decreasing function of $\check{r}$.
%
%
%
\itm With the line element \eqref{e_211} the condition that the
Weyl tensor vanishes takes the form
\begin{equation} \label{e_212}
e^{-2 \beta} (\alpha_{,\check{r}\check{r}}
+ \alpha_{,\check{r}}^2 - \alpha_{,\check{r}} \beta_{,\check{r}})
- e^{-2 \alpha} (\beta_{,\check{t}\check{t}}
+ \beta_{,\check{t}}^2 - \alpha_{,\check{t}} \beta_{,\check{t}})
= \frl{1}{R_{\mbox{\tiny $Q$}}^2}
\mbox{ .}
\end{equation}
Calculating the components of the Einstein tensor from the line element
\eqref{e_211} and using Einstein's field equations it follows that
when equation \eqref{e_212} is fullfilled, the mixed components of the
energy momentum tensor reduce to
\begin{equation} \label{e_216}
T^{\hs{0.3mm} \check{t}}_{\hs{1.5mm} \check{t}} =
T^{\hs{0.3mm} \check{r}}_{\hs{1.5mm} \check{r}} =
-T^{\hs{0.3mm} \theta}_{\hs{1.5mm} \theta} =
-T^{\hs{0.3mm} \phi}_{\hs{1.5mm} \phi} =
- \frl{1}{\kappa R_{\mbox{\tiny $Q$}}^2}
\mbox{ ,}
\end{equation}
which represents a constant radial electric field, as is the case in the
LBR spacetime. This shows that the LBR spacetime does not allow a
non-vanishing cosmological constant.
\itm In the section 4 equation \eqref{e_212} will be solved under
different coordinate conditions. The solutions found is the subsections
4.Ia and 4.Ib will turn out to be special cases of the line element
\begin{equation} \label{e_408}
ds^2 = [R_{\mbox{\tiny $Q$}} / G(x^0,x^1)]^2
[-(dx^0)^2 + (dx^1)^2] + R_{\mbox{\tiny $Q$}}^2 d\Omega^2
\mbox{ ,}
\end{equation}
where $x^0$ is a time coordinate, $x^1$ is a radial coordinate,
$G(x^0,x^1)$ is a function of $x^0$ and $x^1$, and $d\Omega^2$ is
a solid angle element.
\itm In the next section we shall develop a formalism for finding
transformations between the coordinates where the line element takes
the form \eqref{e_408} and the CFS coordinates.
%
%
%
\vspace{10mm} \newline
{\bf 3. Conformally flat spacetime coordinates for the LBR spacetime}
\vspace{3mm} \newline
We want to write the line element \eqref{e_408} of a spacetime with
spherically symmetric space in terms of conformally flat spacetime (CFS)
coordinates $(T,R)$. Then the line element takes the form of a
conformal factor $C(T,R)^2$ times the Minkowski line element,
\begin{equation} \label{e_11}
ds^2 = C(T,R)^2 ds_M^2
= C(T,R)^2 (-dT^2 + dR^2 + R^2 d \Omega^2)
\mbox{ .}
\end{equation}
In order to perform this we shall generalize the method developed in
reference [\ref{r_1}].
\itm We then use transformations of the form
\begin{equation} \label{e_12}
T = \frl{1}{2} \hs{0.6mm} [\hs{0.3mm} f(x^0 + x^1) + g(x^0 - x^1)
\hs{0.3mm}]
\mbox{\hspace{2mm} , \hspace{3mm}}
R = \frl{1}{2} \hs{0.6mm} [\hs{0.3mm} f(x^0 + x^1) - g(x^0 - x^1)
\hs{0.3mm}]
\end{equation}
where $f$ and $g$ are functions that must satisfy an identity deduced below.
A transformation of this form can be described as a composition of three
simple transformations. The first transforms from the coordinates $x^0$
and $x^1$ in the line element \eqref{e_408} to light cone coordinates
(null coordinates) associated with a Minkowski diagram referring to the
$(x^0,x^1)$ coordinate system
\begin{equation} \label{e_409}
u = x^0 + x^1
\mbox{\hspace{2mm} , \hspace{3mm}}
v = x^0 - x^1
\mbox{ .}
\end{equation}
In the Minkowski diagram this rotates the previous coordinate system by
$- \pi / 4$ and scales it by a factor $\sqrt{2}$. The scaling is performed
for later convenience. The coordinate $u$ is constant for light moving in
the negative $x^1$-direction, and $v$ in the positive $x^1$-direction.
The second transforms $u$ and $v$ to the coordinates
\begin{equation} \label{e_410}
U = f(u)
\mbox{\hspace{2mm} , \hspace{3mm}}
V = g(v)
\mbox{ .}
\end{equation}
Finally, we scale and rotate with the inverse of the
transformation \eqref{e_409},
\begin{equation} \label{e_420}
T = \frl{U \p V}{2}
\mbox{\hspace{2mm} , \hspace{3mm}}
R = \frl{U \m V}{2}
\mbox{ .}
\end{equation}
The inverse of the transformation \eqref{e_420} is
\begin{equation} \label{e_2409}
U = T + R
\mbox{\hspace{2mm} , \hspace{3mm}}
V = T - R
\mbox{ ,}
\end{equation}
showing that $U$ and $V$ are light cone coordinates associated with a
Minkowski diagram referring to the CFS coordinate system. The coordinate
$U$ is constant for light moving in the negative $R$-direction and $V$
in the positive $R$-direction. Note that
\begin{equation} \label{e_422}
T^2 - R^2 = UV
\mbox{ .}
\end{equation}
Taking the differentials of $T$ and $R$ we get
\begin{equation} \label{e_421}
-dT^2 + dR^2 = -dU dV
= -f'(u) g'(v) \hs{0.5mm} du \hs{0.5mm} dv
= f'(u) g'(v) (-(d x^0)^2 + (d x^1)^2)
\mbox{ .}
\end{equation}
Comparing the expressions \eqref{e_408} and \eqref{e_11} for the line
element and using the previous formula, we find
\begin{equation} \label{e_390}
C(T,R)^2 = \frl{R_{\mbox{\tiny $Q$}}^2}{f'(u) g'(v) G(x^0,x^1)^2}
\end{equation}
where $x^0$, $x^1$, $u$ and $v$ are functions of $T$ and $R$, and
\begin{equation} \label{e_1381}
C(T,R)^2 = \frl{R_{\mbox{\tiny $Q$}}^2}{R^2}
\mbox{ .}
\end{equation}
From equations \eqref{e_390} and \eqref{e_1381} it follows that
\begin{equation} \label{e_381}
f'(u) g'(v) \hs{0.5mm} G(x^0,x^1)^2 = R^2
\mbox{ .}
\end{equation}
%
%
%
By \eqref{e_12} and \eqref{e_409} equation \eqref{e_381} may be written as
\begin{equation} \label{e_394}
f'(u) g'(v) \hs{0.6mm} G \hs{-0.5mm} \left( \frl{u + v}{2}, \frl{u - v}{2}
\right)^2
= \frl{1}{4} \hs{0.6mm} [ \hs{0.3mm} f(u) - g(v) \hs{0.3mm} ]^2
\mbox{ .}
\end{equation}
Substituting $v = u$ we get the condition
\begin{equation} \label{e_1513}
f'(u) g'(u) \hs{0.6mm} G(u,0)^2
= \frl{1}{4} \hs{0.6mm} [ \hs{0.3mm} f(u) - g(u) \hs{0.3mm} ]^2
\mbox{ .}
\end{equation}
\itm As shown in reference [\ref{r_1}] if $G(u,0) = 0$, the line element
\eqref{e_11} can be written in the form \eqref{e_408} with
$G(x^0,x^1) = S_k (x^1)$, where the function $S_k$ is defined in
equation \eqref{e_801}. Then equation \eqref{e_394} reduces to
\begin{equation} \label{e_513}
f'(u) g'(v) \hs{0.6mm} S_k \hs{-0.5mm} \left( \frl{u - v}{2} \right)^2
= \frl{1}{4} \hs{0.6mm} [ \hs{0.3mm} f(u) - g(v) \hs{0.3mm} ]^2
\mbox{ .}
\end{equation}
Substituting $v = u$ and utilizing that $S_k(0) = 0$, this equation
gives $g(u) = f(u)$. Hence equation \eqref{e_513} may be written
\begin{equation} \label{e_512}
f'(u) f'(v) \hs{0.6mm} S_k \hs{-0.5mm} \left( \frl{u - v}{2} \right)^2
= \frl{1}{4} \hs{0.6mm} [ \hs{0.3mm} f(u) - f(v) \hs{0.3mm} ]^2
\mbox{ ,}
\end{equation}
\begin{equation} \label{e_511}
T = \frl{1}{2} \hs{0.6mm} [\hs{0.3mm} f(x^0 + x^1) + f(x^0 - x^1)
\hs{0.3mm}]
\mbox{\hspace{2mm} , \hspace{3mm}}
R = \frl{1}{2} \hs{0.6mm} [\hs{0.3mm} f(x^0 + x^1) - f(x^0 - x^1)
\hs{0.3mm}]
\mbox{ .}
\end{equation}
With the function [\ref{r_1}]
\begin{equation} \label{e_13}
f(x) =  c \left[ b + I_k \left( \frl{x - a}{2} \right) \right]^{-1} + d
\mbox{ ,}
\end{equation}
\vspace{-1mm} \newline
where $a$, $b$, $c$, $d$ are arbitrary constants and the function
$I_k(x)$ is defined in equation \eqref{e_804}, the transformation
\eqref{e_511} leads from \eqref{e_408} with $G(x^0,x^1) = S_k (x^1)$
to \eqref{e_11} with $C(T,R)$ given by equation \eqref{e_1381} in
the case of the LBR spacetime.
\itm It follows from equations \eqref{e_211}, \eqref{e_11} and
\eqref{e_1381} that the line element of the Minkowski spacetime
inside the domain wall in the different
coordinate systems takes the form
\vspace{-1mm} \newline
\begin{equation} \label{e_1211}
ds_M^2 = \left( \frl{R(\check{t},\check{r})}{R_{\mbox{\tiny $Q$}}}
\right)^2 (- e^{2 \alpha(\check{t},\check{r})} d\check{t}^2
+ e^{2 \beta(\check{t},\check{r})} d\check{r}^2
+ R_{\mbox{\tiny $Q$}}^2 d\Omega^2)
\mbox{ .}
\end{equation}
The equations \eqref{e_211} and \eqref{e_1211} give the general
connection between the form of the line element of the WLBR spacetime
outside the domain wall in an arbitrary coordinate system and the form
of the line element of the flat spacetime inside the domain wall in the
same coordinate system.
%
%
%
%
%
\vspace{6mm} \newline
{\bf 4. The LBR spacetime in different coordinate systems}
\vspace{-2mm} \newline
\itm Equation \eqref{e_212} will now be solved under different
coordinate conditions.
%
%
\vspace{6mm} \newline
{\it Ia. Static metric and coordinates $(\eta,\chi)$ with
$\beta (\chi) = \alpha (\chi)$.}
\vspace{3mm} \newline
In this case equation \eqref{e_212} reduces to
\begin{equation} \label{e_213}
R_{\mbox{\tiny $Q$}}^2 \alpha'' - e^{2 \alpha} = 0
\end{equation}
where the prime means differentiation with respect to the radial
coordinate. This equation may be written
\begin{equation} \label{e_214}
R_{\mbox{\tiny $Q$}}^2 (\alpha' \hs{0.5mm}^2)' = (e^{2 \alpha})'
\mbox{ .}
\end{equation}
Integration gives
\begin{equation} \label{e_215}
R_{\mbox{\tiny $Q$}}^2 \hs{0.5mm} \alpha' \hs{0.5mm}^2
= e^{2 \alpha} - k \hs{0.2mm} c^2 R_{\mbox{\tiny $Q$}}^2
\mbox{ ,}
\end{equation}
where $c > 0$ is an integration constant and $k$ takes the values $1$, $0$
or $-1$. The general solution of \eqref{e_215} is given by
\begin{equation} \label{e_219}
e^{2 \alpha} = c^2 R_{\mbox{\tiny $Q$}}^2 / S_k({\chi}_0 + c \chi)^2
\mbox{ ,}
\end{equation}
where $S_k(x)$ is the function defined in equation \eqref{e_801}
in Appendix A. Here ${\chi}_0$ is an integration constant and $c = 1$
when $k = 0$.
\itm The value $k = 0$ is a very important special case. Then
one can introduce CFS coordinates simply by putting ${\chi}_0 = 0$. The
line element with $(\eta,\chi)$ replaced by $(T,R)$ then takes the form
\begin{equation} \label{e_157}
ds^2 = \frl{R_{\mbox{\tiny $Q$}}^2}{R^2} \hs{0.5mm} ( \hs{0.5mm}
\mbox{$- dT^2$} + dR^2 + R^2 d\Omega^2 \hs{0.5mm})
\end{equation}
with $- \infty < T < \infty$, $R > R_{\mbox{\tiny $Q$}}$ for the WLBR
spacetime, and with $- \infty < T < \infty$, $- \infty < R < \infty$,
$R \ne 0$ for the PLBR spacetime. This
form of the line element is in agreement with equations \eqref{e_11}
and \eqref{e_1381}. Note that the metric is static. This means
that the coordinate clocks go with the same rate at all positions.
The line element has the Minkowski form at the domain wall at
$R = R_{\mbox{\tiny $Q$}}$. At this surface $g_{TT} = -1$, meaning
that the coordinate clocks of the CFS system show the same time as
standard clocks at rest at the domain wall. The fact that there exists
a coordinate system so that the metric is static means that the
LBR spacetime is static, although we will show later that there exist
coordinates so that the metric of this spacetime is time dependent.
This time dependence is due to the motion of the reference frame in
which the coordinates are comoving.
\itm As has been noted by O.\hn J.\hn C.\hn Dias and
J.\hn P.\hn S.\hn Lemos [\ref{r_16}] there is an interesting connection
between the WLBR spacetime and the Reissner-Nordstr\"{o}m spacetime,
which is usually described by the line element
\begin{equation} \label{e_1157}
ds^2 = - \mbox{$\left( \rule[-1.5mm]{0mm}{5.7mm} \right.$} \hs{-0.2mm}
1 - \frl{R_{\mbox{\tiny $S$}}}{r}
+ \frl{R_{\mbox{\tiny $Q$}}^2}{r^2}
\hs{-0.2mm} \mbox{$\left. \rule[-1.5mm]{0mm}{5.7mm} \right)$}
\hs{0.5mm} dt^2 +
\mbox{$\left( \rule[-1.5mm]{0mm}{5.7mm} \right.$} \hs{-0.2mm}
1 - \frl{R_{\mbox{\tiny $S$}}}{r}
+ \frl{R_{\mbox{\tiny $Q$}}^2}{r^2}
\hs{-0.2mm} \mbox{$\left. \rule[-1.5mm]{0mm}{5.7mm} \right)$}
^{\hs{-0.5mm} -1} \hs{0.5mm} dr^2 + r^2 d\Omega^2
\mbox{ ,}
\end{equation}
where $R_{\mbox{\tiny $S$}} = 2 \hs{0.3mm} GM / c^2$ is the Schwarzschild
radius, and $R_{\mbox{\tiny $Q$}}$ is the length corresponding to the
electric charge $Q$. The extremal Reissner-Nordstr\"{o}m spacetime has
$R_{\mbox{\tiny $S$}} = 2 R_{\mbox{\tiny $Q$}}$, and then the line
element takes the form
\begin{equation} \label{e_1158}
ds^2 = - \left( 1 - \frl{R_{\mbox{\tiny $Q$}}}{r} \right)^{\hs{-0.5mm} 2}
\hs{0.5mm} dt^2
+ \left( 1 - \frl{R_{\mbox{\tiny $Q$}}}{r} \right)^{\hs{-0.5mm} -2}
\hs{0.5mm} dr^2 + r^2 d\Omega^2
\mbox{ .}
\end{equation}
A Taylor expansion of $f(r) = (1 - \fr{R_{\mbox{\tiny $Q$}}}{r})^2$
about $r = R_{\mbox{\tiny $Q$}}$ gives to 2.\hn order in $r$,
$f(r) \approx \fr{(r - R_{\mbox{\tiny $Q$}})^2}{R_{\mbox{\tiny $Q$}}^2} \,$.
Hence, the near-horizon limit of the line element for the extremal
Reissner-Nordstr\"{o}m spacetime takes the form
\begin{equation} \label{e_1159}
ds^2 = - \frl{(r - R_{\mbox{\tiny $Q$}})^2}{R_{\mbox{\tiny $Q$}}^2}
\hs{0.5mm} dt^2 +
\frl{R_{\mbox{\tiny $Q$}}^2}{(r - R_{\mbox{\tiny $Q$}})^2} \hs{0.5mm} dr^2
+ R_{\mbox{\tiny $Q$}}^2 d\Omega^2
\mbox{ ,}
\end{equation}
where the angular part is correct only to 0.\hn order in $r$.
Introducing coordinates
\begin{equation} \label{e_1759}
R = (r - R_{\mbox{\tiny $Q$}})^{-1}
\mbox{\hspace{2mm} , \hspace{3mm}}
T = \fr{t}{R_{\mbox{\tiny $Q$}}^2}
\mbox{ ,}
\end{equation}
leads to the form \eqref{e_157} of the line element. Hence the line
element of the near-horizon limit of the Reissner-Nordstr\"{o}m spacetime
has the same form as the line element of the LBR spacetime.
%
%
But the coordinates $R$ and $r$ in equation \eqref{e_1759} increase in
opposite directions. If this is forgotten, gravity seems to be repulsive
in the near-horizon limit of the Reissner-Nordstr\"{o}m spacetime as
expressed in terms of the CFS coordinate $R$, since $\alpha$ is a
decreasing function of $R$. However, gravity is attractive in the
near-horizon limit of the Reissner-Nordstr\"{o}m spacetime. This is a
coordinate independent property of the spacetime. In the LBR spacetime
the CFS coordinate $R$ increases in the direction away from the symmetry
center, and there is repulsive gravity. The LBR spacetime is therefore
very different from the near-horizon limit of the Reissner-Nordstr\"{o}m
spacetime.
\itm We shall define the acceleration of gravity in a coordinate system
with an arbitrary radial coordinate $\check{r}$ as the acceleration of a
free particle instantaneously at rest and measured with standard measuring
rods and clocks. Hence it is the component along the unit radial basis
vector of the second derivative of the radial coordinate with respect to
the proper time of the particle,
\begin{equation} \label{e_357}
a^{\hat{\check{r} \rule[-0.0mm]{0mm}{2.0mm}}}
= (g_{\hs{0.3mm} \check{r} \check{r}})^{1/2} \hs{0.5mm} \ddot{\check{r}}
\mbox{ .}
\end{equation}
In the present case $\ddot{\check{r}} = \ddot{R}$ where $\ddot{R}$ is
given by the geodesic equation
\begin{equation} \label{e_358}
\ddot{R} =
- \Gamma^{\mbox{\tiny $R$}}_{\hs{0.5mm} \mbox{\tiny $T$} \mbox{\tiny $T$}}
\dot{T}^2 =
\frl{\Gamma^{\mbox{\tiny $R$}}_{\hs{0.5mm} \mbox{\tiny $T$}
\mbox{\tiny $T$}}}{g_{\mbox{\tiny $T$} \mbox{\tiny $T$}}}
\mbox{ .}
\end{equation}
For the WLBR spacetime this gives
\begin{equation} \label{e_359}
a^{\hat{R}}
= \sqrt{g_{\mbox{\tiny $R$} \mbox{\tiny $R$}}} \hs{0.8mm} \ddot{R}
= \frl{1}{R_{\mbox{\tiny $Q$}}}
\mbox{ ,}
\end{equation}
i.e. in the CFS system the acceleration of gravity is constant and
directed away from the domain wall.
\itm We will show that the solutions \eqref{e_219} with $k = 1$ and
$k = -1$ represent the same spacetime as the solution with $k = 0$.
This will be shown by demonstrating that there exists a coordinate
transformation that transforms the line elements of the solutions
\eqref{e_219} with $k = 1$ and $k = -1$ to the form \eqref{e_157}.
Putting $c = 1$ and ${\chi}_0 = 0$, the line element \eqref{e_211}
with the solution \eqref{e_219} takes the form
\begin{equation} \label{e_218}
ds^2 = \frl{R_{\mbox{\tiny $Q$}}^2}{S_k(\chi)^2}
(-d \eta^2 + d\chi^2) + R_{\mbox{\tiny $Q$}}^2 d\Omega^2
\mbox{ .}
\end{equation}
In the case $k = 1$ the coordinate clocks showing $\eta$ go at the same
rate as a standard clock at $\chi = \pi / 2$, scaled by the factor
$R_{\mbox{\tiny $Q$}}$. It may be noted that radially moving light
has a coordinate velocity $d\chi / d\eta = \pm 1$ for all values of $k$,
which is due to the condition $\alpha = \beta$.
\itm Note that the form \eqref{e_218} of the line element is valid
for all values of $k$. In the case $k = 0$ the line element reduces
to form \eqref{e_157} with $(T,R)$ replaced by $(\eta,\chi)$.
\itm In order to find a coordinate transformation between the
$(\eta,\chi)$-coordinates and the CFS coordinates we apply
the formalism in section 3. By choosing $a = 0$, $b = 0$, $c = B$
and $d = 0$ in equation \eqref{e_13} we obtain the generating function
\begin{equation} \label{e_339}
f(x) = B \hs{0.8mm} T_k \leftn \fr{x}{2} \rightn
\end{equation}
where $T_k(x)$ is defined in equation \eqref{e_803} and
$B$ is a positive constant satisfying
\begin{equation} \label{e_1139}
(1 - |k|) B = 2(1 - |k|)
\mbox{ .}
\end{equation}
Hence $B$ equals $2$ when $k = 0$, and has an arbitrary positive value
when $k = 1$ and $k = -1$.
Using the generating function \eqref{e_339} as shown in Appendix B, the
transformation \eqref{e_511} between the $(\eta,\chi)$-system and the
CFS system takes the form
\begin{equation} \label{e_279}
T = \frl{B \hs{0.2mm}
S_k(\eta)}{C_k(\eta) \p C_k(\chi)}
\mbox{\hspace{2mm} , \hspace{3mm}}
R = \frl{B \hs{0.2mm}
S_k(\chi)}{C_k(\eta) \p C_k(\chi)}
\mbox{ ,}
\end{equation}
where $C_k(x)$ is defined in equation \eqref{e_802}.
\itm We have shown in Apppendix B how the inverse transformation is
obtained from the generating function
\begin{equation} \label{e_1179}
f(x) = 2 \hs{0.5mm} T_k^{-1} (\fr{x}{B})
\mbox{ ,}
\end{equation}
giving the result
\vspace{-0mm} \newline
\begin{equation} \label{e_346}
I_k(\eta) = \frl{B^2 \m k (T^{2} \m R^{2})}
{2 \hs{0.2mm} B \hs{0.2mm} T}
\mbox{\hspace{2mm} , \hspace{3mm}}
I_k(\chi) = \frl{B^2 \p k (T^{2} \m R^{2})}
{2 \hs{0.2mm} B \hs{0.2mm} R}
\end{equation}
\vspace{-2mm} \newline
when $T \ne 0$, where $I_k(x)$ is defined in equation \eqref{e_804}.
In the case $T = 0$ we have that $\eta = 0$. Note that the formulae
\eqref{e_218} - \eqref{e_346} are valid for all values of $k$.
A special case of the line element \eqref{e_218} with $k = -1$ has
been used by A.\hn C.\hn Ottewill and P. \hn Taylor [\ref{r_13}] in
connection with quantum field theory on the LBR spacetime.
\itm The world lines of points on the domain wall are given by
the second of equations \eqref{e_279} with $R = R_{\mbox{\tiny $Q$}}$,
which leads to
\begin{equation} \label{e_344}
C_k (\eta) = (B / R_{\mbox{\tiny $Q$}}) S_k (\chi) - C_k (\chi)
\mbox{ .}
\end{equation}
Introducing the constant
\begin{equation} \label{e_1184}
{\chi}_{\mbox{\tiny $Q$}}
= I_k^{-1} (B / R_{\mbox{\tiny $Q$}})
\mbox{ ,}
\end{equation}
equation \eqref{e_344} takes the form
\begin{equation} \label{e_1185}
S_k ({\chi}_{\mbox{\tiny $Q$}}) \hs{0.5mm} C_k (\eta)
= S_k (\chi - {\chi}_{\mbox{\tiny $Q$}})
\end{equation}
which can also be written as
\begin{equation} \label{e_1186}
\chi = {\chi}_{\mbox{\tiny $Q$}}
+ S_k^{-1}(S_k ({\chi}_{\mbox{\tiny $Q$}}) \hs{0.5mm} C_k (\eta))
\mbox{ .}
\end{equation}
The point of intersection $(0,{\chi}_0)$ with the $\chi$-axis,
where ${\chi}_0$ is the coordinate radius of the domain wall in the
$(\eta,\chi)$-system at the point of time $\eta = 0$,
is found by inserting $\eta = 0$ in equation \eqref{e_1186}.
Using that $C_k(0) = 1$ for all values of $k$ we then obtain
a physical interpretation of the constant ${\chi}_{\mbox{\tiny $Q$}}$,
\begin{equation} \label{e_1144}
{\chi}_{\mbox{\tiny $Q$}} = {\chi}_0 / 2
\mbox{ .}
\end{equation}
From equation \eqref{e_1184} it then follows that
\begin{equation} \label{e_1167}
B = R_{\mbox{\tiny $Q$}} I_k({\chi}_0 / 2)
\mbox{ .}
\end{equation}
\itm When $k = 1$ equation \eqref{e_344} takes the form
\begin{equation} \label{e_1187}
\cos \eta = (B / R_{\mbox{\tiny $Q$}}) \sin \chi - \cos \chi
\end{equation}
which is plotted in Figure 1 as the left hand boundary of the hatched
region.
It follows that in the case $k = 1$ the WLBR spacetime is represented
in the $(\eta,\chi)$-plane by the hatched region in Figure 1, which is
given by
\begin{equation} \label{e_555}
{\chi}_{\mbox{\tiny $Q$}}
+ \arcsin \left( \sin {\chi}_{\mbox{\tiny $Q$}} \cos \eta \right)
< \chi < \pi - |\eta|
\mbox{\hspace{2mm} , \hspace{3mm}}
- \pi < \eta < \pi
\mbox{ .}
\end{equation}
%
We want to find the corresponding region in the $(\eta,\chi)$-system
representing the PLBR spacetime. From equation \eqref{e_511}
we obtain
\begin{equation} \label{e_669}
T + R = f(\eta + \chi)
\mbox{\hspace{2mm} , \hspace{3mm}}
T - R = f(\eta - \chi)
\mbox{ .}
\end{equation}
Hence $\eta + \chi$ and $\eta - \chi$ must belong to the domain
$(-\pi,\pi)$ of the generator function in equation \eqref{e_339}
with $k = 1$. This gives the region
\begin{equation} \label{e_665}
|\eta| + |\chi| < \pi
\mbox{\hspace{2mm} , \hspace{3mm}}
\chi \ne 0
\end{equation}
%
when $k = 1$, as illustrated in Figure 1.
%
\vspace*{5mm} \newline
\begin{picture}(50,212)(-96,-170)
\qbezier(135.0000, -122.3962)(135.0000, -121.4736)(135.0846, -120.5336)
\qbezier(135.0846, -120.5336)(135.1661, -119.6278)(135.3355, -118.6710)
\qbezier(135.3355, -118.6710)(135.4933, -117.7806)(135.7451, -116.8084)
\qbezier(135.7451, -116.8084)(135.9722, -115.9316)(136.3014, -114.9458)
\qbezier(136.3014, -114.9458)(136.5904, -114.0804)(136.9901, -113.0832)
\qbezier(136.9901, -113.0832)(137.3333, -112.2272)(137.7957, -111.2206)
\qbezier(137.7957, -111.2206)(138.1855, -110.3722)(138.7026, -109.3580)
\qbezier(138.7026, -109.3580)(139.1319, -108.5160)(139.6961, -107.4954)
\qbezier(139.6961, -107.4954)(140.1583, -106.6593)(140.7627, -105.6328)
\qbezier(140.7627, -105.6328)(141.2514, -104.8029)(141.8904, -103.7702)
\qbezier(141.8904, -103.7702)(142.3993, -102.9477)(143.0684, -101.9076)
\qbezier(143.0684, -101.9076)(143.5910, -101.0952)(144.2875, -100.0450)
\qbezier(144.2875, -100.0450)(144.8160, -99.2479)(145.5392, -98.1824)
\qbezier(145.5392, -98.1824)(146.0631, -97.4104)(146.8162, -96.3198)
\qbezier(146.8162, -96.3198)(147.3172, -95.5943)(148.1120, -94.4571)
\qbezier(148.1120, -94.4571)(148.5459, -93.8364)(149.4204, -92.5945)
\qbezier(149.4204, -92.5945)(149.5778, -92.3711)(150.7357, -90.7319)
\qbezier(150.7357, -90.7319)(152.8865, -87.6874)(152.0524, -88.8693)
\qbezier(152.0524, -88.8693)(153.0122, -87.5093)(153.3649, -87.0067)
\qbezier(153.3649, -87.0067)(154.1914, -85.8294)(154.6677, -85.1441)
\qbezier(154.6677, -85.1441)(155.4394, -84.0341)(155.9550, -83.2815)
\qbezier(155.9550, -83.2815)(156.6924, -82.2051)(157.2202, -81.4189)
\qbezier(157.2202, -81.4189)(157.9299, -80.3619)(158.4566, -79.5563)
\qbezier(158.4566, -79.5563)(159.1395, -78.5116)(159.6562, -77.6937)
\qbezier(159.6562, -77.6937)(160.3107, -76.6576)(160.8103, -75.8311)
\qbezier(160.8103, -75.8311)(161.4326, -74.8016)(161.9088, -73.9685)
\qbezier(161.9088, -73.9685)(162.4940, -72.9449)(162.9405, -72.1059)
\qbezier(162.9405, -72.1059)(163.4821, -71.0883)(163.8925, -70.2433)
\qbezier(163.8925, -70.2433)(164.3833, -69.2327)(164.7506, -68.3807)
\qbezier(164.7506, -68.3807)(165.1827, -67.3785)(165.4997, -66.5181)
\qbezier(165.4997, -66.5181)(165.8651, -65.5263)(166.1241, -64.6555)
\qbezier(166.1241, -64.6555)(166.4154, -63.6762)(166.6088, -62.7929)
\qbezier(166.6088, -62.7929)(166.8199, -61.8282)(166.9403, -60.9303)
\qbezier(166.9403, -60.9303)(167.0674, -59.9818)(167.1087, -59.0677)
\qbezier(167.1087, -59.0677)(167.1508, -58.1364)(167.1087, -57.2051)
\qbezier(167.1087, -57.2051)(167.0674, -56.2909)(166.9403, -55.3425)
\qbezier(166.9403, -55.3425)(166.8199, -54.4446)(166.6088, -53.4799)
\qbezier(166.6088, -53.4799)(166.4154, -52.5965)(166.1241, -51.6172)
\qbezier(166.1241, -51.6172)(165.8651, -50.7465)(165.4997, -49.7546)
\qbezier(165.4997, -49.7546)(165.1827, -48.8942)(164.7506, -47.8920)
\qbezier(164.7506, -47.8920)(164.3833, -47.0400)(163.8925, -46.0294)
\qbezier(163.8925, -46.0294)(163.4821, -45.1844)(162.9405, -44.1668)
\qbezier(162.9405, -44.1668)(162.4940, -43.3279)(161.9088, -42.3042)
\qbezier(161.9088, -42.3042)(161.4326, -41.4712)(160.8103, -40.4416)
\qbezier(160.8103, -40.4416)(160.3107, -39.6152)(159.6562, -38.5790)
\qbezier(159.6562, -38.5790)(159.1395, -37.7611)(158.4566, -36.7164)
\qbezier(158.4566, -36.7164)(157.9299, -35.9108)(157.2202, -34.8538)
\qbezier(157.2202, -34.8538)(156.6924, -34.0676)(155.9550, -32.9912)
\qbezier(155.9550, -32.9912)(155.4394, -32.2386)(154.6677, -31.1286)
\qbezier(154.6677, -31.1286)(154.1914, -30.4433)(153.3649, -29.2660)
\qbezier(153.3649, -29.2660)(153.0122, -28.7635)(152.0524, -27.4034)
\qbezier(152.0524, -27.4034)(152.8865, -28.5853)(150.7357, -25.5408)
\qbezier(150.7357, -25.5408)(149.5778, -23.9017)(149.4204, -23.6782)
\qbezier(149.4204, -23.6782)(148.5459, -22.4363)(148.1120, -21.8156)
\qbezier(148.1120, -21.8156)(147.3172, -20.6785)(146.8162, -19.9530)
\qbezier(146.8162, -19.9530)(146.0631, -18.8623)(145.5392, -18.0904)
\qbezier(145.5392, -18.0904)(144.8160, -17.0249)(144.2875, -16.2278)
\qbezier(144.2875, -16.2278)(143.5910, -15.1775)(143.0684, -14.3652)
\qbezier(143.0684, -14.3652)(142.3993, -13.3251)(141.8904, -12.5026)
\qbezier(141.8904, -12.5026)(141.2514, -11.4699)(140.7627, -10.6400)
\qbezier(140.7627, -10.6400)(140.1583,  -9.6134)(139.6961,  -8.7773)
\qbezier(139.6961,  -8.7773)(139.1319,  -7.7567)(138.7026,  -6.9147)
\qbezier(138.7026,  -6.9147)(138.1855,  -5.9005)(137.7957,  -5.0521)
\qbezier(137.7957,  -5.0521)(137.3333,  -4.0455)(136.9901,  -3.1895)
\qbezier(136.9901,  -3.1895)(136.5904,  -2.1923)(136.3014,  -1.3269)
\qbezier(136.3014,  -1.3269)(135.9722,  -0.3411)(135.7451,   0.5357)
\qbezier(135.7451,   0.5357)(135.4933,   1.5079)(135.3355,   2.3983)
\qbezier(135.3355,   2.3983)(135.1661,   3.3550)(135.0846,   4.2609)
\qbezier(135.0846,   4.2609)(135.0000,   5.2008)(135.0000,   6.1235)
\put(135.0846, -121.1472){\line(1, 0){  1.1644}}
\put(135.7451, -117.4220){\line(1, 0){  4.2291}}
\put(136.9901, -113.6968){\line(1, 0){  6.7092}}
\put(138.7026, -109.9716){\line(1, 0){  8.7220}}
\put(140.7627, -106.2464){\line(1, 0){ 10.3871}}
\put(143.0684, -102.5212){\line(1, 0){ 11.8066}}
\put(145.5392, -98.7960){\line(1, 0){ 13.0610}}
\put(148.1120, -95.0708){\line(1, 0){ 14.2134}}
\put(150.7357, -91.3456){\line(1, 0){ 15.3149}}
\put(153.3649, -87.6204){\line(1, 0){ 16.4109}}
\put(155.9550, -83.8952){\line(1, 0){ 17.5461}}
\put(158.4566, -80.1699){\line(1, 0){ 18.7697}}
\put(160.8103, -76.4447){\line(1, 0){ 20.1412}}
\put(162.9405, -72.7195){\line(1, 0){ 21.7362}}
\put(164.7506, -68.9943){\line(1, 0){ 23.6513}}
\put(166.1241, -65.2691){\line(1, 0){ 26.0030}}
\put(166.9403, -61.5439){\line(1, 0){ 28.9120}}
\put(166.6088, -54.0935){\line(1, 0){ 28.6082}}
\put(165.4997, -50.3683){\line(1, 0){ 25.9920}}
\put(163.8925, -46.6431){\line(1, 0){ 23.8741}}
\put(161.9088, -42.9179){\line(1, 0){ 22.1325}}
\put(159.6562, -39.1927){\line(1, 0){ 20.6600}}
\put(157.2202, -35.4674){\line(1, 0){ 19.3707}}
\put(154.6677, -31.7422){\line(1, 0){ 18.1980}}
\put(152.0524, -28.0170){\line(1, 0){ 17.0881}}
\put(149.4204, -24.2918){\line(1, 0){ 15.9949}}
\put(146.8162, -20.5666){\line(1, 0){ 14.8739}}
\put(144.2875, -16.8414){\line(1, 0){ 13.6774}}
\put(141.8904, -13.1162){\line(1, 0){ 12.3493}}
\put(139.6961,  -9.3910){\line(1, 0){ 10.8183}}
\put(137.7957,  -5.6658){\line(1, 0){  8.9935}}
\put(136.3014,  -1.9406){\line(1, 0){  6.7626}}
\put(135.3355,   1.7846){\line(1, 0){  4.0033}}
\put(135.0000,   5.5098){\line(1, 0){  0.6136}}
\put( 47.0455, -58.1364){\vector(1, 0){175.9091}}
\put(135.0000, -146.0909){\vector(0, 1){175.9091}}
\put(135.0000,   6.1235){\line(1, -1){ 64.2598}}
\put(135.0000, -122.3962){\line(1, 1){ 64.2598}}
\put( 70.7402, -58.1364){\line(1, -1){ 64.2598}}
\put( 70.7402, -58.1364){\line(1, 1){ 64.2598}}
\put(230.7273, -61.2045){\makebox(0,0)[]{\footnotesize{$\chi$}}}
\put(128.4545,  30.8409){\makebox(0,0)[]{\footnotesize{$\eta$}}}
\put(205.3962, -64.2727){\makebox(0,0)[]{\footnotesize{$\pi$}}}
\put( 64.6038, -64.2727){\makebox(0,0)[]{\footnotesize{$-\pi$}}}
\put(155.8799, -64.2727){\makebox(0,0)[]{\footnotesize{${\chi}_0$}}}
\put(127.8409,   8.1689){\makebox(0,0)[]{\footnotesize{$\pi$}}}
\put(125.7955, -124.4417){\makebox(0,0)[]{\footnotesize{$-\pi$}}}
\end{picture}
\vspace{0mm} \newline
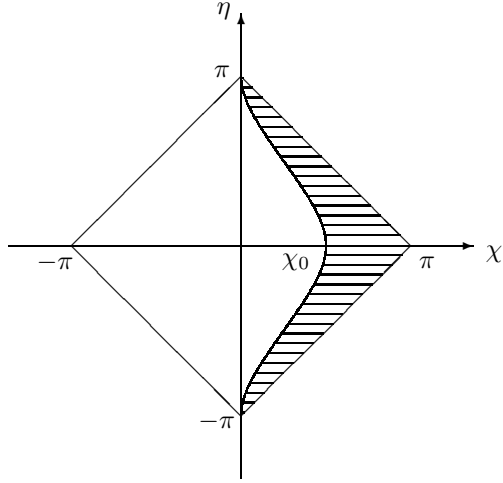
{\footnotesize \sf Figure 1. The square represents the PLBR spacetime
for $k = 1$ in the $(\eta,\chi)$-system given by \eqref{e_665}.
The hatched region represents the WLBR spacetime in the
$(\eta,\chi)$-system given by \eqref{e_555}. The left hand curve
represents the world line of a point on the domain wall as given by
equation \eqref{e_1187} where ${\chi}_0$ is given by \eqref{e_1144}. }
\vspace{2mm} \newline
\itm When $k = -1$ the region representing the PLBR spacetime is the
whole $(\eta,\chi)$ coordinate space except the $\eta$-axis, but $T + R$
and $T - R$ must belong to the range $(-B,B)$ of the generator function
in equation \eqref{e_339} with $k = -1$, which gives the region
\begin{equation} \label{e_1165}
|T| + |R| < B
\mbox{\hspace{2mm} , \hspace{3mm}}
R \ne 0
\mbox{ .}
\end{equation}
%
Hence in this case the $(\eta,\chi)$-system does not cover the whole
PLBR spacetime, but the constant $B$ secures the possibility of choosing
the region given in \eqref{e_1165} to be arbitrarily large.
The WLBR spacetime for $k = -1$ is given by
\begin{equation} \label{e_1555}
\chi > {\chi}_{\mbox{\tiny $Q$}}
+ \mbox{arcsinh} \left( \sinh {\chi}_{\mbox{\tiny $Q$}} \cosh \eta \right)
\mbox{\hspace{2mm} , \hspace{3mm}}
- \infty < \eta < \infty
\mbox{ .}
\end{equation}
\itm The world lines of fixed particles $\chi = {\chi}_1$ in the
$(\eta,\chi)$-system as described in the CFS system is found from
equations \eqref{e_346}, which gives
\begin{equation} \label{e_347}
(R - R_1)^2 - T^2 = R_1^2 + k B^2
\mbox{\hspace{2mm} , \hspace{3mm}}
R_1 = - k B I_k({\chi}_1)
\mbox{ .}
\end{equation}
when $k = 1$ and $k = -1$. For $k = 0$ we get $R = {\chi}_1$.
The corresponding simultaneity curves $\eta = {\eta}_1$ are given by
\begin{equation} \label{e_947}
(T - T_1)^2 - R^2 = T_1^2 + k B^2
\mbox{\hspace{2mm} , \hspace{3mm}}
T_1 = - k B I_k({\eta}_1)
\mbox{ .}
\end{equation}
when $k = 1$ and $k = -1$. For $k = 0$ we get $T = {\eta}_1$.
Note that the $(\eta,\chi)$-coordinates and the CFS coordinates are
comoving in the same reference frame when $k = 0$.
Using the transformation \eqref{e_346} the line element \eqref{e_218}
is given the form \eqref{e_157}. This shows that the solution
\eqref{e_219} represents the LBR spacetime for all values of $k$.
\itm With the line element \eqref{e_218} the coordinate acceleration of
a free particle instantaneously at rest is
\begin{equation} \label{e_342}
a^{\chi} = \ddot{\chi}
= - {\Gamma}^{\hs{0.3mm} \chi}_{\hs{1.5mm} \eta \eta}
\hs{0.8mm} \dot{\eta}^2
= - |g_{\hs{0.3mm} \eta \eta}|^{-1} \hs{0.5mm}
{\Gamma}^{\hs{0.3mm} \chi}_{\hs{1.5mm} \eta \eta}
\mbox{ ,}
\end{equation}
since $\dot{\eta} = |g_{\hs{0.3mm} \eta \eta}|^{-1/2}$ for such a
particle. Calculating the Christoffel symbol
${\Gamma}^{\hs{0.3mm} \chi}_{\hs{1.5mm} \eta \eta}$ from the line
element \eqref{e_218} we obtain
\begin{equation} \label{e_2004}
{\Gamma}^{\hs{0.3mm} \chi}_{\hs{1.5mm} \eta \eta} = -I_k(\chi)
\mbox{ .}
\end{equation}
The acceleration of gravity in the $(\eta,\chi)$-system is
defined as the component of $a^{\chi} {\bf e}_{\chi}$ along
the unit basis vector ${\bf e}_{\hat{\chi}}$, giving
%
\begin{equation} \label{e_360}
a^{\hat{\chi}} = \sqrt{g_{\hs{0.3mm} \chi \chi}} \hs{0.8mm} a^{\chi}
= C_k(\chi) / R_{\mbox{\tiny $Q$}}
\mbox{ .}
\end{equation}
Hence for $k = 1$ the acceleration of gravity in the $(\eta,\chi)$-system
is $a^{\hat{\chi}} = (1 / R_{\mbox{\tiny $Q$}}) \cos \chi$ so that
$a^{\hat{\chi}} > 0$ for $0 < \chi < \pi / 2$ and $a^{\hat{\chi}} < 0$
for $\pi / 2 < \chi < \pi$.
This is different from the situation in the CFS system, where the
acceleration of gravity is directed away from the domain wall everywhere
according to equation \eqref{e_359}. However, in the $(\eta,\chi)$-system
there is a region $\pi / 2 < \chi < \pi$ where the acceleration of gravity
is directed towards the domain wall. This is due to the motion of the
reference frame in which $(\eta,\chi)$ are comoving coordinates, as will be
explained below. The charged domain wall is at rest in the $(T,R)$-system.
Hence the CFS coordinates are those of a static, but not inertial,
reference frame. In this case the world lines are given by equation
\eqref{e_347} with $k = 1$ which represents the hyperbolae shown
in Figure 2.
\itm From this figure it seems that the $(\eta,\chi)$-system covers only
a part of the WLBR spacetime. The worldlines of fixed particles in
$(\eta,\chi)$-system are hyperbolae which never enter the future
region above the asymptotes. This is however not the case because
$R_1$ can have different values depending on ${\chi}_1$. If $R_1$
is moved to the left towards $R = 0$ the hyperbolae are straightened
out. Hence the $(\eta,\chi)$-system covers all of the WLBR spacetime.
\vspace{2mm} \newline
\begin{picture}(50,222)(34,-210)
\qbezier(176.3265, -99.7551)(176.3265, -98.7117)(176.4145, -97.6646)
\qbezier(176.3265, -99.7551)(176.3265, -100.7985)(176.4145, -101.8456)
\qbezier(176.4145, -97.6646)(176.5025, -96.6174)(176.6790, -95.5592)
\qbezier(176.4145, -101.8456)(176.5025, -102.8928)(176.6790, -103.9510)
\qbezier(176.6790, -95.5592)(176.8556, -94.5009)(177.1220, -93.4240)
\qbezier(176.6790, -103.9510)(176.8556, -105.0093)(177.1220, -106.0862)
\qbezier(177.1220, -93.4240)(177.3884, -92.3471)(177.7466, -91.2439)
\qbezier(177.1220, -106.0862)(177.3884, -107.1631)(177.7466, -108.2663)
\qbezier(177.7466, -91.2439)(178.1047, -90.1408)(178.5572, -89.0035)
\qbezier(177.7466, -108.2663)(178.1047, -109.3694)(178.5572, -110.5067)
\qbezier(178.5572, -89.0035)(179.0096, -87.8662)(179.5595, -86.6867)
\qbezier(178.5572, -110.5067)(179.0096, -111.6440)(179.5595, -112.8235)
\qbezier(179.5595, -86.6867)(180.1094, -85.5073)(180.7608, -84.2772)
\qbezier(179.5595, -112.8235)(180.1094, -114.0029)(180.7608, -115.2330)
\qbezier(180.7608, -84.2772)(181.4121, -83.0472)(182.1694, -81.7579)
\qbezier(180.7608, -115.2330)(181.4121, -116.4630)(182.1694, -117.7523)
\qbezier(182.1694, -81.7579)(182.9267, -80.4686)(183.7954, -79.1109)
\qbezier(182.1694, -117.7523)(182.9267, -119.0416)(183.7954, -120.3993)
\qbezier(183.7954, -79.1109)(184.6642, -77.7532)(185.6504, -76.3174)
\qbezier(183.7954, -120.3993)(184.6642, -121.7570)(185.6504, -123.1928)
\qbezier(185.6504, -76.3174)(186.6367, -74.8817)(187.7475, -73.3576)
\qbezier(185.6504, -123.1928)(186.6367, -124.6286)(187.7475, -126.1526)
\qbezier(187.7475, -73.3576)(188.8583, -71.8336)(190.1016, -70.2105)
\qbezier(187.7475, -126.1526)(188.8583, -127.6766)(190.1016, -129.2997)
\qbezier(190.1016, -70.2105)(191.3448, -68.5874)(192.7293, -66.8538)
\qbezier(190.1016, -129.2997)(191.3448, -130.9228)(192.7293, -132.6564)
\qbezier(192.7293, -66.8538)(194.1138, -65.1201)(195.6494, -63.2636)
\qbezier(192.7293, -132.6564)(194.1138, -134.3901)(195.6494, -136.2466)
\qbezier(195.6494, -63.2636)(197.1850, -61.4071)(198.8825, -59.4145)
\qbezier(195.6494, -136.2466)(197.1850, -138.1031)(198.8825, -140.0957)
\qbezier(198.8825, -59.4145)(200.5801, -57.4219)(202.4516, -55.2791)
\qbezier(198.8825, -140.0957)(200.5801, -142.0883)(202.4516, -144.2311)
\qbezier(202.4516, -55.2791)(204.3232, -53.1363)(206.3821, -50.8282)
\qbezier(202.4516, -144.2311)(204.3232, -146.3739)(206.3821, -148.6820)
\qbezier(206.3821, -50.8282)(208.4409, -48.5200)(210.7017, -46.0300)
\qbezier(206.3821, -148.6820)(208.4409, -150.9902)(210.7017, -153.4802)
\qbezier(210.7017, -46.0300)(212.9625, -43.5401)(215.4412, -40.8507)
\qbezier(210.7017, -153.4802)(212.9625, -155.9701)(215.4412, -158.6595)
\qbezier(126.7347, -99.7551)(126.7347, -100.7985)(126.6467, -101.8456)
\qbezier(126.7347, -99.7551)(126.7347, -98.7117)(126.6467, -97.6646)
\qbezier(126.6467, -101.8456)(126.5588, -102.8928)(126.3822, -103.9510)
\qbezier(126.6467, -97.6646)(126.5588, -96.6174)(126.3822, -95.5592)
\qbezier(126.3822, -103.9510)(126.2056, -105.0093)(125.9392, -106.0862)
\qbezier(126.3822, -95.5592)(126.2056, -94.5009)(125.9392, -93.4240)
\qbezier(125.9392, -106.0862)(125.6728, -107.1631)(125.3146, -108.2663)
\qbezier(125.9392, -93.4240)(125.6728, -92.3471)(125.3146, -91.2439)
\qbezier(125.3146, -108.2663)(124.9565, -109.3694)(124.5041, -110.5067)
\qbezier(125.3146, -91.2439)(124.9565, -90.1408)(124.5041, -89.0035)
\qbezier(124.5041, -110.5067)(124.0516, -111.6440)(123.5017, -112.8235)
\qbezier(124.5041, -89.0035)(124.0516, -87.8662)(123.5017, -86.6867)
\qbezier(123.5017, -112.8235)(122.9518, -114.0029)(122.3005, -115.2330)
\qbezier(123.5017, -86.6867)(122.9518, -85.5073)(122.3005, -84.2772)
\qbezier(122.3005, -115.2330)(121.6491, -116.4630)(120.8918, -117.7523)
\qbezier(122.3005, -84.2772)(121.6491, -83.0472)(120.8918, -81.7579)
\qbezier(120.8918, -117.7523)(120.1345, -119.0416)(119.2658, -120.3993)
\qbezier(120.8918, -81.7579)(120.1345, -80.4686)(119.2658, -79.1109)
\qbezier(119.2658, -120.3993)(118.3971, -121.7570)(117.4108, -123.1928)
\qbezier(119.2658, -79.1109)(118.3971, -77.7532)(117.4108, -76.3174)
\qbezier(117.4108, -123.1928)(116.4245, -124.6286)(115.3137, -126.1526)
\qbezier(117.4108, -76.3174)(116.4245, -74.8817)(115.3137, -73.3576)
\qbezier(115.3137, -126.1526)(114.2029, -127.6766)(112.9596, -129.2997)
\qbezier(115.3137, -73.3576)(114.2029, -71.8336)(112.9596, -70.2105)
\qbezier(112.9596, -129.2997)(111.7164, -130.9228)(110.3319, -132.6564)
\qbezier(112.9596, -70.2105)(111.7164, -68.5874)(110.3319, -66.8538)
\qbezier(110.3319, -132.6564)(108.9474, -134.3901)(107.4118, -136.2466)
\qbezier(110.3319, -66.8538)(108.9474, -65.1201)(107.4118, -63.2636)
\qbezier(107.4118, -136.2466)(105.8762, -138.1031)(104.1787, -140.0957)
\qbezier(107.4118, -63.2636)(105.8762, -61.4071)(104.1787, -59.4145)
\qbezier(104.1787, -140.0957)(102.4811, -142.0883)(100.6096, -144.2311)
\qbezier(104.1787, -59.4145)(102.4811, -57.4219)(100.6096, -55.2791)
\qbezier(100.6096, -144.2311)( 98.7380, -146.3739)( 96.6791, -148.6820)
\qbezier(100.6096, -55.2791)( 98.7380, -53.1363)( 96.6791, -50.8282)
\qbezier( 96.6791, -148.6820)( 94.6203, -150.9902)( 92.3595, -153.4802)
\qbezier( 96.6791, -50.8282)( 94.6203, -48.5200)( 92.3595, -46.0300)
\qbezier( 92.3595, -153.4802)( 90.0987, -155.9701)( 87.6200, -158.6595)
\qbezier( 92.3595, -46.0300)( 90.0987, -43.5401)( 87.6200, -40.8507)
\qbezier( 93.6735, -157.6122)( 95.0846, -156.2011)( 96.4958, -154.7899)
\qbezier( 99.3181, -151.9676)(100.7292, -150.5565)(102.1404, -149.1453)
\qbezier(104.9627, -146.3230)(106.3738, -144.9119)(107.7850, -143.5007)
\qbezier(110.6073, -140.6784)(112.0184, -139.2673)(113.4296, -137.8561)
\qbezier(116.2519, -135.0338)(117.6630, -133.6227)(119.0742, -132.2115)
\qbezier(121.8965, -129.3892)(123.3076, -127.9781)(124.7188, -126.5669)
\qbezier(127.5411, -123.7446)(128.9522, -122.3335)(130.3634, -120.9223)
\qbezier(133.1857, -118.1000)(134.5968, -116.6889)(136.0080, -115.2778)
\qbezier(138.8303, -112.4555)(140.2414, -111.0443)(141.6526, -109.6332)
\qbezier(144.4749, -106.8109)(145.8860, -105.3997)(147.2972, -103.9886)
\qbezier(150.1195, -101.1663)(151.5306, -99.7551)(152.9418, -98.3440)
\qbezier(155.7641, -95.5217)(157.1752, -94.1105)(158.5864, -92.6994)
\qbezier(161.4087, -89.8771)(162.8198, -88.4659)(164.2310, -87.0548)
\qbezier(167.0533, -84.2325)(168.4644, -82.8213)(169.8756, -81.4102)
\qbezier(172.6979, -78.5879)(174.1090, -77.1767)(175.5202, -75.7656)
\qbezier(178.3425, -72.9433)(179.7536, -71.5321)(181.1648, -70.1210)
\qbezier(183.9871, -67.2987)(185.3982, -65.8875)(186.8094, -64.4764)
\qbezier(189.6317, -61.6541)(191.0428, -60.2429)(192.4540, -58.8318)
\qbezier(195.2763, -56.0095)(196.6874, -54.5983)(198.0986, -53.1872)
\qbezier(200.9209, -50.3649)(202.3320, -48.9537)(203.7432, -47.5426)
\qbezier(206.5655, -44.7203)(207.9766, -43.3091)(209.3878, -41.8980)
\qbezier( 93.6735, -41.8980)( 95.0846, -43.3091)( 96.4958, -44.7203)
\qbezier( 99.3181, -47.5426)(100.7292, -48.9537)(102.1404, -50.3649)
\qbezier(104.9627, -53.1872)(106.3738, -54.5983)(107.7850, -56.0095)
\qbezier(110.6073, -58.8318)(112.0184, -60.2429)(113.4296, -61.6541)
\qbezier(116.2519, -64.4764)(117.6630, -65.8875)(119.0742, -67.2987)
\qbezier(121.8965, -70.1210)(123.3076, -71.5321)(124.7188, -72.9433)
\qbezier(127.5411, -75.7656)(128.9522, -77.1767)(130.3634, -78.5879)
\qbezier(133.1857, -81.4102)(134.5968, -82.8213)(136.0080, -84.2325)
\qbezier(138.8303, -87.0548)(140.2414, -88.4659)(141.6526, -89.8771)
\qbezier(144.4749, -92.6994)(145.8860, -94.1105)(147.2972, -95.5217)
\qbezier(150.1195, -98.3440)(151.5306, -99.7551)(152.9418, -101.1663)
\qbezier(155.7641, -103.9886)(157.1752, -105.3997)(158.5864, -106.8109)
\qbezier(161.4087, -109.6332)(162.8198, -111.0443)(164.2310, -112.4555)
\qbezier(167.0533, -115.2778)(168.4644, -116.6889)(169.8756, -118.1000)
\qbezier(172.6979, -120.9223)(174.1090, -122.3335)(175.5202, -123.7446)
\qbezier(178.3425, -126.5669)(179.7536, -127.9781)(181.1648, -129.3892)
\qbezier(183.9871, -132.2115)(185.3982, -133.6227)(186.8094, -135.0338)
\qbezier(189.6317, -137.8561)(191.0428, -139.2673)(192.4540, -140.6784)
\qbezier(195.2763, -143.5007)(196.6874, -144.9119)(198.0986, -146.3230)
\qbezier(200.9209, -149.1453)(202.3320, -150.5565)(203.7432, -151.9676)
\qbezier(206.5655, -154.7899)(207.9766, -156.2011)(209.3878, -157.6122)
\put( 52.3469, -99.7551){\vector(1, 0){165.3061}}
\put(135.0000, -182.4082){\vector(0, 1){165.3061}}
\put(225.9184, -103.8878){\makebox(0,0)[]{\footnotesize{$R$}}}
\put(128.1122, -15.7245){\makebox(0,0)[]{\footnotesize{$T$}}}
\put(152.9082, -113.5306){\makebox(0,0)[]{\footnotesize{$R_1$}}}
\put(188.7245, -108.0204){\makebox(0,0)[]{\footnotesize{$R_+$}}}
\put(117.0918, -108.0204){\makebox(0,0)[]{\footnotesize{$R_-$}}}
\put(236.9388, -40.5204){\makebox(0,0)[]{\footnotesize{$\chi = {\chi}_1$}}}
\put( 68.3265, -40.5204){\makebox(0,0)[]{\footnotesize{$\chi = {\chi}_1$}}}
\end{picture}
\begin{picture}(50,222)(-146,-210)
\qbezier(135.0000, -58.4286)(133.9566, -58.4286)(132.9095, -58.3406)
\qbezier(135.0000, -58.4286)(136.0434, -58.4286)(137.0905, -58.3406)
\qbezier(132.9095, -58.3406)(131.8623, -58.2526)(130.8041, -58.0761)
\qbezier(137.0905, -58.3406)(138.1377, -58.2526)(139.1959, -58.0761)
\qbezier(130.8041, -58.0761)(129.7458, -57.8995)(128.6689, -57.6331)
\qbezier(139.1959, -58.0761)(140.2542, -57.8995)(141.3311, -57.6331)
\qbezier(128.6689, -57.6331)(127.5920, -57.3667)(126.4888, -57.0085)
\qbezier(141.3311, -57.6331)(142.4080, -57.3667)(143.5112, -57.0085)
\qbezier(126.4888, -57.0085)(125.3857, -56.6504)(124.2484, -56.1979)
\qbezier(143.5112, -57.0085)(144.6143, -56.6504)(145.7516, -56.1979)
\qbezier(124.2484, -56.1979)(123.1111, -55.7455)(121.9316, -55.1956)
\qbezier(145.7516, -56.1979)(146.8889, -55.7455)(148.0684, -55.1956)
\qbezier(121.9316, -55.1956)(120.7522, -54.6457)(119.5221, -53.9943)
\qbezier(148.0684, -55.1956)(149.2478, -54.6457)(150.4779, -53.9943)
\qbezier(119.5221, -53.9943)(118.2921, -53.3430)(117.0028, -52.5857)
\qbezier(150.4779, -53.9943)(151.7079, -53.3430)(152.9972, -52.5857)
\qbezier(117.0028, -52.5857)(115.7135, -51.8284)(114.3558, -50.9597)
\qbezier(152.9972, -52.5857)(154.2865, -51.8284)(155.6442, -50.9597)
\qbezier(114.3558, -50.9597)(112.9981, -50.0909)(111.5623, -49.1047)
\qbezier(155.6442, -50.9597)(157.0019, -50.0909)(158.4377, -49.1047)
\qbezier(111.5623, -49.1047)(110.1266, -48.1184)(108.6025, -47.0076)
\qbezier(158.4377, -49.1047)(159.8734, -48.1184)(161.3975, -47.0076)
\qbezier(108.6025, -47.0076)(107.0785, -45.8968)(105.4554, -44.6535)
\qbezier(161.3975, -47.0076)(162.9215, -45.8968)(164.5446, -44.6535)
\qbezier(105.4554, -44.6535)(103.8323, -43.4103)(102.0987, -42.0258)
\qbezier(164.5446, -44.6535)(166.1677, -43.4103)(167.9013, -42.0258)
\qbezier(102.0987, -42.0258)(100.3650, -40.6413)( 98.5085, -39.1057)
\qbezier(167.9013, -42.0258)(169.6350, -40.6413)(171.4915, -39.1057)
\qbezier( 98.5085, -39.1057)( 96.6520, -37.5701)( 94.6594, -35.8726)
\qbezier(171.4915, -39.1057)(173.3480, -37.5701)(175.3406, -35.8726)
\qbezier( 94.6594, -35.8726)( 92.6668, -34.1750)( 90.5240, -32.3035)
\qbezier(175.3406, -35.8726)(177.3332, -34.1750)(179.4760, -32.3035)
\qbezier( 90.5240, -32.3035)( 88.3812, -30.4319)( 86.0731, -28.3730)
\qbezier(179.4760, -32.3035)(181.6188, -30.4319)(183.9269, -28.3730)
\qbezier( 86.0731, -28.3730)( 83.7649, -26.3142)( 81.2749, -24.0534)
\qbezier(183.9269, -28.3730)(186.2351, -26.3142)(188.7251, -24.0534)
\qbezier( 81.2749, -24.0534)( 78.7850, -21.7926)( 76.0956, -19.3139)
\qbezier(188.7251, -24.0534)(191.2150, -21.7926)(193.9044, -19.3139)
\qbezier(135.0000, -108.0204)(136.0434, -108.0204)(137.0905, -108.1084)
\qbezier(135.0000, -108.0204)(133.9566, -108.0204)(132.9095, -108.1084)
\qbezier(137.0905, -108.1084)(138.1377, -108.1964)(139.1959, -108.3729)
\qbezier(132.9095, -108.1084)(131.8623, -108.1964)(130.8041, -108.3729)
\qbezier(139.1959, -108.3729)(140.2542, -108.5495)(141.3311, -108.8159)
\qbezier(130.8041, -108.3729)(129.7458, -108.5495)(128.6689, -108.8159)
\qbezier(141.3311, -108.8159)(142.4080, -109.0823)(143.5112, -109.4405)
\qbezier(128.6689, -108.8159)(127.5920, -109.0823)(126.4888, -109.4405)
\qbezier(143.5112, -109.4405)(144.6143, -109.7986)(145.7516, -110.2510)
\qbezier(126.4888, -109.4405)(125.3857, -109.7986)(124.2484, -110.2510)
\qbezier(145.7516, -110.2510)(146.8889, -110.7035)(148.0684, -111.2534)
\qbezier(124.2484, -110.2510)(123.1111, -110.7035)(121.9316, -111.2534)
\qbezier(148.0684, -111.2534)(149.2478, -111.8033)(150.4779, -112.4546)
\qbezier(121.9316, -111.2534)(120.7522, -111.8033)(119.5221, -112.4546)
\qbezier(150.4779, -112.4546)(151.7079, -113.1060)(152.9972, -113.8633)
\qbezier(119.5221, -112.4546)(118.2921, -113.1060)(117.0028, -113.8633)
\qbezier(152.9972, -113.8633)(154.2865, -114.6206)(155.6442, -115.4893)
\qbezier(117.0028, -113.8633)(115.7135, -114.6206)(114.3558, -115.4893)
\qbezier(155.6442, -115.4893)(157.0019, -116.3580)(158.4377, -117.3443)
\qbezier(114.3558, -115.4893)(112.9981, -116.3580)(111.5623, -117.3443)
\qbezier(158.4377, -117.3443)(159.8734, -118.3306)(161.3975, -119.4414)
\qbezier(111.5623, -117.3443)(110.1266, -118.3306)(108.6025, -119.4414)
\qbezier(161.3975, -119.4414)(162.9215, -120.5522)(164.5446, -121.7955)
\qbezier(108.6025, -119.4414)(107.0785, -120.5522)(105.4554, -121.7955)
\qbezier(164.5446, -121.7955)(166.1677, -123.0387)(167.9013, -124.4232)
\qbezier(105.4554, -121.7955)(103.8323, -123.0387)(102.0987, -124.4232)
\qbezier(167.9013, -124.4232)(169.6350, -125.8077)(171.4915, -127.3433)
\qbezier(102.0987, -124.4232)(100.3650, -125.8077)( 98.5085, -127.3433)
\qbezier(171.4915, -127.3433)(173.3480, -128.8789)(175.3406, -130.5764)
\qbezier( 98.5085, -127.3433)( 96.6520, -128.8789)( 94.6594, -130.5764)
\qbezier(175.3406, -130.5764)(177.3332, -132.2740)(179.4760, -134.1455)
\qbezier( 94.6594, -130.5764)( 92.6668, -132.2740)( 90.5240, -134.1455)
\qbezier(179.4760, -134.1455)(181.6188, -136.0171)(183.9269, -138.0760)
\qbezier( 90.5240, -134.1455)( 88.3812, -136.0171)( 86.0731, -138.0760)
\qbezier(183.9269, -138.0760)(186.2351, -140.1348)(188.7251, -142.3956)
\qbezier( 86.0731, -138.0760)( 83.7649, -140.1348)( 81.2749, -142.3956)
\qbezier(188.7251, -142.3956)(191.2150, -144.6564)(193.9044, -147.1351)
\qbezier( 81.2749, -142.3956)( 78.7850, -144.6564)( 76.0956, -147.1351)
\qbezier( 77.1429, -141.0816)( 78.5540, -139.6705)( 79.9652, -138.2593)
\qbezier( 82.7875, -135.4370)( 84.1986, -134.0259)( 85.6098, -132.6147)
\qbezier( 88.4321, -129.7924)( 89.8432, -128.3813)( 91.2544, -126.9701)
\qbezier( 94.0767, -124.1478)( 95.4878, -122.7367)( 96.8990, -121.3255)
\qbezier( 99.7213, -118.5032)(101.1324, -117.0921)(102.5436, -115.6809)
\qbezier(105.3659, -112.8586)(106.7770, -111.4475)(108.1882, -110.0363)
\qbezier(111.0105, -107.2140)(112.4216, -105.8029)(113.8328, -104.3917)
\qbezier(116.6551, -101.5694)(118.0662, -100.1583)(119.4774, -98.7471)
\qbezier(122.2997, -95.9248)(123.7108, -94.5137)(125.1220, -93.1025)
\qbezier(127.9443, -90.2802)(129.3554, -88.8691)(130.7666, -87.4579)
\qbezier(133.5889, -84.6356)(135.0000, -83.2245)(136.4111, -81.8133)
\qbezier(139.2334, -78.9910)(140.6446, -77.5799)(142.0557, -76.1687)
\qbezier(144.8780, -73.3464)(146.2892, -71.9353)(147.7003, -70.5241)
\qbezier(150.5226, -67.7018)(151.9338, -66.2907)(153.3449, -64.8795)
\qbezier(156.1672, -62.0572)(157.5784, -60.6461)(158.9895, -59.2349)
\qbezier(161.8118, -56.4126)(163.2230, -55.0015)(164.6341, -53.5903)
\qbezier(167.4564, -50.7680)(168.8676, -49.3569)(170.2787, -47.9457)
\qbezier(173.1010, -45.1234)(174.5122, -43.7123)(175.9233, -42.3011)
\qbezier(178.7456, -39.4788)(180.1568, -38.0677)(181.5679, -36.6565)
\qbezier(184.3902, -33.8342)(185.8014, -32.4231)(187.2125, -31.0119)
\qbezier(190.0348, -28.1896)(191.4460, -26.7785)(192.8571, -25.3673)
\qbezier( 77.1429, -25.3673)( 78.5540, -26.7785)( 79.9652, -28.1896)
\qbezier( 82.7875, -31.0119)( 84.1986, -32.4231)( 85.6098, -33.8342)
\qbezier( 88.4321, -36.6565)( 89.8432, -38.0677)( 91.2544, -39.4788)
\qbezier( 94.0767, -42.3011)( 95.4878, -43.7123)( 96.8990, -45.1234)
\qbezier( 99.7213, -47.9457)(101.1324, -49.3569)(102.5436, -50.7680)
\qbezier(105.3659, -53.5903)(106.7770, -55.0015)(108.1882, -56.4126)
\qbezier(111.0105, -59.2349)(112.4216, -60.6461)(113.8328, -62.0572)
\qbezier(116.6551, -64.8795)(118.0662, -66.2907)(119.4774, -67.7018)
\qbezier(122.2997, -70.5241)(123.7108, -71.9353)(125.1220, -73.3464)
\qbezier(127.9443, -76.1687)(129.3554, -77.5799)(130.7666, -78.9910)
\qbezier(133.5889, -81.8133)(135.0000, -83.2245)(136.4111, -84.6356)
\qbezier(139.2334, -87.4579)(140.6446, -88.8691)(142.0557, -90.2802)
\qbezier(144.8780, -93.1025)(146.2892, -94.5137)(147.7003, -95.9248)
\qbezier(150.5226, -98.7471)(151.9338, -100.1583)(153.3449, -101.5694)
\qbezier(156.1672, -104.3917)(157.5784, -105.8029)(158.9895, -107.2140)
\qbezier(161.8118, -110.0363)(163.2230, -111.4475)(164.6341, -112.8586)
\qbezier(167.4564, -115.6809)(168.8676, -117.0921)(170.2787, -118.5032)
\qbezier(173.1010, -121.3255)(174.5122, -122.7367)(175.9233, -124.1478)
\qbezier(178.7456, -126.9701)(180.1568, -128.3813)(181.5679, -129.7924)
\qbezier(184.3902, -132.6147)(185.8014, -134.0259)(187.2125, -135.4370)
\qbezier(190.0348, -138.2593)(191.4460, -139.6705)(192.8571, -141.0816)
\put( 52.3469, -99.7551){\vector(1, 0){165.3061}}
\put(135.0000, -182.4082){\vector(0, 1){165.3061}}
\put(225.9184, -103.8878){\makebox(0,0)[]{\footnotesize{$R$}}}
\put(128.1122, -15.7245){\makebox(0,0)[]{\footnotesize{$T$}}}
\put(121.2245, -84.6020){\makebox(0,0)[]{\footnotesize{$T_1$}}}
\put(126.7347, -48.7857){\makebox(0,0)[]{\footnotesize{$T_+$}}}
\put(126.7347, -120.4184){\makebox(0,0)[]{\footnotesize{$T_-$}}}
\put(214.8980, -16.5510){\makebox(0,0)[]{\footnotesize{$\eta = {\eta}_1$}}}
\put(214.8980, -153.4796){\makebox(0,0)[]{\footnotesize{$\eta = {\eta}_1$}}}
\end{picture}
\vspace*{-4mm} \newline
\hspace*{32.0mm} (a) \hspace{73.9mm} (b)
\vspace*{4mm} \newline
%
%
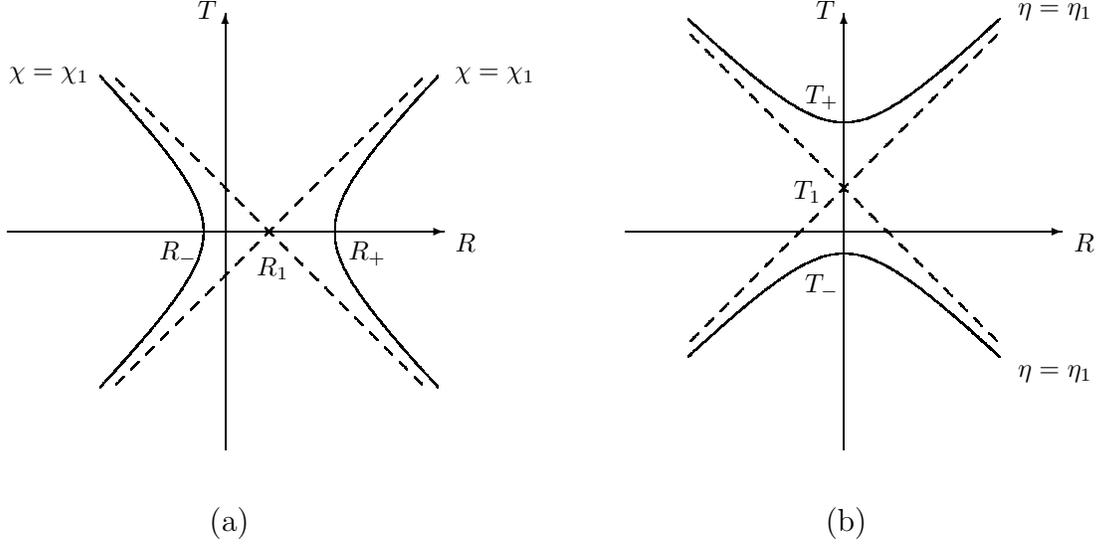
{\footnotesize \sf Figure 2. (a) The world lines of points with $\chi =
{\chi}_1$ as given by equation \eqref{e_347} with $k = 1$. Here
$R_{-} = R_1 - \sqrt{R_1^2 + B^2}$ and $R_{+} = R_1 + \sqrt{R_1^2 + B^2}$.
Note that $R_{+} > 0$. A particle in the WLBR spacetime can only follow
a hyperbola in the region to the right of $R = R_{\mbox{\tiny $Q$}}$.
However, this limitation does not exist in the PLBR spacetime.
(b) The simultaneity curves with $\eta = {\eta}_1$. Here
$T_{-} = T_1 - \sqrt{T_1^2 + B^2}$ and
$T_{+} = T_1 + \sqrt{T_1^2 + B^2}$.}
\vspace{3mm} \newline
\itm It is known from the description of the WLBR spacetime with CFS
coordinates, where the metric is static and the domain wall is at rest,
that there is repulsive gravitation outside the domain wall. Nevertheless
in the region $R > \sqrt{B^2 + T^2}$ in Figure 3
the acceleration of gravity is directed towards the wall in the
$(\eta,\chi)$ coordinate system. This apparent contradiction will be
explained by comparing the acceleration of fixed points in the
$(\eta,\chi)$-system with the acceleration of free particles, both
measured relative to the CFS coordinate system.
\itm The world line of a particle at rest in the $(\eta,\chi)$-system
is given by equation \eqref{e_347}. From this it follows that the
velocity and the acceleration of the particle in the CFS system are
\begin{equation} \label{e_544}
\left( \frl{dR}{dT} \right)_{\chi = {\chi}_1} =
\frl{T_2}{R_2 \m R_1}
\mbox{\hspace{2mm} , \hspace{3mm}}
\left( \frl{d^2 R}{dT^2} \right)_{\chi = {\chi}_1} =
\frl{R_1^2 \p B^2}{(R_2 \m R_1)^3}
\end{equation}
at an arbitrary point $(T_2,R_2)$.
\itm We now consider a free particle with Lagrangian function
\begin{equation} \label{e_545}
L = \frl{R_{\mbox{\tiny $Q$}}^2}{2 R^2} (- \dot{T}^2 + \dot{R}^2)
\mbox{ ,}
\end{equation}
where the dot denotes differentiation with respect to the proper time
of the particle. Since the metric is static,
\begin{equation} \label{e_546}
p_T = \frl{\partial L}{\partial \dot{T}}
= - \frl{R_{\mbox{\tiny $Q$}}^2}{R^2} \frl{dT}{d \tau}
\end{equation}
is a constant of motion. Together with the four-velocity identity
\begin{equation} \label{e_547}
\frl{R_{\mbox{\tiny $Q$}}^2}{R^2} (- \dot{T}^2 + \dot{R}^2) = -1
\end{equation}
this leads to (for a particle moving outwards)
\begin{equation} \label{e_548}
\left( \frl{dR}{dT} \right)_{F} = \frl{\dot{R}}{\dot{T}}
= \frl{1}{R} \sqrt{R^2
- \left( \frl{R_{\mbox{\tiny $Q$}}}{p_T} \right)^2}
\mbox{ ,}
\end{equation}
where $F$ means that the particle is free.
\vspace*{4mm} \newline
\begin{picture}(50,222)(-76,-210)
\qbezier(172.9687, -97.0000)(172.9687, -95.6018)(173.0719, -94.1998)
\qbezier(172.9687, -97.0000)(172.9687, -98.3982)(173.0719, -99.8002)
\qbezier(173.0719, -94.1998)(173.1750, -92.7978)(173.3818, -91.3843)
\qbezier(173.0719, -99.8002)(173.1750, -101.2022)(173.3818, -102.6157)
\qbezier(173.3818, -91.3843)(173.5886, -89.9709)(173.9002, -88.5384)
\qbezier(173.3818, -102.6157)(173.5886, -104.0291)(173.9002, -105.4616)
\qbezier(173.9002, -88.5384)(174.2118, -87.1059)(174.6299, -85.6465)
\qbezier(173.9002, -105.4616)(174.2118, -106.8941)(174.6299, -108.3535)
\qbezier(174.6299, -85.6465)(175.0480, -84.1871)(175.5749, -82.6929)
\qbezier(174.6299, -108.3535)(175.0480, -109.8129)(175.5749, -111.3071)
\qbezier(175.5749, -82.6929)(176.1017, -81.1987)(176.7402, -79.6616)
\qbezier(175.5749, -111.3071)(176.1017, -112.8013)(176.7402, -114.3384)
\qbezier(176.7402, -79.6616)(177.3787, -78.1245)(178.1323, -76.5361)
\qbezier(176.7402, -114.3384)(177.3787, -115.8755)(178.1323, -117.4639)
\qbezier(178.1323, -76.5361)(178.8859, -74.9478)(179.7587, -73.2995)
\qbezier(178.1323, -117.4639)(178.8859, -119.0522)(179.7587, -120.7005)
\qbezier(179.7587, -73.2995)(180.6315, -71.6513)(181.6282, -69.9342)
\qbezier(179.7587, -120.7005)(180.6315, -122.3487)(181.6282, -124.0658)
\qbezier(181.6282, -69.9342)(182.6249, -68.2171)(183.7509, -66.4218)
\qbezier(181.6282, -124.0658)(182.6249, -125.7829)(183.7509, -127.5782)
\qbezier(183.7509, -66.4218)(184.8770, -64.6265)(186.1385, -62.7433)
\qbezier(183.7509, -127.5782)(184.8770, -129.3735)(186.1385, -131.2567)
\qbezier(186.1385, -62.7433)(187.4000, -60.8601)(188.8039, -58.8788)
\qbezier(186.1385, -131.2567)(187.4000, -133.1399)(188.8039, -135.1212)
\qbezier(188.8039, -58.8788)(190.2077, -56.8974)(191.7615, -54.8071)
\qbezier(188.8039, -135.1212)(190.2077, -137.1026)(191.7615, -139.1929)
\qbezier(191.7615, -54.8071)(193.3152, -52.7169)(195.0274, -50.5063)
\qbezier(191.7615, -139.1929)(193.3152, -141.2831)(195.0274, -143.4937)
\qbezier(195.0274, -50.5063)(196.7395, -48.2958)(198.6194, -45.9530)
\qbezier(195.0274, -143.4937)(196.7395, -145.7042)(198.6194, -148.0470)
\qbezier(198.6194, -45.9530)(200.4992, -43.6102)(202.5569, -41.1224)
\qbezier(198.6194, -148.0470)(200.4992, -150.3898)(202.5569, -152.8776)
\qbezier(202.5569, -41.1224)(204.6146, -38.6346)(206.8614, -35.9882)
\qbezier(202.5569, -152.8776)(204.6146, -155.3654)(206.8614, -158.0118)
\qbezier(206.8614, -35.9882)(209.1082, -33.3419)(211.5562, -30.5227)
\qbezier(206.8614, -158.0118)(209.1082, -160.6581)(211.5562, -163.4773)
\qbezier(211.5562, -30.5227)(214.0043, -27.7035)(216.6669, -24.6961)
\qbezier(211.5562, -163.4773)(214.0043, -166.2965)(216.6669, -169.3039)
\qbezier(172.9687, -97.0000)(172.9687, -60.8480)(172.9687, -24.6961)
\qbezier(172.9687, -97.0000)(172.9687, -133.1520)(172.9687, -169.3039)
\qbezier(176.9413, -79.1829)(176.9413, -51.9395)(176.9413, -24.6961)
\qbezier(176.9413, -114.8171)(176.9413, -142.0605)(176.9413, -169.3039)
\qbezier(180.9139, -71.1842)(180.9139, -47.9401)(180.9139, -24.6961)
\qbezier(180.9139, -122.8158)(180.9139, -146.0599)(180.9139, -169.3039)
\qbezier(184.8864, -64.6422)(184.8864, -44.6691)(184.8864, -24.6961)
\qbezier(184.8864, -129.3578)(184.8864, -149.3309)(184.8864, -169.3039)
\qbezier(188.8590, -58.8010)(188.8590, -41.7485)(188.8590, -24.6961)
\qbezier(188.8590, -135.1990)(188.8590, -152.2515)(188.8590, -169.3039)
\qbezier(192.8316, -53.3782)(192.8316, -39.0371)(192.8316, -24.6961)
\qbezier(192.8316, -140.6218)(192.8316, -154.9629)(192.8316, -169.3039)
\qbezier(196.8041, -48.2340)(196.8041, -36.4650)(196.8041, -24.6961)
\qbezier(196.8041, -145.7660)(196.8041, -157.5350)(196.8041, -169.3039)
\qbezier(200.7767, -43.2883)(200.7767, -33.9922)(200.7767, -24.6961)
\qbezier(200.7767, -150.7117)(200.7767, -160.0078)(200.7767, -169.3039)
\qbezier(204.7492, -38.4908)(204.7492, -31.5934)(204.7492, -24.6961)
\qbezier(204.7492, -155.5092)(204.7492, -162.4066)(204.7492, -169.3039)
\qbezier(208.7218, -33.8076)(208.7218, -29.2518)(208.7218, -24.6961)
\qbezier(208.7218, -160.1924)(208.7218, -164.7482)(208.7218, -169.3039)
\qbezier(212.6944, -29.2151)(212.6944, -26.9556)(212.6944, -24.6961)
\qbezier(212.6944, -164.7849)(212.6944, -167.0444)(212.6944, -169.3039)
\qbezier(168.9962, -169.3039)(168.9962, -97.0000)(168.9962, -24.6961)
\qbezier(165.0236, -169.3039)(165.0236, -97.0000)(165.0236, -24.6961)
\qbezier(161.0511, -169.3039)(161.0511, -97.0000)(161.0511, -24.6961)
\thicklines\qbezier(157.0785, -169.3039)(157.0785, -97.0000)(157.0785, -24.6961)
\thinlines\qbezier(172.9687, -97.0000)(194.8178, -97.0000)(216.6669, -97.0000)
\qbezier(172.9687, -97.0000)(194.8178, -97.0000)(216.6669, -97.0000)
\qbezier(173.2062, -92.7468)(194.9366, -92.7468)(216.6669, -92.7468)
\qbezier(173.2062, -101.2532)(194.9366, -101.2532)(216.6669, -101.2532)
\qbezier(173.9099, -88.4937)(195.2884, -88.4937)(216.6669, -88.4937)
\qbezier(173.9099, -105.5063)(195.2884, -105.5063)(216.6669, -105.5063)
\qbezier(175.0554, -84.2405)(195.8611, -84.2405)(216.6669, -84.2405)
\qbezier(175.0554, -109.7595)(195.8611, -109.7595)(216.6669, -109.7595)
\qbezier(176.6060, -79.9873)(196.6365, -79.9873)(216.6669, -79.9873)
\qbezier(176.6060, -114.0127)(196.6365, -114.0127)(216.6669, -114.0127)
\qbezier(178.5185, -75.7341)(197.5927, -75.7341)(216.6669, -75.7341)
\qbezier(178.5185, -118.2659)(197.5927, -118.2659)(216.6669, -118.2659)
\qbezier(180.7476, -71.4810)(198.7073, -71.4810)(216.6669, -71.4810)
\qbezier(180.7476, -122.5190)(198.7073, -122.5190)(216.6669, -122.5190)
\qbezier(183.2495, -67.2278)(199.9582, -67.2278)(216.6669, -67.2278)
\qbezier(183.2495, -126.7722)(199.9582, -126.7722)(216.6669, -126.7722)
\qbezier(185.9838, -62.9746)(201.3254, -62.9746)(216.6669, -62.9746)
\qbezier(185.9838, -131.0254)(201.3254, -131.0254)(216.6669, -131.0254)
\qbezier(188.9154, -58.7214)(202.7912, -58.7214)(216.6669, -58.7214)
\qbezier(188.9154, -135.2786)(202.7912, -135.2786)(216.6669, -135.2786)
\qbezier(192.0138, -54.4683)(204.3404, -54.4683)(216.6669, -54.4683)
\qbezier(192.0138, -139.5317)(204.3404, -139.5317)(216.6669, -139.5317)
\qbezier(195.2532, -50.2151)(205.9601, -50.2151)(216.6669, -50.2151)
\qbezier(195.2532, -143.7849)(205.9601, -143.7849)(216.6669, -143.7849)
\qbezier(198.6122, -45.9619)(207.6396, -45.9619)(216.6669, -45.9619)
\qbezier(198.6122, -148.0381)(207.6396, -148.0381)(216.6669, -148.0381)
\qbezier(202.0727, -41.7087)(209.3698, -41.7087)(216.6669, -41.7087)
\qbezier(202.0727, -152.2913)(209.3698, -152.2913)(216.6669, -152.2913)
\qbezier(205.6199, -37.4556)(211.1434, -37.4556)(216.6669, -37.4556)
\qbezier(205.6199, -156.5444)(211.1434, -156.5444)(216.6669, -156.5444)
\qbezier(209.2412, -33.2024)(212.9541, -33.2024)(216.6669, -33.2024)
\qbezier(209.2412, -160.7976)(212.9541, -160.7976)(216.6669, -160.7976)
\qbezier(212.9265, -28.9492)(214.7967, -28.9492)(216.6669, -28.9492)
\qbezier(212.9265, -165.0508)(214.7967, -165.0508)(216.6669, -165.0508)
\qbezier(172.9687, -24.6961)(172.9687, -97.0000)(172.9687, -169.3039)
\put( 65.3906, -97.0000){\vector(1, 0){196.1719}}
\put(135.0000, -191.9219){\vector(0, 1){189.8437}}
\put(274.2187, -103.3281){\makebox(0,0)[]{\footnotesize{$R$}}}
\put(124.4531,   0.0312){\makebox(0,0)[]{\footnotesize{$T$}}}
\put(240.4687, -20.0078){\makebox(0,0)[]{\footnotesize{$\chi = \fr{\pi}{2}$}}}
\put(148.6410, -109.6563){\makebox(0,0)[]{\footnotesize{$R_{\mbox{\tiny $Q$}}$}}}
\end{picture}
\vspace{0mm} \newline
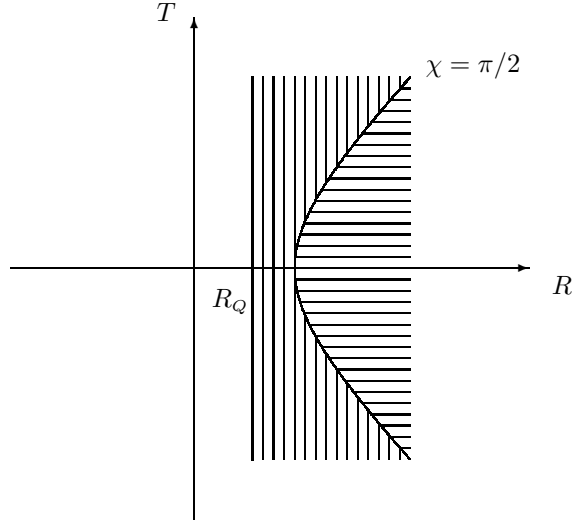
{\footnotesize \sf Figure 3. The WLBR spacetime in the $(T,R)$-system.
The hyperbola $R^2 - T^2 = B^2$ corresponding to $\chi = \pi/2$ separates
the WLBR spacetime in two regions. To the right of this hyperbola an
observer at rest in the $(\eta,\chi)$-system experiences an acceleration
of gravity directed towards the domain wall, and to the left away from
the domain wall. The reason for this is explained in the text.}
\vspace{2mm} \newline
\itm We now demand that the free particle passes through the point $(T_2,R_2)$
with the same velocity as the particle with $\chi = {\chi}_1$. Using
the equations \eqref{e_548} and \eqref{e_347} we obtain
\begin{equation} \label{e_549}
p_T = \frl{R_{\mbox{\tiny $Q$}} (R_2 \m R_1)}
{R_2 \sqrt{R_1^2 \p B^2}}
\mbox{ .}
\end{equation}
Integrating equation \eqref{e_548} we find the equation for the world
line of the particle,
\begin{equation} \label{e_550}
R^2 - (T - T_0)^2 = R_0^2
\mbox{ ,}
\end{equation}
where
\begin{equation} \label{e_559}
R_0 = \frl{R_{\mbox{\tiny $Q$}}}{p_T}
= \frl{R_2 \sqrt{R_1^2 \p B^2}}{R_2 \m R_1}
\mbox{ ,}
\end{equation}
\vspace{-2mm} \newline
and $T_0$ is a constant of integration. With the boundary condition
$R(T_2) = R_2$ it follows that
\begin{equation} \label{e_551}
T_0 = - \frl{R_1 T_2}{R_2 \m R_1}
\mbox{ .}
\end{equation}
Note that $(dR/dT)_F = 0$ for $R = R_0$. Hence the particle falls from
rest at $R = R_0$ at the point of time $T = T_0$. The fact that $T_0$
depends upon $R_1$, i.e.\hspace{1.2mm}on ${\chi}_1$, means that
different reference particles in the $(\eta,\chi)$-system are
instantaneously at rest relative to the CFS system at different points
of time. Differentiating we find that the acceleration of the free
particle at $R = R_2$ is
\begin{equation} \label{e_552}
\left( \frl{d^2 R}{dT^2} \right)_{F} =
\frl{R_1^2 \p B^2}{R_2 (R_2 \m R_1)^2}
\mbox{ .}
\end{equation}
From equations \eqref{e_544} and \eqref{e_552} it follows that the ratio
between the acceleration of a fixed particle in the
$(\eta,\chi)$-system and a free particle is
\begin{equation} \label{e_553}
N = \frl{(\fr{d^2 R}{dT^2})_{\chi = {\chi}_1}}
{(\fr{d^2 R}{dT^2})_{F}} = \frl{R_2}{R_2 \m R_1}
\mbox{ .}
\end{equation}
From the definition of $R_1$ in equation \eqref{e_347} and the
transformation \eqref{e_346} for $k = 1$ it follows that
\begin{equation} \label{e_556}
R_1 = \frl{R_2^2 \m T_2^2 \m B^2}{2 R_2}
\mbox{ .}
\end{equation}
This implies that
\begin{equation} \label{e_557}
N = \frl{2 R_2^2}{R_2^2 \p T_2^2 \p B^2}
\mbox{ .}
\end{equation}
Hence $N > 1$ for $R_2^2 - T_2^2 > B^2$,
i.e.\hspace{1.2mm}to the right of the hyperbola in Figure 3, which
means that the reference particles of the $(\eta,\chi)$-system have a
greater outwards acceleration than a free particle. This is the reason
why an observer at rest in the $(\eta,\chi)$-system experiences that
the acceleration of gravity is directed in the negative $\chi$-direction.
The wall has a decreasing radius in the $(\eta,\chi)$-system. This is,
however, a coordinate effect. In reality the wall is static and the
$(\eta,\chi)$ coordinate system is comoving in an expanding reference
frame.
\itm The acceleration of gravity vanishes in the $(\eta,\chi)$-system
on the hyperbola $\chi = \pi/2$ in Figure 3.
This leads to the following physical interpretation of the constant $B$.
As seen from equation \eqref{e_347} the point $(0,B)$ in the CFS
system corresponds to the point $(0,\pi/2)$ in the $(\eta,\chi)$-system
where the acceleration of gravity vanishes. From equation \eqref{e_360}
it follows that a particle with $\chi = \pi / 2$ moves freely.
Equation \eqref{e_347} gives $R_1 = 0$ for this particle. The coordinate
acceleration of this particle at the point of time $T = 0$ as given by
equation \eqref{e_552} is $1/B$. Hence $B$ is the inverse of the
coordinate acceleration of a free particle at $(0,B)$.
\itm We shall now consider the case $k = -1$. The world lines of
reference particles with $\chi = {\chi}_1$ are given by equation
\eqref{e_347} and are shown in Figure 4. We see that the reference
points in the WLBR spacetime accelerate in the negative $R$ direction.
Hence in this reference frame the acceleration of gravity is directed
outwards and is larger than in the static CFS system. This is verified
by the expression \eqref{e_360} which implies that in this case
$g = (1 / R_{\mbox{\tiny $Q$}}) \cosh \chi \ge
1 / R_{\mbox{\tiny $Q$}}$.
\itm We now consider the flat spacetime inside the shell.
The line element of the Minkowski spacetime in this region has the
following form in the CFS coordinate system,
\begin{equation} \label{e_1913}
ds_M^2 = -dT^2 + dR^2 + R^2 d\Omega^2
\mbox{ .}
\end{equation}
Inserting $e^{\alpha} = R_{\mbox{\tiny $Q$}} / S_k(\chi)$ from the
line element \eqref{e_218} and the expression \eqref{e_279} for $R$
into the line element \eqref{e_1211} we find the form of the line
element \eqref{e_1913} in the $(\eta,\chi)$-system.
\begin{equation} \label{e_1914}
ds_M^2 = \frl{B^2}{[C_k(\eta) \p C_k(\chi)]^2}
[-d\eta^2 + d\chi^2 + S_k(\chi)^2 d\Omega^2]
\mbox{ .}
\end{equation}
Comparing with the line element \eqref{e_218} and using equation
\eqref{e_344} we see that the metric is continuous at the domain wall.
Note that the line element \eqref{e_1914} reduces to \eqref{e_1913} for
$k = 0$ replacing $(\eta,\chi)$ with $(T,R)$. In this case a spherical
surface with radius $R$ has area $4 \pi R^2$ in the region inside the
domain wall. Outside the domain wall, on the other hand, a spherical
surface with radius $R$ has area $4 \pi R_{\mbox{\tiny $Q$}}^2$ which
is independent of $R$. The reason for this strange result is that the
space $T = \mbox{constant}$ is curved outside the domain wall.
\vspace*{5mm} \newline
\begin{picture}(50,257)(-96,-230)
\qbezier(135.0000, -170.3696)(135.9030, -168.4156)(136.7561, -166.4841)
\qbezier(135.0000, -29.5000)(134.0970, -31.4540)(133.2439, -33.3854)
\qbezier(136.7561, -166.4841)(137.6093, -164.5527)(138.4132, -162.6425)
\qbezier(133.2439, -33.3854)(132.3907, -35.3169)(131.5868, -37.2271)
\qbezier(138.4132, -162.6425)(139.2171, -160.7323)(139.9722, -158.8421)
\qbezier(131.5868, -37.2271)(130.7829, -39.1373)(130.0278, -41.0275)
\qbezier(139.9722, -158.8421)(140.7274, -156.9518)(141.4344, -155.0804)
\qbezier(130.0278, -41.0275)(129.2726, -42.9177)(128.5656, -44.7892)
\qbezier(141.4344, -155.0804)(142.1413, -153.2089)(142.8005, -151.3549)
\qbezier(128.5656, -44.7892)(127.8587, -46.6607)(127.1995, -48.5146)
\qbezier(142.8005, -151.3549)(143.4597, -149.5010)(144.0716, -147.6633)
\qbezier(127.1995, -48.5146)(126.5403, -50.3686)(125.9284, -52.2063)
\qbezier(144.0716, -147.6633)(144.6835, -145.8256)(145.2484, -144.0030)
\qbezier(125.9284, -52.2063)(125.3165, -54.0439)(124.7516, -55.8665)
\qbezier(145.2484, -144.0030)(145.8134, -142.1805)(146.3318, -140.3718)
\qbezier(124.7516, -55.8665)(124.1866, -57.6891)(123.6682, -59.4978)
\qbezier(146.3318, -140.3718)(146.8502, -138.5631)(147.3223, -136.7671)
\qbezier(123.6682, -59.4978)(123.1498, -61.3065)(122.6777, -63.1025)
\qbezier(147.3223, -136.7671)(147.7945, -134.9711)(148.2208, -133.1866)
\qbezier(122.6777, -63.1025)(122.2055, -64.8985)(121.7792, -66.6830)
\qbezier(148.2208, -133.1866)(148.6471, -131.4021)(149.0277, -129.6280)
\qbezier(121.7792, -66.6830)(121.3529, -68.4674)(120.9723, -70.2416)
\qbezier(149.0277, -129.6280)(149.4084, -127.8539)(149.7437, -126.0889)
\qbezier(120.9723, -70.2416)(120.5916, -72.0157)(120.2563, -73.7807)
\qbezier(149.7437, -126.0889)(150.0790, -124.3240)(150.3691, -122.5670)
\qbezier(120.2563, -73.7807)(119.9210, -75.5456)(119.6309, -77.3025)
\qbezier(150.3691, -122.5670)(150.6593, -120.8101)(150.9044, -119.0600)
\qbezier(119.6309, -77.3025)(119.3407, -79.0595)(119.0956, -80.8096)
\qbezier(150.9044, -119.0600)(151.1496, -117.3099)(151.3500, -115.5656)
\qbezier(119.0956, -80.8096)(118.8504, -82.5596)(118.6500, -84.3040)
\qbezier(151.3500, -115.5656)(151.5504, -113.8212)(151.7061, -112.0814)
\qbezier(118.6500, -84.3040)(118.4496, -86.0483)(118.2939, -87.7881)
\qbezier(151.7061, -112.0814)(151.8618, -110.3416)(151.9730, -108.6053)
\qbezier(118.2939, -87.7881)(118.1382, -89.5279)(118.0270, -91.2643)
\qbezier(151.9730, -108.6053)(152.0841, -106.8689)(152.1508, -105.1348)
\qbezier(118.0270, -91.2643)(117.9159, -93.0007)(117.8492, -94.7348)
\qbezier(152.1508, -105.1348)(152.2175, -103.4007)(152.2397, -101.6677)
\qbezier(117.8492, -94.7348)(117.7825, -96.4689)(117.7603, -98.2018)
\qbezier(152.2397, -101.6677)(152.2619, -99.9348)(152.2397, -98.2018)
\qbezier(117.7603, -98.2018)(117.7381, -99.9348)(117.7603, -101.6677)
\qbezier(152.2397, -98.2018)(152.2175, -96.4689)(152.1508, -94.7348)
\qbezier(117.7603, -101.6677)(117.7825, -103.4007)(117.8492, -105.1348)
\qbezier(152.1508, -94.7348)(152.0841, -93.0007)(151.9730, -91.2643)
\qbezier(117.8492, -105.1348)(117.9159, -106.8689)(118.0270, -108.6053)
\qbezier(151.9730, -91.2643)(151.8618, -89.5279)(151.7061, -87.7881)
\qbezier(118.0270, -108.6053)(118.1382, -110.3416)(118.2939, -112.0814)
\qbezier(151.7061, -87.7881)(151.5504, -86.0483)(151.3500, -84.3040)
\qbezier(118.2939, -112.0814)(118.4496, -113.8212)(118.6500, -115.5656)
\qbezier(151.3500, -84.3040)(151.1496, -82.5596)(150.9044, -80.8096)
\qbezier(118.6500, -115.5656)(118.8504, -117.3099)(119.0956, -119.0600)
\qbezier(150.9044, -80.8096)(150.6593, -79.0595)(150.3691, -77.3025)
\qbezier(119.0956, -119.0600)(119.3407, -120.8101)(119.6309, -122.5670)
\qbezier(150.3691, -77.3025)(150.0790, -75.5456)(149.7437, -73.7807)
\qbezier(119.6309, -122.5670)(119.9210, -124.3240)(120.2563, -126.0889)
\qbezier(149.7437, -73.7807)(149.4084, -72.0157)(149.0277, -70.2416)
\qbezier(120.2563, -126.0889)(120.5916, -127.8539)(120.9723, -129.6280)
\qbezier(149.0277, -70.2416)(148.6471, -68.4674)(148.2208, -66.6830)
\qbezier(120.9723, -129.6280)(121.3529, -131.4021)(121.7792, -133.1866)
\qbezier(148.2208, -66.6830)(147.7945, -64.8985)(147.3223, -63.1025)
\qbezier(121.7792, -133.1866)(122.2055, -134.9711)(122.6777, -136.7671)
\qbezier(147.3223, -63.1025)(146.8502, -61.3065)(146.3318, -59.4978)
\qbezier(122.6777, -136.7671)(123.1498, -138.5631)(123.6682, -140.3718)
\qbezier(146.3318, -59.4978)(145.8134, -57.6891)(145.2484, -55.8665)
\qbezier(123.6682, -140.3718)(124.1866, -142.1805)(124.7516, -144.0030)
\qbezier(145.2484, -55.8665)(144.6835, -54.0439)(144.0716, -52.2063)
\qbezier(124.7516, -144.0030)(125.3165, -145.8256)(125.9284, -147.6633)
\qbezier(144.0716, -52.2063)(143.4597, -50.3686)(142.8005, -48.5146)
\qbezier(125.9284, -147.6633)(126.5403, -149.5010)(127.1995, -151.3549)
\qbezier(142.8005, -48.5146)(142.1413, -46.6607)(141.4344, -44.7892)
\qbezier(127.1995, -151.3549)(127.8587, -153.2089)(128.5656, -155.0804)
\qbezier(141.4344, -44.7892)(140.7274, -42.9177)(139.9722, -41.0275)
\qbezier(128.5656, -155.0804)(129.2726, -156.9518)(130.0278, -158.8421)
\qbezier(139.9722, -41.0275)(139.2171, -39.1373)(138.4132, -37.2271)
\qbezier(130.0278, -158.8421)(130.7829, -160.7323)(131.5868, -162.6425)
\qbezier(138.4132, -37.2271)(137.6093, -35.3169)(136.7561, -33.3854)
\qbezier(131.5868, -162.6425)(132.3907, -164.5527)(133.2439, -166.4841)
\qbezier(136.7561, -33.3854)(135.9030, -31.4540)(135.0000, -29.5000)
\qbezier(133.2439, -166.4841)(134.0970, -168.4156)(135.0000, -170.3696)
\qbezier(135.0000, -170.3696)(137.6175, -167.4472)(140.0248, -164.7116)
\qbezier(135.0000, -29.5000)(132.3825, -32.4223)(129.9752, -35.1580)
\qbezier(140.0248, -164.7116)(142.4320, -161.9760)(144.6424, -159.4120)
\qbezier(129.9752, -35.1580)(127.5680, -37.8936)(125.3576, -40.4576)
\qbezier(144.6424, -159.4120)(146.8527, -156.8480)(148.8783, -154.4414)
\qbezier(125.3576, -40.4576)(123.1473, -43.0216)(121.1217, -45.4282)
\qbezier(148.8783, -154.4414)(150.9039, -152.0348)(152.7559, -149.7723)
\qbezier(121.1217, -45.4282)(119.0961, -47.8348)(117.2441, -50.0973)
\qbezier(152.7559, -149.7723)(154.6080, -147.5098)(156.2968, -145.3789)
\qbezier(117.2441, -50.0973)(115.3920, -52.3598)(113.7032, -54.4907)
\qbezier(156.2968, -145.3789)(157.9856, -143.2480)(159.5205, -141.2369)
\qbezier(113.7032, -54.4907)(112.0144, -56.6216)(110.4795, -58.6327)
\qbezier(159.5205, -141.2369)(161.0554, -139.2258)(162.4449, -137.3234)
\qbezier(110.4795, -58.6327)(108.9446, -60.6438)(107.5551, -62.5462)
\qbezier(162.4449, -137.3234)(163.8343, -135.4209)(165.0860, -133.6166)
\qbezier(107.5551, -62.5462)(106.1657, -64.4486)(104.9140, -66.2529)
\qbezier(165.0860, -133.6166)(166.3377, -131.8124)(167.4586, -130.0963)
\qbezier(104.9140, -66.2529)(103.6623, -68.0572)(102.5414, -69.7733)
\qbezier(167.4586, -130.0963)(168.5794, -128.3802)(169.5757, -126.7427)
\qbezier(102.5414, -69.7733)(101.4206, -71.4894)(100.4243, -73.1268)
\qbezier(169.5757, -126.7427)(170.5719, -125.1053)(171.4491, -123.5375)
\qbezier(100.4243, -73.1268)( 99.4281, -74.7643)( 98.5509, -76.3321)
\qbezier(171.4491, -123.5375)(172.3262, -121.9697)(173.0891, -120.4628)
\qbezier( 98.5509, -76.3321)( 97.6738, -77.8999)( 96.9109, -79.4068)
\qbezier(173.0891, -120.4628)(173.8519, -118.9559)(174.5048, -117.5017)
\qbezier( 96.9109, -79.4068)( 96.1481, -80.9136)( 95.4952, -82.3679)
\qbezier(174.5048, -117.5017)(175.1576, -116.0474)(175.7040, -114.6378)
\qbezier( 95.4952, -82.3679)( 94.8424, -83.8221)( 94.2960, -85.2318)
\qbezier(175.7040, -114.6378)(176.2504, -113.2281)(176.6934, -111.8551)
\qbezier( 94.2960, -85.2318)( 93.7496, -86.6415)( 93.3066, -88.0144)
\qbezier(176.6934, -111.8551)(177.1364, -110.4822)(177.4784, -109.1385)
\qbezier( 93.3066, -88.0144)( 92.8636, -89.3873)( 92.5216, -90.7311)
\qbezier(177.4784, -109.1385)(177.8204, -107.7947)(178.0634, -106.4727)
\qbezier( 92.5216, -90.7311)( 92.1796, -92.0748)( 91.9366, -93.3968)
\qbezier(178.0634, -106.4727)(178.3064, -105.1507)(178.4516, -103.8431)
\qbezier( 91.9366, -93.3968)( 91.6936, -94.7188)( 91.5484, -96.0264)
\qbezier(178.4516, -103.8431)(178.5968, -102.5356)(178.6452, -101.2352)
\qbezier( 91.5484, -96.0264)( 91.4032, -97.3340)( 91.3548, -98.6344)
\qbezier(178.6452, -101.2352)(178.6935, -99.9348)(178.6452, -98.6344)
\qbezier( 91.3548, -98.6344)( 91.3065, -99.9348)( 91.3548, -101.2352)
\qbezier(178.6452, -98.6344)(178.5968, -97.3340)(178.4516, -96.0264)
\qbezier( 91.3548, -101.2352)( 91.4032, -102.5356)( 91.5484, -103.8431)
\qbezier(178.4516, -96.0264)(178.3064, -94.7188)(178.0634, -93.3968)
\qbezier( 91.5484, -103.8431)( 91.6936, -105.1507)( 91.9366, -106.4727)
\qbezier(178.0634, -93.3968)(177.8204, -92.0748)(177.4784, -90.7311)
\qbezier( 91.9366, -106.4727)( 92.1796, -107.7947)( 92.5216, -109.1385)
\qbezier(177.4784, -90.7311)(177.1364, -89.3873)(176.6934, -88.0144)
\qbezier( 92.5216, -109.1385)( 92.8636, -110.4822)( 93.3066, -111.8551)
\qbezier(176.6934, -88.0144)(176.2504, -86.6415)(175.7040, -85.2318)
\qbezier( 93.3066, -111.8551)( 93.7496, -113.2281)( 94.2960, -114.6378)
\qbezier(175.7040, -85.2318)(175.1576, -83.8221)(174.5048, -82.3679)
\qbezier( 94.2960, -114.6378)( 94.8424, -116.0474)( 95.4952, -117.5017)
\qbezier(174.5048, -82.3679)(173.8519, -80.9136)(173.0891, -79.4068)
\qbezier( 95.4952, -117.5017)( 96.1481, -118.9559)( 96.9109, -120.4628)
\qbezier(173.0891, -79.4068)(172.3262, -77.8999)(171.4491, -76.3321)
\qbezier( 96.9109, -120.4628)( 97.6738, -121.9697)( 98.5509, -123.5375)
\qbezier(171.4491, -76.3321)(170.5719, -74.7643)(169.5757, -73.1268)
\qbezier( 98.5509, -123.5375)( 99.4281, -125.1053)(100.4243, -126.7427)
\qbezier(169.5757, -73.1268)(168.5794, -71.4894)(167.4586, -69.7733)
\qbezier(100.4243, -126.7427)(101.4206, -128.3802)(102.5414, -130.0963)
\qbezier(167.4586, -69.7733)(166.3377, -68.0572)(165.0860, -66.2529)
\qbezier(102.5414, -130.0963)(103.6623, -131.8124)(104.9140, -133.6166)
\qbezier(165.0860, -66.2529)(163.8343, -64.4486)(162.4449, -62.5462)
\qbezier(104.9140, -133.6166)(106.1657, -135.4209)(107.5551, -137.3234)
\qbezier(162.4449, -62.5462)(161.0554, -60.6438)(159.5205, -58.6327)
\qbezier(107.5551, -137.3234)(108.9446, -139.2258)(110.4795, -141.2369)
\qbezier(159.5205, -58.6327)(157.9856, -56.6216)(156.2968, -54.4907)
\qbezier(110.4795, -141.2369)(112.0144, -143.2480)(113.7032, -145.3789)
\qbezier(156.2968, -54.4907)(154.6080, -52.3598)(152.7559, -50.0973)
\qbezier(113.7032, -145.3789)(115.3920, -147.5098)(117.2441, -149.7723)
\qbezier(152.7559, -50.0973)(150.9039, -47.8348)(148.8783, -45.4282)
\qbezier(117.2441, -149.7723)(119.0961, -152.0348)(121.1217, -154.4414)
\qbezier(148.8783, -45.4282)(146.8527, -43.0216)(144.6424, -40.4576)
\qbezier(121.1217, -154.4414)(123.1473, -156.8480)(125.3576, -159.4120)
\qbezier(144.6424, -40.4576)(142.4320, -37.8936)(140.0248, -35.1580)
\qbezier(125.3576, -159.4120)(127.5680, -161.9760)(129.9752, -164.7116)
\qbezier(140.0248, -35.1580)(137.6175, -32.4223)(135.0000, -29.5000)
\qbezier(129.9752, -164.7116)(132.3825, -167.4472)(135.0000, -170.3696)
\put(135.0000, -29.5000){\line(1, -1){ 70.4348}}
\put(135.0000, -170.3696){\line(1, 1){ 70.4348}}
\put( 64.5652, -99.9348){\line(1, -1){ 70.4348}}
\put( 64.5652, -99.9348){\line(1, 1){ 70.4348}}
\qbezier(171.6848, -133.6848)(171.6848, -132.2174)(171.6848, -130.7500)
\qbezier(171.6848, -127.8152)(171.6848, -126.3478)(171.6848, -124.8804)
\qbezier(171.6848, -121.9457)(171.6848, -120.4783)(171.6848, -119.0109)
\qbezier(171.6848, -116.0761)(171.6848, -114.6087)(171.6848, -113.1413)
\qbezier(171.6848, -110.2065)(171.6848, -108.7391)(171.6848, -107.2717)
\qbezier(171.6848, -104.3370)(171.6848, -102.8696)(171.6848, -101.4022)
\qbezier(171.6848, -98.4674)(171.6848, -97.0000)(171.6848, -95.5326)
\qbezier(171.6848, -92.5978)(171.6848, -91.1304)(171.6848, -89.6630)
\qbezier(171.6848, -86.7283)(171.6848, -85.2609)(171.6848, -83.7935)
\qbezier(171.6848, -80.8587)(171.6848, -79.3913)(171.6848, -77.9239)
\qbezier(171.6848, -74.9891)(171.6848, -73.5217)(171.6848, -72.0543)
\qbezier(171.6848, -69.1196)(171.6848, -67.6522)(171.6848, -66.1848)
\put( 32.2826, -99.9348){\vector(1, 0){220.1087}}
\put(135.0000, -202.6522){\vector(0, 1){205.4348}}
\put(261.1957, -104.3370){\makebox(0,0)[]{\footnotesize{$R$}}}
\put(127.6630,   4.2500){\makebox(0,0)[]{\footnotesize{$T$}}}
\put(127.6630, -25.0978){\makebox(0,0)[]{\footnotesize{$B$}}}
\put(211.3043, -107.2717){\makebox(0,0)[]{\footnotesize{$B$}}}
\put(124.7283, -177.7065){\makebox(0,0)[]{\footnotesize{$-B$}}}
\put( 57.2283, -107.2717){\makebox(0,0)[]{\footnotesize{$-B$}}}
\put(164.3478, -108.7391){\makebox(0,0)[]{\footnotesize{$R_{\mbox{\tiny $Q$}}$}}}
\end{picture}
\vspace{0mm} \newline
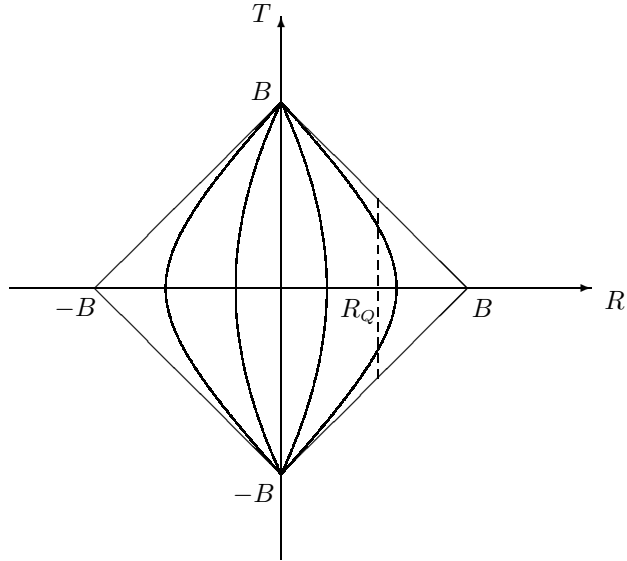
{\footnotesize \sf Figure 4. The square represents that part of the
PLBR spacetime which is described by the $(\eta,\chi)$-system with
$k = -1$. The world lines of points with $\chi = {\chi}_1$ as given by
equation \eqref{e_347}. The region to the right of the vertical line
$R = R_{\mbox{\tiny $Q$}}$ represents a part of the WLBR spacetime
when $B > R_{\mbox{\tiny $Q$}}$ in accordance with
equation \eqref{e_1167}.}
\vspace{3mm} \newline
\itm Calculating the acceleration of gravity inside the shell as
experienced by an observer at rest in the $(\eta,\chi)$-system in the
same way as in equation \eqref{e_360}, we find
\begin{equation} \label{e_1915}
a_M^{\hat{\chi}} = - k S_k(\chi) / B
\mbox{ .}
\end{equation}
In order to find the discontinuity of the acceleration of gravity in
the $(\eta,\chi)$-system at the domain wall, it is sufficient to consider
the point of time $\eta = 0$. Then the domain wall has the position
$\chi = 2 {\chi}_{\mbox{\tiny $Q$}}$, where ${\chi}_{\mbox{\tiny $Q$}}$
is given in equation \eqref{e_1184}. Inserting this into equation
\eqref{e_1915} and using equations \eqref{e_828} and \eqref{e_827}
we find the acceleration of gravity just inside the domain wall,
\begin{equation} \label{e_1916}
a_M^{\hat{\chi}} = - \frl{2k R_{\mbox{\tiny $Q$}}^2}
{B^2 \p k R_{\mbox{\tiny $Q$}}^2} \hs{0.8mm} \frl{1}{R_{\mbox{\tiny $Q$}}}
\mbox{ .}
\end{equation}
We see that the acceleration of gravity depends on the value of $k$.
There is no acceleration if $k = 0$ because in this case the
$(\eta,\chi)$-system is comoving in a static reference frame in
flat spacetime. When $k = 1$ there is an acceleration of gravity
towards the point $\chi = 0$. In this case the $(\eta,\chi)$-system
is comoving in a reference frame accelerating in the outwards direction.
In the case $k = -1$ we must have $B > R_{\mbox{\tiny $Q$}}$
in order that the WLBR spacetime shall exist outside the domain wall
as seen in Figure 4. Then the acceleration of gravity points outwards,
meaning that the reference frame of the $(\eta,\chi)$-system is
accelerating inwards.
\itm The acceleration of gravity just outside the domain wall as given
by equation \eqref{e_360} is found in a similar way using equations
\eqref{e_829} and \eqref{e_827} with the result
\begin{equation} \label{e_1917}
a^{\hat{\chi}} = \frl{B^2 \m k R_{\mbox{\tiny $Q$}}^2}
{B^2 \p k R_{\mbox{\tiny $Q$}}^2} \hs{0.8mm} \frl{1}{R_{\mbox{\tiny $Q$}}}
\mbox{ .}
\end{equation}
When $k = 0$ the $(\eta,\chi)$-system is comoving in the same reference
frame as the CFS coordinates, which is at rest relative to the domain wall.
In this case the acceleration of gravity in the $(\eta,\chi)$-system just
outside the domain wall is equal to $1/R_{\mbox{\tiny $Q$}}$ just as in
the CFS system. When $k = 1$ and $B > R_{\mbox{\tiny $Q$}}$, the
acceleration of gravity is directed away from the domain wall. But
when $B < R_{\mbox{\tiny $Q$}}$ it is directed towards the domain wall.
If $B = R_{\mbox{\tiny $Q$}}$ the acceleration of gravity vanishes.
This behaviour can be understood by considering Figure 3.
In the case $B = R_{\mbox{\tiny $Q$}}$ the hyperbola $\chi = \pi / 2$
touches the domain wall at $R = R_{\mbox{\tiny $Q$}}$ when $T = 0$,
corresponding to $\eta = 0$. For $B > R_{\mbox{\tiny $Q$}}$ the
hyperbola moves to the right, and for $B < R_{\mbox{\tiny $Q$}}$
to the left.
\itm For all values of $k$ and $B$ the discontinuity of the
acceleration of gravity at the domain wall is
\begin{equation} \label{e_1918}
a^{\hat{\chi}} - a_M^{\hat{\chi}} = \frl{1}{R_{\mbox{\tiny $Q$}}}
\mbox{ .}
\end{equation}
This shows that the domain wall produces repulsive gravity.
%
%
%
%
%
\vspace{6mm} \newline
{\it Ib. Time dependent metric and coordinates $(\tau,\rho)$ with
$\beta (\tau) = \alpha (\tau)$.}
\vspace{3mm} \newline
In spite of the fact that the LBR spacetime is static, it may be described
in terms of coordinates comoving with a reference frame expanding in such
a way that the line element takes a time dependent form.
\itm Assuming that the metric functions are independent of the
radial coordinate, equation \eqref{e_212} reduces to
\begin{equation} \label{e_913}
R_{\mbox{\tiny $Q$}}^2 \ddot{\alpha} + e^{2 \alpha} = 0
\mbox{ ,}
\end{equation}
which may be written
\begin{equation} \label{e_914}
R_{\mbox{\tiny $Q$}}^2 (\dot{\alpha} \hs{0.5mm}^2)^{\dot{}}
= - (e^{2 \alpha})^{\dot{}}
\mbox{ .}
\end{equation}
This equation has the general solution
\begin{equation} \label{e_915}
R_{\mbox{\tiny $Q$}}^2 \hs{0.5mm} \dot{\alpha} \hs{0.5mm}^2
= - e^{2 \alpha} + a^2 R_{\mbox{\tiny $Q$}}^2
\mbox{ ,}
\end{equation}
where $a > 0$ is an integration constant. The general solution of
\eqref{e_915} is given by
\begin{equation} \label{e_919}
e^{\alpha} = a R_{\mbox{\tiny $Q$}} / \cosh(a (\tau - {\tau}_0))
\mbox{ ,}
\end{equation}
where ${\tau}_0$ is an integration constant.
Choosing $a = 1$ and ${\tau}_0 = 0$ the line element \eqref{e_211}
takes the form
\begin{equation} \label{e_928}
ds^2 = \frl{R_{\mbox{\tiny $Q$}}^2}{\cosh^2 \tau}
\hs{0.5mm} ( \hs{0.5mm} \mbox{$- d{\tau}^2$} + d{\rho}^2)
+ R_{\mbox{\tiny $Q$}}^2 d\Omega^2 \hs{0.5mm}
\end{equation}
where $-\infty < \tau < \infty$ and $-\infty < \rho < \infty$.
The form of this line element when the proper time of the reference
particles is used as a time coordinate is given in equation \eqref{e_331}.
\itm We want to investigate whether particles with constant $\rho$ are
free, and hence whether their world lines fullfill the geodesic equation.
The radial component of this equation then reduces to
\begin{equation} \label{e_995}
\ddot{\rho} = - {\Gamma}^{\hs{0.3mm} \rho}_{\hs{1.5mm}
\tau \hs{0.5mm} \tau} \dot{\tau}^2
\mbox{ .}
\end{equation}
Calculating the Christoffel symbol from the line element \eqref{e_928}
we find that ${\Gamma}^{\hs{0.3mm} \rho}_{\hs{1.5mm}
\tau \hs{0.5mm} \tau} = 0$. Hence a particle with constant $\rho$
has vanishing acceleration. It is a free particle. Accordingly the
$(\tau,\rho)$-system is comoving with free particles.
\itm From equation \eqref{e_928} it follows that the coordinate clocks
of the $(\tau,\rho)$-system go at a rate
\begin{equation} \label{e_1011}
\dot{\tau} = \frl{d\tau}{ds} = \frl{\cosh \tau}{R_{\mbox{\tiny $Q$}}}
\mbox{ ,}
\end{equation}
which is increasing relative to the rate of standard clocks at rest
in the reference frame where $(\tau,\rho)$ are comoving coordinates.
Note that the coordinate time $\tau$ is not equal to the proper time
$t$ of the reference particles with constant $\rho$. The relationship
between $\tau$ and $t$ will be treated in section IIIb where the proper
time will be used as coordinate time.
\itm In this reference frame the physical distances in the radial direction
are extremely small when $\tau \rightarrow - \infty$. However the space
expands in the radial direction and the radial scale factor has a maximal
value equal to $R_{\mbox{\tiny $Q$}}$ when $\tau = 0$. Then space contracts
in the radial direction towards vanishingly small distances in the
infinitely far future.
\itm We shall find the transformation relating the line elements
\eqref{e_157} and \eqref{e_928}. In this case $G(x^0,x^1) = \cosh (x^0)$
in the line element \eqref{e_408} so that $G(x^0,0) \ne 0$. Hence we need
two generating functions. We introduce the generating functions
\begin{equation} \label{e_980}
f(x) = - B \coth (\fr{x}{2})
\mbox{\hspace{2mm} , \hspace{3mm}}
g(x) = B \tanh (\fr{x}{2})
\mbox{ ,}
\end{equation}
using the same constant $B$ as in equation \eqref{e_339} when $k = 1$
and $k = -1$ in order to simplify the transformations. By means of
equations \eqref{e_12}, using the procedure shown in Appendix B, we
find the following transformation from the $(\tau,\rho)$-coordinates
to the CFS cordinates,
\begin{equation} \label{e_981}
T = - \frl{B \cosh \rho}
{\sinh \tau \p \sinh \rho}
\mbox{\hspace{2mm} , \hspace{3mm}}
R = - \frl{B \cosh \tau}
{\sinh \tau \p \sinh \rho}
\mbox{ .}
\end{equation}
This transforms the region $\tau + \rho < 0$ in the $(\tau,\rho)$-system
to the region $T + R > B$, $|T - R| < B$ in the CFS system, and the region
$\tau + \rho > 0$ in the $(\tau,\rho)$-system to the region
$T + R < - B$, $|T - R| < B$ in the CFS system. As shown in Appendix B
the inverse transformation is found from the generating functions
\begin{equation} \label{e_1980}
f(x) = - 2 \hs{0.5mm} \mbox{arccoth} (\fr{x}{B})
\mbox{\hspace{2mm} , \hspace{3mm}}
g(x) = 2 \hs{0.5mm} \mbox{arctanh} (\fr{x}{B})
\mbox{ ,}
\end{equation}
which gives
\begin{equation} \label{e_946}
\tanh \tau = \frl{(T^{2} \m R^{2}) \m B^2}{2 B R}
\mbox{\hspace{2mm} , \hspace{3mm}}
\tanh \rho = \frl{(R^{2} \m T^{2}) \m B^2}{2 B T}
 \mbox{ .}
\end{equation}
\itm From the second of equations \eqref{e_981} with
$R = R_{\mbox{\tiny $Q$}}$ it follows that the charged domain wall
which represents the inner boundary of the WLBR spacetime, moves
according to
\begin{equation} \label{e_1012}
\sinh \rho = - (B / R_{\mbox{\tiny $Q$}}) \cosh \tau - \sinh \tau
\end{equation}
in the $(\tau,\rho)$-system.
It follows that the WLBR spacetime is represented in the
$(\tau,\rho)$-plane by the hatched region in Figure 5, which is given by
\begin{equation} \label{e_1180}
- \mbox{arcsinh} \hs{0.5mm} ((B / R_{\mbox{\tiny $Q$}})
\cosh \tau + \sinh \tau) < \rho < - \tau
\mbox{ .}
\end{equation}
\vspace{-14mm} \newline
\begin{picture}(50,192)(64,-180)
\qbezier( 88.6814, -134.0560)( 89.3801, -133.3552)( 90.0789, -132.6545)
\qbezier( 90.0789, -132.6545)( 90.7771, -131.9537)( 91.4753, -131.2529)
\qbezier( 91.4753, -131.2529)( 92.2392, -130.4859)( 92.8704, -129.8514)
\qbezier( 92.8704, -129.8514)( 93.6333, -129.0845)( 94.2637, -128.4499)
\qbezier( 94.2637, -128.4499)( 95.0253, -127.6832)( 95.6548, -127.0483)
\qbezier( 95.6548, -127.0483)( 96.4149, -126.2818)( 97.0431, -125.6468)
\qbezier( 97.0431, -125.6468)( 97.8011, -124.8806)( 98.4278, -124.2453)
\qbezier( 98.4278, -124.2453)( 99.1833, -123.4794)( 99.8080, -122.8438)
\qbezier( 99.8080, -122.8438)(100.5602, -122.0783)(101.1824, -121.4422)
\qbezier(101.1824, -121.4422)(101.9306, -120.6773)(102.5496, -120.0407)
\qbezier(102.5496, -120.0407)(103.2927, -119.2764)(103.9077, -118.6392)
\qbezier(103.9077, -118.6392)(104.6444, -117.8757)(105.2543, -117.2376)
\qbezier(105.2543, -117.2376)(105.9830, -116.4752)(106.5865, -115.8361)
\qbezier(106.5865, -115.8361)(107.3052, -115.0749)(107.9007, -114.4346)
\qbezier(107.9007, -114.4346)(108.6071, -113.6750)(109.1926, -113.0331)
\qbezier(109.1926, -113.0331)(109.8837, -112.2753)(110.4568, -111.6315)
\qbezier(110.4568, -111.6315)(111.1293, -110.8760)(111.6871, -110.2300)
\qbezier(111.6871, -110.2300)(112.3370, -109.4772)(112.8760, -108.8285)
\qbezier(112.8760, -108.8285)(113.4989, -108.0788)(114.0151, -107.4269)
\qbezier(114.0151, -107.4269)(114.6058, -106.6809)(115.0946, -106.0254)
\qbezier(115.0946, -106.0254)(115.6480, -105.2834)(116.1042, -104.6239)
\qbezier(116.1042, -104.6239)(116.6144, -103.8863)(117.0324, -103.2223)
\qbezier(117.0324, -103.2223)(117.4939, -102.4895)(117.8679, -101.8208)
\qbezier(117.8679, -101.8208)(118.2752, -101.0929)(118.5993, -100.4193)
\qbezier(118.5993, -100.4193)(118.9472, -99.6964)(119.2158, -99.0178)
\qbezier(119.2158, -99.0178)(119.4999, -98.2999)(119.7078, -97.6162)
\qbezier(119.7078, -97.6162)(119.9247, -96.9033)(120.0676, -96.2147)
\qbezier(120.0676, -96.2147)(120.2146, -95.5066)(120.2891, -94.8132)
\qbezier(120.2891, -94.8132)(120.3647, -94.1098)(120.3686, -93.4116)
\qbezier(120.3686, -93.4116)(120.3726, -92.7130)(120.3048, -92.0101)
\qbezier(120.3048, -92.0101)(120.2380, -91.3162)(120.0988, -90.6086)
\qbezier(120.0988, -90.6086)(119.9633, -89.9195)(119.7541, -89.2070)
\qbezier(119.7541, -89.2070)(119.5531, -88.5229)(119.2762, -87.8055)
\qbezier(119.2762, -87.8055)(119.0141, -87.1264)(118.6730, -86.4040)
\qbezier(118.6730, -86.4040)(118.3547, -85.7299)(117.9538, -85.0025)
\qbezier(117.9538, -85.0025)(117.5850, -84.3333)(117.1292, -83.6009)
\qbezier(117.1292, -83.6009)(116.7157, -82.9365)(116.2106, -82.1994)
\qbezier(116.2106, -82.1994)(115.7583, -81.5394)(115.2094, -80.7979)
\qbezier(115.2094, -80.7979)(114.7240, -80.1420)(114.1370, -79.3963)
\qbezier(114.1370, -79.3963)(113.6236, -78.7441)(113.0041, -77.9948)
\qbezier(113.0041, -77.9948)(112.4673, -77.3457)(111.8202, -76.5933)
\qbezier(111.8202, -76.5933)(111.2643, -75.9470)(110.5941, -75.1918)
\qbezier(110.5941, -75.1918)(110.0225, -74.5477)(109.3332, -73.7902)
\qbezier(109.3332, -73.7902)(108.7490, -73.1481)(108.0441, -72.3887)
\qbezier(108.0441, -72.3887)(107.4497, -71.7482)(106.7321, -70.9872)
\qbezier(106.7321, -70.9872)(106.1295, -70.3480)(105.4017, -69.5856)
\qbezier(105.4017, -69.5856)(104.7925, -68.9475)(104.0566, -68.1841)
\qbezier(104.0566, -68.1841)(103.4421, -67.5468)(102.6996, -66.7826)
\qbezier(102.6996, -66.7826)(102.0811, -66.1459)(101.3333, -65.3810)
\qbezier(101.3333, -65.3810)(100.7115, -64.7449)( 99.9596, -63.9795)
\qbezier( 99.9596, -63.9795)( 99.3352, -63.3438)( 98.5800, -62.5780)
\qbezier( 98.5800, -62.5780)( 97.9536, -61.9427)( 97.1958, -61.1765)
\qbezier( 97.1958, -61.1765)( 96.5677, -60.5414)( 95.8079, -59.7749)
\qbezier( 95.8079, -59.7749)( 95.1785, -59.1401)( 94.4170, -58.3734)
\qbezier( 94.4170, -58.3734)( 93.7867, -57.7387)( 93.0239, -56.9719)
\qbezier( 93.0239, -56.9719)( 92.3928, -56.3373)( 91.6290, -55.5703)
\qbezier( 91.6290, -55.5703)( 90.9309, -54.8696)( 90.2327, -54.1688)
\qbezier( 90.2327, -54.1688)( 89.5340, -53.4680)( 88.8353, -52.7673)
\qbezier( 88.8353, -52.7673)( 88.1362, -52.0665)( 87.4370, -51.3658)
\qbezier( 87.4370, -51.3658)( 86.7375, -50.6650)( 86.0380, -49.9642)
\put( 88.6814, -134.0560){\line(1, 0){ 93.8488}}
\put( 91.4753, -131.2529){\line(1, 0){ 88.2517}}
\put( 94.2637, -128.4499){\line(1, 0){ 82.6603}}
\put( 97.0431, -125.6468){\line(1, 0){ 77.0779}}
\put( 99.8080, -122.8438){\line(1, 0){ 71.5099}}
\put(102.5496, -120.0407){\line(1, 0){ 65.9652}}
\put(105.2543, -117.2376){\line(1, 0){ 60.4575}}
\put(107.9007, -114.4346){\line(1, 0){ 55.0080}}
\put(110.4568, -111.6315){\line(1, 0){ 49.6489}}
\put(112.8760, -108.8285){\line(1, 0){ 44.4266}}
\put(115.0946, -106.0254){\line(1, 0){ 39.4049}}
\put(117.0324, -103.2223){\line(1, 0){ 34.6640}}
\put(118.5993, -100.4193){\line(1, 0){ 30.2941}}
\put(119.7078, -97.6162){\line(1, 0){ 26.3825}}
\put(120.2891, -94.8132){\line(1, 0){ 22.9982}}
\put(120.3048, -92.0101){\line(1, 0){ 20.1794}}
\put(119.7541, -89.2070){\line(1, 0){ 17.9271}}
\put(117.1292, -83.6009){\line(1, 0){ 14.9459}}
\put(115.2094, -80.7979){\line(1, 0){ 14.0626}}
\put(113.0041, -77.9948){\line(1, 0){ 13.4649}}
\put(110.5941, -75.1918){\line(1, 0){ 13.0718}}
\put(108.0441, -72.3887){\line(1, 0){ 12.8187}}
\put(105.4017, -69.5856){\line(1, 0){ 12.6580}}
\put(102.6996, -66.7826){\line(1, 0){ 12.5571}}
\put( 99.9596, -63.9795){\line(1, 0){ 12.4940}}
\put( 97.1958, -61.1765){\line(1, 0){ 12.4548}}
\put( 94.4170, -58.3734){\line(1, 0){ 12.4305}}
\put( 91.6290, -55.5703){\line(1, 0){ 12.4154}}
\put( 88.8353, -52.7673){\line(1, 0){ 12.4061}}
\put( 86.0380, -49.9642){\line(1, 0){ 12.4004}}
\put( 84.9569, -86.5259){\vector(1, 0){100.0862}}
\put(135.0000, -145.8793){\vector(0, 1){109.3966}}
\put( 98.4384, -49.9642){\line(1, -1){ 84.0918}}
\put(189.4655, -88.2716){\makebox(0,0)[]{\footnotesize{$\rho$}}}
\put(131.2759, -35.9009){\makebox(0,0)[]{\footnotesize{$\tau$}}}
\end{picture}
\begin{picture}(50,192)(-36,-180)
\qbezier(134.8040, -134.0560)(134.7917, -133.3218)(134.7790, -132.6545)
\qbezier(134.7790, -132.6545)(134.7650, -131.9203)(134.7507, -131.2529)
\qbezier(134.7507, -131.2529)(134.7349, -130.5188)(134.7188, -129.8514)
\qbezier(134.7188, -129.8514)(134.7010, -129.1173)(134.6828, -128.4499)
\qbezier(134.6828, -128.4499)(134.6628, -127.7158)(134.6422, -127.0483)
\qbezier(134.6422, -127.0483)(134.6196, -126.3143)(134.5964, -125.6468)
\qbezier(134.5964, -125.6468)(134.5710, -124.9128)(134.5448, -124.2453)
\qbezier(134.5448, -124.2453)(134.5161, -123.5113)(134.4866, -122.8438)
\qbezier(134.4866, -122.8438)(134.4542, -122.1098)(134.4209, -121.4422)
\qbezier(134.4209, -121.4422)(134.3844, -120.7083)(134.3469, -120.0407)
\qbezier(134.3469, -120.0407)(134.3057, -119.3069)(134.2634, -118.6392)
\qbezier(134.2634, -118.6392)(134.2170, -117.9054)(134.1693, -117.2376)
\qbezier(134.1693, -117.2376)(134.1170, -116.5040)(134.0632, -115.8361)
\qbezier(134.0632, -115.8361)(134.0042, -115.1026)(133.9437, -114.4346)
\qbezier(133.9437, -114.4346)(133.8772, -113.7013)(133.8089, -113.0331)
\qbezier(133.8089, -113.0331)(133.7340, -112.3000)(133.6571, -111.6315)
\qbezier(133.6571, -111.6315)(133.5728, -110.8988)(133.4862, -110.2300)
\qbezier(133.4862, -110.2300)(133.3913, -109.4977)(133.2937, -108.8285)
\qbezier(133.2937, -108.8285)(133.1870, -108.0967)(133.0772, -107.4269)
\qbezier(133.0772, -107.4269)(132.9573, -106.6958)(132.8338, -106.0254)
\qbezier(132.8338, -106.0254)(132.6992, -105.2950)(132.5603, -104.6239)
\qbezier(132.5603, -104.6239)(132.4094, -103.8945)(132.2535, -103.2223)
\qbezier(132.2535, -103.2223)(132.0845, -102.4942)(131.9096, -101.8208)
\qbezier(131.9096, -101.8208)(131.7208, -101.0942)(131.5248, -100.4193)
\qbezier(131.5248, -100.4193)(131.3144, -99.6946)(131.0952, -99.0178)
\qbezier(131.0952, -99.0178)(130.8613, -98.2953)(130.6168, -97.6162)
\qbezier(130.6168, -97.6162)(130.3576, -96.8965)(130.0854, -96.2147)
\qbezier(130.0854, -96.2147)(129.7995, -95.4982)(129.4976, -94.8132)
\qbezier(129.4976, -94.8132)(129.1835, -94.1005)(128.8498, -93.4116)
\qbezier(128.8498, -93.4116)(128.5067, -92.7033)(128.1396, -92.0101)
\qbezier(128.1396, -92.0101)(127.7670, -91.3066)(127.3651, -90.6086)
\qbezier(127.3651, -90.6086)(126.9631, -89.9104)(126.5258, -89.2070)
\qbezier(126.5258, -89.2070)(126.0952, -88.5146)(125.6222, -87.8055)
\qbezier(125.6222, -87.8055)(125.1642, -87.1189)(124.6562, -86.4040)
\qbezier(124.6562, -86.4040)(124.1727, -85.7234)(123.6310, -85.0025)
\qbezier(123.6310, -85.0025)(123.1241, -84.3278)(122.5507, -83.6009)
\qbezier(122.5507, -83.6009)(122.0228, -82.9319)(121.4201, -82.1994)
\qbezier(121.4201, -82.1994)(120.8740, -81.5356)(120.2449, -80.7979)
\qbezier(120.2449, -80.7979)(119.6829, -80.1388)(119.0304, -79.3963)
\qbezier(119.0304, -79.3963)(118.4550, -78.7415)(117.7822, -77.9948)
\qbezier(117.7822, -77.9948)(117.1956, -77.3437)(116.5055, -76.5933)
\qbezier(116.5055, -76.5933)(115.9095, -75.9453)(115.2048, -75.1918)
\qbezier(115.2048, -75.1918)(114.6013, -74.5464)(113.8844, -73.7902)
\qbezier(113.8844, -73.7902)(113.2747, -73.1471)(112.5478, -72.3887)
\qbezier(112.5478, -72.3887)(111.9331, -71.7473)(111.1981, -70.9872)
\qbezier(111.1981, -70.9872)(110.5794, -70.3473)(109.8378, -69.5856)
\qbezier(109.8378, -69.5856)(109.2159, -68.9469)(108.4690, -68.1841)
\qbezier(108.4690, -68.1841)(107.8445, -67.5464)(107.0933, -66.7826)
\qbezier(107.0933, -66.7826)(106.4669, -66.1456)(105.7123, -65.3810)
\qbezier(105.7123, -65.3810)(105.0842, -64.7447)(104.3269, -63.9795)
\qbezier(104.3269, -63.9795)(103.6976, -63.3436)(102.9381, -62.5780)
\qbezier(102.9381, -62.5780)(102.3078, -61.9425)(101.5467, -61.1765)
\qbezier(101.5467, -61.1765)(100.9155, -60.5413)(100.1531, -59.7749)
\qbezier(100.1531, -59.7749)( 99.5213, -59.1400)( 98.7578, -58.3734)
\qbezier( 98.7578, -58.3734)( 98.0594, -57.6726)( 97.3611, -56.9719)
\qbezier( 97.3611, -56.9719)( 96.6623, -56.2711)( 95.9635, -55.5703)
\qbezier( 95.9635, -55.5703)( 95.2642, -54.8696)( 94.5650, -54.1688)
\qbezier( 94.5650, -54.1688)( 93.8654, -53.4680)( 93.1658, -52.7673)
\qbezier( 93.1658, -52.7673)( 92.4660, -52.0665)( 91.7662, -51.3658)
\qbezier( 91.7662, -51.3658)( 91.0662, -50.6650)( 90.3661, -49.9642)
\put(134.8040, -134.0560){\line(1, 0){ 47.7261}}
\put(134.7507, -131.2529){\line(1, 0){ 44.9764}}
\put(134.6828, -128.4499){\line(1, 0){ 42.2412}}
\put(134.5964, -125.6468){\line(1, 0){ 39.5245}}
\put(134.4866, -122.8438){\line(1, 0){ 36.8313}}
\put(134.3469, -120.0407){\line(1, 0){ 34.1679}}
\put(134.1693, -117.2376){\line(1, 0){ 31.5425}}
\put(133.9437, -114.4346){\line(1, 0){ 28.9651}}
\put(133.6571, -111.6315){\line(1, 0){ 26.4485}}
\put(133.2937, -108.8285){\line(1, 0){ 24.0089}}
\put(132.8338, -106.0254){\line(1, 0){ 21.6658}}
\put(132.2535, -103.2223){\line(1, 0){ 19.4430}}
\put(131.5248, -100.4193){\line(1, 0){ 17.3686}}
\put(130.6168, -97.6162){\line(1, 0){ 15.4736}}
\put(129.4976, -94.8132){\line(1, 0){ 13.7897}}
\put(128.1396, -92.0101){\line(1, 0){ 12.3447}}
\put(126.5258, -89.2070){\line(1, 0){ 11.1554}}
\put(122.5507, -83.6009){\line(1, 0){  9.5244}}
\put(120.2449, -80.7979){\line(1, 0){  9.0272}}
\put(117.7822, -77.9948){\line(1, 0){  8.6867}}
\put(115.2048, -75.1918){\line(1, 0){  8.4611}}
\put(112.5478, -72.3887){\line(1, 0){  8.3150}}
\put(109.8378, -69.5856){\line(1, 0){  8.2220}}
\put(107.0933, -66.7826){\line(1, 0){  8.1634}}
\put(104.3269, -63.9795){\line(1, 0){  8.1267}}
\put(101.5467, -61.1765){\line(1, 0){  8.1039}}
\put( 98.7578, -58.3734){\line(1, 0){  8.0898}}
\put( 95.9635, -55.5703){\line(1, 0){  8.0810}}
\put( 93.1658, -52.7673){\line(1, 0){  8.0756}}
\put( 90.3661, -49.9642){\line(1, 0){  8.0722}}
\put( 84.9569, -86.5259){\vector(1, 0){100.0862}}
\put(135.0000, -145.8793){\vector(0, 1){109.3966}}
\put( 98.4384, -49.9642){\line(1, -1){ 84.0918}}
\put(189.4655, -88.2716){\makebox(0,0)[]{\footnotesize{$\rho$}}}
\put(131.2759, -35.9009){\makebox(0,0)[]{\footnotesize{$\tau$}}}
\end{picture}
\begin{picture}(50,192)(-136,-180)
\qbezier(155.9900, -134.0560)(155.2658, -133.2994)(154.6553, -132.6545)
\qbezier(154.6553, -132.6545)(153.9415, -131.9004)(153.3369, -131.2529)
\qbezier(153.3369, -131.2529)(152.6357, -130.5020)(152.0383, -129.8514)
\qbezier(152.0383, -129.8514)(151.3521, -129.1042)(150.7634, -128.4499)
\qbezier(150.7634, -128.4499)(150.0948, -127.7069)(149.5165, -127.0483)
\qbezier(149.5165, -127.0483)(148.8684, -126.3103)(148.3023, -125.6468)
\qbezier(148.3023, -125.6468)(147.6773, -124.9142)(147.1255, -124.2453)
\qbezier(147.1255, -124.2453)(146.5261, -123.5187)(145.9905, -122.8438)
\qbezier(145.9905, -122.8438)(145.4191, -122.1237)(144.9015, -121.4422)
\qbezier(144.9015, -121.4422)(144.3598, -120.7291)(143.8617, -120.0407)
\qbezier(143.8617, -120.0407)(143.3508, -119.3347)(142.8734, -118.6392)
\qbezier(142.8734, -118.6392)(142.3938, -117.9405)(141.9375, -117.2376)
\qbezier(141.9375, -117.2376)(141.4889, -116.5465)(141.0538, -115.8361)
\qbezier(141.0538, -115.8361)(140.6352, -115.1528)(140.2203, -114.4346)
\qbezier(140.2203, -114.4346)(139.8304, -113.7596)(139.4341, -113.0331)
\qbezier(139.4341, -113.0331)(139.0710, -112.3673)(138.6910, -111.6315)
\qbezier(138.6910, -111.6315)(138.3529, -110.9770)(137.9858, -110.2300)
\qbezier(137.9858, -110.2300)(137.6714, -109.5903)(137.3127, -108.8285)
\qbezier(137.3127, -108.8285)(137.0221, -108.2113)(136.6654, -107.4269)
\qbezier(136.6654, -107.4269)(136.4034, -106.8509)(136.0370, -106.0254)
\qbezier(136.0370, -106.0254)(135.8294, -105.5575)(135.4208, -104.6239)
\qbezier(135.4208, -104.6239)(135.7608, -105.4008)(134.8094, -103.2223)
\qbezier(134.8094, -103.2223)(134.3542, -102.1799)(134.1959, -101.8208)
\qbezier(134.1959, -101.8208)(133.8201, -100.9685)(133.5729, -100.4193)
\qbezier(133.5729, -100.4193)(133.2144, -99.6227)(132.9335, -99.0178)
\qbezier(132.9335, -99.0178)(132.5765, -98.2487)(132.2709, -97.6162)
\qbezier(132.2709, -97.6162)(131.9076, -96.8642)(131.5785, -96.2147)
\qbezier(131.5785, -96.2147)(131.2038, -95.4750)(130.8503, -94.8132)
\qbezier(130.8503, -94.8132)(130.4605, -94.0833)(130.0808, -93.4116)
\qbezier(130.0808, -93.4116)(129.6732, -92.6904)(129.2656, -92.0101)
\qbezier(129.2656, -92.0101)(128.8382, -91.2968)(128.4010, -90.6086)
\qbezier(128.4010, -90.6086)(127.9528, -89.9029)(127.4849, -89.2070)
\qbezier(127.4849, -89.2070)(127.0154, -88.5088)(126.5163, -87.8055)
\qbezier(126.5163, -87.8055)(126.0259, -87.1145)(125.4957, -86.4040)
\qbezier(125.4957, -86.4040)(124.9852, -85.7200)(124.4248, -85.0025)
\qbezier(124.4248, -85.0025)(123.8958, -84.3251)(123.3067, -83.6009)
\qbezier(123.3067, -83.6009)(122.7607, -82.9298)(122.1451, -82.1994)
\qbezier(122.1451, -82.1994)(121.5842, -81.5340)(120.9444, -80.7979)
\qbezier(120.9444, -80.7979)(120.3705, -80.1376)(119.7093, -79.3963)
\qbezier(119.7093, -79.3963)(119.1243, -78.7405)(118.4444, -77.9948)
\qbezier(118.4444, -77.9948)(117.8501, -77.3429)(117.1542, -76.5933)
\qbezier(117.1542, -76.5933)(116.5522, -75.9447)(115.8429, -75.1918)
\qbezier(115.8429, -75.1918)(115.2344, -74.5459)(114.5139, -73.7902)
\qbezier(114.5139, -73.7902)(113.9003, -73.1467)(113.1705, -72.3887)
\qbezier(113.1705, -72.3887)(112.5527, -71.7470)(111.8154, -70.9872)
\qbezier(111.8154, -70.9872)(111.1942, -70.3471)(110.4508, -69.5856)
\qbezier(110.4508, -69.5856)(109.8269, -68.9468)(109.0785, -68.1841)
\qbezier(109.0785, -68.1841)(108.4526, -67.5462)(107.7002, -66.7826)
\qbezier(107.7002, -66.7826)(107.0726, -66.1455)(106.3171, -65.3810)
\qbezier(106.3171, -65.3810)(105.6880, -64.7446)(104.9300, -63.9795)
\qbezier(104.9300, -63.9795)(104.2999, -63.3436)(103.5399, -62.5780)
\qbezier(103.5399, -62.5780)(102.9090, -61.9424)(102.1474, -61.1765)
\qbezier(102.1474, -61.1765)(101.5158, -60.5412)(100.7530, -59.7749)
\qbezier(100.7530, -59.7749)(100.0550, -59.0742)( 99.3571, -58.3734)
\qbezier( 99.3571, -58.3734)( 98.6585, -57.6726)( 97.9599, -56.9719)
\qbezier( 97.9599, -56.9719)( 97.2609, -56.2711)( 96.5619, -55.5703)
\qbezier( 96.5619, -55.5703)( 95.8625, -54.8696)( 95.1631, -54.1688)
\qbezier( 95.1631, -54.1688)( 94.4634, -53.4680)( 93.7637, -52.7673)
\qbezier( 93.7637, -52.7673)( 93.0638, -52.0665)( 92.3638, -51.3658)
\qbezier( 92.3638, -51.3658)( 91.6637, -50.6650)( 90.9636, -49.9642)
\put(155.9900, -134.0560){\line(1, 0){ 26.5401}}
\put(153.3369, -131.2529){\line(1, 0){ 26.3902}}
\put(150.7634, -128.4499){\line(1, 0){ 26.1606}}
\put(148.3023, -125.6468){\line(1, 0){ 25.8186}}
\put(145.9905, -122.8438){\line(1, 0){ 25.3274}}
\put(143.8617, -120.0407){\line(1, 0){ 24.6532}}
\put(141.9375, -117.2376){\line(1, 0){ 23.7742}}
\put(140.2203, -114.4346){\line(1, 0){ 22.6884}}
\put(138.6910, -111.6315){\line(1, 0){ 21.4147}}
\put(137.3127, -108.8285){\line(1, 0){ 19.9899}}
\put(136.0370, -106.0254){\line(1, 0){ 18.4625}}
\put(134.8094, -103.2223){\line(1, 0){ 16.8870}}
\put(133.5729, -100.4193){\line(1, 0){ 15.3205}}
\put(132.2709, -97.6162){\line(1, 0){ 13.8195}}
\put(130.8503, -94.8132){\line(1, 0){ 12.4370}}
\put(129.2656, -92.0101){\line(1, 0){ 11.2187}}
\put(127.4849, -89.2070){\line(1, 0){ 10.1963}}
\put(123.3067, -83.6009){\line(1, 0){  8.7684}}
\put(120.9444, -80.7979){\line(1, 0){  8.3276}}
\put(118.4444, -77.9948){\line(1, 0){  8.0245}}
\put(115.8429, -75.1918){\line(1, 0){  7.8230}}
\put(113.1705, -72.3887){\line(1, 0){  7.6923}}
\put(110.4508, -69.5856){\line(1, 0){  7.6090}}
\put(107.7002, -66.7826){\line(1, 0){  7.5565}}
\put(104.9300, -63.9795){\line(1, 0){  7.5236}}
\put(102.1474, -61.1765){\line(1, 0){  7.5032}}
\put( 99.3571, -58.3734){\line(1, 0){  7.4905}}
\put( 96.5619, -55.5703){\line(1, 0){  7.4826}}
\put( 93.7637, -52.7673){\line(1, 0){  7.4777}}
\put( 90.9636, -49.9642){\line(1, 0){  7.4747}}
\put( 84.9569, -86.5259){\vector(1, 0){100.0862}}
\put(135.0000, -145.8793){\vector(0, 1){109.3966}}
\put( 98.4384, -49.9642){\line(1, -1){ 84.0918}}
\put(189.4655, -88.2716){\makebox(0,0)[]{\footnotesize{$\rho$}}}
\put(131.2759, -35.9009){\makebox(0,0)[]{\footnotesize{$\tau$}}}
\end{picture}
\vspace*{-6mm} \newline
\hspace*{21.5mm} (a) \hspace{45.8mm} (b) \hspace{45.8mm} (c)
\vspace*{4mm} \newline
%
%
\vspace*{-4mm} \newline
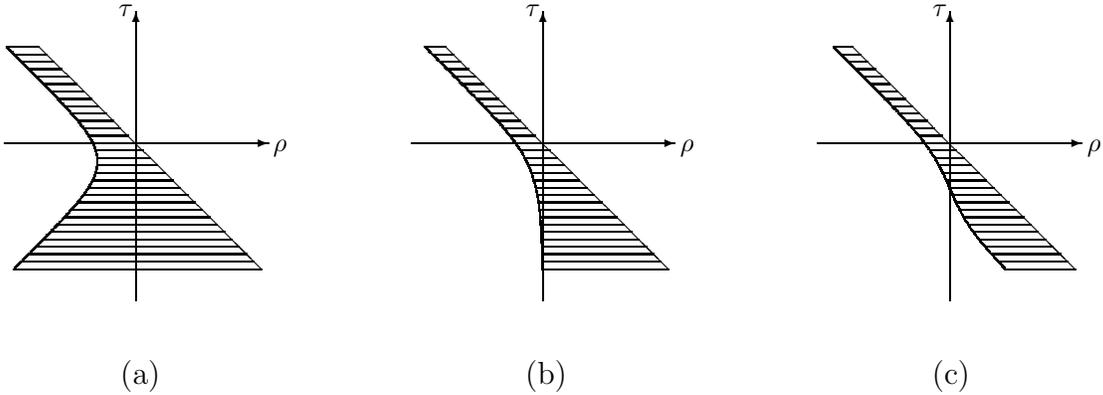
{\footnotesize \sf Figure 5. The hatched region represents the WLBR
spacetime in the $(\tau,\rho)$ coordinate system. The left boundaries
are the world lines of a fixed point on the domain wall
$R = R_{\mbox{\tiny $Q$}}$ as given by equation \eqref{e_1012}.
The cases $B > R_{\mbox{\tiny $Q$}}$, $B = R_{\mbox{\tiny $Q$}}$
and $B < R_{\mbox{\tiny $Q$}}$ are shown in (a), (b) and (c)
respectively.}
\vspace{3mm} \newline
\itm The coordinate velocity of the domain wall is
\begin{equation} \label{e_1196}
\frl{d \rho}{d \tau} = \frl{R_{\mbox{\tiny $Q$}} \p B \tanh \tau}
{\sqrt{(R_{\mbox{\tiny $Q$}} / \cosh \tau)^{2}
\p (B \p R_{\mbox{\tiny $Q$}} \tanh \tau)^2}}
\mbox{ .}
\end{equation}
Using equation \eqref{e_1196} and looking at Figure 5(a) we see that
in the case $B > R_{\mbox{\tiny $Q$}}$ the domain wall initially has
velocity in the positive $\rho$-direction with decelerating motion.
It stops at an event given by
\begin{equation} \label{e_1190}
{\tau}_1 = - \mbox{arccoth} \hs{0.5mm} (B / R_{\mbox{\tiny $Q$}})
\mbox{\hspace{2mm} , \hspace{3mm}}
{\rho}_1 = - \mbox{arcsinh} \hs{0.5mm}
\sqrt{\frl{B \m R_{\mbox{\tiny $Q$}}}{B \p R_{\mbox{\tiny $Q$}}}}
\mbox{ ,}
\end{equation}
and moves in the negative $\rho$-direction. This means that the
$(\tau,\rho)$-system accelerates in the positive $R$-direction.
From equation \eqref{e_1196} it follows that
\begin{equation} \label{e_1197}
\lim_{\tau \rightarrow - \infty} \frl{d \rho}{d \tau}
= \frl{B \m R_{\mbox{\tiny $Q$}}}{|B \m R_{\mbox{\tiny $Q$}}|}
\mbox{\hspace{2mm} , \hspace{3mm}}
\lim_{\tau \rightarrow \infty} \frl{d \rho}{d \tau} = -1
\end{equation}
when $B \ne R_{\mbox{\tiny $Q$}}$.
Hence we see that when $B > R_{\mbox{\tiny $Q$}}$ the domain wall has
initially a velocity close to the velocity of light in the positive
$\rho$-direction, and finally the same velocity in the negative
$\rho$-direction.
\itm Next we consider the case $B = R_{\mbox{\tiny $Q$}}$. Then the
expression for the velocity of the domain wall can be written
\begin{equation} \label{e_1198}
\frl{d \rho}{d \tau} = \frl{e^{\tau}}{\sqrt{1 \p e^{2 \tau}}}
\mbox{ .}
\end{equation}
From this equation it follows that
\begin{equation} \label{e_1199}
\lim_{\tau \rightarrow - \infty} \frl{d \rho}{d \tau} = 0
\mbox{\hspace{2mm} , \hspace{3mm}}
\lim_{\tau \rightarrow \infty} \frl{d \rho}{d \tau} = -1
\end{equation}
In this case the domain wall has initially a vanishing coordinate
velocity, but it accelerates slowly in the negative $\rho$-direction
and ends up approaching the velocity of light.
\itm We finally consider the case $0 < B < R_{\mbox{\tiny $Q$}}$.
In this case the motion of the domain wall is more complicated.
In the limit that $\tau \rightarrow -\infty$ the domain wall moves
nearly with the velocity of light in the negative $\rho$-direction.
Then it decelerates and obtains a minimal velocity
$-\sqrt{1 - (B/R_{\mbox{\tiny $Q$}})^2}$ when it passes $\rho = 0$.
Afterwards it accelerates again and approaches the velocity of light
in the infinite future.
\itm Since the domain wall is at rest in the CFS system, all of this
reflects the motion of the reference frame in which the
$(\tau,\rho)$-coordinates are comoving.
\itm We shall find the transformation relating the line elements
\eqref{e_218} and \eqref{e_928}.
Combining the generating functions \eqref{e_980} with the inverse of
the generating function \eqref{e_339} with $k = 1$, we obtain the
generating functions
\begin{equation} \label{e_834}
f(x) = - 2 \arctan (\coth \frl{x}{2})
\mbox{\hspace{2mm} , \hspace{3mm}}
g(x) = 2 \arctan (\tanh \frl{x}{2})
\end{equation}
which give the transformation
%
\begin{equation} \label{e_835}
\cot \eta = - \frl{\sinh \tau}{\cosh \rho}
\mbox{\hspace{2mm} , \hspace{3mm}}
\cot \chi = - \frl{\sinh \rho}{\cosh \tau}
\mbox{ ,}
\end{equation}
%
as shown in Appendix B. This transforms
the region $\tau + \rho < 0$ in the $(\tau,\rho)$-system to
the region $\pi / 2 < \eta + \chi < \pi$, $|\eta - \chi| < \pi / 2$
in the $(\eta,\chi)$-system, and the region $\tau + \rho > 0$ in the
$(\tau,\rho)$-system to the region $- \pi < \eta + \chi < - \pi / 2 \,$,
$|\eta - \chi| < \pi / 2$ in the $(\eta,\chi)$-system.
\itm The inverse transformation is found in a similar way using the
generating functions
\begin{equation} \label{e_837}
f^{-1}(x) = - 2 \hs{0.9mm} \mbox{arctanh} (\cot \frl{x}{2})
\mbox{\hspace{2mm} , \hspace{3mm}}
g^{-1}(x) = 2 \hs{0.9mm} \mbox{arctanh} (\tan \frl{x}{2})
\end{equation}
which give the transformation
%
\begin{equation} \label{e_838}
\tanh \tau = - \frl{\cos \eta}{\sin \chi}
\mbox{\hspace{2mm} , \hspace{3mm}}
\tanh \rho = - \frl{\cos \chi}{\sin \eta}
\mbox{ .}
\end{equation}
%
In the present case the transformation \eqref{e_835} and its inverse
can also be found from the equations \eqref{e_279}, \eqref{e_346},
\eqref{e_981} and \eqref{e_946}. Combining the first equation in
\eqref{e_946} and \eqref{e_346} for $k = 1$ and substituting for
$R/T$ from \eqref{e_981} we get
%
\begin{equation} \label{e_1938}
\cot \eta = - \frl{R}{T} \tanh \tau = - \frl{\sinh \tau}{\cosh \rho}
\mbox{ .}
\end{equation}
%
Note that the hyperbola $\chi = \pi / 2$ in Figure 3 corresponds to
$\rho = 0$.
\itm As shown above a free particle has constant $\rho$, say
$\rho = {\rho}_1$. Hence it follows from the second of the transformation
equations \eqref{e_838} that the world line of a free particle as
described in the $(\eta,\chi)$-system is given by
%
\begin{equation} \label{e_996}
\cos \chi = k_1 \sin \eta
\mbox{\hspace{2mm} , \hspace{3mm}}
k_1 = - \tanh {\rho}_1
\mbox{ .}
\end{equation}
%
We will now show that this is a solution of the Lagrangian equation
for a free particle moving radially. With the line element \eqref{e_218}
and $k = 1$ the Lagrangian is
%
\begin{equation} \label{e_997}
L = \frl{R_{\mbox{\tiny $Q$}}^2}{2 \hs{0.2mm} \sin^2 \chi}
\hs{0.5mm} (- \dot{\eta}^2 + \dot{\chi}^2)
\mbox{ .}
\end{equation}
%
The conserved momentum conjugate to the time coordinate is
%
\begin{equation} \label{e_998}
p_{\eta} = \frl{\partial L}{\partial \dot{\eta}}
= - \frl{R_{\mbox{\tiny $Q$}}^2}{\sin^2 \chi} \hs{0.8mm} \dot{\eta}
\mbox{ ,}
\end{equation}
%
giving
%
\begin{equation} \label{e_999}
\dot{\eta} = - (p_{\eta} / R_{\mbox{\tiny $Q$}}^2) \sin^2 \chi
\mbox{ .}
\end{equation}
%
The 4-velocity identity then takes the form
%
\begin{equation} \label{e_1000}
\dot{\eta}^2 = (1 / R_{\mbox{\tiny $Q$}}^2) \sin^2 \chi + \dot{\chi}^2
\mbox{ .}
\end{equation}
%
The last two equations lead to
%
\begin{equation} \label{e_1001}
\dot{\chi} - (1 / R_{\mbox{\tiny $Q$}})
\sqrt{(p_{\eta} / R_{\mbox{\tiny $Q$}})^2 \sin^2 \chi - 1}
\hs{0.5mm} \sin \chi = 0
\mbox{ .}
\end{equation}
%
We now transform from differentiation with respect to the proper time of
the particle to differentiation with respect to the coordinate time
$\eta$ by means of equation \eqref{e_999}, and find that the
solution of this differential equation with the initial condition
$\chi (0) = \pi / 2$ is
%
\begin{equation} \label{e_1002}
\cos \chi = - \sqrt{1 - (R_{\mbox{\tiny $Q$}} / p_{\eta})^2} \sin \eta
\mbox{ .}
\end{equation}
%
This is in accordance with equation \eqref{e_996} if the conserved
energy of the particle is
%
\begin{equation} \label{e_1003}
p_{\eta} = - R_{\mbox{\tiny $Q$}} \cosh {\rho}_1
\mbox{ .}
\end{equation}
\vspace{-9mm} \newline
\itm Inserting the expression \eqref{e_919} for $e^{\alpha }$ with
$a = 1$ and ${\tau}_0 = 0$, and \eqref{e_981} for $R$ into the line
element \eqref{e_1211}, we obtain the form of the line element for
the Minkowski spacetime inside the domain wall in the $(\tau,\rho)$
coordinates,
\begin{equation} \label{e_1928}
ds_M^2 = \frl{B^2}{(\sinh \tau \p \sinh \rho)^2}
\hs{0.5mm} ( \hs{0.5mm} \mbox{$- d{\tau}^2$} + d{\rho}^2
+ \cosh^2 \tau \hs{0.7mm} d\Omega^2)
\mbox{ .}
\end{equation}
It follows from equation \eqref{e_928}, the last of equations
\eqref{e_981} with $R = R_{\mbox{\tiny $Q$}}$, and the line element
\eqref{e_1928} that the metric is continuous at the shell.
\itm In the $(\tau,\rho)$-system the acceleration of gravity is
\begin{equation} \label{e_2928}
a^{\hat{\rho}} = \frl{1}{B} \cosh \rho
\end{equation}
which is positive. Hence an observer at rest in this coordinate system
experiences an acceleration of gravity in the outwards direction in the
flat spacetime inside the wall, which means that the reference frame of
these coordinates is accelerating in the inwards direction.
%
%
%
\vspace{6mm} \newline
{\it IIa. Static metric and coordinates $(\tilde{t},\tilde{r})$
with $\beta (\tilde{r}) = - \alpha (\tilde{r})$.}
\vspace{3mm} \newline
In this case equation \eqref{e_212} reduces to
\begin{equation} \label{e_224}
\alpha'' + 2 \hs{0.4mm} \alpha' \hs{0.5mm}^2
= \frl{e^{-2 \alpha}}{R_{\mbox{\tiny $Q$}}^2}
\mbox{ ,}
\end{equation}
which may be written
\begin{equation} \label{e_220}
(e^{2 \alpha})'' = \frl{2}{R_{\mbox{\tiny $Q$}}^2}
\mbox{ .}
\end{equation}
The general solution of this equation can be written as
\begin{equation} \label{e_444}
e^{2 \alpha} = D + \left( \frl{\tilde{r} \m \tilde{r}_0}
{R_{\mbox{\tiny $Q$}}} \right)^2
\mbox{ ,}
\end{equation}
where $D$ and $\tilde{r}_0$ are constants.
The line element \eqref{e_211} then takes the form
\vspace{-1mm} \newline
\begin{equation} \label{e_452}
ds^2 = - \mbox{$\left[ \rule[-1.5mm]{0mm}{5.7mm} \right.$} \hs{-0.2mm}
D + \left( \frl{\tilde{r} \m \tilde{r}_0}
{R_{\mbox{\tiny $Q$}}} \right)^2
\hs{0.0mm} \mbox{$\left. \rule[-1.5mm]{0mm}{5.7mm} \right]$}
\hs{0.8mm} d\tilde{t}^2
+ \mbox{$\left[ \rule[-1.5mm]{0mm}{5.7mm} \right.$} \hs{-0.2mm}
D + \left( \frl{\tilde{r} \m \tilde{r}_0}
{R_{\mbox{\tiny $Q$}}} \right)^2
\hs{0.0mm} \mbox{$\left. \rule[-1.5mm]{0mm}{5.7mm} \right]$}^{-1}
\hs{0.0mm} d\tilde{r}^2
+ R_{\mbox{\tiny $Q$}}^2 d\Omega^2
\mbox{ ,}
\end{equation}
\vspace{-1mm} \newline
This form of the line element with $R_{\mbox{\tiny $Q$}} = 1$,
$\tilde{r}_0 = 0$ and $D = \pm 1$ has been used by Ottewill and
Taylor [\ref{r_13}] and by V.\hn Cardoso, O.\hn J.\hn C.\hn Dias and
J.\hn P.\hn S.\hn Lemos [\ref{r_17}]. It may further be
noted that A.\hn S.\hn Lapedes [\ref{r_18}] has studied particle
creation in the LBR spacetime. He then considered three forms of the
line element \eqref{e_452} with $\tilde{r}_0 = 0$ and $D = - 1,1,0$
respectively, with a rescaling of $\tilde{t}$ and $\tilde{r}$ by
$R_{\mbox{\tiny $Q$}}$, and constructed a Penrose diagram for the LBR
spacetime. The coordinate clocks showing $\tilde{t}$ go at the same rate
as a standard clock at $\tilde{r} = \tilde{r}_0$ scaled by
the factor $\sqrt{D}$ when $D > 0$. For $D = 1$ and $\tilde{r}_0 = 0$
the components $g_{\tilde{t} \tilde{t}}$ and
$g_{\tilde{r} \tilde{r}}$ have the same form as the corresponding
components of the anti De Sitter metric in static coordinates, while
the angular part of the line element represents a spherical surface.
\itm Writing $D = k A^2$ where $k = \mbox{sgn} (D)$ and
\begin{equation} \label{e_692}
A = \left\{ \begin{array}{lcl}
\rule[-2mm]{0mm}{6.0mm} \sqrt{|D|} & \mbox{when} & k \ne 0 \\
   \hs{0.3mm} R_{\mbox{\tiny $Q$}} & \mbox{when} & k = 0
\end{array} \right.
\mbox{ ,}
\end{equation}
the transformation between
the $(\tilde{t},\tilde{r})$- and the $(\eta,\chi)$-system used in
the line element \eqref{e_218} is given by
\begin{equation} \label{e_349}
I_k (\chi) = \frl{\tilde{r}_0 \m \tilde{r}}
{R_{\mbox{\tiny $Q$}} A}
\mbox{\hspace{2mm} , \hspace{3mm}}
\eta = \frl{A}{R_{\mbox{\tiny $Q$}}} \hs{0.6mm} \tilde{t}
\mbox{ .}
\end{equation}
The transformation has been chosen so that $\chi$ and $\tilde{r}$
increases in the same direction. The transformation from $\chi$ to
$\tilde{r}$ is
\begin{equation} \label{e_350}
\tilde{r} = \tilde{r}_0
- R_{\mbox{\tiny $Q$}} A \hs{0.5mm} I_k(\chi)
\mbox{ .}
\end{equation}
It follows that that the $(\tilde{t},\tilde{r})$-system is
comoving with the reference particles of the same reference frame as
the $(\eta,\chi)$-system.
Inserting equation \eqref{e_350} into equation \eqref{e_444} and
using the relation \eqref{e_812} we obtain
\begin{equation} \label{e_351}
e^{2 \alpha} = A^2 \hs{0.5mm} S_k(\chi)^{-2}
\mbox{ .}
\end{equation}
Note that this expression in consistent with the line element
\eqref{e_218} due to the relation between $\eta$ and $\tilde{t}$
in the transformation \eqref{e_349}.
%
%
Differentiating equation \eqref{e_350} we get
\begin{equation} \label{e_278}
d \tilde{r} = R_{\mbox{\tiny $Q$}} \hs{0.5mm} A
\hs{0.5mm} S_k(\chi)^{-2} d \chi
\mbox{ .}
\end{equation}
Using \eqref{e_351} and \eqref{e_278} we see that the line element
\eqref{e_452} takes the form \eqref{e_218}.
\itm It follows from equations \eqref{e_555} and \eqref{e_350} for
$k = 1$ that in this case the WLBR spacetime is represented by the
hatched region in Figure 6 given by
\begin{equation} \label{e_1277}
-\pi R_{\mbox{\tiny $Q$}} / A < \tilde{t} <
\pi R_{\mbox{\tiny $Q$}} / A
\end{equation}
\vspace{-9mm} \newline
and
\vspace{-4mm} \newline
\begin{equation} \label{e_667}
\tilde{r}_0 - R_{\mbox{\tiny $Q$}} A \hs{0.5mm} \cot
\left( {\chi}_{\mbox{\tiny $Q$}} +
\arcsin \left( \sin {\chi}_{\mbox{\tiny $Q$}} \cos \eta \right)
\right) < \tilde{r}
< \tilde{r}_0 + R_{\mbox{\tiny $Q$}} A \hs{0.5mm}
\cot |\eta|
\mbox{ ,}
\end{equation}
where $\eta = (A / R_{\mbox{\tiny $Q$}}) \hspace{0.5mm}
\tilde{t}$ and
${\chi}_{\mbox{\tiny $Q$}} = \arccot (B / R_{\mbox{\tiny $Q$}})$.
\vspace*{5mm} \newline
\begin{picture}(50,212)(-97,-170)
\qbezier( 32.4930, -104.0205)( 46.5732, -103.4904)( 64.5891, -102.4030)
\qbezier( 64.5891, -102.4030)( 73.9694, -101.8368)( 85.6393, -100.7855)
\qbezier( 85.6393, -100.7855)( 92.2352, -100.1912)(100.2724, -99.1679)
\qbezier(100.2724, -99.1679)(105.1151, -98.5513)(110.9236, -97.5504)
\qbezier(110.9236, -97.5504)(114.6078, -96.9155)(118.9730, -95.9329)
\qbezier(118.9730, -95.9329)(121.8609, -95.2828)(125.2496, -94.3154)
\qbezier(125.2496, -94.3154)(127.5719, -93.6524)(130.2760, -92.6978)
\qbezier(130.2760, -92.6978)(132.1855, -92.0238)(134.3949, -91.0803)
\qbezier(134.3949, -91.0803)(135.9958, -90.3966)(137.8383, -89.4628)
\qbezier(137.8383, -89.4628)(139.2040, -88.7706)(140.7682, -87.8453)
\qbezier(140.7682, -87.8453)(141.9514, -87.1453)(143.3003, -86.2277)
\qbezier(143.3003, -86.2277)(144.3398, -85.5206)(145.5188, -84.6102)
\qbezier(145.5188, -84.6102)(146.4437, -83.8961)(147.4861, -82.9927)
\qbezier(147.4861, -82.9927)(148.3187, -82.2713)(149.2492, -81.3752)
\qbezier(149.2492, -81.3752)(150.0067, -80.6457)(150.8435, -79.7576)
\qbezier(150.8435, -79.7576)(151.5396, -79.0188)(152.2958, -78.1401)
\qbezier(152.2958, -78.1401)(152.9413, -77.3900)(153.6262, -76.5226)
\qbezier(153.6262, -76.5226)(154.2292, -75.7588)(154.8493, -74.9051)
\qbezier(154.8493, -74.9051)(155.4157, -74.1252)(155.9751, -73.2875)
\qbezier(155.9751, -73.2875)(156.5078, -72.4898)(157.0095, -71.6700)
\qbezier(157.0095, -71.6700)(157.5087, -70.8543)(157.9548, -70.0525)
\qbezier(157.9548, -70.0525)(158.4175, -69.2209)(158.8096, -68.4350)
\qbezier(158.8096, -68.4350)(159.2304, -67.5914)(159.5691, -66.8174)
\qbezier(159.5691, -66.8174)(159.9412, -65.9674)(160.2260, -65.1999)
\qbezier(160.2260, -65.1999)(160.5418, -64.3492)(160.7706, -63.5824)
\qbezier(160.7706, -63.5824)(161.0230, -62.7366)(161.1920, -61.9649)
\qbezier(161.1920, -61.9649)(161.3753, -61.1286)(161.4800, -60.3473)
\qbezier(161.4800, -60.3473)(161.5905, -59.5240)(161.6263, -58.7298)
\qbezier(161.6263, -58.7298)(161.6629, -57.9211)(161.6263, -57.1123)
\qbezier(161.6263, -57.1123)(161.5905, -56.3182)(161.4800, -55.4948)
\qbezier(161.4800, -55.4948)(161.3753, -54.7135)(161.1920, -53.8772)
\qbezier(161.1920, -53.8772)(161.0230, -53.1055)(160.7706, -52.2597)
\qbezier(160.7706, -52.2597)(160.5418, -51.4929)(160.2260, -50.6422)
\qbezier(160.2260, -50.6422)(159.9412, -49.8747)(159.5691, -49.0247)
\qbezier(159.5691, -49.0247)(159.2304, -48.2507)(158.8096, -47.4071)
\qbezier(158.8096, -47.4071)(158.4175, -46.6212)(157.9548, -45.7896)
\qbezier(157.9548, -45.7896)(157.5087, -44.9878)(157.0095, -44.1721)
\qbezier(157.0095, -44.1721)(156.5078, -43.3523)(155.9751, -42.5546)
\qbezier(155.9751, -42.5546)(155.4157, -41.7169)(154.8493, -40.9370)
\qbezier(154.8493, -40.9370)(154.2292, -40.0833)(153.6262, -39.3195)
\qbezier(153.6262, -39.3195)(152.9413, -38.4521)(152.2958, -37.7020)
\qbezier(152.2958, -37.7020)(151.5396, -36.8233)(150.8435, -36.0845)
\qbezier(150.8435, -36.0845)(150.0067, -35.1964)(149.2492, -34.4669)
\qbezier(149.2492, -34.4669)(148.3187, -33.5709)(147.4861, -32.8494)
\qbezier(147.4861, -32.8494)(146.4437, -31.9461)(145.5188, -31.2319)
\qbezier(145.5188, -31.2319)(144.3398, -30.3215)(143.3003, -29.6144)
\qbezier(143.3003, -29.6144)(141.9514, -28.6968)(140.7682, -27.9968)
\qbezier(140.7682, -27.9968)(139.2040, -27.0715)(137.8383, -26.3793)
\qbezier(137.8383, -26.3793)(135.9958, -25.4455)(134.3949, -24.7618)
\qbezier(134.3949, -24.7618)(132.1855, -23.8183)(130.2760, -23.1443)
\qbezier(130.2760, -23.1443)(127.5719, -22.1897)(125.2496, -21.5267)
\qbezier(125.2496, -21.5267)(121.8609, -20.5593)(118.9730, -19.9092)
\qbezier(118.9730, -19.9092)(114.6078, -18.9266)(110.9236, -18.2917)
\qbezier(110.9236, -18.2917)(105.1151, -17.2908)(100.2724, -16.6742)
\qbezier(100.2724, -16.6742)( 92.2352, -15.6509)( 85.6393, -15.0566)
\qbezier( 85.6393, -15.0566)( 73.9694, -14.0053)( 64.5891, -13.4391)
\qbezier( 64.5891, -13.4391)( 46.5732, -12.3517)( 32.4930, -11.8216)
\put( 32.4930, -109.5836){\line(1, 0){ 54.3602}}
\put( 32.4930, -106.3486){\line(1, 0){ 88.8680}}
\put( 52.2545, -103.1135){\line(1, 0){ 83.2819}}
\put( 94.4461, -99.8785){\line(1, 0){ 49.2263}}
\put(115.6954, -96.6434){\line(1, 0){ 33.5573}}
\put(128.1969, -93.4084){\line(1, 0){ 25.3655}}
\put(136.3972, -90.1733){\line(1, 0){ 20.8085}}
\put(142.2311, -86.9383){\line(1, 0){ 18.2973}}
\put(146.6496, -83.7032){\line(1, 0){ 17.1254}}
\put(150.1621, -80.4682){\line(1, 0){ 17.0062}}
\put(153.0557, -77.2331){\line(1, 0){ 17.9181}}
\put(155.4920, -73.9981){\line(1, 0){ 20.1023}}
\put(157.5506, -70.7630){\line(1, 0){ 24.2267}}
\put(159.2476, -67.5280){\line(1, 0){ 31.9746}}
\put(160.5459, -64.2929){\line(1, 0){ 48.4757}}
\put(161.3706, -61.0579){\line(1, 0){ 88.5706}}
\put(161.3355, -54.5878){\line(1, 0){ 85.7635}}
\put(160.4797, -51.3527){\line(1, 0){ 46.9934}}
\put(159.1553, -48.1177){\line(1, 0){ 31.3391}}
\put(157.4358, -44.8826){\line(1, 0){ 23.8982}}
\put(155.3553, -41.6476){\line(1, 0){ 19.9241}}
\put(152.8942, -38.4125){\line(1, 0){ 17.8304}}
\put(149.9686, -35.1775){\line(1, 0){ 16.9850}}
\put(146.4108, -31.9424){\line(1, 0){ 17.1651}}
\put(141.9239, -28.7074){\line(1, 0){ 18.4070}}
\put(135.9798, -25.4723){\line(1, 0){ 21.0166}}
\put(127.5881, -22.2373){\line(1, 0){ 25.7360}}
\put(114.7217, -19.0022){\line(1, 0){ 34.2366}}
\put( 92.6803, -15.7672){\line(1, 0){ 50.5900}}
\put( 48.4005, -12.5321){\line(1, 0){ 86.5047}}
\put( 32.4930,  -9.2971){\line(1, 0){ 87.6299}}
\put( 32.4930,  -6.0620){\line(1, 0){ 50.5007}}
\qbezier(258.4386, -54.6971)(258.2880, -54.6937)(238.7898, -53.9010)
\qbezier(258.4386, -61.1450)(258.2880, -61.1484)(238.7898, -61.9411)
\qbezier(238.7898, -53.9010)(237.5228, -53.8495)(225.5472, -53.1050)
\qbezier(238.7898, -61.9411)(237.5228, -61.9926)(225.5472, -62.7371)
\qbezier(225.5472, -53.1050)(224.0491, -53.0118)(215.9839, -52.3089)
\qbezier(225.5472, -62.7371)(224.0491, -62.8303)(215.9839, -63.5332)
\qbezier(215.9839, -52.3089)(214.5232, -52.1816)(208.7280, -51.5128)
\qbezier(215.9839, -63.5332)(214.5232, -63.6605)(208.7280, -64.3293)
\qbezier(208.7280, -51.5128)(207.3832, -51.3576)(203.0138, -50.7167)
\qbezier(208.7280, -64.3293)(207.3832, -64.4845)(203.0138, -65.1254)
\qbezier(203.0138, -50.7167)(201.7995, -50.5386)(198.3804, -49.9207)
\qbezier(203.0138, -65.1254)(201.7995, -65.3035)(198.3804, -65.9214)
\qbezier(198.3804, -49.9207)(197.2897, -49.7235)(194.5335, -49.1246)
\qbezier(198.3804, -65.9214)(197.2897, -66.1186)(194.5335, -66.7175)
\qbezier(194.5335, -49.1246)(193.5529, -48.9115)(191.2765, -48.3285)
\qbezier(194.5335, -66.7175)(193.5529, -66.9306)(191.2765, -67.5136)
\qbezier(191.2765, -48.3285)(190.3917, -48.1019)(188.4729, -47.5324)
\qbezier(191.2765, -67.5136)(190.3917, -67.7402)(188.4729, -68.3097)
\qbezier(188.4729, -47.5324)(187.6707, -47.2944)(186.0249, -46.7364)
\qbezier(188.4729, -68.3097)(187.6707, -68.5478)(186.0249, -69.1058)
\qbezier(186.0249, -46.7364)(185.2938, -46.4885)(183.8609, -45.9403)
\qbezier(186.0249, -69.1058)(185.2938, -69.3536)(183.8609, -69.9018)
\qbezier(183.8609, -45.9403)(183.1909, -45.6840)(181.9267, -45.1442)
\qbezier(183.8609, -69.9018)(183.1909, -70.1581)(181.9267, -70.6979)
\qbezier(181.9267, -45.1442)(181.3096, -44.8807)(180.1809, -44.3481)
\qbezier(181.9267, -70.6979)(181.3096, -70.9614)(180.1809, -71.4940)
\qbezier(180.1809, -44.3481)(179.6097, -44.0786)(178.5912, -43.5520)
\qbezier(180.1809, -71.4940)(179.6097, -71.7635)(178.5912, -72.2901)
\qbezier(178.5912, -43.5520)(178.0600, -43.2774)(177.1320, -42.7560)
\qbezier(178.5912, -72.2901)(178.0600, -72.5647)(177.1320, -73.0861)
\qbezier(177.1320, -42.7560)(176.6358, -42.4771)(175.7826, -41.9599)
\qbezier(177.1320, -73.0861)(176.6358, -73.3650)(175.7826, -73.8822)
\qbezier(175.7826, -41.9599)(175.3173, -41.6778)(174.5262, -41.1638)
\qbezier(175.7826, -73.8822)(175.3173, -74.1643)(174.5262, -74.6783)
\qbezier(174.5262, -41.1638)(174.0884, -40.8794)(173.3490, -40.3677)
\qbezier(174.5262, -74.6783)(174.0884, -74.9628)(173.3490, -75.4744)
\qbezier(173.3490, -40.3677)(172.9358, -40.0819)(172.2393, -39.5717)
\qbezier(173.3490, -75.4744)(172.9358, -75.7602)(172.2393, -76.2704)
\qbezier(172.2393, -39.5717)(171.8486, -39.2855)(171.1874, -38.7756)
\qbezier(172.2393, -76.2704)(171.8486, -76.5566)(171.1874, -77.0665)
\qbezier(171.1874, -38.7756)(170.8175, -38.4903)(170.1851, -37.9795)
\qbezier(171.1874, -77.0665)(170.8175, -77.3518)(170.1851, -77.8626)
\qbezier(170.1851, -37.9795)(169.8348, -37.6966)(169.2250, -37.1834)
\qbezier(170.1851, -77.8626)(169.8348, -78.1455)(169.2250, -78.6587)
\qbezier(169.2250, -37.1834)(168.8940, -36.9049)(168.3010, -36.3874)
\qbezier(169.2250, -78.6587)(168.8940, -78.9372)(168.3010, -79.4547)
\qbezier(168.3010, -36.3874)(167.9897, -36.1156)(167.4076, -35.5913)
\qbezier(168.3010, -79.4547)(167.9897, -79.7265)(167.4076, -80.2508)
\qbezier(167.4076, -35.5913)(167.1176, -35.3301)(166.5397, -34.7952)
\qbezier(167.4076, -80.2508)(167.1176, -80.5121)(166.5397, -81.0469)
\qbezier(166.5397, -34.7952)(166.2748, -34.5500)(165.6931, -33.9991)
\qbezier(166.5397, -81.0469)(166.2748, -81.2922)(165.6931, -81.8430)
\qbezier(165.6931, -33.9991)(165.4606, -33.7789)(164.8636, -33.2031)
\qbezier(165.6931, -81.8430)(165.4606, -82.0632)(164.8636, -82.6391)
\qbezier(164.8636, -33.2031)(164.6786, -33.0246)(164.0475, -32.4070)
\qbezier(164.8636, -82.6391)(164.6786, -82.8175)(164.0475, -83.4351)
\qbezier(164.0475, -32.4070)(163.9452, -32.3070)(163.2412, -31.6109)
\qbezier(164.0475, -83.4351)(163.9452, -83.5352)(163.2412, -84.2312)
\qbezier(163.2412, -31.6109)(163.3305, -31.6992)(162.4413, -30.8148)
\qbezier(163.2412, -84.2312)(163.3305, -84.1429)(162.4413, -85.0273)
\qbezier(162.4413, -30.8148)(163.4924, -31.8602)(161.6447, -30.0187)
\qbezier(162.4413, -85.0273)(163.4924, -83.9819)(161.6447, -85.8234)
\qbezier(161.6447, -30.0187)(159.7971, -28.1773)(160.8481, -29.2227)
\qbezier(161.6447, -85.8234)(159.7971, -87.6649)(160.8481, -86.6194)
\qbezier(160.8481, -29.2227)(159.9590, -28.3383)(160.0483, -28.4266)
\qbezier(160.8481, -86.6194)(159.9590, -87.5038)(160.0483, -87.4155)
\qbezier(160.0483, -28.4266)(159.3442, -27.7305)(159.2420, -27.6305)
\qbezier(160.0483, -87.4155)(159.3442, -88.1116)(159.2420, -88.2116)
\qbezier(159.2420, -27.6305)(158.6109, -27.0129)(158.4259, -26.8344)
\qbezier(159.2420, -88.2116)(158.6109, -88.8292)(158.4259, -89.0077)
\qbezier(158.4259, -26.8344)(157.8289, -26.2586)(157.5964, -26.0384)
\qbezier(158.4259, -89.0077)(157.8289, -89.5835)(157.5964, -89.8037)
\qbezier(157.5964, -26.0384)(157.0147, -25.4875)(156.7497, -25.2423)
\qbezier(157.5964, -89.8037)(157.0147, -90.3546)(156.7497, -90.5998)
\qbezier(156.7497, -25.2423)(156.1719, -24.7074)(155.8819, -24.4462)
\qbezier(156.7497, -90.5998)(156.1719, -91.1347)(155.8819, -91.3959)
\qbezier(155.8819, -24.4462)(155.2998, -23.9219)(154.9884, -23.6501)
\qbezier(155.8819, -91.3959)(155.2998, -91.9203)(154.9884, -92.1920)
\qbezier(154.9884, -23.6501)(154.3955, -23.1326)(154.0644, -22.8541)
\qbezier(154.9884, -92.1920)(154.3955, -92.7095)(154.0644, -92.9880)
\qbezier(154.0644, -22.8541)(153.4546, -22.3409)(153.1044, -22.0580)
\qbezier(154.0644, -92.9880)(153.4546, -93.5012)(153.1044, -93.7841)
\qbezier(153.1044, -22.0580)(152.4719, -21.5472)(152.1020, -21.2619)
\qbezier(153.1044, -93.7841)(152.4719, -94.2949)(152.1020, -94.5802)
\qbezier(152.1020, -21.2619)(151.4409, -20.7520)(151.0502, -20.4658)
\qbezier(152.1020, -94.5802)(151.4409, -95.0901)(151.0502, -95.3763)
\qbezier(151.0502, -20.4658)(150.3536, -19.9556)(149.9405, -19.6698)
\qbezier(151.0502, -95.3763)(150.3536, -95.8865)(149.9405, -96.1723)
\qbezier(149.9405, -19.6698)(149.2011, -19.1581)(148.7633, -18.8737)
\qbezier(149.9405, -96.1723)(149.2011, -96.6840)(148.7633, -96.9684)
\qbezier(148.7633, -18.8737)(147.9722, -18.3597)(147.5069, -18.0776)
\qbezier(148.7633, -96.9684)(147.9722, -97.4824)(147.5069, -97.7645)
\qbezier(147.5069, -18.0776)(146.6537, -17.5604)(146.1575, -17.2815)
\qbezier(147.5069, -97.7645)(146.6537, -98.2817)(146.1575, -98.5606)
\qbezier(146.1575, -17.2815)(145.2295, -16.7601)(144.6982, -16.4855)
\qbezier(146.1575, -98.5606)(145.2295, -99.0820)(144.6982, -99.3567)
\qbezier(144.6982, -16.4855)(143.6798, -15.9589)(143.1086, -15.6894)
\qbezier(144.6982, -99.3567)(143.6798, -99.8832)(143.1086, -100.1527)
\qbezier(143.1086, -15.6894)(141.9799, -15.1568)(141.3628, -14.8933)
\qbezier(143.1086, -100.1527)(141.9799, -100.6853)(141.3628, -100.9488)
\qbezier(141.3628, -14.8933)(140.0985, -14.3535)(139.4286, -14.0972)
\qbezier(141.3628, -100.9488)(140.0985, -101.4886)(139.4286, -101.7449)
\qbezier(139.4286, -14.0972)(137.9957, -13.5490)(137.2645, -13.3011)
\qbezier(139.4286, -101.7449)(137.9957, -102.2931)(137.2645, -102.5410)
\qbezier(137.2645, -13.3011)(135.6188, -12.7431)(134.8166, -12.5051)
\qbezier(137.2645, -102.5410)(135.6188, -103.0990)(134.8166, -103.3370)
\qbezier(134.8166, -12.5051)(132.8978, -11.9356)(132.0130, -11.7090)
\qbezier(134.8166, -103.3370)(132.8978, -103.9065)(132.0130, -104.1331)
\qbezier(132.0130, -11.7090)(129.7366, -11.1260)(128.7559, -10.9129)
\qbezier(132.0130, -104.1331)(129.7366, -104.7161)(128.7559, -104.9292)
\qbezier(128.7559, -10.9129)(125.9998, -10.3140)(124.9090, -10.1168)
\qbezier(128.7559, -104.9292)(125.9998, -105.5281)(124.9090, -105.7253)
\qbezier(124.9090, -10.1168)(121.4899,  -9.4989)(120.2756,  -9.3208)
\qbezier(124.9090, -105.7253)(121.4899, -106.3432)(120.2756, -106.5213)
\qbezier(120.2756,  -9.3208)(115.9063,  -8.6799)(114.5615,  -8.5247)
\qbezier(120.2756, -106.5213)(115.9063, -107.1622)(114.5615, -107.3174)
\qbezier(114.5615,  -8.5247)(108.7662,  -7.8559)(107.3056,  -7.7286)
\qbezier(114.5615, -107.3174)(108.7662, -107.9862)(107.3056, -108.1135)
\qbezier(107.3056,  -7.7286)( 99.2404,  -7.0257)( 97.7423,  -6.9325)
\qbezier(107.3056, -108.1135)( 99.2404, -108.8164)( 97.7423, -108.9096)
\qbezier( 97.7423,  -6.9325)( 85.7667,  -6.1880)( 84.4996,  -6.1365)
\qbezier( 97.7423, -108.9096)( 85.7667, -109.6541)( 84.4996, -109.7056)
\qbezier( 84.4996,  -6.1365)( 65.0015,  -5.3438)( 64.8508,  -5.3404)
\qbezier( 84.4996, -109.7056)( 65.0015, -110.4983)( 64.8508, -110.5017)
\qbezier( 64.8508,  -5.3404)( 28.2802,  -4.5078)( 32.4930,  -4.5443)
\qbezier( 64.8508, -110.5017)( 28.2802, -111.3343)( 32.4930, -111.2978)
\put(  5.8483, -57.9211){\vector(1, 0){279.2351}}
\put(135.0000, -152.0658){\vector(0, 1){188.2895}}
\put( 32.4930,  -2.1164){\line(1, 0){225.9456}}
\put( 32.4930, -113.7257){\line(1, 0){225.9456}}
\put(294.8531, -60.5855){\makebox(0,0)[]{\footnotesize{$\widetilde{r}$}}}
\put(152.7632, -65.0263){\makebox(0,0)[]{\footnotesize{$\widetilde{r}_1$}}}
\put(127.5395,  37.1118){\makebox(0,0)[]{\footnotesize{$\widetilde{t}$}}}
\put(116.3487,   4.9888){\makebox(0,0)[]{\scriptsize{$\pi R_{\mbox{\tiny $Q$}} / A$}}}
\put(114.5724, -122.6072){\makebox(0,0)[]{\scriptsize{$-\pi R_{\mbox{\tiny $Q$}} / A$}}}
\end{picture}
\vspace{0mm} \newline
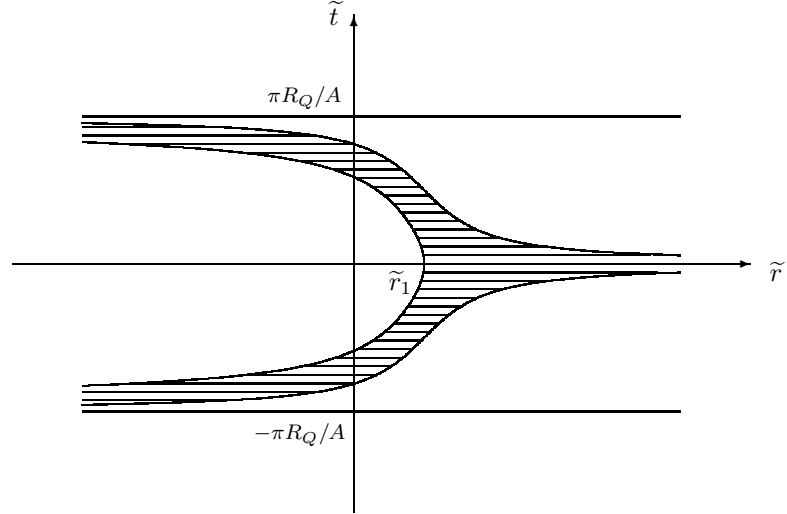
{\footnotesize \sf Figure 6. The hatched region represents
the WLBR spacetime in the $(\tilde{t},\tilde{r})$-system for
$k = 1$. The boundary of the this region follows from the inequalities
\eqref{e_667}, and $\tilde{r}_1 = \tilde{r}_0
- R_{\mbox{\tiny $Q$}} A \hs{0.5mm}
\cot (2 {\chi}_{\mbox{\tiny $Q$}})$ which follows from the left
inequality in \eqref{e_667} with $\eta = 0$.}
\vspace{0mm} \newline
\itm In the PLBR spacetime all the values $k = -1,0,1$ are allowed.
For $k = 1$ the PLBR spacetime is represented by the region between
the horizontal lines in Figure 6 given by
\begin{equation} \label{e_668}
-\pi R_{\mbox{\tiny $Q$}} / A < \tilde{t} <
\pi R_{\mbox{\tiny $Q$}} / A
\mbox{\hspace{2mm} , \hspace{1mm}}
-\infty < \tilde{r} < \infty
\mbox{ .}
\end{equation}
The part to the left of the hatched region corresponds to
$0 < R < R_{\mbox{\tiny $Q$}}$ in the CFS system, while the part
to the right corresponds to $R < 0$.
\itm For $k = -1$ the WLBR spacetime is represented in the
$(\tilde{t},\tilde{r})$-system by a region given by
\begin{equation} \label{e_1267}
-\infty < \tilde{t} < \infty
\mbox{\hspace{0mm} , \hspace{1mm}}
\tilde{r}_0 - R_{\mbox{\tiny $Q$}} A \hs{0.5mm} \coth
\left( {\chi}_{\mbox{\tiny $Q$}} +
\mbox{arcsinh} \left( \sinh {\chi}_{\mbox{\tiny $Q$}} \cosh \eta \right)
\right) < \tilde{r}
< \tilde{r}_0 - R_{\mbox{\tiny $Q$}} A
\mbox{ ,}
\end{equation}
where $\eta = (A / R_{\mbox{\tiny $Q$}}) \hspace{0.5mm}
\tilde{t}$ and
${\chi}_{\mbox{\tiny $Q$}} = \mbox{arccoth} (B / R_{\mbox{\tiny $Q$}})$.
For $k = 0$ it is represented by the region
\begin{equation} \label{e_1287}
-\infty < \tilde{t} < \infty
\mbox{\hspace{0mm} , \hspace{1mm}}
\tilde{r}_0 - R_{\mbox{\tiny $Q$}} < \tilde{r} < \tilde{r}_0
\mbox{ .}
\end{equation}
\vspace{-10mm} \newline
\itm We shall now give a physical interpretation of the constants
$\tilde{r}_0$ and $D$ valid for all values of $k$ by considering
the motion of a free particle. The acceleration of a free particle
instantaneously at rest is here given by
\begin{equation} \label{e_970}
a^{\tilde{r}} = \ddot{\tilde{r}} = - {\Gamma}^{\hs{0.3mm} \tilde{r}}
_{\hs{1.5mm} \tilde{t} \hs{0.5mm} \tilde{t}} \hs{1.5mm}
\dot{\tilde{t}}^{\hs{1.0mm} 2}
\mbox{ .}
\end{equation}
%
Using the line element \eqref{e_452} we get
\begin{equation} \label{e_971}
\dot{\tilde{t}} = \frl{d \tilde{t}}{d \tau}
= |g_{\tilde{t} \hs{0.5mm} \tilde{t}}|^{-1/2}
= \mbox{$\left[ \rule[-1.5mm]{0mm}{5.7mm} \right.$}
\hs{-0.3mm} D + \left( \frl{\tilde{r} \m \tilde{r}_0}
{R_{\mbox{\tiny $Q$}}} \right)^2
\hs{0.0mm} \mbox{$\left. \rule[-1.5mm]{0mm}{5.7mm} \right]$}^{-1/2}
\end{equation}
\vspace{-2mm} \newline
where $\tau$ is the proper time of the particle. Furthermore
\begin{equation} \label{e_972}
{\Gamma}^{\hs{0.3mm} \tilde{r}}_{\hs{1.5mm}
\tilde{t} \hs{0.5mm} \tilde{t}}
= \frl{\tilde{r} - \tilde{r}_0}{R_{\mbox{\tiny $Q$}}^2}
\hs{1.0mm} \mbox{$\left[ \rule[-1.5mm]{0mm}{5.7mm} \right.$}
\hs{-0.3mm} D + \left( \frl{\tilde{r} \m \tilde{r}_0}
{R_{\mbox{\tiny $Q$}}} \right)^2
\hs{0.0mm} \mbox{$\left. \rule[-1.5mm]{0mm}{5.7mm} \right]$}
\mbox{ .}
\end{equation}
%
This gives for the acceleration of gravity in the
$(\tilde{t},\tilde{r})$-system
\begin{equation} \label{e_1010}
a^{\hat{\tilde{r}}}
= \sqrt{g_{\hs{0.3mm} \tilde{r} \tilde{r}}} \hs{0.8mm} a^{\tilde{r}}
= \frl{\tilde{r}_0 - \tilde{r}}{R_{\mbox{\tiny $Q$}}^2}
\hs{1.0mm} \mbox{$\left[ \rule[-1.5mm]{0mm}{5.7mm} \right.$}
\hs{-0.3mm} D + \left( \frl{\tilde{r} \m \tilde{r}_0}
{R_{\mbox{\tiny $Q$}}} \right)^2
\hs{0.0mm} \mbox{$\left. \rule[-1.5mm]{0mm}{5.7mm} \right]$}^{-1/2}
\mbox{ .}
\end{equation}
%
This means that $\tilde{r} = \tilde{r}_0$ is the position
where the acceleration of gravity vanishes in the
$(\tilde{t}, \tilde{r})$-system. In the WLBR spacetime a free particle is
falling towards the domain wall in the region $\tilde{r} > \tilde{r}_0$
and away from the domain wall in the region $\tilde{r} < \tilde{r}_0$.
In section 5 we shall show that this is due to the motion of the reference
frame in which $(\tilde{t}, \tilde{r})$ are comoving coordinates.
In the case $k = 1$ the constant $\tilde{r}_1$ in Figure 6 represents
the position of the shell in the $(\tilde{t}, \tilde{r})$-system
at the point of time $T = 0$. The constant $A = \sqrt{D}$ has the
following physical interpretation. The coordinate clocks $\tilde{t}$
go at a constant rate equal that of the standard clocks at the domain
wall at the point of time $T = 0$ scaled by the factor $A$.
The line element \eqref{e_452} can now be written as
\vspace{-1mm} \newline
\begin{equation} \label{e_1013}
ds^2 = - \left[ D + R_{\mbox{\tiny $Q$}}^2 \hs{0.2mm}
a(\tilde{r})^2 \hs{0.5mm} \right] d\tilde{t}^{\hs{0.9mm} 2}
+ \left[ D + R_{\mbox{\tiny $Q$}}^2 \hs{0.2mm}
a(\tilde{r})^2 \hs{0.5mm} \right]^{-1} d\tilde{r}^{\hs{0.5mm} 2}
+ R_{\mbox{\tiny $Q$}}^2 d\Omega^2
\mbox{ .}
\end{equation}
\vspace{-8mm} \newline
\itm In the previous cases with $k = 1$ the PLBR spacetime corresponds
to only a part of the coordinate region in the
$(\tilde{t},\tilde{r})$-system.
However, when $k = -1$ the $(\tilde{t},\tilde{r})$-system
covers only a part of the PLBR spacetime as shown in Figure 4.
\itm Combining the transformations \eqref{e_349} and \eqref{e_279},
and using the identities \eqref{e_827}, we obtain the coordinate
transformation from $(\tilde{t},\tilde{r})$ to $(T,R)$ in
the following form
\begin{equation} \label{e_228}
T = \frl{B
\sqrt{R_{\mbox{\tiny $Q$}}^2 D \p (\tilde{r} \m \tilde{r}_0)^2}
\hs{0.7mm} S_k (A \hs{0.5mm} \tilde{t} / R_{\mbox{\tiny $Q$}})}
{\sqrt{R_{\mbox{\tiny $Q$}}^2 D \p (\tilde{r} \m \tilde{r}_0)^2}
\hs{0.7mm} C_k (A \hs{0.5mm} \tilde{t} / R_{\mbox{\tiny $Q$}})
\hs{0.5mm} \pm \hs{0.5mm} (\tilde{r}_0 \m \tilde{r})}
\mbox{ ,}
\end{equation}
\begin{equation} \label{e_454}
R = \frl{AB R_{\mbox{\tiny $Q$}}}
{(\tilde{r}_0 \m \tilde{r}) \hs{0.5mm} \pm \hs{0.5mm}
\sqrt{R_{\mbox{\tiny $Q$}}^2 D \p (\tilde{r} \m \tilde{r}_0)^2}
\hs{0.7mm} C_k (A \hs{0.5mm} \tilde{t} / R_{\mbox{\tiny $Q$}})}
\mbox{ .}
\end{equation}
In the cases $k = 0$ and $k = -1$ we use plus when
$\tilde{r} < \tilde{r}_0$ and minus when
$\tilde{r} > \tilde{r}_0$. In the case $k = 1$ we use plus
for all values of $\tilde{r}$.
This generalizes and modifies the corresponding transformation for $k = 1$,
$\tilde{r}_0 = 0$ and $A = B = D = 1$ given by Griffiths and
Podolosky [\ref{r_14}].
\itm When $k = 1$ this transformation maps the region
$-\pi R_{\mbox{\tiny $Q$}} / A < \tilde{t} <
\pi R_{\mbox{\tiny $Q$}} / A$,
$\tilde{r} < \tilde{r}_0 + R_{\mbox{\tiny $Q$}} A
\cot (A |\tilde{t}| / R_{\mbox{\tiny $Q$}})$ in the PLBR
spacetime shown in Figure 6 onto the right half plane in the CFS system,
and the region $-\pi R_{\mbox{\tiny $Q$}} / A < \tilde{t} <
\pi R_{\mbox{\tiny $Q$}} / A$,
$\tilde{r} > \tilde{r}_0 + R_{\mbox{\tiny $Q$}} A
\cot (A |\tilde{t}| / R_{\mbox{\tiny $Q$}})$ in the PLBR
spacetime onto the left half plane in the CFS system.
When $k = -1$ the transformation maps the region
$-\infty < \tilde{t} < \infty$,
$\tilde{r} < \tilde{r}_0 - R_{\mbox{\tiny $Q$}} A$ in the PLBR
spacetime onto the triangle $0 < R < B$, $|T| < B - R$ in the CFS system,
and the region $-\infty < \tilde{t} < \infty$,
$\tilde{r} > \tilde{r}_0 + R_{\mbox{\tiny $Q$}} A$
onto the triangle $-B < R < 0$, $|T| < B + R$ in the CFS system.
The WLBR spacetime described by the inequalities \eqref{e_1267} is
mapped onto the triangle $R_{\mbox{\tiny $Q$}} < R < B$, $|T| < B - R$
in the CFS system.
Using the relationship \eqref{e_810} we find that the inverse
transformation can be written
\begin{equation} \label{e_321}
I_k \mbox{$\left( \rule[-1.5mm]{0mm}{5.7mm} \right.$} \hs{-0.2mm}
\frl{A \hs{0.5mm} \tilde{t}}{R_{\mbox{\tiny $Q$}}}
\hs{-0.2mm} \mbox{$\left. \rule[-1.5mm]{0mm}{5.7mm} \right)$}
= \frl{B^2 - k (T^{2} - R^{2})}{2BT}
\mbox{\hspace{2mm} , \hspace{3mm}}
\frl{\tilde{r}_0 \m \tilde{r}}
{R_{\mbox{\tiny $Q$}} A}
= \frl{B^2 + k (T^{2} - R^{2})}{2BR}
\mbox{ .}
\end{equation}
when $T \ne 0$. In the case $T = 0$, we have $t = 0$. From the last one
of the transformation equations \eqref{e_321} it follows that for $k = 1$
the hyperbola of Figure 3 where the acceleration of gravity vanishes is
given by $\tilde{r} = \tilde{r}_0$, in agreement with equation
\eqref{e_1010}.
\itm In the case $k = 0$ the transformation
\eqref{e_228}, \eqref{e_454} reduces to
\begin{equation} \label{e_455}
T = \tilde{t}
\mbox{\hspace{2mm} , \hspace{3mm}}
R = \frl{R_{\mbox{\tiny $Q$}}^2}{\tilde{r}_0 \m \tilde{r}}
\end{equation}
which has been chosen so that ${\bf e}_{\tilde{r}}$ points in the
same direction as ${\bf e}_R$. This transformation maps the region
$-\infty < \tilde{t} < \infty \,$, $\tilde{r} < \tilde{r}_0$
onto the right half plane in the CFS system, and the region
$-\infty < \tilde{t} < \infty \,$, $\tilde{r} > \tilde{r}_0$
onto the left half plane in the CFS system. The inverse transformation is
\begin{equation} \label{e_456}
\tilde{t} = T
\mbox{\hspace{2mm} , \hspace{3mm}}
\tilde{r} = \tilde{r}_0 -
\frl{R_{\mbox{\tiny $Q$}}^2}{R}
\mbox{ .}
\end{equation}
In this case the line element \eqref{e_452} takes the form
\begin{equation} \label{e_560}
ds^2 = - \left( \frl{\tilde{r} \m \tilde{r}_0}
{R_{\mbox{\tiny $Q$}}} \right)^{\hs{-0.5mm} 2} d\tilde{t}^{\hs{0.5mm} 2}
+ \left( \frl{R_{\mbox{\tiny $Q$}}}{\tilde{r} \m \tilde{r}_0}
\right)^{\hs{-0.5mm} 2} d\tilde{r}^{\hs{0.5mm} 2}
+ R_{\mbox{\tiny $Q$}}^2 d\Omega^2
\mbox{ ,}
\end{equation}
%
which was considered in reference [\ref{r_13}] with $\tilde{r}_0 = 0$.
\itm In these coordinates we shall write down the form of the line
element in the flat spacetime inside the domain wall only for the case
$k = 0$ when the external metric is given by the equation \eqref{e_560}.
Inserting the expression \eqref{e_455} for $R$ in the line element
\eqref{e_1211} then leads to
\begin{equation} \label{e_1241}
ds_M^2 = - d\tilde{t}^{\hs{0.5mm} 2} +
\left( \frl{R_{\mbox{\tiny $Q$}}}{\tilde{r} \m \tilde{r}_0}
\right)^{\hs{-0.5mm} 2} \left[
\left( \frl{R_{\mbox{\tiny $Q$}}}{\tilde{r} \m \tilde{r}_0}
\right)^{\hs{-0.5mm} 2} d\tilde{r}^{\hs{0.5mm} 2}
+ R_{\mbox{\tiny $Q$}}^{\hs{0.5mm} 2} \hs{1.2mm} d\Omega^2 \right]
\mbox{ .}
\end{equation}
It follows from the transformation \eqref{e_456} that
$\fr{(\tilde{r} \m \tilde{r}_0)}{R_{\mbox{\tiny $Q$}}} = 1$
when $R = R_{\mbox{\tiny $Q$}}$, showing that the metric is
continuous at the domain wall.
\itm Calculating the Christoffel symbols from the line element
\eqref{e_1241} shows that there is vanishing acceleration of gravity
inside the domain wall in this coordinate system. The reason is that
for $k = 0$ the $(\tilde{t},\tilde{r})$ coordinates are comoving in a
static reference frame in this region.
%
%
%
\vspace{6mm} \newline
{\it IIb. Time dependent metric and coordinates
$(\overline{t},\overline{r})$ with
$\beta (\overline{t}) = - \alpha (\overline{t})$.}
\vspace{3mm} \newline
In this case equation \eqref{e_212} reduces to
\begin{equation} \label{e_924}
\ddot{\beta} + 2 \hs{0.4mm} \dot{\beta} \hs{0.5mm}^2
= - \frl{e^{-2 \beta}}{R_{\mbox{\tiny $Q$}}^2}
\mbox{ ,}
\end{equation}
\vspace{-6mm} \newline
which may be written
\begin{equation} \label{e_920}
(e^{2 \beta})^{\dot{} \hs{1.0mm} \dot{}} =
- \frl{2}{R_{\mbox{\tiny $Q$}}^2}
\mbox{ .}
\end{equation}
The general solution of this equation can be written as
\begin{equation} \label{e_944}
e^{2 \beta} = D - \left( \frl{ \overline{t} \m \overline{t}_0}
{R_{\mbox{\tiny $Q$}}} \right)^{\hs{-0.5mm} 2}
\mbox{ ,}
\end{equation}
where $D$ and $\overline{t}_0$ are constants. A special case of this
solution with $\overline{t}_0 = 0$ and $D = 1$ has earlier been found
by N.\hn Dadhich [\ref{r_15}].
The line element \eqref{e_211} then takes the form
\vspace{-1mm} \newline
\begin{equation} \label{e_952}
ds^2 = - \mbox{$\left[ \rule[-1.5mm]{0mm}{5.7mm} \right.$} \hs{-0.3mm}
D - \left( \frl{ \overline{t} \m \overline{t}_0}
{R_{\mbox{\tiny $Q$}}} \right)^2
\hs{0.0mm} \mbox{$\left. \rule[-1.5mm]{0mm}{5.7mm} \right]$}^{-1}
\hs{0.0mm} d\overline{t}^2
+ \mbox{$\left[ \rule[-1.5mm]{0mm}{5.7mm} \right.$} \hs{-0.3mm}
D - \left( \frl{ \overline{t} \m \overline{t}_0}
{R_{\mbox{\tiny $Q$}}} \right)^2
\hs{0.0mm} \mbox{$\left. \rule[-1.5mm]{0mm}{5.7mm} \right]$}
\hs{0.8mm} d\overline{r}^2
+ R_{\mbox{\tiny $Q$}}^2 d\Omega^2
\mbox{ .}
\end{equation}
\vspace{-1mm} \newline
From equation \eqref{e_944} we see that the constant $D$ must be positive,
and we introduce the constant $A = \sqrt{D}$ as in section IIa. Here the
standard measuring rods have a time dependent length, and the coordinate
rods have a constant length equal to the length of the standard rods at
the point of time $\overline{t} = \overline{t}_0$ scaled by the
factor $A$. The allowed range of the time $\overline{t}$ is
\begin{equation} \label{e_1016}
\overline{t}_0 - R_{\mbox{\tiny $Q$}} A < \overline{t} <
\overline{t}_0 + R_{\mbox{\tiny $Q$}} A
\mbox{ .}
\end{equation}
\itm The transformation between the $(\overline{t},\overline{r})$- and the
$(\tau,\rho)$-system used in the line element \eqref{e_928} is given by
\begin{equation} \label{e_949}
\tanh \tau = \frl{\overline{t} \m \overline{t}_0}
{R_{\mbox{\tiny $Q$}} A}
\mbox{\hspace{2mm} , \hspace{3mm}}
\rho = \frl{A}{R_{\mbox{\tiny $Q$}}} \hs{0.9mm} \overline{r}
\mbox{ .}
\end{equation}
The transformation has been chosen so that $\tau$ and
$\overline{t}$ increase in the same direction. The second of these
equations shows that the $(\overline{t}, \overline{r})$-system is
comoving with the same reference frame as the $(\tau,\rho)$-system.
Hence particles with $\overline{r} = \mbox{constant}$ are moving
freely. The transformation from $\tau$ to $\overline{t}$ is
\begin{equation} \label{e_950}
\overline{t} = \overline{t}_0
+ R_{\mbox{\tiny $Q$}} A \hs{0.5mm} \tanh \tau
\mbox{ .}
\end{equation}
Inserting equation \eqref{e_950} into equation \eqref{e_944}
we obtain
\begin{equation} \label{e_951}
e^{2 \beta} = D / \cosh^2 \tau
\mbox{ .}
\end{equation}
%
%
%
Differentiating equation \eqref{e_950} we get
\begin{equation} \label{e_978}
d \overline{t} = (R_{\mbox{\tiny $Q$}} A /
\cosh^2 \tau) \hs{1.0mm} d \tau
\mbox{ .}
\end{equation}
Using \eqref{e_951} and \eqref{e_978} we see that the line element
\eqref{e_952} takes the form \eqref{e_928}.
\itm Combining the transformations \eqref{e_981} and \eqref{e_949} we
obtain the coordinate transformation from $(\tilde{t},\tilde{r})$
to $(T,R)$ in the following form
\begin{equation} \label{e_992}
T = \frl{B
\sqrt{R_{\mbox{\tiny $Q$}}^2 D \m (\overline{t} \m \overline{t}_0)^2}
\hs{0.7mm} \cosh (A \hs{0.5mm} \overline{r} / R_{\mbox{\tiny $Q$}})}
{(\overline{t}_0 \m \overline{t}) \m
\sqrt{R_{\mbox{\tiny $Q$}}^2 D \m (\overline{t} \m \overline{t}_0)^2}
\hs{0.7mm} \sinh (A \hs{0.5mm} \overline{r} / R_{\mbox{\tiny $Q$}})}
\mbox{ ,}
\end{equation}
\begin{equation} \label{e_954}
R = \frl{AB R_{\mbox{\tiny $Q$}}}
{(\overline{t}_0 \m \overline{t}) \m
\sqrt{R_{\mbox{\tiny $Q$}}^2 D \m (\overline{t} \m \overline{t}_0)^2}
\hs{0.7mm} \sinh (A \hs{0.5mm} \overline{r} / R_{\mbox{\tiny $Q$}})}
\mbox{ .}
\end{equation}
\vspace{1mm} \newline
The inverse transformation is given by
\vspace{1mm} \newline
\begin{equation} \label{e_991}
\frl{\overline{t} \m \overline{t}_0}
{R_{\mbox{\tiny $Q$}} A}
= \frl{(T^{2} \m R^{2}) \m B^2}{2 BR}
\mbox{\hspace{2mm} , \hspace{3mm}}
\tanh \mbox{$\left( \rule[-1.5mm]{0mm}{5.7mm} \right.$} \hs{-0.2mm}
\frl{A \hs{0.5mm} \overline{r}}{R_{\mbox{\tiny $Q$}}}
\hs{-0.2mm} \mbox{$\left. \rule[-1.5mm]{0mm}{5.7mm} \right)$}
= \frl{(R^{2} \m T^{2}) \m B^2}{2 BT}
\mbox{ .}
\end{equation}
%
%
%
\vspace{6mm} \newline
{\it IIIa. Static metric and coordinates $(\hat{t},\hat{r})$ with
$\alpha = \alpha (\hat{r})$ and $\beta = 0$.}
\vspace{3mm} \newline
In this case the radial coordinate $\hat{r}$ is equal to physical
distance in the radial direction. Equation \eqref{e_212} then reduces to
\begin{equation} \label{e_223}
\alpha''+ \alpha' \hs{0.5mm}^2
= \frl{1}{R_{\mbox{\tiny $Q$}}^2}
\mbox{ ,}
\end{equation}
which may be written
\begin{equation} \label{e_229}
R_{\mbox{\tiny $Q$}}^2 (e^{\alpha})'' - e^{\alpha} = 0
\mbox{ .}
\end{equation}
The general solution of this equation is
\begin{equation} \label{e_225}
e^{\alpha} = c_1 e^{\hat{r} / R_{\mbox{\tiny $Q$}}}
+ c_2 e^{- \hat{r} / R_{\mbox{\tiny $Q$}}}
\end{equation}
\vspace{-9.0mm} \newline
or alternatively
\vspace{-3.0mm} \newline
\begin{equation} \label{e_281}
e^{\alpha} = c_3 \cosh({\hat{r} / R_{\mbox{\tiny $Q$}}})
+ c_4 \sinh({\hat{r} / R_{\mbox{\tiny $Q$}}})
\mbox{ ,}
\end{equation}
where $c_i$, $i = 1,2,3,4$ are constants.
Here the coordinate clocks go with a position independent rate equal to
that of the standard clocks at $\hat{r} = 0$ scaled by the factor
$c_3$. This solution \eqref{e_225} was found already in 1917 by
T.\hn Levi-Civita [\ref{r_4},\ref{r_5}], and was later mentioned
in [\ref{r_15}] and in [\ref{r_22}] with $c_3 = 0$.
\itm We are now going to find the coordinate transformation between
the physical coordinates $(\hat{t},\hat{r})$ and the
CFS coordinates $(T,R)$. In this connection we will also deduce
the transformation between $\chi$ and $\hat{r}$ and between
$\tilde{r}$ and $\hat{r}$.
Since $\hat{r}$ represents the physical radial distance we
have from equation \eqref{e_218} that
\vspace{-2.0mm} \newline
\begin{equation} \label{e_687}
d \hat{r} = \frl{R_{\mbox{\tiny $Q$}}}{|S_k(\chi)|} \hs{0.5mm} d \chi
\mbox{ .}
\end{equation}
\vspace{-2.0mm} \newline
%
%
%
Integration using the identities \eqref{e_808}, \eqref{e_828} and
\eqref{e_825} gives
\begin{equation} \label{e_688}
\hat{r} = \hat{r}_0 - \mbox{sgn} \hs{0.3mm} S_k(\chi)
\hs{0.8mm} R_{\mbox{\tiny $Q$}} \hs{0.3mm}
\ln \left| I_k \leftp \frl{\chi}{1 \p |k|}
\rightp \right|
\mbox{ ,}
\end{equation}
where $\hat{r}_0$ is a constant.
With a suitable scaling of the time coordinate the transformation between
the $(\hat{t},\hat{r})$-system and the $(\eta,\chi)$-system is
given by
\begin{equation} \label{e_689}
\left| I_k \left(\frl{\chi}{1 \p |k|} \right) \right|
= e^{\pm \frac{\hat{r}_0 - \hat{r}}{R_{\mbox{\tiny $Q$}}}}
\mbox{\hspace{2mm} , \hspace{3mm}}
\eta = \frl{A}{R_{\mbox{\tiny $Q$}}} \hs{0.6mm} \hat{t}
\mbox{ ,}
\end{equation}
where $A$ is defined in equation \eqref{e_692}. Comparing with equation
\eqref{e_349} we see that $\hat{t} = \tilde{t}$.
In the case $k = -1$ we use the upper sign when
$\hat{r} < \hat{r}_0$ and the lower sign when
$\hat{r} > \hat{r}_0$. In the cases $k = 0$
and $k = 1$ we use the upper sign for all $\hat{r}$. These rules
mean that $\hat{r}$ increases in the same direction as $\chi$.
Using the identity \eqref{e_832} with $x = \chi/2$ combined with
equations \eqref{e_349} and \eqref{e_689} we obtain
\begin{equation} \label{e_700}
\pm a_{-k} \left(
\frl{\hat{r}_0 - \hat{r}}{R_{\mbox{\tiny $Q$}}} \right)
= I_k (\chi) = \frl{\tilde{r}_0 \m \tilde{r}}
{R_{\mbox{\tiny $Q$}} A}
\end{equation}
with the same rule for choosing the signs as above, meaning that
$\hat{r}$ and $\tilde{r}$ increases in the same direction.
This implies that
\begin{equation} \label{e_701}
\hat{r} = \hat{r}_0 \mp R_{\mbox{\tiny $Q$}} \hs{0.5mm}
a_{-k}^{-1} \left( \frl{\tilde{r}_0 \m \tilde{r}}
{R_{\mbox{\tiny $Q$}} A} \right)
\mbox{ .}
\end{equation}
The inverse transformation is given by
\begin{equation} \label{e_696}
\tilde{r} = \tilde{r}_0 \mp R_{\mbox{\tiny $Q$}} A
\hs{1.0mm} a_{-k} \left(
\frl{\hat{r}_0 - \hat{r}}{R_{\mbox{\tiny $Q$}}} \right)
\mbox{ .}
\end{equation}
Since the relationship between $\tilde{r}$ and $\hat{r}$ is
time-independent, the $\hat{r}$-coordinate is comoving in the same
reference frame as the $\tilde{r}$-coordinate.
\itm It follows from equations \eqref{e_555} and \eqref{e_700} for
$k = 1$ that in this case the WLBR spacetime is given by
\begin{equation} \label{e_1297}
-\pi R_{\mbox{\tiny $Q$}} / A < \hat{t} <
\pi R_{\mbox{\tiny $Q$}} / A
\end{equation}
\vspace{-9mm} \newline
and
\vspace{-4mm} \newline
\begin{equation} \label{e_1667}
\hat{r}_0 - R_{\mbox{\tiny $Q$}} \hs{0.5mm} \mbox{arcsinh} (\cot
\left( {\chi}_{\mbox{\tiny $Q$}} +
\arcsin \left( \sin {\chi}_{\mbox{\tiny $Q$}} \cos \eta \right)
\right) ) < \hat{r}
< \hat{r}_0 + R_{\mbox{\tiny $Q$}} \hs{0.5mm}
\mbox{arcsinh} (\cot |\eta|)
\mbox{ ,}
\end{equation}
where $\eta = (A / R_{\mbox{\tiny $Q$}}) \hspace{0.5mm}
\hat{t}$ and
${\chi}_{\mbox{\tiny $Q$}}$ is given in equation \eqref{e_1144}.
\itm In the PLBR spacetime all the values $k = -1,0,1$ are allowed.
For $k = 1$ the PLBR spacetime is represented by
\begin{equation} \label{e_1668}
-\pi R_{\mbox{\tiny $Q$}} / A < \hat{t} <
\pi R_{\mbox{\tiny $Q$}} / A
\mbox{\hspace{2mm} , \hspace{1mm}}
-\infty < \hat{r} < \infty
\mbox{ .}
\end{equation}
For $k = -1$ the WLBR spacetime is represented in the
$(\hat{t},\hat{r})$-system by a region given by
\begin{equation} \label{e_1367}
-\infty < \hat{t} < \infty
\mbox{\hspace{0mm} , \hspace{1mm}}
\hat{r}_0 - R_{\mbox{\tiny $Q$}} \hs{0.5mm} \mbox{arccosh} (\coth
\left( {\chi}_{\mbox{\tiny $Q$}} +
\mbox{arcsinh} \left( \sinh {\chi}_{\mbox{\tiny $Q$}} \cosh \eta \right)
\right) ) < \hat{r}
< \hat{r}_0
\mbox{ ,}
\end{equation}
where $\eta = (A / R_{\mbox{\tiny $Q$}}) \hspace{0.5mm} \hat{t}$.
For $k = 0$ it is represented by the region
\begin{equation} \label{e_1387}
-\infty < \hat{t} < \infty
\mbox{\hspace{0mm} , \hspace{1mm}}
\hat{r} > \hat{r}_0
+ R_{\mbox{\tiny $Q$}} \ln (R_{\mbox{\tiny $Q$}})
\mbox{ .}
\end{equation}
\vspace{-10mm} \newline
\itm From equations \eqref{e_700}, \eqref{e_808}, \eqref{e_812} and
\eqref{e_833} we obtain
\begin{equation} \label{e_693}
S_k(\chi) = \frr{1}{a_{k} \hs{-0.8mm} \left(
\frl{\hat{r}_0 - \hat{r}}{R_{\mbox{\tiny $Q$}}} \right)}
\mbox{\hspace{2mm} , \hspace{3mm}}
%
%
C_k(\chi)
= \pm \hs{1.1mm} b_{k} \hs{-0.8mm} \left(
\frl{\hat{r}_0 - \hat{r}}{R_{\mbox{\tiny $Q$}}} \right)
\mbox{ .}
\end{equation}
Using the formulae \eqref{e_693} and \eqref{e_689} it follows that
the line element \eqref{e_218} takes the form
\begin{equation} \label{e_1014}
ds^2 = - A^2 \hs{0.5mm} a_{k} \hs{-0.8mm} \left(
\frl{\hat{r}_0 - \hat{r}}{R_{\mbox{\tiny $Q$}}} \right)^2
d \hat{t}^2 + d \hat{r}^2 + R_{\mbox{\tiny $Q$}}^2 d \Omega^2
\mbox{ .}
\end{equation}
The connection between the general solution \eqref{e_225} and the form
\eqref{e_1014} of the line element is given by
\begin{equation} \label{e_1015}
c_1 = \frl{kA}{1 + |k|} \hs{0.5mm}
e^{- \hat{r}_0 / R_{\mbox{\tiny $Q$}}}
\mbox{\hspace{2mm} , \hspace{3mm}}
c_2 = \frl{A}{1 + |k|} \hs{0.5mm}
e^{\hat{r}_0 / R_{\mbox{\tiny $Q$}}}
\mbox{ .}
\end{equation}
Inserting the relations \eqref{e_693} into equations \eqref{e_279} we
obtain the transformation between the physical coordinates
$(\hat{t},\hat{r})$ and the CFS coordinates,
\begin{equation} \label{e_702}
T = \frl{B \hs{1.0mm}
a_k \mbox{$\left( \rule[-1.5mm]{0mm}{5.7mm} \right.$} \hs{-0.2mm}
\frl{\hat{r}_0 \m \hat{r}}{R_{\mbox{\tiny $Q$}}}
\hs{-0.2mm} \mbox{$\left. \rule[-1.5mm]{0mm}{5.7mm} \right)$} \hs{0.5mm}
S_k \mbox{$\left( \rule[-1.5mm]{0mm}{5.7mm} \right.$} \hs{-0.2mm}
\frl{A \hs{0.5mm} \hat{t}}{R_{\mbox{\tiny $Q$}}}
\hs{-0.2mm} \mbox{$\left. \rule[-1.5mm]{0mm}{5.7mm} \right)$}}
{a_{-k} \mbox{$\left( \rule[-1.5mm]{0mm}{5.7mm} \right.$} \hs{-0.2mm}
\frl{\hat{r}_0 \m \hat{r}}{R_{\mbox{\tiny $Q$}}}
\hs{-0.2mm} \mbox{$\left. \rule[-1.5mm]{0mm}{5.7mm} \right)$} \p \hs{0.5mm}
a_k \mbox{$\left( \rule[-1.5mm]{0mm}{5.7mm} \right.$} \hs{-0.2mm}
\frl{\hat{r}_0 \m \hat{r}}{R_{\mbox{\tiny $Q$}}}
\hs{-0.2mm} \mbox{$\left. \rule[-1.5mm]{0mm}{5.7mm} \right)$} \hs{0.5mm}
C_k \mbox{$\left( \rule[-1.5mm]{0mm}{5.7mm} \right.$} \hs{-0.2mm}
\frl{A \hs{0.5mm} \hat{t}}{R_{\mbox{\tiny $Q$}}}
\hs{-0.2mm} \mbox{$\left. \rule[-1.5mm]{0mm}{5.7mm} \right)$}
\rule[-1.5mm]{0mm}{6.9mm}}
\mbox{ ,}
\end{equation}
\begin{equation} \label{e_703}
R = \frl{B}
{a_{-k} \mbox{$\left( \rule[-1.5mm]{0mm}{5.7mm} \right.$} \hs{-0.2mm}
\frl{\hat{r}_0 \m \hat{r}}{R_{\mbox{\tiny $Q$}}}
\hs{-0.2mm} \mbox{$\left. \rule[-1.5mm]{0mm}{5.7mm} \right)$} \p \hs{0.5mm}
a_k \mbox{$\left( \rule[-1.5mm]{0mm}{5.7mm} \right.$} \hs{-0.2mm}
\frl{\hat{r}_0 \m \hat{r}}{R_{\mbox{\tiny $Q$}}}
\hs{-0.2mm} \mbox{$\left. \rule[-1.5mm]{0mm}{5.7mm} \right)$} \hs{0.5mm}
C_k \mbox{$\left( \rule[-1.5mm]{0mm}{5.7mm} \right.$} \hs{-0.2mm}
\frl{A \hs{0.5mm} \hat{t}}{R_{\mbox{\tiny $Q$}}}
\hs{-0.2mm} \mbox{$\left. \rule[-1.5mm]{0mm}{5.7mm} \right)$}
\rule[-1.5mm]{0mm}{6.9mm}}
\mbox{ .}
\end{equation}
\vspace{1mm} \newline
Using equation \eqref{e_700} and the identity
\eqref{e_833} we see that this transformation is consistent
with equations \eqref{e_228} and \eqref{e_454}.
The inverse transformation is given by
\begin{equation} \label{e_322}
I_k \mbox{$\left( \rule[-1.5mm]{0mm}{5.7mm} \right.$} \hs{-0.2mm}
\frl{A \hs{0.5mm} \hat{t}}{R_{\mbox{\tiny $Q$}}}
\hs{-0.2mm} \mbox{$\left. \rule[-1.5mm]{0mm}{5.7mm} \right)$}
= \frl{B^2 - k (T^{2} - R^{2})}{2 \hs{0.2mm} BT}
\mbox{\hspace{2mm} , \hspace{3mm}}
\pm a_{-k} \mbox{$\left( \rule[-1.5mm]{0mm}{5.7mm} \right.$} \hs{-0.2mm}
\frl{\hat{r}_0 \m \hat{r}}{R_{\mbox{\tiny $Q$}}}
\hs{-0.2mm} \mbox{$\left. \rule[-1.5mm]{0mm}{5.7mm} \right)$}
= \frl{B^2 + k (T^{2} - R^{2})}{2 \hs{0.2mm} BR}
\mbox{ .}
\end{equation}
\itm In the case $k = 0$ the transformation
\eqref{e_702}, \eqref{e_703} reduces to
\begin{equation} \label{e_704}
T = \hat{t}
\mbox{\hspace{2mm} , \hspace{3mm}}
R = e^{\frac{\hat{r} - \hat{r}_0}{R_{\mbox{\tiny $Q$}}}}
\end{equation}
which has been chosen so that ${\bf e}_{\hat{r}}$ points in the
same direction as ${\bf e}_R$. The inverse transformation is
\begin{equation} \label{e_705}
\hat{t} = T
\mbox{\hspace{2mm} , \hspace{3mm}}
\hat{r} = \hat{r}_0 +
R_{\mbox{\tiny $Q$}} \hs{0.5mm} \ln R
\mbox{ .}
\end{equation}
%
Then the line element \eqref{e_1014} takes the form
\begin{equation} \label{e_267}
ds^2 = - R_{\mbox{\tiny $Q$}}^2 \hs{0.4mm}
e^{- 2 (\hat{r} - \hat{r}_0) / R_{\mbox{\tiny $Q$}}}
\hs{0.7mm} d\hat{t}^2 + d\hat{r}^{\hs{0.5mm} 2}
+ R_{\mbox{\tiny $Q$}}^2 d\Omega^2
\mbox{ .}
\end{equation}
The line elements \eqref{e_157} and \eqref{e_267} are related by the
transformation \eqref{e_704} with $\hat{r}_0 = R_{\mbox{\tiny $Q$}}$.
In this case the $\hat{r}$-coordinate and the $R$-coordinate
are comoving in the same reference frame. Although different choices of
$c_i$, $i = 1,2,3,4$ all represent conformally flat solutions of
the field equations with the same energy momentum tensor representing a
constant, radial electrical field, the physical properties of the
solutions are different.
\itm This may be most clearly seen by utilizing the geodesic equation.
Inserting the solution \eqref{e_225} in the line element \eqref{e_211}
we find that a free particle instantaneously at rest has a coordinate
acceleration
\vspace{-4mm} \newline
\begin{equation} \label{e_427}
a^{\hat{r}} = \ddot{\hat{r}} = - {\Gamma}^{\hs{0.3mm} \hat{r}}
_{\hs{1.5mm} \hat{t} \hs{0.5mm} \hat{t}} \hs{1.5mm}
\dot{\hat{t}}^{\hs{1.0mm} 2}
\mbox{ .}
\end{equation}
The acceleration of gravity in the $(\hat{t},\hat{r})$-system is the
component of $a^{\hat{r}} {\bf e}_{\hat{r}}$ along the unit basis vector
${\bf e}_{\hat{\hat{r}}}$. Since $g_{\hat{r} \hat{r}} = 1$, we have that
$a^{\hat{\hat{r}}} = a^{\hat{r}}$. A reference particle with given values
of $\hat{r},\theta,\phi$ is at rest relative a reference particle with
given values of $\chi,\theta,\phi$. Hence the transformation from the
$(\eta,\chi)$-system to the $\hat{t},\hat{r}$-system is a so called
internal transformation, i.e. a coordinate transformation inside a
single reference frame. In addition, the unit radial vector in the
$(\eta,\chi)$-system is identical to the unit radial vector in the
$\hat{t},\hat{r}$-system, ${\bf e}_{\hat{r}} = {\bf e}_{\chi}$. These
two conditions mean that $a^{\hat{\hat{r}}} = a^{\hat{\chi}}$. Using
equation \eqref{e_693} we obtain
\vspace{-4mm} \newline
\begin{equation} \label{e_2011}
a^{\hat{\hat{r}}} = \pm \hs{0.3mm}
\frl{1}{R_{\mbox{\tiny $Q$}}} \hs{1.1mm} b_{k} \hs{-0.8mm} \left(
\frl{\hat{r}_0 - \hat{r}}{R_{\mbox{\tiny $Q$}}} \right)
%
\mbox{ .}
\end{equation}
%
This expression for the acceleration of gravity can be positive or
negative. We shall here discuss these possibilities for the WLBR
spacetime.
\itm The reason for these differences is found in the different motions
of the $(\hat{t},\hat{r})$-systems relative to the CFS system for
different values of $k$ and $\hat{r}$.
The world line of a reference point $\hat{r} = \hat{r}_1$ as
described with reference to the $(T,R)$-system is given by
equation \eqref{e_347}. By means of equation \eqref{e_700} the constant
$R_1$ in equation \eqref{e_347} can be expressed in terms of the
coordinate $\hat{r}_1$ as
$R_1 = \mp k B a_{-k} ((\hat{r}_0 - \hat{r}_1)
/ R_{\mbox{\tiny $Q$}}) \,$. The world line is shown for $k = 1$ in
Figure 1.
\itm The form \eqref{e_267} of the line element for the WLBR spacetime in
a uniform electric field outside a charged domain wall shows that the time
does not proceed infinitely far from the domain wall. The coordinate
velocity of light moving radially outwards is
\begin{equation} \label{e_227}
\frl{d \hat{r}}{d \hat{t}} =
e^{- (\hat{r} - R_{\mbox{\tiny $Q$}}) /
R_{\mbox{\tiny $Q$}}}
\mbox{ .}
\end{equation}
Hence $\lim_{\hat{r} \rightarrow \infty} \fr{d \hat{r}}
{d \hat{t}} = 0$.
There is, however, no horizon at a finite distance from the wall.
\itm Again we shall write down the form of the line element in the
flat spacetime inside the domain wall only for the case $k = 0$.
Inserting $e^{\alpha}$ from the external line element \eqref{e_267}
and the expression \eqref{e_704} for $R$ we obtain the internal line
element in the $(\hat{t},\hat{r})$ coordinates,
\begin{equation} \label{e_1767}
ds_M^2 = - d\hat{t}^2 + (\fr{1}{R_{\mbox{\tiny $Q$}}^2}) \hs{0.4mm}
e^{2 (\hat{r} - \hat{r}_0) / R_{\mbox{\tiny $Q$}}}
\hs{0.7mm} (d\hat{r}^{\hs{0.5mm} 2}
+ R_{\mbox{\tiny $Q$}}^2 d\Omega^2)
\mbox{ .}
\end{equation}
It follows from equation \eqref{e_705} that the $\hat{r}$ coordinate
of the shell is
\begin{equation} \label{e_2267}
\hat{r}_{\mbox{\tiny $Q$}} = \hat{r}_0 +
R_{\mbox{\tiny $Q$}} \hs{0.5mm} \ln R_{\mbox{\tiny $Q$}}
\mbox{ ,}
\end{equation}
showing that $e^{2 (\hat{r} - \hat{r}_0) / R_{\mbox{\tiny $Q$}}}
= R_{\mbox{\tiny $Q$}}^2$ at $\hat{r} = \hat{r}_{\mbox{\tiny $Q$}}$.
Inserting this in the line elements \eqref{e_267} and \eqref{e_1767}
shows that metric is continuous at the shell.
\itm As in the $(\tilde{t},\tilde{r})$ coordinates there is vanishing
acceleration of gravity inside the domain wall in the $(\hat{t},\hat{r})$
coordinate system for the case that $k = 0$ because then it is comoving
in a static reference frame in this region.
%
%
%
%
%
\newpage
{\it IIIb. Time dependent metric and coordinates $(t,r)$ with $\alpha = 0$,
$\beta = \beta (t)$.}
\vspace{3mm} \newline
With $e^{\beta (t)} = a(t)$ the line element then takes the form
\begin{equation} \label{e_328}
ds^2 = - d{t}^2 + a(t)^2 dr^2 + R_{\mbox{\tiny $Q$}}^2 d\Omega^2
\mbox{ .}
\end{equation}
Here $t$ corresponds to the cosmic time of the FRW universe models,
i.e. it is the proper time of clocks with fixed spatial coordinates,
and $a(t)$ is a scale factor describing the expansion of space in
the radial direction. There is no expansion in the directions orthogonal
to the radius vector.
\itm Calculating the Christoffel symbols from this line element we find
that
\begin{equation} \label{e_326}
\Gamma^{r}_{\hs{0.5mm} t t} =
\Gamma^{\theta}_{\hs{0.5mm} t t} =
\Gamma^{\phi}_{\hs{0.5mm} t t} = 0
\mbox{ .}
\end{equation}
From the geodesic equation it follows that a free particle
instantaneously at rest will remain at rest in this coordinate system.
Hence the coordinates $r,\theta,\phi$ are comoving with free particles.
Therefore $(t,r,\theta,\phi)$ are the coordinates of an inertial
reference frame. They may be called inertial coordinates in the PLBR
spacetime. These coordinates are analogous to the standard coordinates
used in the FRW universe models.
\itm The Einstein-Maxwell equations then take the form
\begin{equation} \label{e_329}
R_{\mbox{\tiny $Q$}}^2 \ddot{a} + a = 0
\mbox{ ,}
\end{equation}
where the dots denote differentiation with respect to the proper time
of the reference particles.
With the line element \eqref{e_328} this is also the condition that the
Weyl tensor vanishes. The general solution of equation \eqref{e_329} is
\begin{equation} \label{e_330}
a(t) = d_1 \cos (t / R_{\mbox{\tiny $Q$}})
+ d_2 \sin (t / R_{\mbox{\tiny $Q$}})
\end{equation}
where $d_1$ and $d_2$ are constants.
\itm We shall find the transformation relating this line element to
the line element \eqref{e_157} of the LBR spactime in CFS coordinates.
In this connection we will also deduce the transformation between
$\tau$ and $t$ and between $\overline{t}$ and $t$.
Since $t$ represents the proper time of clocks with fixed spatial
coordinates it follows from the line element \eqref{e_928} that
\begin{equation} \label{e_1028}
dt = \frl{R_{\mbox{\tiny $Q$}}}{\cosh \tau} \hs{0.5mm} d \tau
= \frl{R_{\mbox{\tiny $Q$}} \cosh \tau }{1 \p \sinh^2 \tau} \hs{0.5mm}
d \tau
\mbox{ .}
\end{equation}
Integration gives
\begin{equation} \label{e_1029}
t = t_0 + R_{\mbox{\tiny $Q$}} \arctan (\sinh \tau)
\mbox{ ,}
\end{equation}
where $t_0$ is a constant.
With a suitable scaling of the radial coordinate the transformation from
the $(t,r)$-system to the $(\tau,\rho)$-system is given by
\begin{equation} \label{e_1030}
\sinh \tau = \tan \left( \frl{t - t_0}{R_{\mbox{\tiny $Q$}}} \right)
\mbox{\hspace{2mm} , \hspace{3mm}}
\rho = \frl{A}{R_{\mbox{\tiny $Q$}}} \hs{0.6mm} r
\mbox{ ,}
\end{equation}
transforming the region $t_0 - R_{\mbox{\tiny $Q$}} \hs{0.3mm} \pi / 2 < t
< t_0 + R_{\mbox{\tiny $Q$}} \hs{0.3mm} \pi / 2 \,$,
$-\infty < r < \infty$ in the inertial system to the region
$-\infty < \tau < \infty$, $-\infty < \rho < \infty$ in the
$(\tau,\rho)$-system. It also follows that
\begin{equation} \label{e_1021}
\cosh \tau = 1 / \cos \left( \frl{t - t_0}{R_{\mbox{\tiny $Q$}}} \right)
\mbox{ .}
\end{equation}
Combining this with equation \eqref{e_949} we obtain
\begin{equation} \label{e_1049}
\sin \left( \frl{t - t_0}{R_{\mbox{\tiny $Q$}}} \right)
= \tanh \tau = \frl{\overline{t} \m \overline{t}_0}
{R_{\mbox{\tiny $Q$}} A}
\mbox{ ,}
\end{equation}
which implies that
\begin{equation} \label{e_1024}
t = t_0 + R_{\mbox{\tiny $Q$}} \hs{0.5mm}
\arcsin \left( \frl{\overline{t} \m \overline{t}_0}
{R_{\mbox{\tiny $Q$}} A} \right)
\mbox{ .}
\end{equation}
The inverse transformation is given by
\begin{equation} \label{e_1025}
\overline{t} = \overline{t}_0 + R_{\mbox{\tiny $Q$}} A
\hs{1.0mm} \sin \left(
\frl{t - t_0}{R_{\mbox{\tiny $Q$}}} \right)
\mbox{ .}
\end{equation}
Using the formulae \eqref{e_1028}, \eqref{e_1021} and \eqref{e_1030} it
follows that the line element \eqref{e_928} takes the form
\begin{equation} \label{e_331}
ds^2 = - d{t}^2 + A^2 \hs{0.1mm} \cos^2 \hs{-1.2mm}
\left( \frl{t - t_0}{R_{\mbox{\tiny $Q$}}} \right) \hs{0.5mm} dr^2
+ R_{\mbox{\tiny $Q$}}^2 d\Omega^2
\mbox{ .}
\end{equation}
In these coordinates the line element has a form similar to that of a
Friedmann Robertson Walker universe model with radial scale factor
$a(t) = A \hs{0.1mm} \cos (\fr{(t - t_0)}{R_{\mbox{\tiny $Q$}}})$.
The coordinate time $t$ corresponds to the cosmic time as measured
by clocks comoving with free particles. There is initially an expansion
in the radial direction, turning to contraction at the point of time
$t = t_0$. Hence $t_0$ is the point of time with maximal physical
distances.
\itm The connection between the general solution \eqref{e_330} and the
form \eqref{e_331} of the line element is given by
\begin{equation} \label{e_1115}
d_1 = A \hs{0.5mm} \cos \left( \frl{t_0}{R_{\mbox{\tiny $Q$}}} \right)
\mbox{\hspace{2mm} , \hspace{3mm}}
d_2 = A \hs{0.5mm} \sin \left( \frl{t_0}{R_{\mbox{\tiny $Q$}}} \right)
\mbox{ .}
\end{equation}
Inserting the relations \eqref{e_1030} and \eqref{e_1021} into equations
\eqref{e_981} we obtain the transformation between the inertial
coordinates $(t,r)$ and the CFS coordinates,
\begin{equation} \label{e_332}
T = \frl{B \hs{0.5mm}
\cos \mbox{$\left( \rule[-1.5mm]{0mm}{5.7mm} \right.$} \hs{-0.2mm}
\frl{t \m t_0}{R_{\mbox{\tiny $Q$}}}
\hs{-0.2mm} \mbox{$\left. \rule[-1.5mm]{0mm}{5.7mm} \right)$} \hs{0.5mm}
\cosh \mbox{$\left( \rule[-1.5mm]{0mm}{5.7mm} \right.$} \hs{-0.2mm}
\frl{A \hs{0.5mm} r}{R_{\mbox{\tiny $Q$}}}
\hs{-0.2mm} \mbox{$\left. \rule[-1.5mm]{0mm}{5.7mm} \right)$}}
{\sin \mbox{$\left( \rule[-1.5mm]{0mm}{5.7mm} \right.$} \hs{-0.2mm}
\frl{t_0 \m t}{R_{\mbox{\tiny $Q$}}}
\hs{-0.2mm} \mbox{$\left. \rule[-1.5mm]{0mm}{5.7mm} \right)$} \m \hs{0.5mm}
\cos \mbox{$\left( \rule[-1.5mm]{0mm}{5.7mm} \right.$} \hs{-0.2mm}
\frl{t \m t_0}{R_{\mbox{\tiny $Q$}}}
\hs{-0.2mm} \mbox{$\left. \rule[-1.5mm]{0mm}{5.7mm} \right)$} \hs{0.5mm}
\sinh \mbox{$\left( \rule[-1.5mm]{0mm}{5.7mm} \right.$} \hs{-0.2mm}
\frl{A \hs{0.5mm} r}{R_{\mbox{\tiny $Q$}}}
\hs{-0.2mm} \mbox{$\left. \rule[-1.5mm]{0mm}{5.7mm} \right)$}
\rule[-1.5mm]{0mm}{6.9mm}}
\mbox{ ,}
\end{equation}
\begin{equation} \label{e_333}
R = \frl{B}
{\sin \mbox{$\left( \rule[-1.5mm]{0mm}{5.7mm} \right.$} \hs{-0.2mm}
\frl{t_0 \m t}{R_{\mbox{\tiny $Q$}}}
\hs{-0.2mm} \mbox{$\left. \rule[-1.5mm]{0mm}{5.7mm} \right)$} \m \hs{0.5mm}
\cos \mbox{$\left( \rule[-1.5mm]{0mm}{5.7mm} \right.$} \hs{-0.2mm}
\frl{t \m t_0}{R_{\mbox{\tiny $Q$}}}
\hs{-0.2mm} \mbox{$\left. \rule[-1.5mm]{0mm}{5.7mm} \right)$} \hs{0.5mm}
\sinh \mbox{$\left( \rule[-1.5mm]{0mm}{5.7mm} \right.$} \hs{-0.2mm}
\frl{A \hs{0.5mm} r}{R_{\mbox{\tiny $Q$}}}
\hs{-0.2mm} \mbox{$\left. \rule[-1.5mm]{0mm}{5.7mm} \right)$}
\rule[-1.5mm]{0mm}{6.9mm}}
\mbox{ ,}
\end{equation}
\vspace{1mm} \newline
transforming the region $t_0 - R_{\mbox{\tiny $Q$}} \hs{0.3mm} \pi / 2 < t
< t_0 + R_{\mbox{\tiny $Q$}} \hs{0.3mm} \pi / 2 \,$,
$-\infty < r < \infty$ in the inertial system to the region
$|T + R| > B$, $|T - R| < B$ in the CFS system (see Figure 7).
Using equation \eqref{e_1049} we see that this transformation is
consistent with equations \eqref{e_992} and \eqref{e_954}.
The inverse transformation is given by
\begin{equation} \label{e_348}
\sin \mbox{$\left( \rule[-1.5mm]{0mm}{5.7mm} \right.$} \hs{-0.2mm}
\frl{t \m t_0}{R_{\mbox{\tiny $Q$}}}
\hs{-0.2mm} \mbox{$\left. \rule[-1.5mm]{0mm}{5.7mm} \right)$}
= \frl{(T^{2} \m R^{2}) \m B^2}{2 \hs{0.2mm} B \hs{0.2mm} R}
\mbox{\hspace{2mm} , \hspace{3mm}}
\tanh \mbox{$\left( \rule[-1.5mm]{0mm}{5.7mm} \right.$} \hs{-0.2mm}
\frl{A \hs{0.5mm} r}{R_{\mbox{\tiny $Q$}}}
\hs{-0.2mm} \mbox{$\left. \rule[-1.5mm]{0mm}{5.7mm} \right)$}
= \frl{(R^2 \m T^2) \m B^2}{2 \hs{0.2mm} B \hs{0.2mm} T}
\mbox{ .}
\end{equation}
Note that the denominators cannot vanish in the regions specified
above.
\itm The world lines of fixed points $r = r_1$ in the inertial frame
with reference to the CFS system are given by
\begin{equation} \label{e_361}
R^2 - (T - T_1)^2 = B^2 - T_1^2
\mbox{\hspace{2mm} , \hspace{3mm}}
T_1 = - B
\tanh \mbox{$\left( \rule[-1.5mm]{0mm}{5.7mm} \right.$} \hs{-0.2mm}
\frl{A \hs{0.5mm} r_1}{R_{\mbox{\tiny $Q$}}}
\hs{-0.2mm} \mbox{$\left. \rule[-1.5mm]{0mm}{5.7mm} \right)$}
\mbox{ ,}
\end{equation}
which represents hyperbolae as shown in Figure 7. This form of the
world line of a fixed point $r = r_1$ is in accordance with
equation \eqref{e_550} for the world line of a free particle.
\vspace*{2mm} \newline
\begin{picture}(50,257)(-96,-230)
\qbezier(173.1522, -99.9348)(173.4828, -99.3975)(173.8288, -98.8506)
\qbezier( 96.8478, -99.9348)( 96.5172, -100.4721)( 96.1712, -101.0190)
\qbezier(173.8288, -98.8506)(174.1747, -98.3037)(174.5362, -97.7469)
\qbezier( 96.1712, -101.0190)( 95.8253, -101.5658)( 95.4638, -102.1226)
\qbezier(174.5362, -97.7469)(174.8977, -97.1901)(175.2750, -96.6229)
\qbezier( 95.4638, -102.1226)( 95.1023, -102.6795)( 94.7250, -103.2467)
\qbezier(175.2750, -96.6229)(175.6523, -96.0556)(176.0457, -95.4776)
\qbezier( 94.7250, -103.2467)( 94.3477, -103.8139)( 93.9543, -104.3920)
\qbezier(176.0457, -95.4776)(176.4392, -94.8995)(176.8490, -94.3101)
\qbezier( 93.9543, -104.3920)( 93.5608, -104.9701)( 93.1510, -105.5595)
\qbezier(176.8490, -94.3101)(177.2589, -93.7207)(177.6856, -93.1195)
\qbezier( 93.1510, -105.5595)( 92.7411, -106.1489)( 92.3144, -106.7501)
\qbezier(177.6856, -93.1195)(178.1122, -92.5183)(178.5560, -91.9049)
\qbezier( 92.3144, -106.7501)( 91.8878, -107.3512)( 91.4440, -107.9647)
\qbezier(178.5560, -91.9049)(178.9997, -91.2915)(179.4609, -90.6653)
\qbezier( 91.4440, -107.9647)( 91.0003, -108.5781)( 90.5391, -109.2043)
\qbezier(179.4609, -90.6653)(179.9222, -90.0391)(180.4012, -89.3997)
\qbezier( 90.5391, -109.2043)( 90.0778, -109.8305)( 89.5988, -110.4699)
\qbezier(180.4012, -89.3997)(180.8802, -88.7603)(181.3775, -88.1071)
\qbezier( 89.5988, -110.4699)( 89.1198, -111.1093)( 88.6225, -111.7625)
\qbezier(181.3775, -88.1071)(181.8747, -87.4539)(182.3906, -86.7865)
\qbezier( 88.6225, -111.7625)( 88.1253, -112.4157)( 87.6094, -113.0831)
\qbezier(182.3906, -86.7865)(182.9064, -86.1190)(183.4413, -85.4368)
\qbezier( 87.6094, -113.0831)( 87.0936, -113.7505)( 86.5587, -114.4328)
\qbezier(183.4413, -85.4368)(183.9761, -84.7545)(184.5304, -84.0570)
\qbezier( 86.5587, -114.4328)( 86.0239, -115.1150)( 85.4696, -115.8126)
\qbezier(184.5304, -84.0570)(185.0847, -83.3594)(185.6588, -82.6459)
\qbezier( 85.4696, -115.8126)( 84.9153, -116.5102)( 84.3412, -117.2236)
\qbezier(185.6588, -82.6459)(186.2330, -81.9325)(186.8275, -81.2025)
\qbezier( 84.3412, -117.2236)( 83.7670, -117.9371)( 83.1725, -118.6670)
\qbezier(186.8275, -81.2025)(187.4220, -80.4726)(188.0373, -79.7256)
\qbezier( 83.1725, -118.6670)( 82.5780, -119.3970)( 81.9627, -120.1439)
\qbezier(188.0373, -79.7256)(188.6525, -78.9787)(189.2891, -78.2141)
\qbezier( 81.9627, -120.1439)( 81.3475, -120.8909)( 80.7109, -121.6555)
\qbezier(189.2891, -78.2141)(189.9257, -77.4495)(190.5841, -76.6666)
\qbezier( 80.7109, -121.6555)( 80.0743, -122.4201)( 79.4159, -123.2029)
\qbezier(190.5841, -76.6666)(191.2424, -75.8838)(191.9231, -75.0821)
\qbezier( 79.4159, -123.2029)( 78.7576, -123.9858)( 78.0769, -124.7875)
\qbezier(191.9231, -75.0821)(192.6038, -74.2804)(193.3073, -73.4592)
\qbezier( 78.0769, -124.7875)( 77.3962, -125.5892)( 76.6927, -126.4104)
\qbezier(193.3073, -73.4592)(194.0109, -72.6380)(194.7378, -71.7967)
\qbezier( 76.6927, -126.4104)( 75.9891, -127.2316)( 75.2622, -128.0729)
\qbezier(194.7378, -71.7967)(195.4648, -70.9553)(196.2157, -70.0932)
\qbezier( 75.2622, -128.0729)( 74.5352, -128.9142)( 73.7843, -129.7764)
\qbezier(196.2157, -70.0932)(196.9667, -69.2310)(197.7422, -68.3474)
\qbezier( 73.7843, -129.7764)( 73.0333, -130.6385)( 72.2578, -131.5222)
\qbezier(197.7422, -68.3474)(198.5178, -67.4637)(199.3185, -66.5579)
\qbezier( 72.2578, -131.5222)( 71.4822, -132.4058)( 70.6815, -133.3117)
\qbezier(199.3185, -66.5579)(200.1192, -65.6520)(200.9458, -64.7232)
\qbezier( 70.6815, -133.3117)( 69.8808, -134.2175)( 69.0542, -135.1463)
\qbezier(200.9458, -64.7232)(201.7724, -63.7945)(202.6255, -62.8420)
\qbezier( 69.0542, -135.1463)( 68.2276, -136.0751)( 67.3745, -137.0275)
\qbezier(202.6255, -62.8420)(203.4785, -61.8896)(204.3588, -60.9128)
\qbezier( 67.3745, -137.0275)( 66.5215, -137.9799)( 65.6412, -138.9568)
\qbezier(204.3588, -60.9128)(205.2390, -59.9359)(206.1471, -58.9339)
\qbezier( 65.6412, -138.9568)( 64.7610, -139.9336)( 63.8529, -140.9356)
\qbezier(206.1471, -58.9339)(207.0553, -57.9319)(207.9920, -56.9039)
\qbezier( 63.8529, -140.9356)( 62.9447, -141.9377)( 62.0080, -142.9657)
\qbezier(207.9920, -56.9039)(208.9287, -55.8759)(209.8947, -54.8211)
\qbezier( 62.0080, -142.9657)( 61.0713, -143.9937)( 60.1053, -145.0485)
\qbezier(209.8947, -54.8211)(210.8607, -53.7663)(211.8569, -52.6838)
\qbezier( 60.1053, -145.0485)( 59.1393, -146.1033)( 58.1431, -147.1857)
\qbezier(211.8569, -52.6838)(212.8530, -51.6014)(213.8801, -50.4905)
\qbezier( 58.1431, -147.1857)( 57.1470, -148.2682)( 56.1199, -149.3791)
\qbezier(213.8801, -50.4905)(214.9071, -49.3795)(215.9658, -48.2392)
\qbezier( 56.1199, -149.3791)( 55.0929, -150.4900)( 54.0342, -151.6304)
\qbezier(215.9658, -48.2392)(217.0246, -47.0989)(218.1158, -45.9283)
\qbezier( 54.0342, -151.6304)( 52.9754, -152.7707)( 51.8842, -153.9413)
\qbezier(218.1158, -45.9283)(219.2071, -44.7577)(220.3318, -43.5559)
\qbezier( 51.8842, -153.9413)( 50.7929, -155.1118)( 49.6682, -156.3136)
\qbezier(220.3318, -43.5559)(221.4565, -42.3541)(222.6155, -41.1202)
\qbezier( 49.6682, -156.3136)( 48.5435, -157.5154)( 47.3845, -158.7494)
\qbezier(222.6155, -41.1202)(223.7745, -39.8862)(224.9687, -38.6191)
\qbezier( 47.3845, -158.7494)( 46.2255, -159.9834)( 45.0313, -161.2505)
\qbezier(224.9687, -38.6191)(226.1630, -37.3520)(227.3934, -36.0507)
\qbezier( 45.0313, -161.2505)( 43.8370, -162.5176)( 42.6066, -163.8188)
\qbezier(227.3934, -36.0507)(228.6238, -34.7495)(229.8913, -33.4130)
\qbezier( 42.6066, -163.8188)( 41.3762, -165.1201)( 40.1087, -166.4565)
\put(246.0326, -29.0109){\makebox(0,0)[]{\scriptsize{$r_1 > 0$}}}
\put( 26.9022, -173.7935){\makebox(0,0)[]{\scriptsize{$r_1 < 0$}}}
\qbezier(173.1522, -99.9348)(173.1522, -99.2293)(173.1783, -98.5234)
\qbezier( 96.8478, -99.9348)( 96.8478, -100.6402)( 96.8217, -101.3461)
\qbezier(173.1783, -98.5234)(173.2044, -97.8175)(173.2566, -97.1102)
\qbezier( 96.8217, -101.3461)( 96.7956, -102.0520)( 96.7434, -102.7594)
\qbezier(173.2566, -97.1102)(173.3088, -96.4028)(173.3873, -95.6930)
\qbezier( 96.7434, -102.7594)( 96.6912, -103.4668)( 96.6127, -104.1766)
\qbezier(173.3873, -95.6930)(173.4657, -94.9832)(173.5704, -94.2701)
\qbezier( 96.6127, -104.1766)( 96.5343, -104.8863)( 96.4296, -105.5995)
\qbezier(173.5704, -94.2701)(173.6752, -93.5569)(173.8064, -92.8394)
\qbezier( 96.4296, -105.5995)( 96.3248, -106.3127)( 96.1936, -107.0302)
\qbezier(173.8064, -92.8394)(173.9376, -92.1218)(174.0954, -91.3990)
\qbezier( 96.1936, -107.0302)( 96.0624, -107.7477)( 95.9046, -108.4706)
\qbezier(174.0954, -91.3990)(174.2532, -90.6761)(174.4379, -89.9469)
\qbezier( 95.9046, -108.4706)( 95.7468, -109.1935)( 95.5621, -109.9227)
\qbezier(174.4379, -89.9469)(174.6226, -89.2177)(174.8343, -88.4811)
\qbezier( 95.5621, -109.9227)( 95.3774, -110.6519)( 95.1657, -111.3884)
\qbezier(174.8343, -88.4811)(175.0461, -87.7446)(175.2853, -86.9997)
\qbezier( 95.1657, -111.3884)( 94.9539, -112.1250)( 94.7147, -112.8698)
\qbezier(175.2853, -86.9997)(175.5245, -86.2549)(175.7913, -85.5006)
\qbezier( 94.7147, -112.8698)( 94.4755, -113.6147)( 94.2087, -114.3689)
\qbezier(175.7913, -85.5006)(176.0582, -84.7464)(176.3532, -83.9818)
\qbezier( 94.2087, -114.3689)( 93.9418, -115.1232)( 93.6468, -115.8878)
\qbezier(176.3532, -83.9818)(176.6482, -83.2172)(176.9716, -82.4411)
\qbezier( 93.6468, -115.8878)( 93.3518, -116.6524)( 93.0284, -117.4285)
\qbezier(176.9716, -82.4411)(177.2951, -81.6650)(177.6475, -80.8765)
\qbezier( 93.0284, -117.4285)( 92.7049, -118.2045)( 92.3525, -118.9931)
\qbezier(177.6475, -80.8765)(177.9999, -80.0879)(178.3817, -79.2858)
\qbezier( 92.3525, -118.9931)( 92.0001, -119.7816)( 91.6183, -120.5838)
\qbezier(178.3817, -79.2858)(178.7635, -78.4837)(179.1752, -77.6669)
\qbezier( 91.6183, -120.5838)( 91.2365, -121.3859)( 90.8248, -122.2027)
\qbezier(179.1752, -77.6669)(179.5869, -76.8501)(180.0292, -76.0175)
\qbezier( 90.8248, -122.2027)( 90.4131, -123.0195)( 89.9708, -123.8521)
\qbezier(180.0292, -76.0175)(180.4714, -75.1849)(180.9447, -74.3354)
\qbezier( 89.9708, -123.8521)( 89.5286, -124.6847)( 89.0553, -125.5342)
\qbezier(180.9447, -74.3354)(181.4181, -73.4859)(181.9231, -72.6183)
\qbezier( 89.0553, -125.5342)( 88.5819, -126.3837)( 88.0769, -127.2513)
\qbezier(181.9231, -72.6183)(182.4282, -71.7507)(182.9657, -70.8638)
\qbezier( 88.0769, -127.2513)( 87.5718, -128.1189)( 87.0343, -129.0058)
\qbezier(182.9657, -70.8638)(183.5033, -69.9769)(184.0740, -69.0695)
\qbezier( 87.0343, -129.0058)( 86.4967, -129.8927)( 85.9260, -130.8001)
\qbezier(184.0740, -69.0695)(184.6447, -68.1621)(185.2493, -67.2330)
\qbezier( 85.9260, -130.8001)( 85.3553, -131.7074)( 84.7507, -132.6365)
\qbezier(185.2493, -67.2330)(185.8540, -66.3039)(186.4934, -65.3518)
\qbezier( 84.7507, -132.6365)( 84.1460, -133.5657)( 83.5066, -134.5178)
\qbezier(186.4934, -65.3518)(187.1328, -64.3997)(187.8079, -63.4233)
\qbezier( 83.5066, -134.5178)( 82.8672, -135.4699)( 82.1921, -136.4463)
\qbezier(187.8079, -63.4233)(188.4830, -62.4469)(189.1947, -61.4448)
\qbezier( 82.1921, -136.4463)( 81.5170, -137.4227)( 80.8053, -138.4248)
\qbezier(189.1947, -61.4448)(189.9064, -60.4427)(190.6556, -59.4137)
\qbezier( 80.8053, -138.4248)( 80.0936, -139.4268)( 79.3444, -140.4559)
\qbezier(190.6556, -59.4137)(191.4049, -58.3846)(192.1927, -57.3271)
\qbezier( 79.3444, -140.4559)( 78.5951, -141.4850)( 77.8073, -142.5425)
\qbezier(192.1927, -57.3271)(192.9805, -56.2696)(193.8080, -55.1822)
\qbezier( 77.8073, -142.5425)( 77.0195, -143.6000)( 76.1920, -144.6873)
\qbezier(193.8080, -55.1822)(194.6355, -54.0949)(195.5037, -52.9762)
\qbezier( 76.1920, -144.6873)( 75.3645, -145.7747)( 74.4963, -146.8934)
\qbezier(195.5037, -52.9762)(196.3720, -51.8575)(197.2822, -50.7059)
\qbezier( 74.4963, -146.8934)( 73.6280, -148.0121)( 72.7178, -149.1637)
\qbezier(197.2822, -50.7059)(198.1925, -49.5543)(199.1459, -48.3682)
\qbezier( 72.7178, -149.1637)( 71.8075, -150.3153)( 70.8541, -151.5014)
\qbezier(199.1459, -48.3682)(200.0994, -47.1821)(201.0974, -45.9600)
\qbezier( 70.8541, -151.5014)( 69.9006, -152.6874)( 68.9026, -153.9096)
\qbezier(201.0974, -45.9600)(202.0954, -44.7379)(203.1393, -43.4780)
\qbezier( 68.9026, -153.9096)( 67.9046, -155.1317)( 66.8607, -156.3916)
\qbezier(203.1393, -43.4780)(204.1832, -42.2181)(205.2744, -40.9187)
\qbezier( 66.8607, -156.3916)( 65.8168, -157.6515)( 64.7256, -158.9509)
\qbezier(205.2744, -40.9187)(206.3656, -39.6193)(207.5056, -38.2787)
\qbezier( 64.7256, -158.9509)( 63.6344, -160.2502)( 62.4944, -161.5909)
\qbezier(207.5056, -38.2787)(208.6456, -36.9381)(209.8360, -35.5544)
\qbezier( 62.4944, -161.5909)( 61.3544, -162.9315)( 60.1640, -164.3152)
\qbezier(209.8360, -35.5544)(211.0264, -34.1706)(212.2688, -32.7419)
\qbezier( 60.1640, -164.3152)( 58.9736, -165.6989)( 57.7312, -167.1276)
\qbezier(212.2688, -32.7419)(213.5112, -31.3132)(214.8073, -29.8376)
\qbezier( 57.7312, -167.1276)( 56.4888, -168.5563)( 55.1927, -170.0320)
\qbezier(214.8073, -29.8376)(216.1034, -28.3620)(217.4550, -26.8374)
\qbezier( 55.1927, -170.0320)( 53.8966, -171.5076)( 52.5450, -173.0322)
\qbezier(217.4550, -26.8374)(218.8065, -25.3128)(220.2154, -23.7372)
\qbezier( 52.5450, -173.0322)( 51.1935, -174.5568)( 49.7846, -176.1324)
\put(236.3567, -19.3350){\makebox(0,0)[]{\scriptsize{$r_1 = 0$}}}
\put( 36.5781, -183.4694){\makebox(0,0)[]{\scriptsize{$r_1 = 0$}}}
\qbezier(173.1522, -99.9348)(172.1337, -98.6107)(171.2047, -97.3543)
\qbezier( 96.8478, -99.9348)( 97.8663, -101.2588)( 98.7953, -102.5153)
\qbezier(171.2047, -97.3543)(170.2758, -96.0978)(169.4319, -94.9028)
\qbezier( 98.7953, -102.5153)( 99.7242, -103.7718)(100.5681, -104.9667)
\qbezier(169.4319, -94.9028)(168.5880, -93.7079)(167.8252, -92.5687)
\qbezier(100.5681, -104.9667)(101.4120, -106.1617)(102.1748, -107.3009)
\qbezier(167.8252, -92.5687)(167.0623, -91.4295)(166.3767, -90.3406)
\qbezier(102.1748, -107.3009)(102.9377, -108.4400)(103.6233, -109.5290)
\qbezier(166.3767, -90.3406)(165.6912, -89.2517)(165.0797, -88.2078)
\qbezier(103.6233, -109.5290)(104.3088, -110.6179)(104.9203, -111.6618)
\qbezier(165.0797, -88.2078)(164.4681, -87.1639)(163.9277, -86.1599)
\qbezier(104.9203, -111.6618)(105.5319, -112.7057)(106.0723, -113.7096)
\qbezier(163.9277, -86.1599)(163.3872, -85.1560)(162.9152, -84.1872)
\qbezier(106.0723, -113.7096)(106.6128, -114.7135)(107.0848, -115.6823)
\qbezier(162.9152, -84.1872)(162.4433, -83.2184)(162.0374, -82.2801)
\qbezier(107.0848, -115.6823)(107.5567, -116.6511)(107.9626, -117.5894)
\qbezier(162.0374, -82.2801)(161.6316, -81.3418)(161.2901, -80.4294)
\qbezier(107.9626, -117.5894)(108.3684, -118.5278)(108.7099, -119.4402)
\qbezier(161.2901, -80.4294)(160.9485, -79.5170)(160.6695, -78.6262)
\qbezier(108.7099, -119.4402)(109.0515, -120.3525)(109.3305, -121.2434)
\qbezier(160.6695, -78.6262)(160.3905, -77.7353)(160.1727, -76.8617)
\qbezier(109.3305, -121.2434)(109.6095, -122.1342)(109.8273, -123.0079)
\qbezier(160.1727, -76.8617)(159.9550, -75.9881)(159.7974, -75.1275)
\qbezier(109.8273, -123.0079)(110.0450, -123.8815)(110.2026, -124.7421)
\qbezier(159.7974, -75.1275)(159.6398, -74.2669)(159.5416, -73.4152)
\qbezier(110.2026, -124.7421)(110.3602, -125.6026)(110.4584, -126.4543)
\qbezier(159.5416, -73.4152)(159.4435, -72.5635)(159.4043, -71.7166)
\qbezier(110.4584, -126.4543)(110.5565, -127.3061)(110.5957, -128.1530)
\qbezier(159.4043, -71.7166)(159.3651, -70.8696)(159.3846, -70.0234)
\qbezier(110.5957, -128.1530)(110.6349, -128.9999)(110.6154, -129.8462)
\qbezier(159.3846, -70.0234)(159.4042, -69.1771)(159.4826, -68.3274)
\qbezier(110.6154, -129.8462)(110.5958, -130.6925)(110.5174, -131.5421)
\qbezier(159.4826, -68.3274)(159.5610, -67.4778)(159.6987, -66.6206)
\qbezier(110.5174, -131.5421)(110.4390, -132.3918)(110.3013, -133.2489)
\qbezier(159.6987, -66.6206)(159.8363, -65.7635)(160.0339, -64.8947)
\qbezier(110.3013, -133.2489)(110.1637, -134.1061)(109.9661, -134.9749)
\qbezier(160.0339, -64.8947)(160.2314, -64.0259)(160.4898, -63.1413)
\qbezier(109.9661, -134.9749)(109.7686, -135.8437)(109.5102, -136.7283)
\qbezier(160.4898, -63.1413)(160.7482, -62.2566)(161.0687, -61.3519)
\qbezier(109.5102, -136.7283)(109.2518, -137.6129)(108.9313, -138.5176)
\qbezier(161.0687, -61.3519)(161.3892, -60.4472)(161.7734, -59.5181)
\qbezier(108.9313, -138.5176)(108.6108, -139.4223)(108.2266, -140.3515)
\qbezier(161.7734, -59.5181)(162.1575, -58.5889)(162.6072, -57.6308)
\qbezier(108.2266, -140.3515)(107.8425, -141.2807)(107.3928, -142.2388)
\qbezier(162.6072, -57.6308)(163.0568, -56.6727)(163.5741, -55.6811)
\qbezier(107.3928, -142.2388)(106.9432, -143.1968)(106.4259, -144.1885)
\qbezier(163.5741, -55.6811)(164.0914, -54.6894)(164.6789, -53.6594)
\qbezier(106.4259, -144.1885)(105.9086, -145.1802)(105.3211, -146.2102)
\qbezier(164.6789, -53.6594)(165.2663, -52.6294)(165.9268, -51.5561)
\qbezier(105.3211, -146.2102)(104.7337, -147.2401)(104.0732, -148.3134)
\qbezier(165.9268, -51.5561)(166.5872, -50.4828)(167.3239, -49.3610)
\qbezier(104.0732, -148.3134)(103.4128, -149.3868)(102.6761, -150.5085)
\qbezier(167.3239, -49.3610)(168.0605, -48.2392)(168.8769, -47.0635)
\qbezier(102.6761, -150.5085)(101.9395, -151.6303)(101.1231, -152.8060)
\qbezier(168.8769, -47.0635)(169.6933, -45.8879)(170.5933, -44.6526)
\qbezier(101.1231, -152.8060)(100.3067, -153.9817)( 99.4067, -155.2170)
\qbezier(170.5933, -44.6526)(171.4934, -43.4174)(172.4814, -42.1166)
\qbezier( 99.4067, -155.2170)( 98.5066, -156.4522)( 97.5186, -157.7530)
\qbezier(172.4814, -42.1166)(173.4695, -40.8158)(174.5503, -39.4432)
\qbezier( 97.5186, -157.7530)( 96.5305, -159.0538)( 95.4497, -160.4263)
\qbezier(174.5503, -39.4432)(175.6311, -38.0706)(176.8100, -36.6196)
\qbezier( 95.4497, -160.4263)( 94.3689, -161.7989)( 93.1900, -163.2499)
\qbezier(176.8100, -36.6196)(177.9888, -35.1686)(179.2713, -33.6322)
\qbezier( 93.1900, -163.2499)( 92.0112, -164.7009)( 90.7287, -166.2373)
\qbezier(179.2713, -33.6322)(180.5538, -32.0958)(181.9461, -30.4666)
\qbezier( 90.7287, -166.2373)( 89.4462, -167.7738)( 88.0539, -169.4030)
\qbezier(181.9461, -30.4666)(183.3385, -28.8373)(184.8475, -27.1074)
\qbezier( 88.0539, -169.4030)( 86.6615, -171.0323)( 85.1525, -172.7622)
\qbezier(184.8475, -27.1074)(186.3564, -25.3774)(187.9892, -23.5385)
\qbezier( 85.1525, -172.7622)( 83.6436, -174.4921)( 82.0108, -176.3311)
\qbezier(187.9892, -23.5385)(189.6220, -21.6995)(191.3865, -19.7426)
\qbezier( 82.0108, -176.3311)( 80.3780, -178.1701)( 78.6135, -180.1270)
\qbezier(191.3865, -19.7426)(193.1511, -17.7857)(195.0559, -15.7015)
\qbezier( 78.6135, -180.1270)( 76.8489, -182.0838)( 74.9441, -184.1681)
\qbezier(195.0559, -15.7015)(196.9606, -13.6173)(199.0148, -11.3957)
\qbezier( 74.9441, -184.1681)( 73.0394, -186.2523)( 70.9852, -188.4739)
\qbezier(199.0148, -11.3957)(201.0691,  -9.1741)(203.2826,  -6.8043)
\qbezier( 70.9852, -188.4739)( 68.9309, -190.6955)( 66.7174, -193.0652)
\put(219.4239,  -2.4022){\makebox(0,0)[]{\scriptsize{$r_1 < 0$}}}
\put( 53.5109, -200.4022){\makebox(0,0)[]{\scriptsize{$r_1 > 0$}}}
\put(135.0000, -61.7826){\line(1, -1){ 38.1522}}
\put(135.0000, -61.7826){\line(1, 1){ 61.6304}}
\put(173.1522, -99.9348){\line(1, 1){ 61.6304}}
\put( 96.8478, -99.9348){\line(1, -1){ 38.1522}}
\put( 73.3696, -199.7174){\line(1, 1){ 61.6304}}
\put( 35.2174, -161.5652){\line(1, 1){ 61.6304}}
\put(164.3478, -102.5761){\line(0, 1){  5.2826}}
\qbezier(164.3478, -91.1304)(164.3478, -89.8544)(164.3478, -88.5784)
\qbezier(164.3478, -86.0265)(164.3478, -84.7505)(164.3478, -83.4745)
\qbezier(164.3478, -80.9225)(164.3478, -79.6465)(164.3478, -78.3705)
\qbezier(164.3478, -75.8185)(164.3478, -74.5425)(164.3478, -73.2665)
\qbezier(164.3478, -70.7146)(164.3478, -69.4386)(164.3478, -68.1626)
\qbezier(164.3478, -65.6106)(164.3478, -64.3346)(164.3478, -63.0586)
\qbezier(164.3478, -60.5066)(164.3478, -59.2306)(164.3478, -57.9546)
\qbezier(164.3478, -55.4026)(164.3478, -54.1267)(164.3478, -52.8507)
\qbezier(164.3478, -50.2987)(164.3478, -49.0227)(164.3478, -47.7467)
\qbezier(164.3478, -45.1947)(164.3478, -43.9187)(164.3478, -42.6427)
\qbezier(164.3478, -40.0907)(164.3478, -38.8147)(164.3478, -37.5388)
\qbezier(164.3478, -34.9868)(164.3478, -33.7108)(164.3478, -32.4348)
\put( 32.2826, -99.9348){\vector(1, 0){220.1087}}
\put(135.0000, -202.6522){\vector(0, 1){205.4348}}
\put(261.1957, -104.3370){\makebox(0,0)[]{\footnotesize{$R$}}}
\put(127.6630,   4.2500){\makebox(0,0)[]{\footnotesize{$T$}}}
\put(126.1957, -63.2500){\makebox(0,0)[]{\footnotesize{$B$}}}
\put(174.6196, -108.7391){\makebox(0,0)[]{\footnotesize{$B$}}}
\put(159.9457, -108.7391){\makebox(0,0)[]{\footnotesize{$R_{\mbox{\tiny $Q$}}$}}}
\put(149.6739, -139.5543){\makebox(0,0)[]{\footnotesize{$-B$}}}
\put( 98.3152, -94.0652){\makebox(0,0)[]{\footnotesize{$-B$}}}
\end{picture}
\vspace{-3mm} \newline
%
{\footnotesize \sf Figure 7. The world lines of freely falling particles
with $r = r_1$ as shown in the CFS system for different values of $r_1$.
The comoving coordinates $(t,r)$ cover a part of the PLBR spacetime
given by $|T - R| < B$, $|T + R| > B$. The region to the right of the
vertical line $R = R_{\mbox{\tiny $Q$}}$ represents a part of the WLBR
spacetime.}
\vspace{3mm} \newline
\itm Differentiating equation \eqref{e_361} we find the coordinate
velocity of a particle with $r = r_1$ in the CFS system. The initial
velocity of the particle at $T = 0$, $R = B$ is
\begin{equation} \label{e_1361}
\left( \frl{dR}{dT} \right)_{T = 0} = - \frl{T_1}{B} =
\tanh \mbox{$\left( \rule[-1.5mm]{0mm}{5.7mm} \right.$} \hs{-0.2mm}
\frl{A \hs{0.5mm} r_1}{R_{\mbox{\tiny $Q$}}}
\hs{-0.2mm} \mbox{$\left. \rule[-1.5mm]{0mm}{5.7mm} \right)$}
\mbox{ .}
\end{equation}
\vspace{-9mm} \newline
\itm We want to find the region in the $(t,r)$-system corresponding
to WLBR spacetime. This region is given by $R > R_{\mbox{\tiny $Q$}}$.
From equation \eqref{e_333} it follows that this corresponds to
\begin{equation} \label{e_1333}
0 < \sin \mbox{$\left( \rule[-1.5mm]{0mm}{5.7mm} \right.$} \hs{-0.2mm}
\frl{t_0 \m t}{R_{\mbox{\tiny $Q$}}}
\hs{-0.2mm} \mbox{$\left. \rule[-1.5mm]{0mm}{5.7mm} \right)$} \m
\cos \mbox{$\left( \rule[-1.5mm]{0mm}{5.7mm} \right.$} \hs{-0.2mm}
\frl{t \m t_0}{R_{\mbox{\tiny $Q$}}}
\hs{-0.2mm} \mbox{$\left. \rule[-1.5mm]{0mm}{5.7mm} \right)$} \hs{0.5mm}
\sinh \mbox{$\left( \rule[-1.5mm]{0mm}{5.7mm} \right.$} \hs{-0.2mm}
\frl{A \hs{0.5mm} r}{R_{\mbox{\tiny $Q$}}}
\hs{-0.2mm} \mbox{$\left. \rule[-1.5mm]{0mm}{5.7mm} \right)$}
< \frl{B}{R_{\mbox{\tiny $Q$}}}
\mbox{ ,}
\end{equation}
\vspace{-7mm} \newline
which gives
\begin{equation} \label{e_1335}
t_0 - R_{\mbox{\tiny $Q$}} \hs{0.3mm} \pi / 2 < t
< t_0 + R_{\mbox{\tiny $Q$}} \hs{0.3mm} \pi / 2
\end{equation}
and
\begin{equation} \label{e_1334}
\frl{R_{\mbox{\tiny $Q$}}
\sin \mbox{$\left( \rule[-1.5mm]{0mm}{5.7mm} \right.$} \hs{-0.2mm}
\frl{t \m t_0}{R_{\mbox{\tiny $Q$}}}
\hs{-0.2mm} \mbox{$\left. \rule[-1.5mm]{0mm}{5.7mm} \right)$} \m B}
{R_{\mbox{\tiny $Q$}}
\cos \mbox{$\left( \rule[-1.5mm]{0mm}{5.7mm} \right.$} \hs{-0.2mm}
\frl{t \m t_0}{R_{\mbox{\tiny $Q$}}}
\hs{-0.2mm} \mbox{$\left. \rule[-1.5mm]{0mm}{5.7mm} \right)$}}
< \sinh \mbox{$\left( \rule[-1.5mm]{0mm}{5.7mm} \right.$} \hs{-0.2mm}
\frl{A \hs{0.5mm} r}{R_{\mbox{\tiny $Q$}}}
\hs{-0.2mm} \mbox{$\left. \rule[-1.5mm]{0mm}{5.7mm} \right)$}
< \tan \mbox{$\left( \rule[-1.5mm]{0mm}{5.7mm} \right.$} \hs{-0.2mm}
\frl{t \m t_0}{R_{\mbox{\tiny $Q$}}}
\hs{-0.2mm} \mbox{$\left. \rule[-1.5mm]{0mm}{5.7mm} \right)$}
\mbox{ .}
\end{equation}
\vspace{1mm} \newline
The WLBR spacetime is shown as the hatched region in Figure 8.
The part to the left of the hatched region corresponds to
$0 < R < R_{\mbox{\tiny $Q$}}$ in the CFS system, while the part
to the right corresponds to $R < 0$.
\newpage
\begin{picture}(50,212)(-97,-170)
\qbezier( 12.4368, -112.1558)( 12.6321, -112.1613)( 39.2843, -110.1690)
\qbezier( 39.2843, -110.1690)( 42.2743, -109.9455)( 54.8101, -108.1823)
\qbezier( 54.8101, -108.1823)( 57.7352, -107.7708)( 64.9866, -106.1955)
\qbezier( 64.9866, -106.1955)( 67.3525, -105.6815)( 72.0286, -104.2088)
\qbezier( 72.0286, -104.2088)( 73.8610, -103.6316)( 77.0345, -102.2220)
\qbezier( 77.0345, -102.2220)( 78.4337, -101.6005)( 80.6355, -100.2352)
\qbezier( 80.6355, -100.2352)( 81.6936, -99.5791)( 83.2257, -98.2485)
\qbezier( 83.2257, -98.2485)( 84.0146, -97.5633)( 85.0651, -96.2617)
\qbezier( 85.0651, -96.2617)( 85.6387, -95.5510)( 86.3316, -94.2749)
\qbezier( 86.3316, -94.2749)( 86.7301, -93.5410)( 87.1508, -92.2882)
\qbezier( 87.1508, -92.2882)( 87.4044, -91.5327)( 87.6132, -90.3014)
\qbezier( 87.6132, -90.3014)( 87.7448, -89.5256)( 87.7857, -88.3147)
\qbezier( 87.7857, -88.3147)( 87.8126, -87.5194)( 87.7182, -86.3279)
\qbezier( 87.7182, -86.3279)( 87.6537, -85.5137)( 87.4485, -84.3411)
\qbezier( 87.4485, -84.3411)( 87.3028, -83.5084)( 87.0058, -82.3544)
\qbezier( 87.0058, -82.3544)( 86.7867, -81.5034)( 86.4123, -80.3676)
\qbezier( 86.4123, -80.3676)( 86.1259, -79.4986)( 85.6857, -78.3809)
\qbezier( 85.6857, -78.3809)( 85.3364, -77.4940)( 84.8394, -76.3941)
\qbezier( 84.8394, -76.3941)( 84.4307, -75.4897)( 83.8839, -74.4073)
\qbezier( 83.8839, -74.4073)( 83.4183, -73.4857)( 82.8273, -72.4206)
\qbezier( 82.8273, -72.4206)( 82.3066, -71.4822)( 81.6756, -70.4338)
\qbezier( 81.6756, -70.4338)( 81.1010, -69.4792)( 80.4332, -68.4470)
\qbezier( 80.4332, -68.4470)( 79.8055, -67.4769)( 79.1031, -66.4603)
\qbezier( 79.1031, -66.4603)( 78.4226, -65.4754)( 77.6870, -64.4735)
\qbezier( 77.6870, -64.4735)( 76.9536, -63.4746)( 76.1855, -62.4868)
\qbezier( 76.1855, -62.4868)( 75.3987, -61.4747)( 74.5982, -60.5000)
\qbezier( 74.5982, -60.5000)( 73.7571, -59.4756)( 72.9238, -58.5132)
\qbezier( 72.9238, -58.5132)( 72.0269, -57.4773)( 71.1598, -56.5265)
\qbezier( 71.1598, -56.5265)( 70.2052, -55.4797)( 69.3029, -54.5397)
\qbezier( 69.3029, -54.5397)( 68.2882, -53.4826)( 67.3488, -52.5530)
\qbezier( 67.3488, -52.5530)( 66.2708, -51.4861)( 65.2919, -50.5662)
\qbezier( 65.2919, -50.5662)( 64.1466, -49.4899)( 63.1256, -48.5794)
\qbezier( 63.1256, -48.5794)( 61.9082, -47.4939)( 60.8417, -46.5927)
\qbezier( 60.8417, -46.5927)( 59.5463, -45.4980)( 58.4305, -44.6059)
\qbezier( 58.4305, -44.6059)( 57.0499, -43.5021)( 55.8804, -42.6191)
\qbezier( 55.8804, -42.6191)( 54.4058, -41.5059)( 53.1775, -40.6324)
\qbezier( 53.1775, -40.6324)( 51.5982, -39.5093)( 50.3052, -38.6456)
\qbezier( 50.3052, -38.6456)( 48.6081, -37.5120)( 47.2435, -36.6589)
\qbezier( 47.2435, -36.6589)( 45.4121, -35.5139)( 43.9679, -34.6721)
\qbezier( 43.9679, -34.6721)( 41.9819, -33.5145)( 40.4486, -32.6853)
\qbezier( 40.4486, -32.6853)( 38.2819, -31.5136)( 36.6484, -30.6986)
\qbezier( 36.6484, -30.6986)( 34.2675, -29.5107)( 32.5204, -28.7118)
\qbezier( 32.5204, -28.7118)( 29.8811, -27.5051)( 28.0045, -26.7251)
\qbezier( 28.0045, -26.7251)( 25.0471, -25.4959)( 23.0218, -24.7383)
\qbezier( 23.0218, -24.7383)( 19.6631, -23.4819)( 17.4664, -22.7515)
\qbezier( 17.4664, -22.7515)( 13.5855, -21.4612)( 11.1906, -20.7648)
\put( 22.6132, -111.5482){\line(1, 0){231.6890}}
\put( 58.3534, -107.5747){\line(1, 0){182.2564}}
\put( 73.7402, -103.6012){\line(1, 0){144.9464}}
\put( 81.5209, -99.6276){\line(1, 0){121.8606}}
\put( 85.5066, -95.6541){\line(1, 0){105.9810}}
\put( 87.3266, -91.6806){\line(1, 0){ 94.3230}}
\put( 87.7886, -87.7071){\line(1, 0){ 85.3748}}
\put( 87.3303, -83.7335){\line(1, 0){ 78.2831}}
\put( 86.2036, -79.7600){\line(1, 0){ 72.5271}}
\put( 84.5584, -75.7865){\line(1, 0){ 67.7700}}
\put( 82.4849, -71.8130){\line(1, 0){ 63.7834}}
\put( 80.0356, -67.8395){\line(1, 0){ 60.4068}}
\put( 77.2369, -63.8659){\line(1, 0){ 57.5234}}
\put( 74.0955, -59.8924){\line(1, 0){ 55.0465}}
\put( 70.6020, -55.9189){\line(1, 0){ 52.9104}}
\put( 66.7310, -51.9454){\line(1, 0){ 51.0641}}
\put( 62.4400, -47.9718){\line(1, 0){ 49.4677}}
\put( 57.6659, -43.9983){\line(1, 0){ 48.0898}}
\put( 52.3179, -40.0248){\line(1, 0){ 46.9050}}
\put( 46.2657, -36.0513){\line(1, 0){ 45.8932}}
\put( 39.3182, -32.0777){\line(1, 0){ 45.0380}}
\put( 31.1837, -28.1042){\line(1, 0){ 44.3265}}
\put( 21.3895, -24.1307){\line(1, 0){ 43.7483}}
\put( 11.1906, -20.1572){\line(1, 0){ 41.2019}}
\put( 11.1906, -16.1837){\line(1, 0){ 24.3680}}
\put( 11.1906, -12.2101){\line(1, 0){  1.0146}}
\qbezier(254.9146, -109.3702)(253.4696, -109.2392)(242.0476, -107.7795)
\qbezier(242.0476, -107.7795)(240.1082, -107.5316)(231.8376, -106.1887)
\qbezier(231.8376, -106.1887)(229.7911, -105.8564)(223.3579, -104.5979)
\qbezier(223.3579, -104.5979)(221.3417, -104.2035)(216.0928, -103.0072)
\qbezier(216.0928, -103.0072)(214.1558, -102.5657)(209.7255, -101.4164)
\qbezier(209.7255, -101.4164)(207.8827, -100.9383)(204.0476, -99.8256)
\qbezier(204.0476, -99.8256)(202.2998, -99.3185)(198.9144, -98.2348)
\qbezier(198.9144, -98.2348)(197.2569, -97.7043)(194.2212, -96.6441)
\qbezier(194.2212, -96.6441)(192.6474, -96.0944)(189.8902, -95.0533)
\qbezier(189.8902, -95.0533)(188.3928, -94.4879)(185.8613, -93.4625)
\qbezier(185.8613, -93.4625)(184.4335, -92.8842)(182.0877, -91.8717)
\qbezier(182.0877, -91.8717)(180.7229, -91.2827)(178.5318, -90.2810)
\qbezier(178.5318, -90.2810)(177.2244, -89.6833)(175.1630, -88.6902)
\qbezier(175.1630, -88.6902)(173.9080, -88.0856)(171.9562, -87.0994)
\qbezier(171.9562, -87.0994)(170.7490, -86.4895)(168.8902, -85.5087)
\qbezier(168.8902, -85.5087)(167.7272, -84.8950)(165.9472, -83.9179)
\qbezier(165.9472, -83.9179)(164.8253, -83.3020)(163.1119, -82.3271)
\qbezier(163.1119, -82.3271)(162.0285, -81.7106)(160.3711, -80.7363)
\qbezier(160.3711, -80.7363)(159.3243, -80.1210)(157.7131, -79.1456)
\qbezier(157.7131, -79.1456)(156.7018, -78.5333)(155.1279, -77.5548)
\qbezier(155.1279, -77.5548)(154.1518, -76.9480)(152.6062, -75.9640)
\qbezier(152.6062, -75.9640)(151.6661, -75.3656)(150.1399, -74.3733)
\qbezier(150.1399, -74.3733)(149.2382, -73.7870)(147.7216, -72.7825)
\qbezier(147.7216, -72.7825)(146.8628, -72.2137)(145.3445, -71.1917)
\qbezier(145.3445, -71.1917)(144.5366, -70.6480)(143.0023, -69.6009)
\qbezier(143.0023, -69.6009)(142.2596, -69.0941)(140.6891, -68.0102)
\qbezier(140.6891, -68.0102)(140.0375, -67.5604)(138.3995, -66.4194)
\qbezier(138.3995, -66.4194)(137.8913, -66.0654)(136.1281, -64.8286)
\qbezier(136.1281, -64.8286)(135.8895, -64.6613)(133.8699, -63.2378)
\qbezier(133.8699, -63.2378)(134.3426, -63.5710)(131.6200, -61.6471)
\qbezier(131.6200, -61.6471)(130.4967, -60.8517)(129.3735, -60.0563)
\qbezier(129.3735, -60.0563)(128.2496, -59.2609)(127.1257, -58.4655)
\qbezier(127.1257, -58.4655)(124.8947, -56.8909)(124.8717, -56.8748)
\qbezier(124.8717, -56.8748)(123.0195, -55.5725)(122.6068, -55.2840)
\qbezier(122.6068, -55.2840)(120.9231, -54.1072)(120.3258, -53.6932)
\qbezier(120.3258, -53.6932)(118.7305, -52.5873)(118.0237, -52.1024)
\qbezier(118.0237, -52.1024)(116.4764, -51.0409)(115.6951, -50.5117)
\qbezier(115.6951, -50.5117)(114.1717, -49.4798)(113.3341, -48.9209)
\qbezier(113.3341, -48.9209)(111.8183, -47.9094)(110.9349, -47.3301)
\qbezier(110.9349, -47.3301)(109.4142, -46.3330)(108.4907, -45.7394)
\qbezier(108.4907, -45.7394)(106.9548, -44.7521)(105.9945, -44.1486)
\qbezier(105.9945, -44.1486)(104.4341, -43.1680)(103.4384, -42.5578)
\qbezier(103.4384, -42.5578)(101.8448, -41.5813)(100.8137, -40.9670)
\qbezier(100.8137, -40.9670)( 99.1780, -39.9925)( 98.1109, -39.3763)
\qbezier( 98.1109, -39.3763)( 96.4236, -38.4018)( 95.3189, -37.7855)
\qbezier( 95.3189, -37.7855)( 93.5698, -36.8096)( 92.4253, -36.1947)
\qbezier( 92.4253, -36.1947)( 90.6029, -35.2157)( 89.4157, -34.6039)
\qbezier( 89.4157, -34.6039)( 87.5069, -33.6203)( 86.2735, -33.0132)
\qbezier( 86.2735, -33.0132)( 84.2629, -32.0234)( 82.9792, -31.4224)
\qbezier( 82.9792, -31.4224)( 80.8482, -30.4247)( 79.5095, -29.8316)
\qbezier( 79.5095, -29.8316)( 77.2355, -28.8242)( 75.8363, -28.2409)
\qbezier( 75.8363, -28.2409)( 73.3913, -27.2215)( 71.9254, -26.6501)
\qbezier( 71.9254, -26.6501)( 69.2737, -25.6163)( 67.7345, -25.0593)
\qbezier( 67.7345, -25.0593)( 64.8295, -24.0080)( 63.2099, -23.4685)
\qbezier( 63.2099, -23.4685)( 59.9894, -22.3958)( 58.2824, -21.8778)
\qbezier( 58.2824, -21.8778)( 54.6608, -20.7787)( 52.8604, -20.2870)
\qbezier( 52.8604, -20.2870)( 48.7151, -19.1550)( 46.8191, -18.6962)
\qbezier( 46.8191, -18.6962)( 41.9667, -17.5222)( 39.9822, -17.1054)
\qbezier( 39.9822, -17.1054)( 34.1313, -15.8767)( 32.0884, -15.5147)
\qbezier( 32.0884, -15.5147)( 24.7398, -14.2125)( 22.7261, -13.9239)
\qbezier( 22.7261, -13.9239)( 12.9317, -12.5204)( 11.1906, -12.3331)
\put(-25.3225, -79.7600){\vector(1, 0){316.7503}}
\put(130.0000, -150.1458){\vector(0, 1){179.2917}}
\put( 11.1906,  -6.3403){\line(1, 0){243.7240}}
\put( 11.1906, -114.6597){\line(1, 0){243.7240}}
\put( 78.8816, -114.6597){\line(0, 1){108.3195}}
\put(302.3817, -86.3594){\makebox(0,0)[]{\footnotesize{$r$}}}
\put(122.2105,  30.8698){\makebox(0,0)[]{\footnotesize{$t$}}}
\put(159.2105,   1.2452){\makebox(0,0)[]{\scriptsize{$t_0 + R_{\mbox{\tiny $Q$}} \hs{0.3mm} \pi / 2$}}}
\put(159.2105, -124.3139){\makebox(0,0)[]{\scriptsize{$t_0 - R_{\mbox{\tiny $Q$}} \hs{0.3mm} \pi / 2$}}}
\put(140.2237, -58.7760){\makebox(0,0)[]{\scriptsize{$t_0$}}}
\put( 72.7961, -85.2767){\makebox(0,0)[]{\scriptsize{$r_1$}}}
\put( 72.0658, -98.0340){\makebox(0,0)[]{\scriptsize{$P_1$}}}
\put( 72.0658, -66.6579){\makebox(0,0)[]{\scriptsize{$P_2$}}}
\put( 87.4013, -25.9725){\makebox(0,0)[]{\scriptsize{$P_3$}}}
\end{picture}
\vspace{-2mm} \newline
%
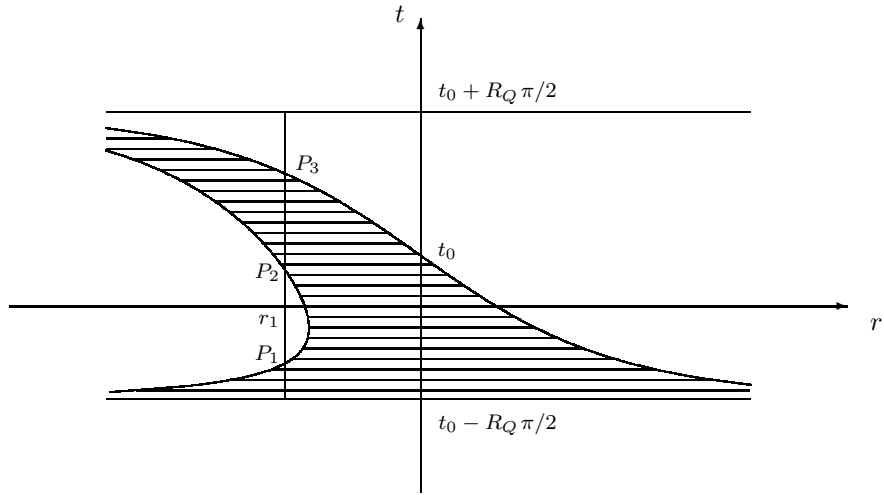
{\footnotesize \sf Figure 8. The region between the horizontal lines
in this figure represents that part of the PLBR spacetime covered by
the $(t,r)$ coordinate system. The hatched region in this figure
represents the WLBR spacetime in the $(t,r)$-system, with
$0 < R < R_{\mbox{\tiny $Q$}}$ to the left of this region and
$R < 0$ to the right. Consider the vertical line $r = r_1$ where
$r_1 < 0$. This is the world line of a free particle with $r_1 < 0$
in Figure 7. The initial point with
$t = t_0 - R_{\mbox{\tiny $Q$}} \hs{0.3mm} \pi / 2$ corresponds
to an event with coordinates $(0,B)$ in the CFS system.
According to equation \eqref{e_1361} the initial velocity of a particle
with $r_1 < 0$ is directed inwards. The world line between the events
$P_1$ and $P_2$ is only possible in the PLBR spacetime which has no
domain wall. At the event $P_2$ the particle enters the WLBR spacetime,
and at $P_3$ it leaves the WLBR spacetime as $R$ and $T$ approach
infinity. Then it appears in the PLBR spacetime as $R$ and $T$ comes from
minus infinity. Finally it arrives at $(0,-B)$ in the CFS system when
$t = t_0 + R_{\mbox{\tiny $Q$}} \hs{0.3mm} \pi / 2$.}
\vspace{3mm} \newline
\itm Let us consider a particle falling freely with outwards directed
initial velocity from $R = R_{\mbox{\tiny $Q$}}$ at the event $P_2$.
This particle follows the world line $r = r_1 < 0$ as shown in Figure 7.
As observed in the CFS system it accelerates away from the wall as
seen from equation \eqref{e_359}. Hence there is repulsive gravitation.
However, it follows from the line element \eqref{e_331} that as observed
by freely falling observers, the 3-space $t = \mbox{constant}$ first
expands and then contracts in the radial direction. This strange
behaviour can be understood by considering the equation of geodesic
deviation.
\itm In comoving coordinates with tangent vector ${\bf u} = (1,0,0,0)$
for the geodesic curves the equation takes the form
\begin{equation} \label{e_706}
\frl{d^2 s^i}{d t^2} + R^i_{\hs{1.0mm} 0 j 0} s^j = 0
\mbox{ .}
\end{equation}
With the line element \eqref{e_331} this equation reduces to
\begin{equation} \label{e_707}
\frl{d^2 s^r}{d t^2} + \frl{1}{R_{\mbox{\tiny $Q$}}^2} s^r = 0
\mbox{ ,}
\end{equation}
having the solution
\begin{equation} \label{e_708}
s^r = A \hs{0.1mm} \cos \hs{-1.2mm}
\left( \frl{t - t_0}{R_{\mbox{\tiny $Q$}}} \right)
\mbox{ ,}
\end{equation}
which is equal to the scale factor in the line element \eqref{e_331}.
This then provides an explanation for the surprising contraction of
the space between the events $(t_0,r_1)$ and $P_3$ in Figure 8.
The transition from expansion to contractions happens at $t = t_0$,
corresponding to the simultaneity curve
$T^2 - R^2 = B^2$ as seen from the first of the
equations \eqref{e_348}. The world line of an observer with
$r = r_1$ will intersect this simultaneity curve only when $r_1 < 0$.
Hence only these observers will experience contraction.
\itm A simple special case of the line element
\eqref{e_331} is obtained by choosing
$t_0 = R_{\mbox{\tiny $Q$}} \hs{0.3mm} \pi / 2$ and $A = 1$, giving
\begin{equation} \label{e_1331}
ds^2 = - d{t}^2 + \sin^2 \hs{-1.2mm}
\left( \fr{t}{R_{\mbox{\tiny $Q$}}} \right) \hs{0.5mm} dr^2
+ R_{\mbox{\tiny $Q$}}^2 d\Omega^2
\mbox{ .}
\end{equation}
\vspace{-9mm} \newline
\itm A deeper understanding of the $t$ coordinate may be obtained by
giving a parametric description of a free particle in the PLBR spacetime
with the proper time $t$ of the particle as parameter. We consider a
particle with $r = r_1$ in the inertial coordinate system, with world
line given in equation \eqref{e_361}. The particle is instantaneously
at rest at the point $P$ with CFS coordinates $(T_1,R_1)$ where
$R_1 = \sqrt{B^2 - T_1^2}$. We shall now apply
Lagrangian dynamics in the CFS system to this particle. Putting the
velocity \eqref{e_548} equal to zero at the point $P$ shows that the
constant of motion $p_T$ for this particle is
\begin{equation} \label{e_562}
p_T = - \frl{R_{\mbox{\tiny $Q$}}}{R_1}
\mbox{ ,}
\end{equation}
where the minus sign has been chosen in order that
\begin{equation} \label{e_563}
\dot{T} = - \frl{R^2}{R_{\mbox{\tiny $Q$}}^2} p_T
= \frl{R^2}{R_1 R_{\mbox{\tiny $Q$}}} > 0
\mbox{ .}
\end{equation}
The first equality follows from equation \eqref{e_546}.
Inserting this into the four-velocity identity \eqref{e_547} and
integrating leads to
\begin{equation} \label{e_564}
R = R_1 / \sin \left( \frl{{t}_1 \m t}{R_{\mbox{\tiny $Q$}}} \right)
\mbox{ ,}
\end{equation}
where ${t}_1$ is a constant of integration. Inserting the expression
\eqref{e_564} into \eqref{e_563} and integrating gives
\begin{equation} \label{e_565}
T = T_2 + R_1 \cot \left( \frl{{t}_1 \m t}{R_{\mbox{\tiny $Q$}}}
\right)
\mbox{ ,}
\end{equation}
where $T_2$ is a new constant of integration. Demanding that equations
\eqref{e_564} and \eqref{e_565} is a parametric representation of the
hyperbola \eqref{e_361} gives $T_2 = T_1$. From equations
\eqref{e_564} and \eqref{e_565} we then have
\begin{equation} \label{e_566}
\cos \left( \frl{{t}_1 \m t}{R_{\mbox{\tiny $Q$}}} \right)
= \frl{T - T_1}{R}
\mbox{ .}
\end{equation}
The constant ${t}_1$ is now determined by eliminating $t$ from
equations \eqref{e_566} and the first of the transformation formulae
\eqref{e_348} at the point $P$. This gives
\begin{equation} \label{e_567}
t_1 - t_0 = R_{\mbox{\tiny $Q$}}
\left( \frl{\pi}{2} - \arcsin \frl{R_1}{B} \right)
= R_{\mbox{\tiny $Q$}} \arcsin \frl{T_1}{B}
\mbox{ .}
\end{equation}
The equations \eqref{e_564} to \eqref{e_567} give a parametric
representation of the world lines of free particles with $r = r_1$
as shown in Figure 7.
\itm We shall now show that this parametric description of the path
of a free particle with the proper time of the particle as parameter
is in agreement with the transformation \eqref{e_348} from the CFS
coordinates to the comoving coordinates of the particle. We have that
\begin{equation} \label{e_569}
\sin \left( \frl{t \m t_0}{R_{\mbox{\tiny $Q$}}} \right)
= \sin \left( \arcsin \frl{T_1}{B}
- \frl{t_1 \m t}{R_{\mbox{\tiny $Q$}}} \right)
= \frl{T_1}{B} \hs{0.5mm}
\cos \left( \frl{t_1 \m t}{R_{\mbox{\tiny $Q$}}} \right)
- \frl{R_1}{B} \hs{0.5mm}
\sin \left( \frl{t_1 \m t}{R_{\mbox{\tiny $Q$}}} \right)
\mbox{ .}
\end{equation}
Inserting equations \eqref{e_564} and \eqref{e_566} gives
\begin{equation} \label{e_570}
\sin \left( \frl{t \m t_0}{R_{\mbox{\tiny $Q$}}} \right)
= \frl{T_1 (T \m T_1) \m R_1^2}{BR}
= \frl{T_1 T \m B^2}{BR}
\mbox{ .}
\end{equation}
From equation \eqref{e_361} and the second of the equations \eqref{e_348}
it follows that
\begin{equation} \label{e_571}
T_1 = \frl{(T^2 \m R^2) \p B^2}{2 T}
\mbox{ .}
\end{equation}
Inserting this into equation \eqref{e_570} we finally obtain the first
of the transformation equations \eqref{e_348}.
\itm In Figure 7 we have drawn the world lines of particles with different
values of $r_1$. All of the particles come from the point $(B,0)$ and
move so that $T$ and $R$ approach infinity when $t$ increases towards
${t}_0$ as seen from equations \eqref{e_564} and \eqref{e_565}. When $t$
passes $t_1$ the values of $T$ and $R$ switch to minus infinity, and all
particles approach the point $(-B,0)$ when $t$ increases towards
$t_0 + \pi R_{\mbox{\tiny $Q$}} \hs{0.3mm} / 2$. This highly surprising
behaviour may be understood by noting that according to the line element
\eqref{e_157} the physical distances in the PLBR spacetime
approach zero when $|R|$ approaches infinity.
\itm The reference particles of the $(t,r)$-system are freely falling.
Their world lines are hyperbolae corresponding to particles with constant
proper acceleration. This is in accordance with the fact that the
acceleration of gravity as given in equation \eqref{e_359} is constant
in the LBR spacetime.
\itm In the final part of this section we shall present the
transformations between the previous coordinate systems and the
inertial one.
Combining equation \eqref{e_1049} with the transformation \eqref{e_949}
we obtain
\begin{equation} \label{e_1004}
\overline{t} = \overline{t}_0 + R_{\mbox{\tiny $Q$}} A \hs{0.5mm}
\sin \left( \frl{t \m t_0}{R_{\mbox{\tiny $Q$}}} \right)
\mbox{\hspace{2mm} , \hspace{3mm}}
\overline{r} = r
\mbox{ .}
\end{equation}
\vspace{-1mm} \newline
The inverse transformation is
\begin{equation} \label{e_1005}
\sin \left( \frl{t \m t_0}{R_{\mbox{\tiny $Q$}}} \right)
= \frl{\overline{t} - \overline{t}_0}{R_{\mbox{\tiny $Q$}} A}
\mbox{\hspace{2mm} , \hspace{3mm}}
r = \overline{r}
\mbox{ .}
\end{equation}
\vspace{-6mm} \newline
Hence
\vspace{-1mm} \newline
\begin{equation} \label{e_1009}
\frl{d \overline{t}}{dt} = A \hs{0.5mm}
\cos \left( \frl{t \m t_0}{R_{\mbox{\tiny $Q$}}} \right)
= A / \cosh \tau
\mbox{ ,}
\end{equation}
\vspace{-1mm} \newline
which means that the coordinate clocks showing $\overline{t}$ go at an
increasingly slower rate than the standard clocks showing $t$.
\itm We shall find the transformation relating the line element
\eqref{e_211} of the LBR spacetime in inertial and physical coordinates
respectively. Here $\alpha$ in the line element \eqref{e_211} is given
by equation \eqref{e_225} and $\beta = 0$.
Combining the transformation \eqref{e_835} with the equations
\eqref{e_1030} and \eqref{e_1021} we obtain the transformation
%
\begin{equation} \label{e_836}
\cot \eta = \frr{
\tan \mbox{$\left( \rule[-1.5mm]{0mm}{5.7mm} \right.$} \hs{-0.2mm}
\frl{t_0 \m t}{R_{\mbox{\tiny $Q$}}}
\hs{-0.2mm} \mbox{$\left. \rule[-1.5mm]{0mm}{5.7mm} \right)$}}
{\cosh \mbox{$\left( \rule[-1.5mm]{0mm}{5.7mm} \right.$} \hs{-0.2mm}
\frl{A \hs{0.5mm} r}{R_{\mbox{\tiny $Q$}}}
\hs{-0.2mm} \mbox{$\left. \rule[-1.5mm]{0mm}{5.7mm} \right)$}
\rule[-1.5mm]{0mm}{6.9mm}}
\mbox{\hspace{2mm} , \hspace{3mm}}
\cot \chi = -
\cos \mbox{$\left( \rule[-1.5mm]{0mm}{5.7mm} \right.$} \hs{-0.2mm}
\frl{t \m t_0}{R_{\mbox{\tiny $Q$}}}
\hs{-0.2mm} \mbox{$\left. \rule[-1.5mm]{0mm}{5.7mm} \right)$}
\sinh \mbox{$\left( \rule[-1.5mm]{0mm}{5.7mm} \right.$} \hs{-0.2mm}
\frl{A \hs{0.5mm} r}{R_{\mbox{\tiny $Q$}}}
\hs{-0.2mm} \mbox{$\left. \rule[-1.5mm]{0mm}{5.7mm} \right)$}
\mbox{ .}
\end{equation}
%
Now using
$\eta = (A / R_{\mbox{\tiny $Q$}}) \hs{0.6mm} \hat{t}$
and \eqref{e_700} we obtain the transformation
\vspace{-2mm} \newline
\begin{equation} \label{e_334}
\cot \mbox{$\left( \rule[-1.5mm]{0mm}{5.7mm} \right.$} \hs{-0.2mm}
\frl{A \hs{0.6mm} \hat{t}}{R_{\mbox{\tiny $Q$}}}
\hs{0.2mm} \mbox{$\left. \rule[-1.5mm]{0mm}{5.7mm} \right)$}
= \frr{
\tan \mbox{$\left( \rule[-1.5mm]{0mm}{5.7mm} \right.$} \hs{-0.2mm}
\frl{t_0 \m t}{R_{\mbox{\tiny $Q$}}}
\hs{-0.2mm} \mbox{$\left. \rule[-1.5mm]{0mm}{5.7mm} \right)$}}
{\cosh \mbox{$\left( \rule[-1.5mm]{0mm}{5.7mm} \right.$} \hs{-0.2mm}
\frl{A \hs{0.5mm} r}{R_{\mbox{\tiny $Q$}}}
\hs{-0.2mm} \mbox{$\left. \rule[-1.5mm]{0mm}{5.7mm} \right)$}
\rule[-1.5mm]{0mm}{6.9mm}}
\mbox{\hspace{0mm} , \hspace{1mm}}
\sinh \mbox{$\left( \rule[-1.5mm]{0mm}{5.7mm} \right.$} \hs{-0.2mm}
\frl{\hat{r} \m \hat{r}_0}{R_{\mbox{\tiny $Q$}}}
\hs{-0.2mm} \mbox{$\left. \rule[-1.5mm]{0mm}{5.7mm} \right)$}
= \cos \mbox{$\left( \rule[-1.5mm]{0mm}{5.7mm} \right.$} \hs{-0.2mm}
\frl{t \m t_0}{R_{\mbox{\tiny $Q$}}}
\hs{-0.2mm} \mbox{$\left. \rule[-1.5mm]{0mm}{5.7mm} \right)$}
\sinh \mbox{$\left( \rule[-1.5mm]{0mm}{5.7mm} \right.$} \hs{-0.2mm}
\frl{A \hs{0.5mm} r}{R_{\mbox{\tiny $Q$}}}
\hs{-0.2mm} \mbox{$\left. \rule[-1.5mm]{0mm}{5.7mm} \right)$}
\mbox{ .}
\end{equation}
%
\itm The inverse transformation is found in a similar way by combining
the transformation \eqref{e_838} with equation \eqref{e_1049} which gives
%
\begin{equation} \label{e_839}
\sin \mbox{$\left( \rule[-1.5mm]{0mm}{5.7mm} \right.$} \hs{-0.2mm}
\frl{t_0 \m t}{R_{\mbox{\tiny $Q$}}}
\hs{-0.2mm} \mbox{$\left. \rule[-1.5mm]{0mm}{5.7mm} \right)$}
= \frl{\cos \eta}{\sin \chi}
\mbox{\hspace{2mm} , \hspace{3mm}}
\tanh \mbox{$\left( \rule[-1.5mm]{0mm}{5.7mm} \right.$} \hs{-0.2mm}
\frl{A \hs{0.5mm} r}{R_{\mbox{\tiny $Q$}}}
\hs{-0.2mm} \mbox{$\left. \rule[-1.5mm]{0mm}{5.7mm} \right)$}
= - \frl{\cos \chi}{\sin \eta}
\mbox{ .}
\end{equation}
%
Introducing $\hat{t}$ and using the equations \eqref{e_693}
we obtain the transformation
%
\begin{equation} \label{e_336}
\sin \mbox{$\left( \rule[-1.5mm]{0mm}{5.7mm} \right.$} \hs{-0.2mm}
\frl{t_0 \m t}{R_{\mbox{\tiny $Q$}}}
\hs{-0.2mm} \mbox{$\left. \rule[-1.5mm]{0mm}{5.7mm} \right)$}
= \cosh \mbox{$\left( \rule[-1.5mm]{0mm}{5.7mm} \right.$} \hs{-0.2mm}
\frl{\hat{r} \m \hat{r}_0}{R_{\mbox{\tiny $Q$}}}
\hs{-0.2mm} \mbox{$\left. \rule[-1.5mm]{0mm}{5.7mm} \right)$}
\cos \mbox{$\left( \rule[-1.5mm]{0mm}{5.7mm} \right.$} \hs{-0.2mm}
\frl{A \hs{0.6mm} \hat{t}}{R_{\mbox{\tiny $Q$}}}
\hs{0.2mm} \mbox{$\left. \rule[-1.5mm]{0mm}{5.7mm} \right)$}
\mbox{\hspace{0mm} , \hspace{1mm}}
\tanh \mbox{$\left( \rule[-1.5mm]{0mm}{5.7mm} \right.$} \hs{-0.2mm}
\frl{A \hs{0.5mm} r}{R_{\mbox{\tiny $Q$}}}
\hs{-0.2mm} \mbox{$\left. \rule[-1.5mm]{0mm}{5.7mm} \right)$}
= \frr{
\tanh \mbox{$\left( \rule[-1.5mm]{0mm}{5.7mm} \right.$} \hs{-0.2mm}
\frl{\hat{r} \m \hat{r}_0}{R_{\mbox{\tiny $Q$}}}
\hs{-0.2mm} \mbox{$\left. \rule[-1.5mm]{0mm}{5.7mm} \right)$}}
{\sin \mbox{$\left( \rule[-1.5mm]{0mm}{5.7mm} \right.$} \hs{-0.2mm}
\frl{A \hs{0.6mm} \hat{t}}{R_{\mbox{\tiny $Q$}}}
\hs{0.2mm} \mbox{$\left. \rule[-1.5mm]{0mm}{5.7mm} \right)$}
\rule[-1.5mm]{0mm}{6.9mm}}
\mbox{ .}
\end{equation}
Hence the world line of a free particle with $r = r_1$ as described with
reference to the $(\hat{t},\hat{r})$-system is given by
\begin{equation} \label{e_337}
\tanh \mbox{$\left( \rule[-1.5mm]{0mm}{5.7mm} \right.$} \hs{-0.2mm}
\frl{\hat{r} \m \hat{r}_0}{R_{\mbox{\tiny $Q$}}}
\hs{-0.2mm} \mbox{$\left. \rule[-1.5mm]{0mm}{5.7mm} \right)$}
= a_1 \sin \mbox{$\left( \rule[-1.5mm]{0mm}{5.7mm} \right.$} \hs{-0.2mm}
\frl{A \hs{0.6mm} \hat{t}}{R_{\mbox{\tiny $Q$}}}
\hs{0.2mm} \mbox{$\left. \rule[-1.5mm]{0mm}{5.7mm} \right)$}
\mbox{\hspace{2mm} , \hspace{3mm}}
a_1 = \tanh \mbox{$\left( \rule[-1.5mm]{0mm}{5.7mm} \right.$} \hs{-0.2mm}
\frl{A \hs{0.5mm} r_1}{R_{\mbox{\tiny $Q$}}}
\hs{-0.2mm} \mbox{$\left. \rule[-1.5mm]{0mm}{5.7mm} \right)$}
\mbox{ .}
\end{equation}
\vspace{-9mm} \newline
\itm The coordinate transformations from the CFS system to the inertial
system and the $(\hat{t},\hat{r})$-system with $k = -1$ are
defined on the disjoint domains
$|T + R| > B$, $|T - R| < B$ and
$|T + R| < B$, $|T - R| < B$, $R \ne 0$
respectively. There is therefore no coordinate transformation from
the inertial system to the $(\hat{t},\hat{r})$-system in
this case.
For $k = 0$ we have
%
\begin{equation} \label{e_362}
\hat{t} = \frl{B
\cos \mbox{$\left( \rule[-1.5mm]{0mm}{5.7mm} \right.$} \hs{-0.2mm}
\frl{t \m t_0}{R_{\mbox{\tiny $Q$}}}
\hs{-0.2mm} \mbox{$\left. \rule[-1.5mm]{0mm}{5.7mm} \right)$} \hs{0.5mm}
\cosh \mbox{$\left( \rule[-1.5mm]{0mm}{5.7mm} \right.$} \hs{-0.2mm}
\frl{A \hs{0.5mm} r}{R_{\mbox{\tiny $Q$}}}
\hs{-0.2mm} \mbox{$\left. \rule[-1.5mm]{0mm}{5.7mm} \right)$}}
{\sin \mbox{$\left( \rule[-1.5mm]{0mm}{5.7mm} \right.$} \hs{-0.2mm}
\frl{t_0 \m t}{R_{\mbox{\tiny $Q$}}}
\hs{-0.2mm} \mbox{$\left. \rule[-1.5mm]{0mm}{5.7mm} \right)$} \m \hs{0.5mm}
\cos \mbox{$\left( \rule[-1.5mm]{0mm}{5.7mm} \right.$} \hs{-0.2mm}
\frl{t \m t_0}{R_{\mbox{\tiny $Q$}}}
\hs{-0.2mm} \mbox{$\left. \rule[-1.5mm]{0mm}{5.7mm} \right)$} \hs{0.5mm}
\sinh \mbox{$\left( \rule[-1.5mm]{0mm}{5.7mm} \right.$} \hs{-0.2mm}
\frl{A \hs{0.5mm} r}{R_{\mbox{\tiny $Q$}}}
\hs{-0.2mm} \mbox{$\left. \rule[-1.5mm]{0mm}{5.7mm} \right)$}
\rule[-1.5mm]{0mm}{6.9mm}}
\mbox{ ,}
\end{equation}
\vspace{-4mm} \newline
\begin{equation} \label{e_363}
B \hs{0.5mm} e^{(\hat{r}_0 \m \hat{r}) / R_{\mbox{\tiny $Q$}}} =
\sin \mbox{$\left( \rule[-1.5mm]{0mm}{5.7mm} \right.$} \hs{-0.2mm}
\frl{t_0 \m t}{R_{\mbox{\tiny $Q$}}}
\hs{-0.2mm} \mbox{$\left. \rule[-1.5mm]{0mm}{5.7mm} \right)$}
- \cos \mbox{$\left( \rule[-1.5mm]{0mm}{5.7mm} \right.$} \hs{-0.2mm}
\frl{t \m t_0}{R_{\mbox{\tiny $Q$}}}
\hs{-0.2mm} \mbox{$\left. \rule[-1.5mm]{0mm}{5.7mm} \right)$} \hs{0.5mm}
\sinh \mbox{$\left( \rule[-1.5mm]{0mm}{5.7mm} \right.$} \hs{-0.2mm}
\frl{A \hs{0.5mm} r}{R_{\mbox{\tiny $Q$}}}
\hs{-0.2mm} \mbox{$\left. \rule[-1.5mm]{0mm}{5.7mm} \right)$}
\rule[-1.5mm]{0mm}{6.9mm}
\mbox{ .}
\end{equation}
\vspace{-6mm} \newline
\itm We can also find the transformation between the
$(\tilde{t},\tilde{r})$-system and the inertial system.
Combining the transformations \eqref{e_836} and \eqref{e_349} we obtain
%
\begin{equation} \label{e_934}
\cot \mbox{$\left( \rule[-1.5mm]{0mm}{5.7mm} \right.$} \hs{-0.2mm}
\frl{A \hs{0.6mm} \tilde{t}}{R_{\mbox{\tiny $Q$}}}
\hs{0.2mm} \mbox{$\left. \rule[-1.5mm]{0mm}{5.7mm} \right)$}
= \frr{
\tan \mbox{$\left( \rule[-1.5mm]{0mm}{5.7mm} \right.$} \hs{-0.2mm}
\frl{t_0 \m t}{R_{\mbox{\tiny $Q$}}}
\hs{-0.2mm} \mbox{$\left. \rule[-1.5mm]{0mm}{5.7mm} \right)$}}
{\cosh \mbox{$\left( \rule[-1.5mm]{0mm}{5.7mm} \right.$} \hs{-0.2mm}
\frl{A \hs{0.5mm} r}{R_{\mbox{\tiny $Q$}}}
\hs{-0.2mm} \mbox{$\left. \rule[-1.5mm]{0mm}{5.7mm} \right)$}
\rule[-1.5mm]{0mm}{6.9mm}}
\mbox{\hspace{2mm} , \hspace{3mm}}
\tilde{r} = \tilde{r}_0 + R_{\mbox{\tiny $Q$}} A
\cos \mbox{$\left( \rule[-1.5mm]{0mm}{5.7mm} \right.$} \hs{-0.2mm}
\frl{t \m t_0}{R_{\mbox{\tiny $Q$}}}
\hs{-0.2mm} \mbox{$\left. \rule[-1.5mm]{0mm}{5.7mm} \right)$}
\sinh \mbox{$\left( \rule[-1.5mm]{0mm}{5.7mm} \right.$} \hs{-0.2mm}
\frl{A \hs{0.5mm} r}{R_{\mbox{\tiny $Q$}}}
\hs{-0.2mm} \mbox{$\left. \rule[-1.5mm]{0mm}{5.7mm} \right)$}
\end{equation}
%
The inverse transformation is given by
%
\begin{equation} \label{e_993}
\sin \mbox{$\left( \rule[-1.5mm]{0mm}{5.7mm} \right.$} \hs{-0.2mm}
\frl{t_0 \m t}{R_{\mbox{\tiny $Q$}}}
\hs{-0.2mm} \mbox{$\left. \rule[-1.5mm]{0mm}{5.7mm} \right)$}
= \frl{1}{R_{\mbox{\tiny $Q$}}}
\cos \mbox{$\left( \rule[-1.5mm]{0mm}{5.7mm} \right.$} \hs{-0.2mm}
\frl{A \hs{0.6mm} \tilde{t}}{R_{\mbox{\tiny $Q$}}}
\hs{0.2mm} \mbox{$\left. \rule[-1.5mm]{0mm}{5.7mm} \right)$}
\sqrt{R_{\mbox{\tiny $Q$}}^2 D + (\tilde{r} - \tilde{r}_0)^2}
\mbox{ ,}
\end{equation}
%
\begin{equation} \label{e_994}
\tanh \mbox{$\left( \rule[-1.5mm]{0mm}{5.7mm} \right.$} \hs{-0.2mm}
\frl{A \hs{0.5mm} r}{R_{\mbox{\tiny $Q$}}}
\hs{-0.2mm} \mbox{$\left. \rule[-1.5mm]{0mm}{5.7mm} \right)$}
= \frl{\tilde{r} - \tilde{r}_0}
{\sin \mbox{$\left( \rule[-1.5mm]{0mm}{5.7mm} \right.$} \hs{-0.2mm}
\frac{A \hs{0.6mm} \tilde{t}}{R_{\mbox{\tiny $Q$}}}
\hs{0.2mm} \mbox{$\left. \rule[-1.5mm]{0mm}{5.7mm} \right)$}
\sqrt{R_{\mbox{\tiny $Q$}}^2 D + (\tilde{r} - \tilde{r}_0)^2}}
\mbox{ .}
\end{equation}
\vspace{-2mm} \newline
\itm Equation \eqref{e_1010} gives the following equation of motion for
a free particle in the $(\tilde{t}, \tilde{r})$-system,
\begin{equation} \label{e_1035}
R_{\mbox{\tiny $Q$}}^2 \frl{d^2 \tilde{r}}{dt^2}
+ \tilde{r} - \tilde{r}_0 = 0
\mbox{ ,}
\end{equation}
where $t$ is the proper time of the particle.
This is the equation of harmonic motion about the position
$\tilde{r} = \tilde{r}_0$ as noted by Dadhich [\ref{r_15}]. His
interpretation is that a free particle would execute harmonic
oscillation about $\tilde{r} = \tilde{r}_0$. He has given an
explanation of this motion in terms of electrostatic energy
filling the LBR spacetime.
\itm In our opinion, however, there is another explanation for this
motion. Equation \eqref{e_1035} has the general solution
\begin{equation} \label{e_1345}
\tilde{r} = \tilde{r}_0 + A_1
\cos \mbox{$\left( \rule[-1.5mm]{0mm}{5.7mm} \right.$} \hs{-0.2mm}
\frl{t \m t_0}{R_{\mbox{\tiny $Q$}}}
\hs{-0.2mm} \mbox{$\left. \rule[-1.5mm]{0mm}{5.7mm} \right)$}
\mbox{ ,}
\end{equation}
where $A_1$ and $t_0$ are integration constants.
According to equation \eqref{e_934} a fixed point $r = r_1$
in a freely moving reference frame has a radial coordinate given by
the above equation with $A_1 = R_{\mbox{\tiny $Q$}} A \sinh
(\fr{A r_1}{R_{\mbox{\tiny $Q$}}})$. Hence $(\tilde{t},\tilde{r})$
are comoving coordinates in a reference frame that performs harmonic
motion relatively to a freely falling reference frame. This is the reason
for the oscillating motion of a free particle in the
$(\tilde{t},\tilde{r})$-system which was noted by Dadhich, assuming
that $-\infty < t < \infty$.
\itm In the context of the LBR as interpreted in the present article,
the situation is different. The coordinate region in $(t,r)$-system
of the LBR spacetime is given by the inequalities \eqref{e_1335} and
$-\infty < r < \infty$. This restriction of the time interval means
that the oscillating character of the motion of a free particle in
the $(\tilde{t},\tilde{r})$-system as given by equation \eqref{e_1345}
vanishes.
\itm Choosing $A = 1$, $B = R_{\mbox{\tiny $Q$}}$, $t_0 = 0$ and
introducing the coordinates $\tilde{\tau} = t \,$, $x = r$,
$y = R_{\mbox{\tiny $Q$}} \phi$ and
$z = R_{\mbox{\tiny $Q$}} (\theta - \pi / 2)$ in equation \eqref{e_331},
the PLBR line element takes the form [\ref{r_13}]
\begin{equation} \label{e_364}
ds^2 = - d \tilde{\tau}^2
+ \cos^2 (\tilde{\tau} / R_{\mbox{\tiny $Q$}}) \hs{0.5mm} dx^2
+ \cos^2 (z / R_{\mbox{\tiny $Q$}}) \hs{0.5mm} dy^2 + dz^2
\mbox{ .}
\end{equation}
Using the formulae \eqref{e_332} and \eqref{e_333} we see that
this form of the line element is obtained from \eqref{e_157} by the
transformation
%
\begin{equation} \label{e_237}
T = \frl{R_{\mbox{\tiny $Q$}} \cos (\tilde{\tau} / R_{\mbox{\tiny $Q$}})
\cosh (x / R_{\mbox{\tiny $Q$}})}
{\sin (\tilde{\tau} / R_{\mbox{\tiny $Q$}})
\m \cos (\tilde{\tau} / R_{\mbox{\tiny $Q$}})
\sinh (x / R_{\mbox{\tiny $Q$}}) \rule[-1.5mm]{0mm}{4.9mm}}
\mbox{ ,}
\end{equation}
\begin{equation} \label{e_238}
R = \frl{R_{\mbox{\tiny $Q$}}}
{\sin (\tilde{\tau} / R_{\mbox{\tiny $Q$}})
\m \cos (\tilde{\tau} / R_{\mbox{\tiny $Q$}})
\sinh (x / R_{\mbox{\tiny $Q$}}) \rule[-1.5mm]{0mm}{4.9mm}}
\mbox{ ,}
\end{equation}
\vspace{-5.5mm} \newline
\begin{equation} \label{e_239}
\hspace{-4.5mm}
\theta = \frl{z}{R_{\mbox{\tiny $Q$}}} + \frl{\pi}{2}
\mbox{\hspace{2mm} , \hspace{3mm}}
\phi = \frl{y}{R_{\mbox{\tiny $Q$}}}
\hspace{15.5mm}
\end{equation}
\vspace{-2mm} \newline
Note that $x$, $y$ and $z$ are not to be interpreted as Cartesian
coordinates.
%
%
\vspace{6mm} \newline
{\it IV. A new type of coordinate systems for the case $k = -1$.}
\vspace{3mm} \newline
When $k = -1$ in equation \eqref{e_218} there exist different types of
coordinates for the LBR spacetime obeying the coordinate conditions
$\beta = \alpha$, $\beta = -\alpha$ and $\beta = 0$. Here we will
introduce coordinates $(\eta',\chi')$ with $\beta = \alpha$,
$(\tilde{t}',\tilde{r}')$ with $\beta = -\alpha$ and $(\hat{t}',\hat{r}')$
with $\beta = 0$ different from the coordinates $(\eta,\chi)$,
$(\tilde{t},\tilde{r})$ and $(\hat{t},\hat{r})$ respectively.
\itm In order to find the transformation between the line element
\eqref{e_157} and the line element \eqref{e_218} with marked coordinates,
\begin{equation} \label{e_1274}
ds^2 = \frl{R_{\mbox{\tiny $Q$}}^2}{\sinh^2 \chi'}
(-d {\eta'}^2 + d{\chi'}^2) + R_{\mbox{\tiny $Q$}}^2 d\Omega^2
\mbox{ ,}
\end{equation}
we replace the generating function \eqref{e_339} by
\begin{equation} \label{e_1268}
f(x) = B \hs{0.2mm} e^x
\mbox{ .}
\end{equation}
Like the function \eqref{e_339} it satisfies the condition \eqref{e_512}.
It is obtained from equation \eqref{e_13} with $a = 0$, $b = -1$,
$c = 2\hs{0.2mm} B$ and $d = B$. This leads to the transformation
\begin{equation} \label{e_1269}
T = B \hs{0.2mm} e^{\eta'} \cosh \chi'
\mbox{\hspace{2mm} , \hspace{3mm}}
R = B \hs{0.2mm} e^{\eta'} \sinh \chi'
\mbox{ .}
\end{equation}
The inverse transformation is
\begin{equation} \label{e_1270}
B \hs{0.2mm} e^{\eta'} = \sqrt{T^2 - R^2}
\mbox{\hspace{2mm} , \hspace{3mm}}
\tanh \chi' = \frl{R}{T}
\mbox{ .}
\end{equation}
\itm In the $(T,R)$-system, each reference particle with
$\chi' = \mbox{constant}$ in the coordinate system has
a constant velocity
\begin{equation} \label{e_2065}
V = \frl{R}{T} = \tanh \chi'
\end{equation}
which is less than $1$. According to this equation $\chi'$ is the
rapidity of a reference particle with radial coordinate $\chi'$.
\itm Figure 9 shows the $(\eta',\chi')$-system in a Minkowski
diagram referring to the CFS system of the observer at
$\chi' = 0$. It follows from equations \eqref{e_1270} that the world
lines of the reference particles with $\chi' = \mbox{constant}$ are
straight lines, and the curves of the space $\eta' = \mbox{constant}$
are hyperbolae with centre at the origin as shown in the diagram in
\mbox{Figure 9.}
\vspace*{-4mm} \newline
\begin{picture}(50,192)(-76,-182)
\qbezier(125.7955, -87.7955)(127.3448, -87.7955)(128.8996, -87.6648)
\qbezier(128.8996, -87.6648)(130.4544, -87.5342)(132.0258, -87.2720)
\qbezier(132.0258, -87.2720)(133.5971, -87.0099)(135.1961, -86.6143)
\qbezier(135.1961, -86.6143)(136.7952, -86.2187)(138.4332, -85.6869)
\qbezier(138.4332, -85.6869)(140.0713, -85.1551)(141.7600, -84.4833)
\qbezier(141.7600, -84.4833)(143.4487, -83.8115)(145.2000, -82.9949)
\qbezier(145.2000, -82.9949)(146.9513, -82.1784)(148.7777, -81.2113)
\qbezier(148.7777, -81.2113)(150.6041, -80.2442)(152.5185, -79.1197)
\qbezier(152.5185, -79.1197)(154.4329, -77.9952)(156.4489, -76.7052)
\qbezier(156.4489, -76.7052)(158.4649, -75.4153)(160.5968, -73.9509)
\qbezier(160.5968, -73.9509)(162.7288, -72.4864)(164.9917, -70.8370)
\qbezier(164.9917, -70.8370)(167.2546, -69.1876)(169.6647, -67.3416)
\qbezier(169.6647, -67.3416)(172.0747, -65.4956)(174.6489, -63.4398)
\qbezier(174.6489, -63.4398)(177.2231, -61.3840)(179.9798, -59.1039)
\qbezier(179.9798, -59.1039)(182.7365, -56.8238)(185.6952, -54.3032)
\qbezier(185.6952, -54.3032)(188.6539, -51.7826)(191.8356, -49.0036)
\qbezier(191.8356, -49.0036)(195.0173, -46.2246)(198.4446, -43.1675)
\qbezier(198.4446, -43.1675)(201.8718, -40.1104)(205.5690, -36.7535)
\qbezier(205.5690, -36.7535)(209.2662, -33.3966)(213.2595, -29.7161)
\put(125.7955, -124.6136){\line(1, 1){ 95.1136}}
\qbezier(125.7955, -124.6136)(156.2436, -77.6823)(186.6918, -30.7509)
\put( 88.9773, -124.6136){\vector(1, 0){128.8636}}
\put(125.7955, -161.4318){\vector(0, 1){144.2045}}
\put(227.0455, -129.2159){\makebox(0,0)[]{\footnotesize{$R$}}}
\put(118.1250, -15.6932){\makebox(0,0)[]{\footnotesize{$T$}}}
\put(222.4432, -20.2955){\makebox(0,0)[]{\footnotesize{$\eta' = {\eta}'_1$}}}
\put(182.5568, -20.2955){\makebox(0,0)[]{\footnotesize{$\chi' = {\chi}'_1$}}}
\put(133.7727, -132.2841){\makebox(0,0)[]{\footnotesize{$O$}}}
\put(151.8750, -71.8409){\makebox(0,0)[]{\footnotesize{$P$}}}
\end{picture}
\vspace{-2mm} \newline
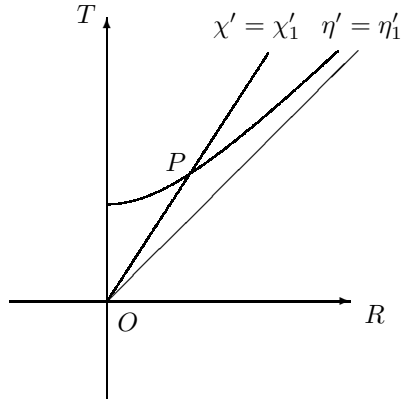
{\footnotesize \sf Figure 9. Minkowski diagram for the LBR spacetime
with reference to the CFS coordinates $(T,R)$. Here the line OP is the
world line of a reference particle with $\chi' = {\chi}_1'$.
The hyperbola represents a simultaneity curve $\eta' = {\eta}_1'$.}
\vspace{5mm} \newline
One rather sublime point should be noted. Since the line element of the
LBR spacetime has similar coordinate expressions in \eqref{e_218} and
\eqref{e_1274}, one might think that for instance the kinematics of free
particles are identical in these coordinate systems. Calculating the
acceleration of a free particle from the geodesic equation, one finds
identical coordinate expressions for the Christoffel symbols. Hence it
seems that the coordinate acceleration of a free particle at a given event
is the same in the marked and unmarked coordinate systems. This is, however,
not the case since the transformations \eqref{e_346} and \eqref{e_1270}
imply that the $\chi$- and $\chi'$-coordinates of an event with
coordinates $(T_1,R_1)$ are different.
%
%
\itm Inserting the expression
$e^{\alpha} = \fr{R_{\mbox{\tiny $Q$}}}{\sinh \chi'}$ from the line
element \eqref{e_1274} and the expression \eqref{e_1269} for $R$, we
find the line element of the flat spacetime inside the domain wall
in the $(\eta',\chi')$ coordinates,
\begin{equation} \label{e_2274}
ds_M^2 = B^2 e^{2 \eta'}
(-d {\eta'}^2 + d{\chi'}^2 + \sinh^2 \chi' d\Omega^2)
\mbox{ .}
\end{equation}
From equations \eqref{e_1274}, \eqref{e_1269} with
$R = R_{\mbox{\tiny $Q$}}$ and \eqref{e_2274} it is seen that
the metric is continuous at the domain wall.
\itm Combining the transformations \eqref{e_1269} and \eqref{e_349} we
obtain the transformation
\begin{equation} \label{e_1271}
T = e^{A \tilde{t} / R_{\mbox{\tiny $Q$}}}
\frl{B (\tilde{r}_0' \m \tilde{r}')}
{\sqrt{(\tilde{r}' \m \tilde{r}_0')^2 \m R_{\mbox{\tiny $Q$}}^2 A^2}}
\mbox{ ,}
\end{equation}
\begin{equation} \label{e_1272}
R = e^{A \tilde{t}' / R_{\mbox{\tiny $Q$}}} \frl{AB R_{\mbox{\tiny $Q$}}}
{\sqrt{(\tilde{r}' \m \tilde{r}_0')^2 \m R_{\mbox{\tiny $Q$}}^2 A^2}}
\mbox{ ,}
\end{equation}
with inverse transformation given by
\begin{equation} \label{e_1273}
B \hs{0.4mm} e^{A \tilde{t}' / R_{\mbox{\tiny $Q$}}} = \sqrt{T^2 - R^2}
\mbox{\hspace{2mm} , \hspace{3mm}}
\tilde{r}' = \tilde{r}_0' - R_{\mbox{\tiny $Q$}} A T / R
\mbox{ .}
\end{equation}
Combining the transformations \eqref{e_1269} and \eqref{e_693} we
obtain the transformation
\begin{equation} \label{e_1276}
T = B \hs{0.4mm} e^{A \hat{t}' / R_{\mbox{\tiny $Q$}}} \coth \left(
\frl{\hat{r}_0' - \hat{r}'}{R_{\mbox{\tiny $Q$}}} \right)
\mbox{\hspace{2mm} , \hspace{3mm}}
R = B \hs{0.4mm} e^{A \hat{t}' / R_{\mbox{\tiny $Q$}}} / \sinh \left(
\frl{\hat{r}_0' - \hat{r}'}{R_{\mbox{\tiny $Q$}}} \right)
\mbox{ .}
\end{equation}
with inverse transformation given by
\begin{equation} \label{e_1275}
B \hs{0.4mm} e^{A \hat{t}' / R_{\mbox{\tiny $Q$}}} = \sqrt{T^2 - R^2}
\mbox{\hspace{2mm} , \hspace{3mm}}
\cosh \left( \frl{\hat{r}_0' - \hat{r}'}{R_{\mbox{\tiny $Q$}}} \right)
= \frl{T}{R}
\mbox{ .}
\end{equation}
This transformation has earlier be considered by Zaslavskii [\ref{r_25}].
%
%
%
\vspace{6mm} \newline
{\it V. Static metric and coordinates $(\hat{\eta},\hat{\chi})$ with
$e^{2 \beta(\hat{\chi})} - e^{2 \alpha(\hat{\chi})} =
R_{\mbox{\tiny $Q$}}^2$.}
\vspace{3mm} \newline
In this section we shall introduce new coordinates
$(\hat{\eta},\hat{\chi})$ which will be useful when we introduce lightlike
coordinates in section VII. These coordinates are assumed to obey the
coordinate condition
\begin{equation} \label{e_2275}
e^{2 \beta(\hat{\chi})} - e^{2 \alpha(\hat{\chi})} =
R_{\mbox{\tiny $Q$}}^2
\mbox{ .}
\end{equation}
Introducing the function $f(\hat{\chi}) = e^{\alpha(\hat{\chi})}$
the differential equation \eqref{e_212} takes the form
\begin{equation} \label{e_2276}
\left( \frl{f'}{\sqrt{f^2 \p R_{\mbox{\tiny $Q$}}^2}} \right)'
\frl{1}{\sqrt{f^2 \p R_{\mbox{\tiny $Q$}}^2}}
= \frl{f}{R_{\mbox{\tiny $Q$}}^2}
\mbox{ .}
\end{equation}
In order to solve this differential equation we introduce
a function $y(\hat{\chi})$ defined by
\begin{equation} \label{e_2277}
R_{\mbox{\tiny $Q$}} \tan y = f
\mbox{ .}
\end{equation}
This transforms the equation \eqref{e_2276} to
\begin{equation} \label{e_2278}
y'' = (1 - y'^2) \tan y
\mbox{ .}
\end{equation}
The general solution of this differential equation is
\begin{equation} \label{e_2283}
\int \frl{dy}{\sqrt{1 \m a \cos^2 y}} = \pm (\hat{\chi} - \hat{\chi}_0)
\mbox{ ,}
\end{equation}
where $a$ and $\hat{\chi}_0$ are integration constants. There are two
special cases where the solution can be expressed in terms of elementary
functions. The first is $a = 0$. Choosing $\hat{\chi}_0 = - \pi/2$ we
obtain
\begin{equation} \label{e_2284}
y = \pm (\hat{\chi} + \pi / 2)
\mbox{ ,}
\end{equation}
giving
\begin{equation} \label{e_2279}
e^{2 \alpha} = R_{\mbox{\tiny $Q$}}^2 \cot^2 \hat{\chi}
\mbox{\hspace{0mm} , \hspace{1mm}}
e^{2 \beta} = \frl{R_{\mbox{\tiny $Q$}}^2}{\sin^2 \hat{\chi}}
\end{equation}
where $-\infty < \hat{\eta} < \infty$ and $-\pi/2 < \hat{\chi} < \pi/2$,
$\hat{\chi} \ne 0$.
With these coordinates the line element of the LBR spacetime takes
the form
\begin{equation} \label{e_2280}
ds^2 = \frl{R_{\mbox{\tiny $Q$}}^2}{\sin^2 \hat{\chi}}
(- \cos^2 \hat{\chi} \hs{0.9mm} d \hat{\eta}^2 + d \hat{\chi}^2)
+ R_{\mbox{\tiny $Q$}}^2 d \Omega^2
\mbox{ .}
\end{equation}
The transformation between the $(\hat{\eta},\hat{\chi})$- and
the $(\eta,\chi)$-system used in the line element \eqref{e_218}
is given by
\begin{equation} \label{e_2384}
\eta = \hat{\eta}
\mbox{\hspace{2mm} , \hspace{3mm}}
\tanh \chi = \sin \hat{\chi}
\mbox{ .}
\end{equation}
This transformation shows that $(\hat{\eta},\hat{\chi})$ and
$(\eta,\chi)$ are comoving coordinates in the same reference frame.
The transformation represents a rescaling of the radial coordinate such
that the infinite interval $-\infty < \chi < \infty$ is transformed
into the finite interval $-\pi/2 < \hat{\chi} < \pi/2$, where
$\chi \ne 0$ and $\hat{\chi} \ne 0$.
\itm Combining the transformation \eqref{e_2384} and \eqref{e_279} with
$k = -1$, we obtain the coordinate transformation from
$(\hat{\eta},\hat{\chi})$ to $(T,R)$ in the following form
\begin{equation} \label{e_2289}
T = \frl{B \hs{0.2mm} \cos \hat{\chi} \sinh \hat{\eta}}
{1 \p \cos \hat{\chi} \cosh \hat{\eta}}
\mbox{\hspace{2mm} , \hspace{3mm}}
R = \frl{B \hs{0.2mm} \sin \hat{\chi}}
{1 \p \cos \hat{\chi} \cosh \hat{\eta}}
\mbox{ .}
\end{equation}
The inverse transformation is
\vspace{-0mm} \newline
\begin{equation} \label{e_2290}
\tanh \hat{\eta} =
\frl{2 \hs{0.2mm} B \hs{0.2mm} T}{B^2 \p T^{2} \m R^{2}}
\mbox{\hspace{2mm} , \hspace{3mm}}
\sin \hat{\chi} =
\frl{2 \hs{0.2mm} B \hs{0.2mm} R}{B^2 \m T^{2} \p R^{2}}
\mbox{ .}
\end{equation}
\vspace{-7mm} \newline
\itm On the other hand, choosing $\hat{\chi}_0 = 0$ we obtain
\begin{equation} \label{e_2484}
y = \pm \hat{\chi}
\mbox{ ,}
\end{equation}
giving
\begin{equation} \label{e_2379}
e^{2 \alpha} = R_{\mbox{\tiny $Q$}}^2 \tan^2 \hat{\chi}
\mbox{\hspace{0mm} , \hspace{1mm}}
e^{2 \beta} = \frl{R_{\mbox{\tiny $Q$}}^2}{\cos^2 \hat{\chi}}
\end{equation}
where $-\infty < \hat{\eta} < \infty$ and $0 < \hat{\chi} < \pi$,
$\hat{\chi} \ne \pi/2$.
With these coordinates the line element of the LBR spacetime takes
the form
\begin{equation} \label{e_2380}
ds^2 = \frl{R_{\mbox{\tiny $Q$}}^2}{\cos^2 \hat{\chi}}
(- \sin^2 \hat{\chi} \hs{0.9mm} d \hat{\eta}^2 + d \hat{\chi}^2)
+ R_{\mbox{\tiny $Q$}}^2 d \Omega^2
\mbox{ .}
\end{equation}
The transformation between the $(\hat{\eta},\hat{\chi})$- and
the $(\eta,\chi)$-system used in the line element \eqref{e_218}
is given by
\begin{equation} \label{e_2584}
\eta = \hat{\eta}
\mbox{\hspace{2mm} , \hspace{3mm}}
\tanh \chi = -\cos \hat{\chi}
\mbox{ .}
\end{equation}
\vspace{-7mm} \newline
\itm From the line element \eqref{e_2380} it follows that the coordinate
velocity of light moving in the radial direction is
\begin{equation} \label{e_2984}
\frl{d\hat{\chi}}{d\hat{\eta}} = \pm \sin \hat{\chi}
\mbox{ .}
\end{equation}
%
Integrating we obtain the equation of the world line of light in the
$(\hat{\eta},\hat{\chi})$-system
\begin{equation} \label{e_2985}
e^{\pm \hat{\eta}} \cot \frl{\hat{\chi}}{2} = e^{\pm \hat{\eta}_0}
\mbox{ ,}
\end{equation}
\vspace{-7mm} \newline
where $\hat{\eta}_0$ is a constant.
\itm Combining the transformation \eqref{e_2584} and \eqref{e_279} with
$k = -1$, we obtain the coordinate transformation from
$(\hat{\eta},\hat{\chi})$ to $(T,R)$ in the following form
\begin{equation} \label{e_2389}
T = \frl{B \hs{0.2mm} \sin \hat{\chi} \sinh \hat{\eta}}
{1 \p \sin \hat{\chi} \cosh \hat{\eta}}
\mbox{\hspace{2mm} , \hspace{3mm}}
R = - \frl{B \hs{0.2mm} \cos \hat{\chi}}
{1 \p \sin \hat{\chi} \cosh \hat{\eta}}
\mbox{ .}
\end{equation}
The inverse transformation is
\vspace{-0mm} \newline
\begin{equation} \label{e_2390}
\tanh \hat{\eta} =
\frl{2 \hs{0.2mm} B \hs{0.2mm} T}{T^{2} \m R^{2} \p B^2}
\mbox{\hspace{2mm} , \hspace{3mm}}
\cos \hat{\chi} =
\frl{2 \hs{0.2mm} B \hs{0.2mm} R}{T^{2} \m R^{2} \m B^2}
\mbox{ .}
\end{equation}
\vspace{-7mm} \newline
\itm The second case is $a = 1$. Choosing $\hat{\chi}_0 = 0$ the
solution of the differential equation \eqref{e_2283} can then be written
\begin{equation} \label{e_2684}
\tan y = \mp \frl{1}{\sinh \hat{\chi}}
\mbox{ ,}
\end{equation}
giving
\begin{equation} \label{e_2479}
e^{2 \alpha} = \frl{R_{\mbox{\tiny $Q$}}^2}{\sinh^2 \hat{\chi}}
\mbox{\hspace{0mm} , \hspace{1mm}}
e^{2 \beta} = R_{\mbox{\tiny $Q$}}^2 \coth^2 \hat{\chi}
\mbox{ .}
\end{equation}
With these coordinates the line element of the LBR spacetime takes
the form
\begin{equation} \label{e_2480}
ds^2 = \frl{R_{\mbox{\tiny $Q$}}^2}{\sinh^2 \hat{\chi}}
(- d \hat{\eta}^2 + \cosh^2 \hat{\chi} \hs{0.9mm} d \hat{\chi}^2)
+ R_{\mbox{\tiny $Q$}}^2 d \Omega^2
\mbox{ .}
\end{equation}
The transformation between the $(\hat{\eta},\hat{\chi})$- and
the CFS system used in the line element \eqref{e_157}
is given by
\begin{equation} \label{e_2294}
T = \hat{\eta}
\mbox{\hspace{2mm} , \hspace{3mm}}
R = \sinh \hat{\chi}
\mbox{ .}
\end{equation}
\vspace{-7mm} \newline
\itm From the line element \eqref{e_2480} it follows that the coordinate
velocity of light moving in the radial direction is
\begin{equation} \label{e_2984}
\frl{d\hat{\chi}}{d\hat{\eta}} = \pm \frl{1}{\cosh \hat{\chi}}
\mbox{ .}
\end{equation}
%
Integrating with the initial condition $\hat{\chi}(0) = 0$ we obtain the
equation of the world line of light in the $(\hat{\eta},\hat{\chi})$-system
\begin{equation} \label{e_2985}
\sinh \hat{\chi} = \pm \hat{\eta}
\mbox{ .}
\end{equation}
%
According to equation \eqref{e_2294} this corresponds to $R = \pm T$,
which is the equation of radially moving light in the CFS system as seen
from the line element \eqref{e_157}.
\itm Combining the transformation \eqref{e_2294} and \eqref{e_346}, we
obtain the coordinate transformation from
$(\hat{\eta},\hat{\chi})$ to $(\eta,\chi)$ in the following form
\begin{equation} \label{e_2299}
I_k(\eta) = \frl{B^2 \m k (\hat{\eta}^{2} \m \sinh^{2} \hat{\chi})}
{2 \hs{0.2mm} B \hs{0.2mm} \hat{\eta}}
\mbox{\hspace{2mm} , \hspace{3mm}}
I_k(\chi) = \frl{B^2 \p k (\hat{\eta}^{2} \m \sinh^{2} \hat{\chi})}
{2 \hs{0.2mm} B \hs{0.2mm} \sinh^{2} \hat{\chi}}
\end{equation}
\vspace{-2mm} \newline
when $\hat{\eta} \ne 0$. In the case $\hat{\eta} = 0$ we have that
$\eta = 0$. The inverse transformation is
\vspace{-0mm} \newline
\begin{equation} \label{e_2300}
\hat{\eta} = \frl{B \hs{0.2mm}
S_k(\eta)}{C_k(\eta) \p C_k(\chi)}
\mbox{\hspace{2mm} , \hspace{3mm}}
\sinh \hat{\chi} = \frl{B \hs{0.2mm}
S_k(\chi)}{C_k(\eta) \p C_k(\chi)}
\mbox{ .}
\end{equation}
\vspace{-7mm} \newline
%
%
%
\vspace{6mm} \newline
{\it VI. Cylindrical coordinates.}
\vspace{3mm} \newline
We shall now consider an axially symmetric space using cylindrical
coordinates $\rho$, $\theta$, $z$, assuming that the line element
has the form
\begin{equation} \label{e_284}
ds^2 = R_{\mbox{\tiny $Q$}}^2 \hs{0.5mm} [ \hs{0.5mm}
\mbox{$-f$} \hs{0.7mm} dT^2
+ \frl{1}{f} (d{\rho}^2 + dz^2 + {\rho}^2 d{\theta}^2)
\hs{0.5mm}]
\end{equation}
where $f = f(\rho,z)$. Demanding that the Weyl tensor vanishes,
we find that
\begin{equation} \label{e_285}
f(\rho,z) = a (\rho^2 + z^2) + bz + c
\end{equation}
where $a$, $b$ and $c$ are constants.
\itm In general the energy momentum tensor has the following physical
interpretation. Since the tensor is symmetrical, the eigenvectors of
the tensor can be chosen to be orthonormal with one timelike and three
spacelike vectors. These vectors will then represent an orthonormal
basis that may be associated with an observer with four velocity equal
to the timelike eigenvector ${\bf u} = {\bf e}_{\hs{0.5mm} 0}$. The
eigenvalue ${\lambda}_{\hs{0.5mm} 0}$ is interpreted as the energy density
measured by this observer, and the eigenvalues ${\lambda}_{\hs{0.5mm} i}$
are interpreted as the stresses he measures. For $a = 1$, $b = c = 0$ the
vectors of the observer's orthonormal basis are
\begin{equation} \label{e_286}
{\bf e}_0 = \frn{1}{R_{\mbox{\tiny $Q$}} \hs{0.2mm} \sqrt{\rho^2 + z^2}}
\hs{1.0mm} {\bf e}_t
\mbox{\hspace{0mm} , \hspace{1mm}}
{\bf e}_1 = \frm{\sqrt{\rho^2 + z^2}}{R_{\mbox{\tiny $Q$}} \hs{0.2mm} \rho}
\hs{1.0mm} {\bf e}_{\theta}
\mbox{\hspace{0mm} , \hspace{1mm}}
{\bf e}_2 = \frm{1}{R_{\mbox{\tiny $Q$}}}
(z \hs{0.5mm} {\bf e}_z + \rho \hs{0.5mm} {\bf e}_{\rho})
\mbox{\hspace{0mm} , \hspace{1mm}}
{\bf e}_3 = \frm{1}{R_{\mbox{\tiny $Q$}}}
(\rho \hs{0.5mm} {\bf e}_z - z \hs{0.5mm} {\bf e}_{\rho})
\mbox{ .}
\end{equation}
The corresponding eigenvalues of the energy momentum tensor are
\begin{equation} \label{e_283}
{\lambda}_{\hs{0.5mm} 0} = {\lambda}_{\hs{0.5mm} 2} =
- \frl{1}{\kappa R_{\mbox{\tiny $Q$}}^2}
\mbox{\hspace{2mm} , \hspace{3mm}}
{\lambda}_{\hs{0.5mm} 1} = {\lambda}_{\hs{0.5mm} 3} =
\frl{1}{\kappa R_{\mbox{\tiny $Q$}}^2}
\mbox{ .}
\end{equation}
These eigenvalues are recognized as those of the energy momentum tensor
of an electrical field. The line element then takes the form
\begin{equation} \label{e_235}
ds^2 = R_{\mbox{\tiny $Q$}}^2 \hs{0.5mm} [ \hs{0.5mm}
\mbox{$- ({\rho}^2 + z^2)$} \hs{0.7mm} dT^2
+ \frl{1}{{\rho}^2 + z^2} (d{\rho}^2 + dz^2 + {\rho}^2 d{\theta}^2)
\hs{0.5mm}]
\mbox{ .}
\end{equation}
As shown by D.\hn Garfinkle and E.\hn N.\hn Glass [\ref{r_19}] this
may also be found by transforming the line element \eqref{e_157} to
cylindrical coordinates by means of
\begin{equation} \label{e_233}
\rho = \frl{\sin \theta}{R}
\mbox{\hspace{2mm} , \hspace{3mm}}
z = \frl{\cos \theta}{R}
\mbox{ ,}
\end{equation}
or inversely
\begin{equation} \label{e_234}
R = \frl{1}{\sqrt{{\rho}^2 + z^2}}
\mbox{\hspace{2mm} , \hspace{3mm}}
\tan \theta = \frl{\rho}{z}
\mbox{ ,}
\end{equation}
Note that the charged domain wall defining the inner boundary of
the WLBR spacetime according to our interpretation is now given by
${\rho}^2 + z^2 = R_{\mbox{\tiny $Q$}}^{-2}$.
\itm From equations \eqref{e_1211}, \eqref{e_235} and \eqref{e_234}
we find that in the cylinder coordinates the line element of the
Minkowski spacetime inside the domain wall takes the form
\begin{equation} \label{e_1235}
ds_M^2 = -dT^2 + \frl{1}{({\rho}^2 + z^2)^2} \hs{0.4mm}
(d{\rho}^2 + dz^2 + {\rho}^2 d{\theta}^2)
\mbox{ .}
\end{equation}
It follows from equations \eqref{e_235}, \eqref{e_234} with
$R = R_{\mbox{\tiny $Q$}}$ and \eqref{e_1235} that the metric is
continuous at the domain wall.
\itm There is no acceleration of gravity in the reference frame in which
these coordinates are comoving. The unusual form of the spatial part of
the line element is a coordinate effect. The space is defined by
$T = \mbox{constant}$ just as in the CFS coordinate system. Hence it is
a Euclidean space described by using non-standard coordinate measuring
rods that are related to the standard rods by the transformation
\eqref{e_233}. In these coordinates the space looks like a curved,
but conformally flat space. In Cartesian and spherical coordinates,
respectively, this line element takes the form
\begin{equation} \label{e_1835}
ds^2 = -dT^2 + \frl{dx^2 + dy^2 + dz^2}{(x^2 + y^2 + z^2)^2}
= -dT^2 + \frl{1}{r^4} \hs{0.4mm} (dr^2 + r^2 d\Omega^2)
\mbox{ .}
\end{equation}
%
%
%
%
\vspace{-3mm} \newline
{\it VII. Light cone coordinates.}
\vspace{3mm} \newline
In spherical coordinates the line element on a 2-sphere with radius
$R_{\mbox{\tiny $Q$}}$ has the form
\begin{equation} \label{e_2258}
ds_2^2 = R_{\mbox{\tiny $Q$}}^2 d \Omega^2
= R_{\mbox{\tiny $Q$}}^2 (d \theta^2 + \sin^2 \theta d \phi^2)
\mbox{ .}
\end{equation}
One can project the spherical surface from the north pole onto the
equatorial plane by means of stereographic coordinates given by
\begin{equation} \label{e_1238}
\zeta = \cot \frl{\theta}{2} \hs{1.5mm} e^{i \phi}
\mbox{\hspace{2mm} , \hspace{3mm}}
\overline{\zeta} = \cot \frl{\theta}{2} \hs{1.5mm} e^{-i \phi}
\mbox{ ,}
\end{equation}
with inverse transformation given by
\begin{equation} \label{e_1239}
\cot \frl{\theta}{2} = \sqrt{\zeta \hs{0.2mm} \overline{\zeta}}
\mbox{\hspace{2mm} , \hspace{3mm}}
\cos 2 \phi = \mbox{Re}
\mbox{$\left( \rule[-1.5mm]{0mm}{5.7mm} \right.$} \hs{-0.2mm}
\frl{\zeta}{\overline{\zeta}
\rule[-0.0mm]{0mm}{3.7mm}}
\hs{-0.2mm} \mbox{$\left. \rule[-1.5mm]{0mm}{5.7mm} \right)$}
\mbox{ .}
\end{equation}
\vspace{-2mm} \newline
Taking the differentials and inserting into equation \eqref{e_2258}
we find the line element of the 2-sphere parametrized with the
stereographic coordinate $\zeta$ representing two real coordinates,
i.e. the real and imaginary part of $\zeta$,
\begin{equation} \label{e_2233}
ds_{\mbox{\tiny $S$}}^2 = \frl{4 R_{\mbox{\tiny $Q$}}^{2} \hs{0.3mm}
d\zeta \hs{0.3mm} d\overline{\zeta}}
{(1 \p \zeta \overline{\zeta})^2}
\mbox{ ,}
\end{equation}
where $\overline{\zeta}$ is the complex conjugate of $\zeta$.
\itm We shall now deduce a corresponding form for the line element of
the 2-dimensional anti de Sitter spacetime. For this purpose we
introduce new coordinates $U$ and $V$ for the anti de Sitter part of
the LBR spacetime in analogy with stereographic coordinates for the
spherical part,
\begin{equation} \label{e_2288}
U = \cot \frl{\hat{\chi}}{2} \hs{1.5mm} e^{\hat{\eta}}
\mbox{\hspace{2mm} , \hspace{3mm}}
V = \cot \frl{\hat{\chi}}{2} \hs{1.5mm} e^{- \hat{\eta}}
\mbox{ ,}
\end{equation}
where $(\hat{\eta}, \hat{\chi})$ are the coordinates introduced in
section 4.V. From equation \eqref{e_2985} it follows that
$U = \mbox{constant}$ for light moving in the positive
$\hat{\chi}$-direction, and $V = \mbox{constant}$ for light moving in
the negative $\hat{\chi}$-direction. Hence $(U,V)$ are light cone
coordinates.
\itm The inverse of the transformation \eqref{e_2288} is given by
\begin{equation} \label{e_2489}
\cot \frl{\hat{\chi}}{2} = \sqrt{UV}
\mbox{\hspace{2mm} , \hspace{3mm}}
e^{2 \hat{\eta}} =
\frl{U}{V \rule[-0.0mm]{0mm}{3.7mm}}
\mbox{ .}
\end{equation}
\vspace{-2mm} \newline
In the same way as for the spherical part we find
\begin{equation} \label{e_1290}
ds_{\mbox{\tiny $A$}}^2 =
\frl{4 R_{\mbox{\tiny $Q$}}^{2} \hs{0.3mm} dU \hs{0.3mm} dV}
{(1 \m UV)^2}
\mbox{ .}
\end{equation}
In terms of the light cone coordinates $(U,V)$ and the stereographic
coordinates the line element of the LBR spacetime takes the form
\begin{equation} \label{e_1233}
ds^2 = \frl{4 R_{\mbox{\tiny $Q$}}^{2} \hs{0.3mm} dU \hs{0.3mm} dV}
{(1 \m UV)^2} +
\frl{4 R_{\mbox{\tiny $Q$}}^{2} \hs{0.3mm} d\zeta \hs{0.3mm}
d\overline{\zeta}}{(1 \p \zeta \overline{\zeta})^2}
\mbox{ .}
\end{equation}
This form of the line element has earlier been considered by
M.\hn Ortaggio [\ref{r_33}] and later mentioned by Ortaggio and
Podolsk\'{y} [\ref{r_34}] and Griffiths and Podolsk\'{y} [\ref{r_14}]
with a different scaling of the coordinates.
\itm In order to find the coordinate transformation between the light
cone coordinates $(U,V)$ and the CFS coordinates, we will utilize
the transformation \eqref{e_279} between the $(\eta,\chi)$ and the
CFS coordinates. The $(U,V)$ coordinates are related to the
$(\hat{\eta},\hat{\chi})$ coordinates by the transformation
\eqref{e_2288}. Using the transformation \eqref{e_2584} we get
\begin{equation} \label{e_2306}
\cot \frl{\hat{\chi}}{2} = \frl{\sin \hat{\chi}}{1 \m \cos \hat{\chi}}
= \frl{1}{\cosh \chi} \cdot \frl{1}{1 \p \tanh \chi} = e^{- \chi}
\mbox{ .}
\end{equation}
Inserting this in equation \eqref{e_2288} we obtain the transformation
from the $(\eta,\chi)$- to the $(U,V)$-system,
\begin{equation} \label{e_2307}
U = e^{-\chi + \eta}
\mbox{\hspace{2mm} , \hspace{3mm}}
V = e^{-\chi - \eta}
\mbox{ .}
\end{equation}
The inverse transformation is given by
\begin{equation} \label{e_2308}
e^{\eta} = \sqrt{\frl{U}{V}}
\mbox{\hspace{2mm} , \hspace{3mm}}
e^{\chi} = \frl{1}{\sqrt{UV}}
\mbox{ .}
\end{equation}
From this we also obtain
\begin{equation} \label{e_2309}
\sinh \eta = \frl{U \m V}{2 \sqrt{UV}}
\mbox{\hspace{2mm} , \hspace{3mm}}
\cosh \eta = \frl{U \p V}{2 \sqrt{UV}}
\end{equation}
and
\begin{equation} \label{e_2310}
\sinh \chi = \frl{1 \m UV}{2 \sqrt{UV}}
\mbox{\hspace{2mm} , \hspace{3mm}}
\cosh \chi = \frl{1 \p UV}{2 \sqrt{UV}}
\mbox{ .}
\end{equation}
Inserting these expressions into the transformation \eqref{e_279}
we find
\vspace{-0mm} \newline
\begin{equation} \label{e_1234}
T = \frl{B}{2} \left(
\frl{1 \m V}{1 \p V} - \frl{1 \m U}{1 \p U} \right)
\mbox{\hspace{2mm} , \hspace{3mm}}
R = \frl{B}{2} \left(
\frl{1 \m V}{1 \p V} + \frl{1 \m U}{1 \p U} \right)
\mbox{ .}
\end{equation}
The inverse transformation is
\begin{equation} \label{e_1236}
U = \frl{B \m (R \m T)}{B \p (R \m T)}
\mbox{\hspace{2mm} , \hspace{3mm}}
V = \frl{B \m (R \p T)}{B \p (R \p T)}
\mbox{ .}
\end{equation}
We now introduce coordinates $(\tilde{u},\tilde{v})$ by the coordinate
transformation
\begin{equation} \label{e_2247}
\tilde{u} = - \frl{1}{U}
\mbox{\hspace{2mm} , \hspace{3mm}}
\tilde{v} = - \frl{U}{1 \m UV}
\mbox{ .}
\end{equation}
The inverse transformation is
\begin{equation} \label{e_2377}
U = - \frl{1}{\tilde{u}}
\mbox{\hspace{2mm} , \hspace{3mm}}
V = \frl{1}{\tilde{v}} - \tilde{u}
\mbox{ .}
\end{equation}
Taking the differentials of $U$ and $V$ and substituting into
the line element \eqref{e_1233} gives
\begin{equation} \label{e_2243}
ds^2 = - 4 R_{\mbox{\tiny $Q$}}^2 \hs{0.5mm} (\tilde{v}^2 d\tilde{u}^2
+ d \tilde{u} \hs{0.5mm} d\tilde{v}) +
\frl{4 R_{\mbox{\tiny $Q$}}^{2} \hs{0.3mm} d\zeta \hs{0.3mm}
d\overline{\zeta}}{(1 \p \zeta \overline{\zeta})^2}
\mbox{ .}
\end{equation}
This line element has earlier been studied by Podolsk\'{y} and
Ortaggio [\ref{r_32}] with a different scaling of the coordinates.
Inserting the formulae \eqref{e_2377} for $U$ and $V$ into
\eqref{e_1234}, we obtain the transformation between the
$(\tilde{u}, \tilde{v})$ coordinates and the CFS coordinates,
\begin{equation} \label{e_2643}
T = \frl{B \hs{0.2mm} [(\tilde{u}^2 \m 1) \tilde{v} \m \tilde{u}]}
{(\tilde{u} \m 1)[1 \m (\tilde{u} \m 1) \tilde{v}]}
\mbox{\hspace{2mm} , \hspace{3mm}}
R = \frl{B}{(\tilde{u} \m 1)[1 \m (\tilde{u} \m 1) \tilde{v}]}
\mbox{ .}
\end{equation}
%
This transformation corresponds to the transformation immediately
preceding (A1) in the appendix of reference [\ref{r_32}], but with
a different scaling of the coordinates.
The inverse transformation is found by inserting the exressions for
$U$ and $V$ in \eqref{e_1236} into equation \eqref{e_2247}, giving
\vspace{-3mm} \newline
\begin{equation} \label{e_2645}
\tilde{u}  =  \frl{(R \m T) \p B}{(R \m T) \m B}
\mbox{\hspace{2mm} , \hspace{3mm}}
\tilde{v}  = \frl{R^2 \m (T \p B)^2}{4BR}
\mbox{ .}
\end{equation}
%
%
%
%
%
\vspace{6mm} \newline
{\bf 5. A Milne-LBR universe model}
\vspace{3mm} \newline
We consider the flat spacetime inside the domain wall. In the
$(\eta',\chi')$-system the line element is given by \eqref{e_2274}.
Introducing the proper time $\tau'$ of the reference particles in the
$(\eta',\chi')$-system as a time coordinate we have
\begin{equation} \label{e_2237}
d \tau' = B e^{\eta'} d\eta'
\mbox{ .}
\end{equation}
Integrating with the initial condition $\tau'(0) = 0$, we obtain
\begin{equation} \label{e_2238}
\tau' = B e^{\eta'}
\mbox{ ,}
\end{equation}
and the line element of the flat spacetime inside the domain wall
takes the form
\begin{equation} \label{e_2376}
ds_M^2 = -d {\tau'}^2 + a({\tau'})^2 (d{\chi'}^2 + \sinh^2 \chi' d\Omega^2)
\end{equation}
where the scale factor is
\begin{equation} \label{e_2476}
a({\tau'}) = \tau'
\mbox{ .}
\end{equation}
This line element represents the Milne spacetime, which is simply the
flat Minkowski spacetime as described from a uniformly expanding
reference frame. The coordinates $(\tau',\chi')$ will here be called
the Milne coordinates. The transformation between the CFS coordinates
and the Milne coordinates is obtained from equation \eqref{e_1269}
with the substitution $B e^{\eta'} = \tau'$, giving
\begin{equation} \label{e_2269}
T = \tau' \cosh \chi'
\mbox{\hspace{2mm} , \hspace{3mm}}
R = \tau' \sinh \chi'
\mbox{ .}
\end{equation}
The inverse transformation is
\begin{equation} \label{e_2270}
\tau' = \sqrt{T^2 - R^2}
\mbox{\hspace{2mm} , \hspace{3mm}}
\tanh \chi' = \frl{R}{T}
\mbox{ .}
\end{equation}
In these coordinates the line element of the WLBR spacetime outside
the domain wall takes the form
\begin{equation} \label{e_2374}
ds^2 = \frl{R_{\mbox{\tiny $Q$}}^2}
{\sinh^2 \hs{-0.4mm} \chi' \rule[-0mm]{0mm}{3.95mm}} \left(
- \frl{d {\tau'}^2}{{\tau'}^2 \rule[-0mm]{0mm}{3.95mm}}
+ d{\chi'}^2 \right)
+ R_{\mbox{\tiny $Q$}}^2 d\Omega^2
\mbox{ ,}
\end{equation}
\vspace{-6mm} \newline
\itm It follows from the transformation \eqref{e_2270} that the world
lines of particles with $\chi' = \mbox{constant}$ are straight lines
both inside and outside the domain wall as illustrated in Figure 9.
Imagine observers with constant value of $\chi'$. The coordinate time
$\tau'$ both inside and outside the domain wall is equal to the proper
time of these observers. Note from the form of the line element
\eqref{e_2374} that $\tau'$ is not equal to the proper time of standard
clocks with $\chi' = \mbox{constant}$ outside the domain wall. The rate
of the proper time of these clocks is given by
\begin{equation} \label{e_2227}
d \tau_{\mbox{\tiny $W$}}'
= \frl{R_{\mbox{\tiny $Q$}}}
{\tau' \sinh \hs{-0.4mm} \chi'} \hs{0.6mm} d \tau'
\mbox{ .}
\end{equation}
This formula shows that a standard clock outside the domain wall
with $\chi' = \mbox{constant}$ goes at an increasingly slower rate
than a standard clock inside the domain wall. The reason is that
the clocks with $\chi' = \mbox{constant}$ move in the outwards
direction in the static CFS system. This does not change the rate
of the clocks inside the domain wall because they have constant
velocity and there is no gravitational field in this region. However,
outside the domain wall there is an outwards directed gravitational
field. Hence a clock with constant $\chi'$ comes lower in this field,
and therefore its rate decreases.
\itm We will now consider clocks in the WLBR region with a fixed
physical distance from the domain wall, $R = R_1$. It follows from
the transformation \eqref{e_2269} that the $\chi'$ coordinate of
this clock is given by $\tau' \sinh \chi' = R_1$. Hence equation
\eqref{e_2227} shows that these clocks go at a constant rate.
\itm Observers comoving with the reference particles inside the
domain wall, $\chi' = \mbox{constant}$, will observe that the
domain wall collapses towards them. The physical distance from an
observer at the origin to an object with coordinate $\chi'$ is
\begin{equation} \label{e_2570}
l = a(\tau') \hs{0.3mm} \chi'
\mbox{ .}
\end{equation}
The physical velocity of the object relative to the observer is
\begin{equation} \label{e_2571}
\dot{l} = \dot{a} \chi' + a \dot{\chi}'
= Hl  + a \dot{\chi}'
\mbox{ ,}
\end{equation}
where $H = \dot{a} / a$ is the Hubble parameter. The first term is the
velocity of the Hubble flow as given by Hubble's law, i.e. in the
present case the velocity of "the river of space" [\ref{r_26}] in the
Milne universe, and the second term represents the so-called peculiar
velocity of the object, i.e. its velocity through space.
\itm The physical velocity of the domain wall is found by inserting
$R = R_{\mbox{\tiny $Q$}}$ in the second of the transformation
equations \eqref{e_2269}, which gives
\begin{equation} \label{e_2572}
\sinh \chi' = \fr{R_{\mbox{\tiny $Q$}}}{\tau'}
\mbox{ .}
\end{equation}
%
Hence the coordinate velocity of the domain wall is
\begin{equation} \label{e_2573}
\dot{\chi}' = - \frl{\fr{R_{\mbox{\tiny $Q$}}}{\tau'}}
{\sqrt{R_{\mbox{\tiny $Q$}}^2 + \tau'^2}}
\mbox{ ,}
\end{equation}
and its physical velocity is
\begin{equation} \label{e_2574}
\dot{l}_{\mbox{\tiny $Q$}} = \mbox{arcsinh}
(\fr{R_{\mbox{\tiny $Q$}}}{\tau'})
- \frl{R_{\mbox{\tiny $Q$}}}
{\sqrt{R_{\mbox{\tiny $Q$}}^2 + \tau'^2}}
\mbox{ .}
\end{equation}
%
Surprisingly the domain wall has a non-vanishing physical velocity in the
Milne universe inside the wall which is even infinitely great initially,
and then decreases to zero in an infinitely far future. Hence the Hubble
flow dominates over the peculiar motion all the time. Integrating with the
initial condition $l(0) = 0$ we find the physical distance from the
observer at the center to the domain wall,
\begin{equation} \label{e_2575}
l = \tau' \mbox{arcsinh}
(\fr{R_{\mbox{\tiny $Q$}}}{\tau'})
= \tau' \chi'
\mbox{ .}
\end{equation}
The chosen initial condition is necessary in order to obtain a result in
accordance with the expression for $l$ in equation \eqref{e_2570}. Taking
the limit when $\tau' \rightarrow \infty$ we find that the final distance
of the domain wall from the observer at the center is
$l = R_{\mbox{\tiny $Q$}}$.
\itm The physical velocity of the domain wall in the CFS system inside the
wall vanishes. Hence, as described by an observer at the center of these
coordinates, which coincides with the center of the Milne coordinates, the
domain wall is at rest. Since we talk about physical velocity and physical
distance one might think that these quantities should be coordinate
invariant. The reason that this is not so, is that the spaces of the
CFS system and the Milne universe are different simultaneity spaces.
%
%
%
\vspace{6mm} \newline
{\bf 6. The Killing vector field defining the motion of the reference
frames}
\vspace{3mm} \newline
The LBR spacetime has a timelike Killing vector field which is most easily
seen in the coordinate systems in which the metric is static. Then the
timelike coordinate basis vector is a timelike Killing vector [\ref{r_24}].
\itm From the line element \eqref{e_218} it follows that
${\bf K} = {\bf e}_{\eta} = \partial / \partial \eta$ is a Killing vector.
In order to make it explicit that there are three different cases we
define unit vectors in the direction of ${\bf K}$ by
\begin{equation} \label{e_2001}
{\bf V} = \frl{{\bf K}}{\sqrt{K_{\mu} K^{\mu}}}
= \frl{{\bf K}}{\sqrt{- g_{\eta \eta}}}
= \frl{S_k(\chi)}{R_{\mbox{\tiny $Q$}}} \hs{0.5mm} {\bf e}_{\eta}
\mbox{ .}
\end{equation}
These timelike unit vectors ${\bf V}$ can be interpreted as the 4-velocity
of reference particles following trajectories of the Killing vector field
${\bf K}$, i.e. it is the 4-velocity of the reference particles defining
the reference frame in which the $(\eta,\chi)$-coordinates are comoving.
The vectors ${\bf V}$ given in equation \eqref{e_2001} for $k = -1,0,1$
may be distinguished by the magnitude of the 4-accelerations of the
particles. The 4-acceleration of a particle with a world line having
${\bf V}$ as a unit tangent vector is
\begin{equation} \label{e_2002}
{\bf A} = A^{\chi} {\bf e}_{\chi}
= V_{;\nu}^{\chi} V^{\nu} {\bf e}_{\chi}
= (V_{,\nu}^{\chi} V^{\nu} +
{\Gamma}^{\hs{0.3mm} \chi}_{\hs{1.5mm} \alpha \beta}
\hs{0.8mm} V^{\alpha} V^{\beta}) \hs{0.5mm} {\bf e}_{\chi}
\mbox{ .}
\end{equation}
Since the only non-vanishing component of ${\bf V}$ is $V^{\eta}$,
this expression reduces to
\begin{equation} \label{e_2003}
{\bf A} = {\Gamma}^{\hs{0.3mm} \chi}_{\hs{1.5mm} \eta \eta}
\hs{0.8mm} (V^{\eta})^2 {\bf e}_{\chi}
\mbox{ .}
\end{equation}
Using the expression \eqref{e_2004} for the Christoffel symbol we obtain
\begin{equation} \label{e_2005}
{\bf A} = - I_k(\chi) \hs{0.5mm} \frl{S_k(\chi)^2}{R_{\mbox{\tiny $Q$}}^2}
\hs{0.7mm} {\bf e}_{\chi}
= - \frl{S_k(2 \chi)}{2 R_{\mbox{\tiny $Q$}}^2} \hs{0.7mm} {\bf e}_{\chi}
\mbox{ .}
\end{equation}
The square of the acceleration scalar of this reference particle is
\begin{equation} \label{e_2006}
A^2 = A_{\mu} A^{\mu} = \frl{C_k(\chi)^2}{R_{\mbox{\tiny $Q$}}^2}
\mbox{ .}
\end{equation}
\vspace{-8mm} \newline
\itm The physical meaning of the acceleration scalar of an arbitrary
particle is that it represents the ordinary acceleration of the particle
as measured with standard clocks and measuring rods relative to a local
inertial frame in which the particle is instantaneously at rest. In
other words it represents the acceleration of the particle relative to
a free particle. This is called the {\it proper acceleration} of the
particle and will here be denoted by $A_k$ for reasons that will be
apparent below. It follows that the acceleration of gravity as defined
in equation \eqref{e_357} is equal to minus the proper acceleration of
the reference particles defining the motion of a reference frame.
\itm The acceleration of a free particle instantaneously at rest in the
$(\eta,\chi)$-system is given by equation \eqref{e_360}. This acceleration
is due to the non-inertial character of the reference frame. Hence the
proper acceleration of a reference particle in the $(\eta,\chi)$-system
is given by
\begin{equation} \label{e_2007}
A_k = - \frl{C_k(\chi)}{R_{\mbox{\tiny $Q$}}}
\end{equation}
%
for $k = 1,0,-1$.
\itm In the $(\tilde{t},\tilde{r})$-system the Killing vector is
${\bf K} = (R_{\mbox{\tiny $Q$}} / A) \hs{0.5mm} {\bf e}_{\tilde{t}}
= (R_{\mbox{\tiny $Q$}} / A) \hs{0.5mm} \partial / \partial \tilde{t}$.
Using the transformation \eqref{e_349} and the formula \eqref{e_812} we
find that the proper acceleration of the reference particle is given by
\begin{equation} \label{e_2008}
A_k = \frl{\tilde{r} \m \tilde{r}_0}{R_{\mbox{\tiny $Q$}}
\sqrt{(\tilde{r} \m \tilde{r}_0)^2 \p k A^2 R_{\mbox{\tiny $Q$}}^2}}
\mbox{ .}
\end{equation}
This expression has earlier been deduced by Lapedes [\ref{r_18}] with
$\tilde{r}_0 = 0$ and $A = 1 / R_{\mbox{\tiny $Q$}}$. We shall here
use equation \eqref{e_2008} to discuss the motion of the reference frame
in which the $(\eta,\chi)$, $(\tilde{t},\tilde{r})$ and
$(\hat{t},\hat{r})$ coordinate systems are comoving in the WLBR
spacetime. The equation shows how the reference particles move in the
radial direction. We first consider the case $k = 0$. Equation
\eqref{e_456} implies that $\tilde{r} < \tilde{r}_0 - R_{\mbox{\tiny $Q$}}$
in the WLBR spacetime. Hence $\tilde{r} < \tilde{r}_0$ in this region.
Equation \eqref{e_2008} shows that in this case the reference particles
have a constant acceleration $A_0 = -1 / R_{\mbox{\tiny $Q$}}$ which is
directed towards the domain wall relative to a free particle, with just the
magnitude that keeps it at rest relative to the domain wall.
\itm We then consider the case $k = 1$. At $\tilde{r} = \tilde{r}_0$ the
reference particles have vanishing proper acceleration, i.e. they are
freely falling. When $\tilde{r} < \tilde{r}_0$ in the region given by the
inequalities \eqref{e_667} we then have
$- 1 / R_{\mbox{\tiny $Q$}} \le A_1 < 0$. This means that a reference
particle in this region accelerates away from the domain wall, but with
a smaller acceleration than that of a free particle. Hence in this region
the reference frame accelerates inwards relative to a local inertial frame.
When $\tilde{r} > \tilde{r}_0$ in the region given by the inequalities
\eqref{e_667} we have that $0 < A_1 \le 1 / R_{\mbox{\tiny $Q$}}$, and
the reference frame accelerates outwards relative to a local inertial
reference frame.
\itm Finally we consider the case $k = -1$. In this case the proper
acceleration of the reference particles is directed towards the domain
wall and has a magnitude greater than $1 / R_{\mbox{\tiny $Q$}}$, i.e.
greater than that of a free particle. Thus the reference frame accelerates
towards the domain wall.
\itm The proper acceleration of the reference particles depends upon $k$
in the following way
\begin{equation} \label{e_2010}
|A_1| \le 1 / R_{\mbox{\tiny $Q$}}
\mbox{\hspace{2mm} , \hspace{3mm}}
|A_0| = 1 / R_{\mbox{\tiny $Q$}}
\mbox{\hspace{2mm} and \hspace{3mm}}
|A_{-1}| \ge 1 / R_{\mbox{\tiny $Q$}}
\mbox{ .}
\end{equation}
This gives a physical meaning of the constant $k$. It tells whether the
magnitude of the proper acceleration of the reference particles is smaller
than, equal to, or greater than that of a free particle. If $k = 1$ the
reference frame accelerates away from the domain wall, if $k = 0$ it is
at rest relative to the domain wall, and if $k = -1$ it accelerates
towards the domain wall.
\itm Finally, in the $(\hat{t},\hat{r})$-system the Killing vector is
${\bf K} = (R_{\mbox{\tiny $Q$}} / A) \hs{0.5mm} {\bf e}_{\hat{t}}
= (R_{\mbox{\tiny $Q$}} / A) \hs{0.5mm} \partial / \partial \hat{t}$.
From equation \eqref{e_2007} and the transformation \eqref{e_700} we get
\begin{equation} \label{e_2009}
A_k = \mp \hs{0.3mm} \frl{1}{R_{\mbox{\tiny $Q$}}} \hs{1.1mm}
b_{k} \hs{-0.8mm} \left(
\frl{\hat{r}_0 - \hat{r}}{R_{\mbox{\tiny $Q$}}} \right)
\mbox{ ,}
\end{equation}
which is consistent with the expression \eqref{e_2011} for the
acceleration of gravity in the $(\hat{t},\hat{r})$-system.
%
%
\vspace{6mm} \newline
{\bf 7. Embedding of the LBR spacetime in a flat six-dimensional manifold}
\vspace{3mm} \newline
In order to exhibit the topological structure of the LBR spacetime Dias
and Lemos [\ref{r_16}] considered the embedding of the LBR spacetime in a
flat six-dimensional manifold $M^{2,4}$.
\itm We shall here show how LBR spacetime is parametrized in $M^{2,4}$ in
the six main coordinate systems that we have considered in this paper.
The coordinates in $M^{2,4}$ are denoted by $(z_0,z_1,z_2,z_3,z_4,z_5)$,
and for $k = \pm 1$ the line element of $M^{2,4}$ has the form
\begin{equation} \label{e_1937}
ds^2 = -d z_0^2 + k \hs{0.5mm} d z_1^2 - k \hs{0.5mm} d z_2^2
+ d z_3^2 + d z_4^2 + d z_5^2
\mbox{ .}
\end{equation}
Note that $z_1$ and $z_2$ are exchanged when $k$ changes sign. The LBR
4-submanifold is determined by the two constraints
\begin{equation} \label{e_1948}
z_0^2 - k \hs{0.5mm} z_1^2 + k \hs{0.5mm} z_2^2 = R_{\mbox{\tiny $Q$}}^2
\mbox{ ,}
\end{equation}
\begin{equation} \label{e_1939}
z_3^2 + z_4^2 + z_5^2 = R_{\mbox{\tiny $Q$}}^2
\mbox{ .}
\end{equation}
The first of these constraints defines the $\mbox{AdS}_{\hs{0.2mm} 2}$
hyperboloid, and the second defines the 2-sphere of radius $
R_{\mbox{\tiny $Q$}}$.
\itm From equation \eqref{e_211} it follows that the spherical part of
the LBR submanifold is invariant. Hence the parametrization of the
2-sphere takes the same form in all the coordinate systems,
\begin{equation} \label{e_1940}
z_3 = R_{\mbox{\tiny $Q$}} \sin \theta \cos \phi
\mbox{\hspace{2mm} , \hspace{3mm}}
z_4 = R_{\mbox{\tiny $Q$}} \sin \theta \sin \phi
\mbox{\hspace{2mm} , \hspace{3mm}}
z_5 = R_{\mbox{\tiny $Q$}} \cos \theta
\mbox{ .}
\end{equation}
This satisfies the constraint \eqref{e_1939} and gives the last three
terms of equation \eqref{e_1937}.
\itm We shall now consider the different parametrizations of the AdS
hyperboloid satisfying the constraint \eqref{e_1948} and giving the first
three terms at the right hand side of equation \eqref{e_1937} using the
coordinate systems mentioned above. In CFS coordinates the parametrization
takes the form
\begin{equation} \label{e_1941}
z_0 = R_{\mbox{\tiny $Q$}} \frl{T}{R}
\mbox{\hspace{2mm} , \hspace{3mm}}
z_1 = R_{\mbox{\tiny $Q$}} \frl{B^2 \p k (T^{2} \m R^{2})}
{2 \hs{0.2mm} B \hs{0.2mm} R}
\mbox{\hspace{2mm} , \hspace{3mm}}
z_2 = R_{\mbox{\tiny $Q$}} \frl{B^2 \m k (T^{2} \m R^{2})}
{2 \hs{0.2mm} B \hs{0.2mm} R}
\mbox{ ,}
\end{equation}
giving
\begin{equation} \label{e_1942}
-d z_0^2 + k \hs{0.5mm} d z_1^2 - k \hs{0.5mm} d z_2^2
= \frl{R_{\mbox{\tiny $Q$}}^2}{R^2} \hs{0.5mm} ( \hs{0.5mm}
\mbox{$- dT^2$} + dR^2 \hs{0.5mm})
\mbox{ .}
\end{equation}
Note that the cases $k = 1$ and $k = -1$ give the same parametrization,
but with the coordinates $z_1$ and $z_2$ exchanged.
A special case of this parametrization has earlier been considered by
O.\hn B.\hn Zaslavskii [\ref{r_25}].
\itm From the transformation \eqref{e_346} it follows that with the
$(\eta,\chi)$-coordinates the parametri- zation of the AdS hyperboloid
in $M^{2,4}$ takes the form
\begin{equation} \label{e_1943}
z_0 = R_{\mbox{\tiny $Q$}} \frl{S_k(\eta)}{S_k(\chi)}
\mbox{\hspace{2mm} , \hspace{3mm}}
z_1 = R_{\mbox{\tiny $Q$}} I_k(\chi)
\mbox{\hspace{2mm} , \hspace{3mm}}
z_2 = R_{\mbox{\tiny $Q$}} \frl{C_k(\eta)}{S_k(\chi)}
\mbox{ .}
\end{equation}
Using equations \eqref{e_810}, \eqref{e_812}, \eqref{e_824} and
\eqref{e_825} one may show that this parametrization fullfills
equation \eqref{e_1937} and the constraint \eqref{e_1948}.
\itm Using the transformation \eqref{e_946} we find that the
parametrization that transforms between the line element \eqref{e_928}
with $(\tau,\rho)$-coordinates and the first three terms of \eqref{e_1937}
with $k = -1$ is
\begin{equation} \label{e_1944}
z_0 = R_{\mbox{\tiny $Q$}} \frl{\cosh \rho}{\cosh \tau}
\mbox{\hspace{2mm} , \hspace{3mm}}
z_1 = - R_{\mbox{\tiny $Q$}} \tanh \tau
\mbox{\hspace{2mm} , \hspace{3mm}}
z_2 = - R_{\mbox{\tiny $Q$}} \frl{\sinh \rho}{\cosh \tau}
\mbox{ .}
\end{equation}
With the $(\tilde{t},\tilde{r})$-coordinates equation \eqref{e_321}
gives
\begin{equation} \label{e_1945}
z_0 = \mbox{$\left[ \rule[-1.5mm]{0mm}{5.7mm} \right.$} \hs{-0.2mm}
k R_{\mbox{\tiny $Q$}}^2
+ \left(\frl{\tilde{r} \m \tilde{r}_0}{A} \right)^2
\hs{-0.2mm} \mbox{$\left. \rule[-1.5mm]{0mm}{5.7mm} \right]$}^{1/2}
\hs{0.7mm} S_k \left( \frl{A \hs{0.5mm} \tilde{t}}{R_{\mbox{\tiny $Q$}}}
\right)
\mbox{ ,}
\end{equation}
\begin{equation} \label{e_1946}
z_1 = \frl{\tilde{r}_0 \m \tilde{r}}{A}
\mbox{ ,} \hs{46.5mm}
\end{equation}
\begin{equation} \label{e_1947}
z_2 = \mbox{$\left[ \rule[-1.5mm]{0mm}{5.7mm} \right.$} \hs{-0.2mm}
k R_{\mbox{\tiny $Q$}}^2
+ \left(\frl{\tilde{r} \m \tilde{r}_0}{A} \right)^2
\hs{-0.2mm} \mbox{$\left. \rule[-1.5mm]{0mm}{5.7mm} \right]$}^{1/2}
\hs{0.7mm} C_k \left( \frl{A \hs{0.5mm} \tilde{t}}{R_{\mbox{\tiny $Q$}}}
\right)
\mbox{ .}
\end{equation}
\vspace{-1mm} \newline
From equation \eqref{e_991} it follows that the parametrization with
the $(\overline{t},\overline{r})$-coordinates has the form
\begin{equation} \label{e_1949}
z_0 = \mbox{$\left[ \rule[-1.5mm]{0mm}{5.7mm} \right.$} \hs{-0.2mm}
R_{\mbox{\tiny $Q$}}^2
- \left(\frl{\overline{t} \m \overline{t}_0}{A} \right)^2
\hs{-0.2mm} \mbox{$\left. \rule[-1.5mm]{0mm}{5.7mm} \right]$}^{1/2}
\hs{0.7mm} \cosh \left(
\frl{A \hs{0.5mm} \overline{r}}{R_{\mbox{\tiny $Q$}}} \right)
\mbox{ ,}
\end{equation}
\begin{equation} \label{e_1950}
z_1 = \frl{\overline{t}_0 \m \overline{t}}{A}
\mbox{ ,} \hs{49.5mm}
\end{equation}
\begin{equation} \label{e_1951}
\hs{2.5mm}
z_2 = - \mbox{$\left[ \rule[-1.5mm]{0mm}{5.7mm} \right.$} \hs{-0.2mm}
R_{\mbox{\tiny $Q$}}^2
- \left(\frl{\overline{t} \m \overline{t}_0}{A} \right)^2
\hs{-0.2mm} \mbox{$\left. \rule[-1.5mm]{0mm}{5.7mm} \right]$}^{1/2}
\hs{0.7mm} \sinh \left(
\frl{A \hs{0.5mm} \overline{r}}{R_{\mbox{\tiny $Q$}}} \right)
\mbox{ ,}
\end{equation}
\vspace{-1mm} \newline
corresponding to $k = -1$ in equations \eqref{e_1937} and \eqref{e_1948}.
With $(\hat{t},\hat{r})$-coordinates equation \eqref{e_322} leads to
the parametrization
\begin{equation} \label{e_1952}
z_0 = R_{\mbox{\tiny $Q$}} \hs{0.5mm}
a_{k} \mbox{$\left( \rule[-1.5mm]{0mm}{5.7mm} \right.$} \hs{-0.2mm}
\frl{\hat{r}_0 \m \hat{r}}{R_{\mbox{\tiny $Q$}}}
\hs{-0.2mm} \mbox{$\left. \rule[-1.5mm]{0mm}{5.7mm} \right)$}
\hs{0.7mm} S_k \left( \frl{A \hs{0.5mm} \hat{t}}{R_{\mbox{\tiny $Q$}}}
\right)
\mbox{ ,}
\end{equation}
\begin{equation} \label{e_1953}
z_1 = R_{\mbox{\tiny $Q$}} \hs{0.5mm}
a_{-k} \mbox{$\left( \rule[-1.5mm]{0mm}{5.7mm} \right.$} \hs{-0.2mm}
\frl{\hat{r}_0 \m \hat{r}}{R_{\mbox{\tiny $Q$}}}
\hs{-0.2mm} \mbox{$\left. \rule[-1.5mm]{0mm}{5.7mm} \right)$}
\mbox{ ,} \hs{15.5mm}
\end{equation}
\begin{equation} \label{e_1954}
z_2 = R_{\mbox{\tiny $Q$}} \hs{0.5mm}
a_{k} \mbox{$\left( \rule[-1.5mm]{0mm}{5.7mm} \right.$} \hs{-0.2mm}
\frl{\hat{r}_0 \m \hat{r}}{R_{\mbox{\tiny $Q$}}}
\hs{-0.2mm} \mbox{$\left. \rule[-1.5mm]{0mm}{5.7mm} \right)$}
\hs{0.7mm} C_k \left( \frl{A \hs{0.5mm} \hat{t}}{R_{\mbox{\tiny $Q$}}}
\right)
\mbox{ .}
\end{equation}
\vspace{-1mm} \newline
In order to show that this parametrization fullfills the constraint
\eqref{e_1948}, one has to use equation \eqref{e_833}.
Equation \eqref{e_348} leads to the following parametrization in
$(t,r)$-coordinates,
\begin{equation} \label{e_1955}
z_0 = R_{\mbox{\tiny $Q$}}
\cos \mbox{$\left( \rule[-1.5mm]{0mm}{5.7mm} \right.$} \hs{-0.2mm}
\frl{t \m t_0}{R_{\mbox{\tiny $Q$}}}
\hs{-0.2mm} \mbox{$\left. \rule[-1.5mm]{0mm}{5.7mm} \right)$}
\hs{0.7mm} \cosh \left( \frl{A \hs{0.5mm} r}{R_{\mbox{\tiny $Q$}}}
\right)
\mbox{ ,}
\end{equation}
\begin{equation} \label{e_1956}
z_1 = - R_{\mbox{\tiny $Q$}}
\sin \mbox{$\left( \rule[-1.5mm]{0mm}{5.7mm} \right.$} \hs{-0.2mm}
\frl{t \m t_0}{R_{\mbox{\tiny $Q$}}}
\hs{-0.2mm} \mbox{$\left. \rule[-1.5mm]{0mm}{5.7mm} \right)$}
\mbox{ ,} \hs{19.5mm}
\end{equation}
\begin{equation} \label{e_1957}
\hs{2.5mm} z_2 = - R_{\mbox{\tiny $Q$}}
\cos \mbox{$\left( \rule[-1.5mm]{0mm}{5.7mm} \right.$} \hs{-0.2mm}
\frl{t \m t_0}{R_{\mbox{\tiny $Q$}}}
\hs{-0.2mm} \mbox{$\left. \rule[-1.5mm]{0mm}{5.7mm} \right)$}
\hs{0.7mm} \sinh \left( \frl{A \hs{0.5mm} r}{R_{\mbox{\tiny $Q$}}}
\right)
\mbox{ ,}
\end{equation}
\vspace{-1mm} \newline
again corresponding to $k = -1$ in equations \eqref{e_1937} and
\eqref{e_1948}.
\itm The parametrization of the AdS hyperboloid in
$(\eta',\chi')$-coordinates takes the form
\begin{equation} \label{e_1983}
z_0 = R_{\mbox{\tiny $Q$}} \coth \chi'
\mbox{\hspace{2mm} , \hspace{3mm}}
z_1 = k R_{\mbox{\tiny $Q$}} \frl{a_{k}(\eta')}{\sinh \chi'}
\mbox{\hspace{2mm} , \hspace{3mm}}
z_2 = k R_{\mbox{\tiny $Q$}} \frl{a_{-k}(\eta')}{\sinh \chi'}
\mbox{ .}
\end{equation}
With $(\tilde{t}',\tilde{r}')$-coordinates the parametrization of the AdS
hyperboloid is
\begin{equation} \label{e_1986}
z_0 = \frl{\tilde{r}_0' \m \tilde{r}'}{A}
\mbox{ ,} \hs{46.5mm}
\end{equation}
\begin{equation} \label{e_1985}
z_1 = k \hs{0.2mm} \mbox{$\left[ \rule[-1.5mm]{0mm}{5.7mm} \right.$}
\hs{-1.2mm} \left( \frl{\tilde{r}' \m \tilde{r}_0'}{A} \right)^2
- R_{\mbox{\tiny $Q$}}^2
\hs{-0.2mm} \mbox{$\left. \rule[-1.5mm]{0mm}{5.7mm} \right]$}^{1/2}
\hs{0.7mm} a_k \left( \frl{A \hs{0.5mm} \tilde{t}'}{R_{\mbox{\tiny $Q$}}}
\right)
\mbox{ ,}
\end{equation}
\begin{equation} \label{e_1987}
\hs{2.5mm}
z_2 = k \hs{0.2mm}\mbox{$\left[ \rule[-1.5mm]{0mm}{5.7mm} \right.$}
\hs{-1.2mm} \left( \frl{\tilde{r}' \m \tilde{r}_0'}{A} \right)^2
- R_{\mbox{\tiny $Q$}}^2
\hs{-0.2mm} \mbox{$\left. \rule[-1.5mm]{0mm}{5.7mm} \right]$}^{1/2}
\hs{0.7mm} a_{-k} \left( \frl{A \hs{0.5mm} \tilde{t}'}{R_{\mbox{\tiny $Q$}}}
\right)
\mbox{ .}
\end{equation}
\vspace{-1mm} \newline
With $(\hat{t}',\hat{r}')$-coordinates the parametrization is
\begin{equation} \label{e_1993}
z_0 = R_{\mbox{\tiny $Q$}} \hs{0.5mm}
\cosh \mbox{$\left( \rule[-1.5mm]{0mm}{5.7mm} \right.$} \hs{-0.2mm}
\frl{\hat{r}_0' \m \hat{r}'}{R_{\mbox{\tiny $Q$}}}
\hs{-0.2mm} \mbox{$\left. \rule[-1.5mm]{0mm}{5.7mm} \right)$}
\mbox{ ,} \hs{19.5mm}
\end{equation}
\begin{equation} \label{e_1992}
z_1 = k R_{\mbox{\tiny $Q$}} \hs{0.5mm}
\sinh \mbox{$\left( \rule[-1.5mm]{0mm}{5.7mm} \right.$} \hs{-0.2mm}
\frl{\hat{r}_0' \m \hat{r}'}{R_{\mbox{\tiny $Q$}}}
\hs{-0.2mm} \mbox{$\left. \rule[-1.5mm]{0mm}{5.7mm} \right)$}
\hs{0.7mm} a_k \left( \frl{A \hs{0.5mm} \hat{t}'}{R_{\mbox{\tiny $Q$}}}
\right)
\mbox{ ,}
\end{equation}
\begin{equation} \label{e_1994}
\hs{2.5mm} z_2 = k R_{\mbox{\tiny $Q$}} \hs{0.5mm}
\sinh \mbox{$\left( \rule[-1.5mm]{0mm}{5.7mm} \right.$} \hs{-0.2mm}
\frl{\hat{r}_0' \m \hat{r}'}{R_{\mbox{\tiny $Q$}}}
\hs{-0.2mm} \mbox{$\left. \rule[-1.5mm]{0mm}{5.7mm} \right)$}
\hs{0.7mm} a_{-k} \left( \frl{A \hs{0.5mm} \hat{t}'}{R_{\mbox{\tiny $Q$}}}
\right)
\mbox{ .}
\end{equation}
%
\itm We shall now consider the case $k = 0$. Then the line element of
$M^{2,4}$ has the form
\begin{equation} \label{e_1958}
ds^2 = -d z_0^2 - d z_1^2 + d z_2^2
+ d z_3^2 + d z_4^2 + d z_5^2
\mbox{ .}
\end{equation}
The LBR 4-submanifold is determined by the constraint \eqref{e_1939}
and
\begin{equation} \label{e_1959}
z_0^2 + z_1^2 - z_2^2 = R_{\mbox{\tiny $Q$}}^2
\mbox{ .}
\end{equation}
The $(\eta,\chi)$-system with $k = 0$ coincides with the CFS system,
and the line element takes the form \eqref{e_157}. In this case the
parametrization of the AdS hyperboloid is given by \eqref{e_1941}
with $T = \eta$, $R = \chi$ and $k = -1$,
\begin{equation} \label{e_1960}
z_0 = R_{\mbox{\tiny $Q$}} \frl{\eta}{\chi}
\mbox{\hspace{2mm} , \hspace{3mm}}
z_1 = R_{\mbox{\tiny $Q$}} \frl{B^2 \m (\eta^{2} \m \chi^{2})}
{2 \hs{0.2mm} B \hs{0.2mm} \chi}
\mbox{\hspace{2mm} , \hspace{3mm}}
z_2 = R_{\mbox{\tiny $Q$}} \frl{B^2 \p (\eta^{2} \m \chi^{2})}
{2 \hs{0.2mm} B \hs{0.2mm} \chi}
\mbox{ .}
\end{equation}
The reason for inserting $k = -1$ instead of $k = 0$ is that
the case $k = 0$ concerns the type of coordinate system which we
consider, while the $k = -1$ value in equation \eqref{e_1941} concerns
the parametrization.
\itm These parametrizations of the AdS hyperboloid in the $M^{2,4}$
manifold makes it clear that the line elements \eqref{e_157},
\eqref{e_218}, \eqref{e_928}, \eqref{e_452}, \eqref{e_952},
\eqref{e_1014} and \eqref{e_331} describe the same LBR spacetime.
\itm With $(\hat{\eta},\hat{\chi})$-coordinates the parametrization of
the AdS hyperboloid is
\begin{equation} \label{e_2952}
z_0 = -R_{\mbox{\tiny $Q$}} \hs{0.5mm}
\tan \hat{\chi} \sinh \hat{\eta}
\mbox{\hspace{2mm} , \hspace{3mm}}
z_1 = -R_{\mbox{\tiny $Q$}} \hs{0.5mm}
\tan \hat{\chi} \cosh \hat{\eta}
\mbox{\hspace{2mm} , \hspace{3mm}}
z_2 = - \frl{R_{\mbox{\tiny $Q$}}}{\cos \hat{\chi}}
\mbox{ .}
\end{equation}
\vspace{-9mm} \newline
\itm From the embedding parametrization \eqref{e_1943} and
equations \eqref{e_2309} and \eqref{e_2310} we obtain the
following embedding parametrization of the LBR spacetime in the
coordinates introduced in section 4.VII,
\begin{equation} \label{e_2960}
z_0 = R_{\mbox{\tiny $Q$}} \frl{U \m V}{1 \m UV}
\mbox{\hspace{2mm} , \hspace{3mm}}
z_1 = R_{\mbox{\tiny $Q$}} \frl{U \p V}{1 \m UV}
\mbox{\hspace{2mm} , \hspace{3mm}}
z_2 = R_{\mbox{\tiny $Q$}} \frl{1 \p UV}{1 \m UV}
\mbox{ ,}
\end{equation}
\begin{equation} \label{e_2961}
z_3 = R_{\mbox{\tiny $Q$}} \frl{\zeta \p \overline{\zeta}}
{1 \p \zeta\overline{\zeta}}
\mbox{\hspace{2mm} , \hspace{3mm}}
z_4 = -i R_{\mbox{\tiny $Q$}} \frl{\zeta \m \overline{\zeta}}
{1 \p \zeta\overline{\zeta}}
\mbox{\hspace{2mm} , \hspace{3mm}}
z_5 = R_{\mbox{\tiny $Q$}}
\frl{1 \m \zeta\overline{\zeta}}{1 \p \zeta\overline{\zeta}}
\mbox{ ,}
\end{equation}
corresponding to $k = -1$ in equations \eqref{e_1937} and \eqref{e_1948}.
This is in agreement with the embedding parametrization of the LBR
spacetime used by Ortaggio and Podolsk\'{y} [\ref{r_34}] with a different
scaling of the coordinates.
%
%
%
\vspace{6mm} \newline
{\bf 8. Conclusion}
\vspace{3mm} \newline
The LBR solution of Einstein's field equations was found more than 90
years ago by T.\hn Levi-Civita [\ref{r_4},\ref{r_5}] and rediscovered in
1959 by B.\hn Bertotti [\ref{r_6}] and I.\hn Robinson [\ref{r_7}]. The
solution was interpreted physically as a spacetime with an electric or
a magnetic field with constant field strength. However the source of
the electrical field remained rather obscure. We recently used Israel's
formalism [\ref{r_27}] for describing singular shells in general
relativity to investigate the physical properties of a shell with LBR
spacetime outside the shell and flat spacetime inside it, and found
[\ref{r_12}] that the source then had to be a charged domain wall with
a radius equal to the distance corresponding to its charge. From equation
(44) in reference [\ref{r_12}] we see that the radius of the shell is
one half of its Schwarzschild radius.
\itm We have found different coordinate representations of the LBR
spacetime by taking a general form \eqref{e_211} of a spherically
symmetric line element as our point of departure, permitting the metric
functions to depend upon the radial and the time coordinate. The
differential equation \eqref{e_212} obtained from the requirement that
the spacetime is conformally flat, i.e. that the Weyl curvature tensor
vanishes, was then solved under different coordinate conditions.
Remarkably, with the general form \eqref{e_211} of the line element and
the requirement that the Weyl tensor vanishes, Einstein's field equations
restrict the energy-momentum tensor to be of a form \eqref{e_216}
representing a constant electric or magnetic field. In the present
article we have only discussed the case of an electric field.
\itm Next we have given a general prescription for finding coordinate
transformations between the "canonical" CFS coordinate system in which
the line element of the LBR spacetime is equal to a conformal factor
times the Minkowski line element, and the coordinate representations
obtained by the method based on solving equation \eqref{e_211}. In
sections 4 and 6 of this article we have given a detailed discussion
of the kinematical properties of the reference frames both outside and
inside the domain wall, in which the coordinate systems are comoving.
\itm We have found that in several coordinate systems there are three
cases which we have parameterized by the constant $k$ having the values
$1$, $0$ or $-1$. The corresponding reference frames have different
motions. In the case $k = 0$ the $(\eta,\chi)$ coordinate system is
comoving in the same referenc frame as that of the CFS coordinates.
The domain wall at $R = R_{\mbox{\tiny $Q$}}$ of the WLBR spacetime
is static in this reference frame, and the acceleration of gravity
is constant and equal to $1 / R_{\mbox{\tiny $Q$}}$. In the case
$k = 1$ the $(\eta,\chi)$ coordinate system is comoving with a reference
frame that accelerates away from the domain wall in the WLBR spacetime.
Then the acceleration of gravity is smaller that that in the static case
($k = 0$), and even directed towards the domain wall for
$R > \sqrt{B^2 + T^2}$. In the case $k = - 1$ the $(\eta,\chi)$ coordinate
system is comoving with a reference frame that accelerates towards the
domain wall in the WLBR spacetime. Hence observers in this reference frame
will experience an acceleration of gravity directed away from the domain
wall larger than $1 / R_{\mbox{\tiny $Q$}}$.
\itm In section 5 we have presented a Milne-LBR universe model with a
part of the Milne universe inside the domain wall and an infinitely
extended LBR spacetime outside it.
\itm Finally we have considered embedding of the LBR spacetime in a flat,
6-dimensional manifold, $M^{2,4}$, and deduced the parameterizations of
this embedding for the main coordinate systems considered in the
present article.
%
%
\vspace{10mm} \newline
{\bf \parbox[t]{140mm}{Appendix A. Calculus of k-functions}}
\vspace{-2mm} \newline
\setcounter{equation}{0}
\appendix
\setcounter{chapter}{1}
\numberwithin{equation}{chapter}
In this appendix we shall define functions which we call k-functions
and deduce their main properties. Motivated by the angular part of the
Robertson-Walker line element in standard coordinates it is natural to
introduce the function
\begin{equation} \label{e_801}
S_k (x) = \left\{ \begin{array}{lcl}
\sin x  & \mbox{for} & k = 1  \\
x       & \mbox{for} & k = 0  \\
\sinh x & \mbox{for} & k = -1
\end{array} \right.
\mbox{ .}
\end{equation}
In the present paper we shall need several functions of similar type
defined by
\begin{equation} \label{e_802}
C_k(x) = \left\{ \begin{array}{lcl}
\cos x  & \mbox{for} & k = 1  \\
1       & \mbox{for} & k = 0  \\
\cosh x & \mbox{for} & k = -1
\end{array} \right.
\mbox{ ,}
\end{equation}
\begin{equation} \label{e_803}
T_k (x) = \left\{ \begin{array}{lcl}
\tan x  & \mbox{for} & k = 1  \\
x       & \mbox{for} & k = 0  \\
\tanh x & \mbox{for} & k = -1
\end{array} \right.
\end{equation}
and
\begin{equation} \label{e_804}
I_k(x) =
\left\{ \begin{array}{lcl}
\cot x  & \mbox{for} & k = 1  \\
1 / x   & \mbox{for} & k = 0  \\
\coth x & \mbox{for} & k = -1
\end{array} \right.
\mbox{ .}
\end{equation}
\vspace{0mm} \newline
Motivated by the scale factor of the DeSitter line element we also
introduce
\begin{equation} \label{e_805}
a_k(x) =
\left\{ \begin{array}{lcl}
\cosh x & \mbox{for} & k = 1  \\
e^x     & \mbox{for} & k = 0  \\
\sinh x & \mbox{for} & k = -1
\end{array} \right.
\end{equation}
and
\begin{equation} \label{e_1805}
b_k(x) =
\left\{ \begin{array}{lcl}
\tanh x & \mbox{for} & k = 1  \\
1       & \mbox{for} & k = 0  \\
\coth x & \mbox{for} & k = -1
\end{array} \right.
\mbox{ .}
\end{equation}
Note that
\begin{equation} \label{e_1806}
b_k(x) = \frl{a_{-k}(x)}{a_{k}(x)}
\mbox{ .}
\end{equation}
\vspace{-2mm} \newline
The series expansions for the function $S_k(x)$ and $C_k(x)$ are
\begin{equation} \label{e_806}
S_k(x) = x + \sum_{k = 1}^{\infty} \frl{(-k)^n}{(2n \p 1)!} x^{2n + 1}
\end{equation}
\vspace{-5mm} \newline
and
\begin{equation} \label{e_807}
C_k(x) = 1 + \sum_{k = 1}^{\infty} \frl{(-k)^n}{(2n)!} x^{2n}
\mbox{ .}
\end{equation}
Furthermore
\vspace{-2mm} \newline
\begin{equation} \label{e_808}
T_k(x) = \frl{S_k(x)}{C_k(x)}
\mbox{\hspace{2mm} , \hspace{3mm}}
I_k(x) = \frl{C_k(x)}{S_k(x)}
\end{equation}
and
\begin{equation} \label{e_810}
C_k(x)^2 + k S_k(x)^2 = 1
\mbox{ ,}
\end{equation}
which implies that
\begin{equation} \label{e_811}
1 + k T_k(x)^2 = C_k(x)^{-2}
\end{equation}
and
\begin{equation} \label{e_812}
I_k(x)^2 + k = S_k(x)^{-2}
\mbox{\hspace{2mm} , \hspace{3mm}}
C_k(x)^{\hs{0.3mm} 2} = \frl{I_k(x)^2}{I_k(x)^2 \p k}
\mbox{ .}
\end{equation}
\vspace{-2mm} \newline
We have the following addition formulae
\begin{equation} \label{e_813}
S_k(x + y) = S_k(x) \hs{0.5mm} C_k(y) + C_k(x) S_k(y)
\mbox{ ,}
\end{equation}
\begin{equation} \label{e_814}
C_k(x + y) = C_k(x) \hs{0.5mm} C_k(y) - k S_k(x) S_k(y)
\mbox{ ,}
\end{equation}
\begin{equation} \label{e_815}
T_k(x + y) = \frl{T_k(x) \p T_k(y)}{1 \m k T_k(x) T_k(y)}
\end{equation}
and
\begin{equation} \label{e_816}
I_k(x + y) = \frl{I_k(x) I_k(y) \m k}{I_k(x) \p I_k(y)}
\mbox{ .}
\end{equation}
With $y = x$ this gives
\begin{equation} \label{e_828}
S_k(2x) = 2 S_k(x) \hs{0.5mm} C_k(x)
\mbox{ ,}
\end{equation}
\begin{equation} \label{e_829}
C_k(2x) = C_k(x)^2 - k S_k(x)^2
\mbox{ ,}
\end{equation}
\begin{equation} \label{e_830}
T_k(2x) = \frl{2 T_k(x)}{1 \m k T_k(x)^2}
\end{equation}
and
\begin{equation} \label{e_831}
I_k(2x) = \frl{I_k(x)^2 \m k}{2 I_k(x)}
\mbox{ .}
\end{equation}
From equations \eqref{e_815} and \eqref{e_816} we also obtain
\begin{equation} \label{e_1815}
T_k^{-1}(x) + T_k^{-1}(y) = T_k^{-1} \left( \frl{x \p y}{1 \m kxy} \right)
\end{equation}
and
\begin{equation} \label{e_1816}
I_k^{-1}(x) + I_k^{-1}(y) = I_k^{-1} \left( \frl{xy \m k}{x \p y} \right)
\mbox{ .}
\end{equation}
Furthermore
\begin{equation} \label{e_817}
S_k(-x) = -S_k(x)
\mbox{\hspace{2mm} , \hspace{3mm}}
C_k(-x) = C_k(x)
\end{equation}
and
\begin{equation} \label{e_818}
T_k(-x) = -T_k(x)
\mbox{\hspace{2mm} , \hspace{3mm}}
I_k(-x) = -I_k(x)
\mbox{ .}
\end{equation}
\vspace{-2mm} \newline
Combining equations \eqref{e_813} - \eqref{e_816} we also have that
\begin{equation} \label{e_819}
S_k(x) + S_k(y) = 2 S_k(\frl{x \p y}{2}) \hs{0.5mm} C_k(\frl{x \m y}{2})
\mbox{ ,}
\end{equation}
\begin{equation} \label{e_820}
S_k(x) - S_k(y) = 2 S_k(\frl{x \m y}{2}) \hs{0.5mm} C_k(\frl{x \p y}{2})
\mbox{ ,}
\end{equation}
\begin{equation} \label{e_821}
C_k(x) + C_k(y) = 2 \hs{0.5mm} C_k(\frl{x \p y}{2}) \hs{0.5mm}
C_k(\frl{x \m y}{2})
\mbox{ ,}
\end{equation}
\begin{equation} \label{e_822}
C_k(x) - C_k(y) = -2k S_k(\frl{x \p y}{2}) \hs{0.5mm} S_k(\frl{x \m y}{2})
\end{equation}
and
\begin{equation} \label{e_823}
T_k(x/2) = \frl{S_k(x)}{1 \p C_k(x)}
\mbox{\hspace{2mm} , \hspace{3mm}}
I_k(x/2) = \frl{1 \p C_k(x)}{S_k(x)}
\mbox{ .}
\end{equation}
\vspace{-1mm} \newline
Using \eqref{e_808}, \eqref{e_828} and \eqref{e_829} we obtain
\begin{equation} \label{e_832}
I_k(2x) = \frl{C_k(2x)}{S_k(2x)}
= \frl{C_k(x)^2 - k S_k(x)^2}{2 S_k(x) \hs{0.5mm} C_k(x)}
= \frl{1}{2} \left[ \hs{0.5mm}
I_k(x) - k \hs{0.5mm} I_k(x)^{-1} \right]
\mbox{ .}
\end{equation}
The derivatives of the k-functions are
\begin{equation} \label{e_824}
S_k'(x) = C_k(x)
\mbox{\hspace{2mm} , \hspace{3mm}}
C_k'(x) = -k S_k(x)
\end{equation}
\vspace{-8mm} \newline
and
\begin{equation} \label{e_825}
T_k'(x) = C_k(x)^{-2}
\mbox{\hspace{2mm} , \hspace{3mm}}
I_k'(x) = -S_k(x)^{-2}
\mbox{ .}
\end{equation}
\vspace{-2mm} \newline
The following identities will also be needed
\begin{equation} \label{e_827}
|S_k(I_k^{-1}(x))| = \frl{1}{\sqrt{x^2 + k}}
\mbox{\hspace{2mm} , \hspace{3mm}}
C_k(I_k^{-1}(x)) = x S_k(I_k^{-1}(x))
\mbox{ .}
\end{equation}
From the definition \eqref{e_805} it follows that
\begin{equation} \label{e_833}
a_k(x)^2 - a_{-k}(x)^2 = k
\end{equation}
and
\begin{equation} \label{e_826}
a_k'(x) = a_{-k}(x)
\mbox{ .}
\end{equation}
%
%
%
\vspace{10mm} \newline
{\bf \parbox[t]{140mm}{Appendix B. From generating functions
to transformations}}
\vspace{-2mm} \newline
\setcounter{equation}{0}
\appendix
\setcounter{chapter}{2}
\numberwithin{equation}{chapter}
We shall here show how the transformation \eqref{e_279} is deduced
from the generating function \eqref{e_339}. From equation \eqref{e_12}
with $g = f$, $x^0 = \eta$ and $x^1 = \chi$ and using equations
\eqref{e_808}, \eqref{e_819} and \eqref{e_821} it follows that
\begin{equation} \label{e_1827}
T + R = f(\eta + \chi)
= B \hs{0.8mm} T_k \leftn \frl{\eta + \chi}{2} \rightn
= B \hs{0.8mm} \frl{2 \hs{0.4mm} S_k \leftn \frac{\eta + \chi}{2} \rightn
\hs{0.4mm} C_k \leftn \frac{\eta - \chi}{2} \rightn}
{2 \hs{0.4mm} C_k \leftn \frac{\eta + \chi}{2} \rightn
\hs{0.4mm} C_k \leftn \frac{\eta - \chi}{2} \rightn}
= B \hs{0.8mm} \frl{S_k (\eta) \p S_k(\chi)}{C_k (\eta) \p C_k(\chi)}
\mbox{ .}
\end{equation}
In the same way we find
\begin{equation} \label{e_1857}
T - R = B \hs{0.8mm} \frl{S_k (\eta) \m S_k(\chi)}{C_k (\eta) \p C_k(\chi)}
\mbox{ ,}
\end{equation}
which gives the transformation \eqref{e_279}.
\itm Next we show how the transformation \eqref{e_346} is deduced
from the generating function \eqref{e_1179}. Using this generating
function and equation \eqref{e_12}, replacing $T$ by $\eta$,
$x^0$ by $T$ and $x^1$ by $R$, we obtain
\begin{equation} \label{e_1867}
\eta = \frl{1}{2} \hs{0.6mm}
[\hs{0.3mm} f(T + R) + f(T - R) \hs{0.3mm}]
= T_k^{-1} \left( \frm{T + R}{B} \right)
+ T_k^{-1} \left( \frm{T - R}{B} \right)
\mbox{ .}
\end{equation}
From equation \eqref{e_1815} it then follows that
\begin{equation} \label{e_1878}
\eta = T_k^{-1} \left( \frl{2BT}{B^2 \m k (T^2 \m R^2)} \right)
\hs{0.3mm}
\mbox{ .}
\end{equation}
Hence we obtain the first of equations \eqref{e_346}. The second
is found in the same way.
\itm We shall now deduce the transformation \eqref{e_981} between the
$(\tau,\rho)$-koordinates and the CFS coordinates.  From equation
\eqref{e_12} with $x^0 = \tau$ and $x^1 = \rho$ and using the generating
functions \eqref{e_980}, it follows that
\begin{equation} \label{e_1327}
T + R
= -B \hs{0.8mm} \coth \leftn \frl{\tau + \rho}{2} \rightn
= -B \hs{0.8mm} \frl{2 \hs{0.4mm} \cosh \leftn \frac{\tau + \rho}{2}
\rightn \hs{0.4mm} \cosh \leftn \frac{\tau - \rho}{2} \rightn}
{2 \hs{0.4mm} \sinh \leftn \frac{\tau + \rho}{2} \rightn
\hs{0.4mm} \cosh \leftn \frac{\tau - \rho}{2} \rightn}
= -B \hs{0.8mm} \frl{\cosh \tau \p \cosh \rho}
{\sinh \tau \p \sinh \rho}
\mbox{ .}
\end{equation}
In the same way we find
\begin{equation} \label{e_1357}
T - R = B \hs{0.8mm} \frl{\cosh \tau \m \cosh \rho}
{\sinh \tau \p \sinh \rho}
\mbox{ ,}
\end{equation}
which gives the transformation \eqref{e_981}.
\itm Next we show how the transformation \eqref{e_946} is deduced
from the generating functions \eqref{e_1980}. Using these generating
functions and equation \eqref{e_12}, replacing $T$ by $\tau$,
$x^0$ by $T$ and $x^1$ by $R$, we obtain
\begin{equation} \label{e_1377}
\tau = \frl{1}{2} \hs{0.6mm}
[\hs{0.3mm} f(T + R) + g(T - R) \hs{0.3mm}]
= - \mbox{arctanh} \left( \frm{B}{T + R} \right)
+ \mbox{arctanh} \left( \frm{T - R}{B} \right)
\mbox{ .}
\end{equation}
Hence we find
\begin{equation} \label{e_1868}
\tau = \mbox{arctanh} \left( \frl{\frm{T - R}{B} \m \frm{B}{T + R}}
{1 \m \left( \frm{T - R}{B} \right)
\left( \frm{B}{T + R} \right) \rule[-0mm]{0mm}{4.75mm}}
\cdot \frl{B \hs{0.5mm} (T + R \hs{0.5mm} )}
{B \hs{0.5mm} (T + R \hs{0.5mm} )} \right)
= \mbox{arctanh} \left( \frl{(T^2 \m R^2) \m B^2}{2BR} \right)
\mbox{ .}
\end{equation}
Hence we obtain the first of equations \eqref{e_946}. The second equation
is found in the same way.
\itm We shall now deduce the transformation \eqref{e_835} between the
$(\tau,\rho)$- and the $(\eta,\chi)$-coordinates. From equation
\eqref{e_12} with $x^0 = \tau$ and $x^1 = \rho$ and using the generating
functions \eqref{e_834}, it follows that
\begin{equation} \label{e_1477}
\eta = \frl{1}{2} \hs{0.6mm}
[\hs{0.3mm} f(\tau + \rho) + g(\tau - \rho) \hs{0.3mm}]
= - \arctan \left( \coth \frm{\tau + \rho}{2} \right)
+ \arctan \left( \coth \frm{\tau - \rho}{2} \right)
\mbox{ ,}
\end{equation}
which may be rewritten as
\begin{equation} \label{e_1968}
\eta = - \arctan \left( \frl{\coth \frm{\tau + \rho}{2} \m
\tanh \frm{\tau - \rho}{2}}
{1 \p \coth \frm{\tau + \rho}{2} \tanh \frm{\tau - \rho}{2}
\rule[-0mm]{0mm}{4.45mm}}
\cdot \frl{\sinh \frm{\tau + \rho}{2} \cosh \frm{\tau - \rho}{2}}
{\sinh \frm{\tau + \rho}{2} \cosh \frm{\tau - \rho}{2}
\rule[-0mm]{0mm}{4.45mm}} \right)
\mbox{ .}
\end{equation}
Multiplication and using the addition formulae for hyperbolic functions
give
\begin{equation} \label{e_1869}
\eta = - \arctan \left(
\frl{\cosh \frm{\tau + \rho}{2} \cosh \frm{\tau - \rho}{2} \m
\sinh \frm{\tau + \rho}{2} \sinh \frm{\tau - \rho}{2}}
{\sinh \frm{\tau + \rho}{2} \cosh \frm{\tau - \rho}{2} \p
\cosh \frm{\tau + \rho}{2} \sinh \frm{\tau - \rho}{2}
\rule[-0mm]{0mm}{4.45mm}} \right)
= - \arctan \left( \frl{\cosh \rho}{\sinh \tau} \right)
\mbox{ .}
\end{equation}
\vspace{-2mm} \newline
Hence we obtain the first of equations \eqref{e_835}. The second equation
is found in a similar way.
\vspace{5mm} \newline
{\bf Acknowledgement}
\vspace{3mm} \newline
We would like to thank Marcello Ortaggio for providing us with the
references \ref{r_31}, \ref{r_32}, \ref{r_33} and \ref{r_34}.
\vspace{5mm} \newline
\newpage
{\bf References}
%
\begin{enumerate}
\item \O .\hn Gr\o n and S.\hn Johannesen, \textit{FRW Universe Models in
Conformally Flat Spacetime Coordinates. I: General Formalism},
Eur.\hn Phys.\hn J.\hn Plus \textbf{126}: 28 (2011).
\label{r_1}

\item \O .\hn Gr\o n and S.\hn Johannesen, \textit{FRW Universe Models in
Conformally Flat Spacetime Coordinates. II: Universe models with
negative and vanishing spatial curvature},
Eur.\hn Phys.\hn J.\hn Plus \textbf{126}: 29 (2011).
\label{r_2}

\item \O .\hn Gr\o n and S.\hn Johannesen, \textit{FRW Universe Models in
Conformally Flat Spacetime Coordinates. III: Universe models with positive
spatial curvature}, Eur.\hn Phys.\hn J.\hn Plus \textbf{126}: 30 (2011).
\label{r_3}

\item T.\hn Levi-Civita, \textit{Realt\'{a} fisica di alcuni spazi normali
del Bianchi}, Rediconti della Reale Accademia dei Lincei \textbf{26},
519 - 531 (1917).
\label{r_4}

\item T.\hn Levi-Civita, \textit{The physical reality of some normal spaces
of Bianchi}, Gen.\hn Rel.\hn Grav.\hn \textbf{43}, 2307 - 2320 (2011).
\label{r_5}

\item B.\hn Bertotti, \textit{Uniform Electromagnetic Field in the Theory
of General Relativity}, Phys.\hn Rev. \textbf{116}, 1331 - 1333 (1959).
\label{r_6}

\item I.\hn Robinson, \textit{A Solution of the Maxwell-Einstein
Equations}, Bull.\hn Acad.\hn Pol.\hn Sci.\hn Ser.\hn Sci.\hn
Math.\hn Astr.\hn Phys. \textbf{7}, 351 - 352 (1959).
\label{r_7}

\item N.\hn Tariq and B.\hn O.\hn J.\hn Tupper, \textit{The
uniqueness of the Bertotti-Robinson electromagnetic universe},
J.\hn Math.\hn Phys. \textbf{15}, 2232 - 2235 (1974).
\label{r_20}

\item N.\hn Tariq and R.\hn G.\hn McLenaghan, \textit{Note on the
Bertotti-Robinson electromagnetic universe},
J.\hn Math.\hn Phys. \textbf{19}, 349 - 351 (1978).
\label{r_21}

\item H.\hn Stephani, D.\hn Kramer, M.\hn MacCallum, C.\hn Hoenselaers
and E.\hn Herlt, \textit{Exact Solutions of Einstein's Field Equations},
Cambridge University Press (2009).
\label{r_8}

\item D.\hn Lovelock, \textit{Weakened Field Equations in General
Relativity Admitting an 'Unphysical' Metric}, Commun.\hn math.\hn Phys.
\textbf{5}, 205 - 214 (1967).
\label{r_9}

\item D.\hn Lovelock, \textit{A Spherically Symmetric Solution of the
Maxwell-Einstein Equations}, Commun.\hn math.\hn Phys. \textbf{5},
257 - 261 (1967).
\label{r_10}

\item P.\hn Dolan, \textit{A Singularity Free Solution of the
Maxwell-Einstein Equations}, Commun.\hn math.\hn Phys. \textbf{9},
161 - 168 (1968).
\label{r_11}

\item \O .\hn Gr\o n and S.\hn Johannesen, \textit{A solution of the
Einstein-Maxwell equations describing conformally flat spacetime
outside a charged domain wall},
Eur.\hn Phys.\hn J.\hn Plus \textbf{126}: 89 (2011).
\label{r_12}

\item M.\hn G\"{u}rzes and \"{O}.\hn Sario\u{g}lu, \textit{Accelerated
Levi-Civita-Bertotti-Robinson Metric in D-Dimensions},
Gen.\hn Rel.\hn Grav.\hn \textbf{37}, 2015 - 2022 (2005).
\label{r_23}

\item V.\hn I.\hn Khlebnikov and \'{E}.\hn Shelkovenko, \textit{On exact
solutions of the Einstein-Maxwell equations in the Newman-Penrose formalism.
II. Conformally plane metrics}, Sov.\hn Phys.\hn J.\hn \textbf{19},
960 - 962 (1976).
\label{r_31}

\item J.\hn Podolsk\'{y} and M.\hn Ortaggio, \textit{Explicit Kundt type II
and N solutions as gravitational waves in various type D and O universes},
Class.\hn Quant.\hn Grav.\hn \textbf{20}, 1685 - 1701 (2003).
\label{r_32}

\vspace{-5mm}
\item O.\hn J.\hn C.\hn Dias and J.\hn P.\hn S.\hn Lemos,
\textit{The external limits of the C-metric: Nariai, Bertotti - Robinson
and anti - Nariai C-metrics}, Phys.\hn Rev.\hn \textbf{D68}, 104010 (2003).
\label{r_16}

\item A.\hn C.\hn Ottewill and P.\hn Taylor, \textit{Quantum field theory
on the Bertotti-Robinson space-time}, arXiv:1209.6080 (2012).
\label{r_13}

\item V.\hn Cardoso, O.\hn J.\hn C.\hn Dias and J.\hn P.\hn S.\hn Lemos,
\textit{Nariai, Bertotti - Robinson and anti - Nariai solutions in
higher dimensions}, Phys.\hn Rev.\hn \textbf{D70}, 024002 (2004).
\label{r_17}

\item A.\hn S.\hn Lapedes, \textit{Euclidean quantum field theory and
the Hawking effect}, Phys.\hn Rev.\hn \textbf{D17}, 2556 - 2566 (1978).
\label{r_18}

\item J.\hn B.\hn Griffiths and J.\hn Podolsk\'{y}, \textit{Exact
space-times in Einstein's general relativity}. Cambridge
University Press, section 7.1 (2009).
\label{r_14}

\item N.\hn Dadhich, \textit{On product spacetime with 2-sphere of
constant curvature}, arXiv:0003026 (2000).
\label{r_15}

\item G.\hn A.\hn Gonz\'{a}lez and R.\hn Vera, \textit{A local
characterization for static charged black holes},
Class.\hn Quant.\hn Grav.\hn \textbf{28}, 025008 (2011).
\label{r_22}

\item D.\hn Garfinkle and E.\hn N.\hn Glass, \textit{Bertotti-Robinson
and Melvin Spacetimes}, Class.\hn Quant.\hn Grav.\hn \textbf{28}, 215012
(2011).
\label{r_19}

\item M.\hn Ortaggio, \textit{Impulsive waves in the Nariai universe},
Phys.\hn Rev.\hn \textbf{D65}, 084046 (2002).
\label{r_33}

\vspace{-5mm}
\item M.\hn Ortaggio and J.\hn Podolsk\'{y}, \textit{Impulsive waves in
electrovac direct product spacetimes with $\Lambda$},
Class.\hn Quant.\hn Grav.\hn \textbf{19}, 5221 - 5227 (2002).
\label{r_34}

\item S.\hn Braeck and \O .\hn Gr\o n, \textit{A river model of space},
Eur.\hn Phys.\hn J.\hn Plus \textbf{128}: 24 (2013).
\label{r_26}

\item \O .\hn Gr\o n and S.\hn Hervik, \textit{Einstein's General Theory
of Relativity}, Springer, p.147 (2007).
\label{r_24}

\item O.\hn B.\hn Zaslavskii, \textit{Geometry of nonextreme black holes
near the extreme state}, Phys.\hn Rev.\hn \textbf{D56}, 2188 - 2191 (1997).
\label{r_25}

\item W.\hn Israel, \textit{Singular hypersurfaces and thin shells in
general relativity}, Nuovo Cimento \textbf{B44}, 1-14 (1966).
\label{r_27}

\end{enumerate}
\end{document}